\documentclass[journal,twocolumn]{IEEEtran}

\usepackage[utf8]{inputenc}
\usepackage[T1]{fontenc}
\usepackage{textcomp}
\usepackage{fixltx2e}
\usepackage{microtype}
\usepackage{calc}
\usepackage{footnote}
\usepackage{csquotes}
\usepackage{subcaption}
\usepackage{todonotes}

\usepackage[font=footnotesize]{caption}

\usepackage{mathtools}
\usepackage{amsmath,amsfonts}
\usepackage{algorithmic}

\usepackage{soul} 

\usepackage{pgfplots}
\pgfplotsset{compat=1.18}
\usepackage{xcolor} 
\usepackage{graphicx}
\usepackage{tikz}
\definecolor{myred}{HTML}{D95319}
\definecolor{myyellow}{HTML}{EDB120}
\definecolor{myblue}{HTML}{0072BD}
\definecolor{mypurple}{HTML}{7E2F8E}
\definecolor{mygreen}{HTML}{77AC30}

\usepackage{tabularx}
\usepackage{booktabs}
\usepackage{rotating}

\usepackage[skip=0.33\baselineskip]{caption}
\newcommand\mycell[1]{%
   \begin{tabular}[t]{@{}l@{}} #1 \end{tabular}}

\usepackage[flushleft]{threeparttable}
\usepackage{multirow}
\usepackage{xltabular}
\usepackage{pdflscape}
\usepackage{bigstrut}
\usepackage{pifont} 
\usepackage{stfloats} 

\usepackage{import} %

\usepackage[
  range-phrase=\text{ to },
  per-mode=symbol-or-fraction,
  range-units=single,
  list-units=single,
  detect-all,
  list-final-separator={, },
]{siunitx}
\usepackage{silence}
\DeclareSIUnit\samples{samples}
\DeclareSIUnit\molecule{molecule}
\DeclareSIUnit\param{Parameters}
\DeclareSIUnit\w{weights}
\usepackage{acro}
\DeclareAcronym{MC}{short=MC, long=molecular communication}
\DeclareAcronym{QS}{short=QS, long=quorum sensing}
\DeclareAcronym{IMU}{short=IMU, long=inertial measurement unit}
\DeclareAcronym{PCR}{short=PCR, long=polymerase chain reaction}
\DeclareAcronym{NAHL}{short=NAHL, long=N-Acyl homoserine lactone}
\DeclareAcronym{AHL}{short=AHL, long=Acyl homoserine lactone}
\DeclareAcronym{ELISA}{short=ELISA, long=enzyme linked immuno sorbent assay}
\DeclareAcronym{SDM}{short=SDM, long=software-defined metamaterials}
\DeclareAcronym{CSK}{short=CSK, long=concentration shift keying}
\DeclareAcronym{SPION}{short=SPION, long=superparamagnetic iron oxide nanoparticle}
\DeclareAcronym{DNA}{short=DNA, long=deoxyribonucleic acid}
\DeclareAcronym{GRN}{short=GRN, long=gene regulatory network}
\DeclareAcronym{RNA}{short=RNA, long=ribonucleic acid}
\DeclareAcronym{IoBNT}{short=IoBNT, long=Internet of Bio-Nano Things}
\DeclareAcronym{IoNT}{short=IoNT, long=Internet of Nano Things}
\DeclareAcronym{BVS}{short=BVS, long=BloodVoyagerS}
\DeclareAcronym{CGM}{short=CGM, long=continuous glucose monitoring}
\DeclareAcronym{PEG}{short=PEG, long=polyethylene glycol}
\DeclareAcronym{BRET}{short=BRET, long=bioluminescence resonance energy transfer}
\DeclareAcronym{ATP}{short=ATP, long=adenosine triphosphate}
\DeclareAcronym{LU}{short=LU, long=luciferase}
\DeclareAcronym{IPTG}{short=IPTG, long=isopropyl thiogalactopyranoside}
\DeclareAcronym{MM}{short=MM, long=media modulation}
\DeclareAcronym{Tcell}{short=T-cell, long=T-lymphocyte cell}
\DeclareAcronym{MoSK}{short=MoSK, long=molecule shift keying}
\DeclareAcronym{GQD}{short=GQD, long=graphene quantum dot}
\DeclareAcronym{ITR}{short=ITR, long=interference-to-total received molecule ratio}
\DeclareAcronym{ISK}{short=ISK, long=interfacial shift keying}
\DeclareAcronym{CRN}{short=CRN, long=chemical reaction networks}
\DeclareAcronym{AWGN}{short=AWGN, long=additive white Gaussian noise}

\DeclareAcronym{PHY}{short=PHY, long=physical layer}
\DeclareAcronym{CIR}{short=CIR, long=channel impulse response}
\DeclareAcronym{MIMO}{short=MIMO, long=multiple-input multiple-output}
\DeclareAcronym{MISO}{short=MISO, long=multiple-input single-output}
\DeclareAcronym{RTT}{short=RTT, long=round trip time}
\DeclareAcronym{ISI}{short=ISI, long=inter-symbol interference}
\DeclareAcronym{ILI}{short=ILI, long=inter-link interference}
\DeclareAcronym{MAI}{short=MAI, long=multiple-access interference}
\DeclareAcronym{BER}{short=BER, long=bit error rate}
\DeclareAcronym{OOK}{short=OOK, long=on-off keying}
\DeclareAcronym{ASK}{short=ASK, long=amplitude shift keying}
\DeclareAcronym{SNR}{short=SNR, long=signal-to-noise ratio}
\DeclareAcronym{CSI}{short=CSI, long=channel state information}
\DeclareAcronym{TDMA}{short=TDMA, long=time division multiple access}
\DeclareAcronym{MAC}{short=MAC, long=medium access control}
\DeclareAcronym{RF}{short=RF, long=radio frequency}

\DeclareAcronym{STO}{short=STO, long=symbol time offset}
\DeclareAcronym{MAP}{short=MAP, long=maximum a posteriori}
\DeclareAcronym{MLE}{short=MLE, long=maximum likelihood estimation}
\DeclareAcronym{FBMLE}{short=FBMLE, long=filter-based maximum likelihood estimation}
\DeclareAcronym{RMSE}{short=RMSE, long=root mean squared error}
\DeclareAcronym{PCA}{short=PCA, long=principal component analysis}
\DeclareAcronym{PIV}{short=PIV, long=particle image velocimetry}
\DeclareAcronym{PLIF}{short=PLIF, long=planar laser-induced fluorescence}
\DeclareAcronym{PSO}{short=PSO, long=particle swarm optimization}

\DeclareAcronym{IIR}{short=IIR, long=infinite impulse response}

\DeclareAcronym{HCS}{short=HCS, long=human circulatory system}
 
\DeclareAcronym{AI}{short=AI, long=artificial intelligence}
\DeclareAcronym{XAI}{short=XAI, long=explainable artificial intelligence}
\DeclareAcronym{ML}{short=ML, long=machine learning}
\DeclareAcronym{RL}{short=RL, long=reinforcement learning}
\DeclareAcronym{DRL}{short=DRL, long=deep reinforcement learning}
\DeclareAcronym{DQN}{short=DQN, long=deep Q-network}
\DeclareAcronym{PPO}{short=PPO, long=proximal policy optimization}
\DeclareAcronym{NN}{short=NN, long=neural network}
\DeclareAcronym{ANN}{short=ANN, long=artificial neural network}
\DeclareAcronym{RNN}{short=RNN, long=recurrent neural network}
\DeclareAcronym{SBiRNN}{short=SBRNN, long=sliding bidirectional recurrent neural network}
\DeclareAcronym{BiRNN}{short=BiRNN, long=bidirectional recurrent neural network}
\DeclareAcronym{BRNN}{short=BiRNN, long=bidirectional recurrent neural network}
\DeclareAcronym{CNN}{short=CNN, long=convolutional neural network}
\DeclareAcronym{GNN}{short=GNN, long=graph neural network}
\DeclareAcronym{MLP}{short=MLP, long=multilayer perceptron}
\DeclareAcronym{FC}{short=FC, long=fully connected}
\DeclareAcronym{ReLU}{short=ReLU, long=rectified linear unit}
\DeclareAcronym{MSE}{short=MSE, long=mean square error}
\DeclareAcronym{LIME}{short=LIME, long=local interpretable model-agnostic explanation}
\DeclareAcronym{SHAP}{short=SHAP, long=shapley additive explanation}
\DeclareAcronym{LRP}{short=LRP, long=layer-wise relevance propagation}
\DeclareAcronym{DeepLIFT}{short=DeepLIFT, long=deep learning important features}
\DeclareAcronym{OMP}{short=OMP, long=group orthogonal matching pursuit}
\DeclareAcronym{PDP}{short=PDP, long=partial dependence plot}
\DeclareAcronym{ICE}{short=ICE, long=individual conditional expectation}
\DeclareAcronym{GD}{short=GD, long=gradient descent}
\DeclareAcronym{LSTM}{short=LSTM, long=long short-term memory}
\DeclareAcronym{BiLSTM}{short=Bi-LSTM, long=bidirectional long short-term memory}
\DeclareAcronym{TCN}{short=TCN, long=temporal convolutional neural network}
\DeclareAcronym{DNN}{short=DNN, long=deep neural network}
\DeclareAcronym{HGT}{short=HGT, long=heterogeneous graph transformers}
\DeclareAcronym{ARX}{short=ARX, long=auto-regressive exogenous}
\DeclareAcronym{SGD}{short=SGD, long=stochastic gradient descent}
\DeclareAcronym{Adam}{short=Adam, long=adaptive moment estimation}
\DeclareAcronym{BFGS}{short=BFGS, long=Broyden–Fletcher–Goldfarb–Shanno}
\DeclareAcronym{RMSprop}{short=RMSprop, long=root mean square propagation}
\DeclareAcronym{ANFIS}{short=ANFIS, long=adaptive-network-based fuzzy inference system}
\DeclareAcronym{TPR}{short=TPR, long=true positive rate}
\DeclareAcronym{FPR}{short=FPR, long=false positive rate}
\DeclareAcronym{AEC}{short=AEC, long=autoencoder}
\DeclareAcronym{FiLM}{short=FiLM, long=feature-wise linear modulation}
\DeclareAcronym{AAEC}{short=AAEC, long=asymmetric autoencoder}
\DeclareAcronym{BCE}{short=BCE, long=binary cross entropy}
\DeclareAcronym{FF}{short=FF, long=feedforward}
\DeclareAcronym{MNIST}{short=MNIST, long=modified National Institute of Standards and Technology}
\DeclareAcronym{GA}{short=GA, long=genetic algorithm}
\DeclareAcronym{GRU}{short=GRU, long=gated recurrent unit }
\DeclareAcronym{MPI}{short=MPI, long=manual permutation importance}
\DeclareAcronym{PINN}{short=PINN, long=physics-informed neural network}
\DeclareAcronym{PDE}{short=PDE, long=partial differential equation}
\DeclareAcronym{MIP}{short=MIP, long=manual permutation importance}
\DeclareAcronym{MDP}{short=MIP, long=Markov decision process}

\DeclareAcronym{CFD}{short=CFD, long=computational fluid dynamics}
\DeclareAcronym{TRL}{short=TRL, long=technology readiness level}

\DeclareAcronym{MoHANET}{short=MoHANET, long=mobile human ad hoc network}
\DeclareAcronym{PBS}{short=PBS, long=particle-based simulation}
\DeclareAcronym{GPU}{short=GPU, long=graphical processing unit}
\DeclareAcronym{CUDA}{short=CUDA, long=compute unified device architecture}

\DeclareAcronym{FIR}{short=FIR, long=finite impulse response}
\DeclareAcronym{DSP}{short=DSP, long=digital signal processing}
\DeclareAcronym{ADC}{short=ADC, long=analog-to-digital converter}

\usepackage{lipsum}

\usepackage{chemformula}

\usepackage{csquotes}
\usepackage[backend=biber,style=ieee,doi=false,isbn=false,maxnames=6,mincitenames=1,maxcitenames=2,sorting=none]{biblatex}
\DeclareFieldFormat{sentencecase}{#1} 
\DeclareFieldFormat{titlecase}{#1} 
\usepackage{xpatch}
\xpatchbibmacro{textcite}{\addspace}{\addnbspace}{}{}
\xpatchbibmacro{Textcite}{\addspace}{\addnbspace}{}{}

\DefineBibliographyStrings{english}{
	andothers = et~al\adddot\addspace
}

\usepackage[hidelinks]{hyperref}
\usepackage[capitalise,noabbrev]{cleveref}
\Crefformat{figure}{#2Fig.~#1#3}
\Crefformat{section}{#2Sec.~#1#3}
\Crefformat{equation}{#2Eq.~(#1)#3}
\Crefformat{appendices}{#2Appendix~(#1)#3}

\usepackage{stfloats}
\usepackage{tikz}
\usetikzlibrary{shadows}
\usetikzlibrary{mindmap,backgrounds}

\usepackage{titlesec}
\titlespacing\subsection{0pt}{2pt plus 1pt minus 2pt}{2pt plus 0pt minus 1pt}

\begin{document}

\title{Communicating Smartly in Molecular Communication Environments: \\ Neural Networks in the Internet of Bio-Nano Things}

\author{
    Jorge Torres G\'omez,~\IEEEmembership{Senior~Member,~IEEE},
    Pit Hofmann,~\IEEEmembership{Graduate~Student~Member,~IEEE},
    \\
    Lisa Y. Debus,~\IEEEmembership{Graduate~Student~Member,~IEEE},
    Osman Tugay Başaran,~\IEEEmembership{Graduate~Student~Member,~IEEE},\\
    Sebastian Lotter,
    Roya Khanzadeh, ~\IEEEmembership{Member,~IEEE},
    Stefan Angerbauer, Bige~Deniz~Unluturk,~\IEEEmembership{Member,~IEEE},
    \\
    Sergi Abadal,
    Werner Haselmayr,
    Frank H.P. Fitzek,~\IEEEmembership{Fellow,~IEEE},
    \\
    Robert Schober,~\IEEEmembership{Fellow,~IEEE},
    and Falko Dressler,~\IEEEmembership{Fellow,~IEEE}%
    \thanks{J. Torres G\'omez, L. Debus, O. Tugay Başaran, and F. Dressler are with the School of Electrical Engineering and Computer Science, TU Berlin, Germany, Email: \{torres-gomez,debus,basaran,dressler\}@ccs-labs.org.
    R. Khanzadeh, S. Angerbauer, and W. Haselmayr are with the Johannes Kepler University Linz, Austria, Email: \{roya.khanzadeh,stefan.angerbauer,werner.haselmayr\}@jku.at.P. Hofmann and F. H.P. Fitzek are with the Deutsche Telekom Chair of Communication Networks, Dresden University of Technology, Germany, and the Centre for Tactile Internet with Human-in-the-Loop (CeTI), Dresden, Germany, Email: \{pit.hofmann,frank.fitzek\}@tu-dresden.de. B. D. Unluturk is with Michigan State University, 775 Woodlot Dr, East Lansing, MI, USA, e-mail: unluturk@msu.edu.S. Abadal is with Universitat Politècnica de Catalunya, Spain, Email: abadal@ac.upc.edu.S. Lotter and R. Schober are with the Friedrich-Alexander-Universität Erlangen-Nürnberg (FAU), Germany, Email: \{sebastian.g.lotter,robert.schober\}@fau.de. This work was supported by the German Federal Ministry of Research, Technology and Space (BMFTR) through the project IoBNT, grant numbers 16KIS1986K \& 16KIS1994, the project 6G-life, grant number 16KIS2413K, and the project CommUnity, grant number 16KISS012K.
    This work was also supported by the German Research Foundation (DFG) through the project NaBoCom, grant number DR 639/21-1, SmartSynch grant number TO 1422/4-1, and the Excellence Strategy – EXC 2050/2, Project ID 390696704 – Cluster of Excellence "Centre for Tactile Internet with Human-in-the-Loop" (CeTI).
    Further support has been given by the "University SAL Labs" initiative of Silicon Austria Labs (SAL) and its Austrian partner universities for applied fundamental research for electronic-based systems'.
    Stefan Angerbauer and Werner Haselmayr acknowledge the Linz Institute of Technology (LIT), Johannes Kepler University at Linz, and the State of Upper Austria (Project LIT-2024-13-SEE-115) for the funds.}%
}%

\maketitle

%
%

\begin{abstract}
Recent developments in the \ac{IoBNT} are laying the foundation for innovative healthcare applications that envision a network of remotely coordinated nanodevices within the human body to monitor and actuate over potential diseases.
However, interconnecting such nanodevices requires communication strategies that can cope with \ac{MC} channels, whose complex, stochastic, and dynamic behavior often makes accurate physical modeling infeasible.
To explore the limits of nanodevice interconnectivity under these conditions, this survey focuses on data-driven communication strategies for \ac{MC} systems, with particular emphasis on \ac{ML} methods and \ac{NN} architectures for a robust and adaptive communication scheme at the nanoscale.
Research on \ac{NN}-enabled \ac{MC} spans several aspects covered in this survey, including \acp{NN} for communication in \ac{IoBNT} networks, the feasibility of biocompatible NN realization, explainable approaches, and the generation of training datasets.
We also include open-source code examples to support reproducible research across key \ac{MC} scenarios.
Finally, we identify emerging challenges, including the need for robust \ac{NN} architectures, biologically integrated \ac{NN} modules, and scalable training strategies.
\end{abstract}

\begin{IEEEkeywords}
Machine learning, neural networks, deep learning, molecular communication, Internet of Bio-Nano Things
\end{IEEEkeywords}

\acresetall

%

\section{Introduction}

Inspired by Schrödinger's reflections about the question ``What is life?'' \cite{schroedinger1944what}, the physics community joined biology to describe constituent components on the boundaries between inanimate matter and life.
Today, more communities are joining the realm of biology, including electrical engineers and computer scientists, to enable medical, sensing, and communication applications. 
Frameworks like the \ac{IoBNT} \cite{akyildiz2020panacea} can only be realized through truly interdisciplinary research leveraging expertise from engineering, computer science, and life sciences.
Focusing on recent advancements in \ac{AI} to realize {IoBNT} applications, this survey takes the journey one step further. 
We explore the research progress at the intersection of computer science and biology, specifically the use of \acp{NN} to enable \ac{MC} links and nanonetworks.

The constituent {MC} links of the {IoBNT} framework serve as synthetic tools with promising applications in healthcare and industry, though they pose new challenges for practical deployment.
As in the case of wireless links, communication over {MC} channels is unreliable, albeit for distinct reasons.
The combined impacts of random molecular mobility, reactive environments, and system geometry render analytical modeling impractical for most realistic scenarios.
As such, model-based \ac{MC} design becomes infeasible in many practical settings, motivating self-learning mechanisms, such as \ac{ML}, and particularly \ac{NN} architectures, that can account for unknown parameters and changing environmental conditions.

\subsection{Prelude}
\begin{figure}
    \centering
    \includegraphics[width=0.8\columnwidth]{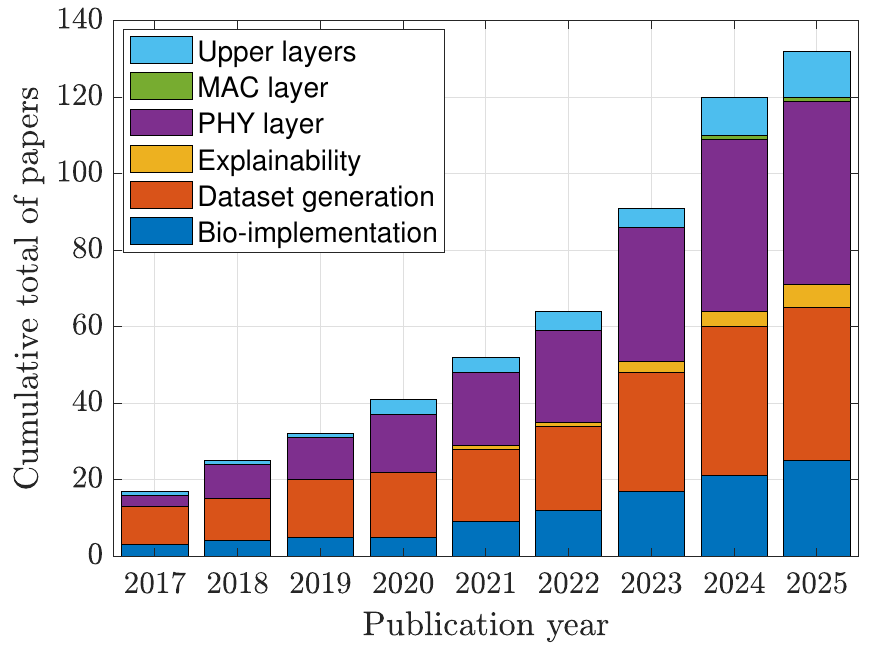}
    \caption{Trends in contributions related to {NN} research for the {IoBNT}.}
    \label{fig_trends}
\end{figure}

Today, the research community contributes to the joint field of {ML} and {IoBNT} networks in various directions.
As indicated in \Cref{fig_trends}, a growing number of works focus on developing the \ac{PHY} of {IoBNT}, while other contributions focus on the \ac{MAC} and upper layers.
Alongside these, other relevant directions arise: The transparency of these models for sensitive applications, such as the healthcare sector; the representativeness of the datasets used to train these models; and the feasibility of {NN} deployment at the nanoscale level.
These ongoing research activities lay the groundwork for numerous new developments, as we unveil in this survey.

This research landscape also distinguishes \ac{IoBNT} from intelligent wireless networks.
In wireless systems, \ac{ML} was introduced after decades of mature communication design; in \ac{IoBNT}, by contrast, communication architectures are being developed in an era where \ac{ML} is already a foundational tool.
Nevertheless, several lessons transfer: purely modular designs may be suboptimal, and joint optimization of communication blocks can improve performance.
Unlike wireless systems, \ac{MC} links are shaped by diffusion, flow, reactions, and biological variability, requiring models and learning strategies tailored to molecular environments.
Together, these advances and challenges frame the central question of this survey: how \acp{NN} can support reliable, explainable, and deployable communication in \ac{MC}-based \ac{IoBNT} networks.

%

\subsection{Literature Review Strategy}

Tracing the above question within the literature, this survey followed a three-stage search process.
First, we searched paper abstracts using the keywords ``Neural Network'' \texttt{AND} (``Molecular Communication'' \texttt{OR} ``{IoBNT}'').
Second, we expanded this set by examining both the references cited by these papers and the works that cited these works.
Finally, we curated the resulting collection by selecting papers relevant to biocompatible implementations of \acp{NN} and dataset generation.
In the first stage, we conducted a comprehensive search across the leading databases IEEE Xplore, ACM Digital Library, ScienceDirect (Elsevier), Springer, and Wiley Online Library.
In all cases, our inclusion criteria refer to: (i) The use of {NNs} as enablers of communication over {MC} links, which includes the communication design across all layers of the {MC} channel, i.e., from the \ac{PHY} layer to the application layer, (ii)~biocompatible implementation of {NN} models, (iii)~explainable methods for {NNs} as applied to {MC}, and (iv)~dataset creation.
We listed a total of \num{397} papers, and adhering to the inclusion criteria outlined above, we filtered these references to \num{133} published papers, spanning from \num{2017} to \num{2025}.
Our selection aimed to identify representative works that illustrate the evolution of {NN} applications relevant to the \ac{IoBNT} framework, ensuring coverage of both foundational and recent contributions across all communication layers.

%

\subsection{Contributions}
Focusing on {NN} architectures, this survey aims to outline the potential of {NNs} in {IoBNT} applications from various perspectives.
This paper not only covers the architectures and deployment of {NNs} but also complementary directions, such as implementation, explainability approaches, and dataset generation for {NN} model training.
The contributions are summarized as follows:
\begin{itemize}
    \item Deployment of {NN} architectures for {MC} links:
    We discuss in detail the {NN} architectures reported for the various {IoBNT} layers, with a primary focus on the \ac{PHY} layer, as it remains the most mature research area.
    We aim to provide readers with a holistic view of problems, environments, and {NN} architectures in each communication layer.
    %
    %
    \item Code developments for {NN}-based designs in {MC} links:
    We provide illustrative code examples for training and the use of {NNs} in {MC} systems.
    This open-access code encompasses a range of {NN} architectures across various {MC} environments, including free-diffusion, flow-based, and vessel-like channels, using both synthetic and testbed-generated datasets.
    We provide practical insight into the integration of {NNs} in {MC} channels.\footnote{We provide the code on two platforms: (i) In the Ocean Code platform, we provide access to the cell-to-cell example developed in \Cref{sec_code_dist} under the link~\url{https://codeocean.com/capsule/6777864/tree/v1}, and (ii) in the GitHub platform we provide access to all code on this paper related to synchronization, decoding, and autoencoding under the link \url{https://github.com/tkn-tub/NN_molecular_communications}.
    Furthermore, we provide the database used to train and test the reported {NN} modules in \cite{cell2cell2024data}.}
    \item Review of potential implementations at the nanoscale level:
    We summarize state-of-the-art technologies for implementing {NN} architectures at the nanoscale level.
    We describe biocompatible technologies such as microfluidic circuits and \ac{DNA} chemical reactions, which provide means to run {NN} modules.
    We discuss the feasibility of deploying reported {NN}-solutions in the nanoscale domain.
    \item Summary of explainable approaches to describe the operation of {NNs}:
    We provide a tutorial-style description of explainable methods, offering insights into their application in \ac{MC} channels.
    We survey the latest explainable methods studied for \ac{MC} applications.
    %
    \item Comprehensive summary of synthetic and testbed-based generation of datasets for training {NNs}:
    We elaborate in detail on {MC}-related datasets used to train {NN} modules.
    We also review dataset accessibility, documentation, and usability based on the published code and documentation per dataset.
    \item New upcoming challenges:
    We summarize new research directions related to the convergence of {NN} architectures and {MC}.
    We identify open problems in developing, deploying, and training robust {NN}s, as well as challenges in explaining their operation.
\end{itemize}

%

\subsection{Reader's Itinerary}

\begin{figure}
    \centering
    \includegraphics[width=\columnwidth]{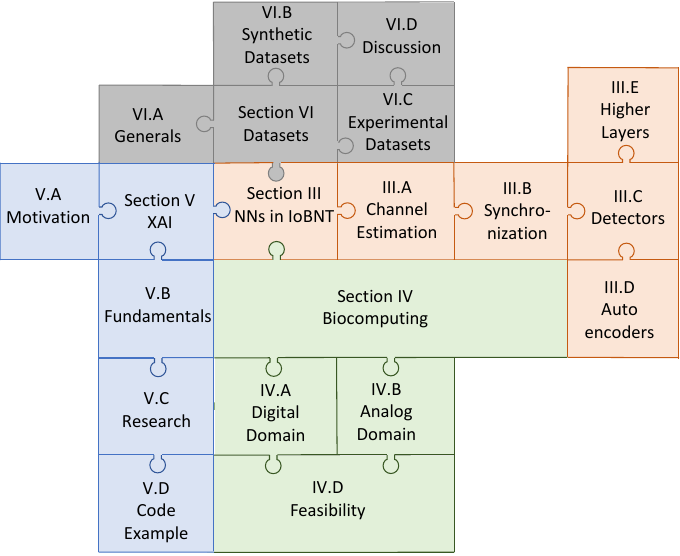}
    \caption{Mosaic representation of the survey content.}
    \label{fig_itinerary_2}
\end{figure}

We structured the content to guide the reader for an in-depth examination of the joint topic \ac{NN} and \ac{MC} as follows:
In \Cref{sec_communication}, we discuss mechanisms by which \acp{NN} enable \ac{IoBNT}  networks, which constitutes the central theme of the paper.
In \Cref{sec_bio_AI}, we examine the synthesis of \ac{NN} modules using biological components.
In \Cref{sec_XAI}, we review explainability approaches specifically tailored to \ac{AI} in \ac{MC} channels.
In \Cref{sec_data}, we present synthetic and testbed-generated datasets for training and evaluating \ac{NN} modules.
Among these sections, the most extensive part is \Cref{sec_communication}, which examines standard communication components (supported by {NNs}) such as channel estimation, synchronization, detection, and autoencoder designs.

Besides, this survey is organized to accommodate readers with different levels of familiarity with \ac{NN} and \ac{MC}.
\Cref{fig_itinerary_2} provides a visual roadmap highlighting the relationships among the main sections and possible independent reading paths.\footnote{Here, we invite the reader to follow the collage model for reading this survey, as first suggested by Julio Cortázar within the novel \textit{Rayuela}.}
\Cref{sec_communication} forms the core and examines \ac{NN} architectures for channel estimation, synchronization, detection, and autoencoder design.
Readers primarily interested in recent developments may begin with~\Cref{sec_bio_AI,sec_XAI}, which focus on biological implementations and explainability, respectively.
Those seeking a gradual introduction can start with \Cref{sec_communication,sec_data} for foundational background before moving on to these more specialized topics.
Besides, all sections link back to the core discussion in~\Cref{sec_communication}, ensuring continuity across the themes of {NN}s in \ac{MC}, biocomputing, explainable {AI}, and dataset generation.

%

\section{Most Recent Developments in Molecular Communications: A Survey-based Perspective}
\label{sec_SoA}

This section contextualizes recent developments in molecular communication and nanonetworked~\ac{IoBNT} systems within the landscape of related surveys published between \num{2019} and~\num{2025}.
Across these surveys, a common trend is the growing need for adaptive communication methods that can operate in complex, dynamic, and analytically intractable molecular environments.
Tracing this trend, we can group the most recent survey papers into four categories:
{(i}) {IoBNT} frameworks and applications, a field that also integrates {MC} and electromagnetics;
{(ii})~theoretical and technical developments in {MC}, which links communication theory with {MC} schemes; and
{(iii})~recent {AI} innovations in the {MC} field as summarized within the first surveyed materials on the topic.

\noindent
\textbf{IoBNT/MC Frameworks and Applications:} Grounded in applications for the early detection and treatment of diseases in the human body, published surveys described the {IoBNT} architecture to fulfill a futuristic paradigm: To connect human cells to the internet \cite{akyildiz2020panacea}.
The {IoBNT} architecture comprises a nanonetwork within the human body, in which nanobiological devices are the main components, e.g., engineered bacteria, human cells, miniaturized biosensors, and wearables.
The nanonetwork is connected to wearables attached to the skin surface, which are then linked to healthcare providers via the internet; see an illustration in \cite[Figures 7 and 9]{etemadi2023abnormality}.
These conceptual developments and early-stage technologies embody the {IoBNT} in a symbiotic relationship with the collective intelligence of the body~\cite{lagasse2023future}.


Towards this vision, existing surveys have reviewed a variety of interfaces between nanodevices and the internet, enabled by heterogeneous links that exploit the biological and electromagnetic domains, which are among the most relevant; see examples in \cite{kuscu2021internet} and \cite{aktas2024odorbased}, respectively.
Such interfaces enable groundbreaking healthcare applications, such as reducing the detection time for bacterial infections, as illustrated in \cite{etemadi2023abnormality}.
Applications also span smart agriculture and environmental monitoring, tracking the health status of animals and plants, as summarized in \cite{kuscu2021internet,lu2020wireless,marzo2019nanonetworks}.
In industrial environments, nanodevices can be deployed inside underground pipes to detect corrosion and damage \cite[Sec. III.b]{lu2020wireless}.

In the electromagnetic domain, surveys have addressed analog front-end units in the terahertz ($\si{\tera\hertz}$) band as a possible interface between implanted nanosensors and the outside world, as in~\cite{yang2020comprehensive}.
In graphene materials, the plasmonic effect observed at this frequency band is of particular applicability due to the existence of signal generators~\cite{dong2020recent} and antennas~\cite{saeed2021body-centric, abadal2019graphenebased} that can be miniaturized to a few microns in length and width.
Besides the $\si{\tera\hertz}$ band, recent surveys also pointed to optical bands exploiting nano-lasers and nano-antennas, megahertz ($\si{\mega\hertz}$) frequencies with magnetoelectric antennas based on nano-electromechanical devices, and to non-radiative techniques, either coupling-based or wave-based, such as galvanic coupling or ultrasound~\cite{abadal2024electromagnetic}.

Furthermore, several works have summarized potential healthcare-related applications of $\si{\tera\hertz}$ networks \cite{saeed2021body-centric, yin2022biomedical}, including {(i)} ultra-precise detection and localization of diseases with nanomachines flowing in the human blood system \cite{canovascarrasco2020understanding, lemic2021survey}, and {(ii)} brain-computer interfaces enabled by a wireless interaction with the human brain \cite{moioli2021neurosciences} even down to single neurons \cite{jornet2023nanonetworking}.
Notably, however, none of the existing work on the electromagnetic side of nanonetworks identifies {AI}/{ML} as a potential tool for modeling communication or protocol design.

Recent surveys on {IoBNT} also highlight the self-power capabilities of nanodevices in maintaining continuous operation of nanonetworks.
Micro-batteries built with micro-electromechanical and nano-electromechanical devices are powered by energy harvested from the environment or through wireless power transfer, as described in \cite{kuscu2021internet,zafar2021systematic}.
Besides, due to the highly sensitive applications of {IoBNT} networks, especially in healthcare, surveys also note the challenges and risks of {MC} \cite{qiu2024review} and wireless channels; see \cite[Sec.~IV]{zafar2021systematic}.
Likely attacks include eavesdropping, spoofing attacks, and jamming; see a full description in \cite[Table 6]{zafar2021systematic}.
These attacks exploit the limited computational capacities of nano nodes, which preclude the implementation of sophisticated protection algorithms.
Mitigating strategies use fingerprinting based on channel identification (a domain well-suited for {AI}) or cryptographic mechanisms to avoid eavesdropping. 

%

\noindent
\textbf{Theoretical and Technical Developments in {MC} Channel Modeling:} {MC} channel models are a common research topic in the scientific community across different scales: {(i)} molecular scale as for free-diffusion or junction {MC} channels between cells; {(ii)} cellular scale, which includes cell-to-cell communication and information processing; and {(iii)} larger scales, which span human vessels, organs, and tissues~\cite{akyildiz2019moving}.
These theoretical models provide the analytical basis for evaluating the feasibility of applications such as drug delivery, disease monitoring (e.g., cancer initiation and progression), and the deployment of body-network infrastructures \cite{lu2020wireless}.
The authors of \cite{jamali2019channel} summarize the end-to-end \acp{CIR} for various {MC} geometries at the molecular scale, irrespective of healthcare or industrial applications.
{MC} channels include free-diffusion, advection, and chemical-degradation channels~\cite[Table 1]{jamali2019channel}.
At the tissue scale, mobility models for passive and active bio-nano machines are summarized in~\cite{nakano2019methods}.
Theoretical work also addresses more practical problems arising in {MC} systems.
Examples include methods for channel parameter estimation~\cite{huang2022survey}, modulation and demodulation~\cite{kuran2021survey}, coding~\cite{hofmann2023coding}, and the formulation of spatial domain resources through \ac{MIMO} schemes~\cite{koo2021mimo}.
When analytical approaches become intractable, theoretical modeling is complemented by simulators that enable the numerical study of \ac{MC} links, as listed in~\cite[Sec. V.A]{jamali2019channel}, \cite[Sec. II.A.5]{lu2020wireless}. This progression from theory to simulation is further linked to experimental validation, with recent surveys categorizing testbeds by their expected near-future technological readiness~\cite {lotter2023experimental,lotter2023experimental2}.

%
\noindent
\textbf{Recent {AI} Innovations in the {MC} Field:} A prevalent viewpoint expressed in some {MC}-related surveys is that data-driven detectors can complement model-based schemes that are limited to represent realistic \ac{MC} channels~\cite{huang2021signal}, yet experimental validation remains indispensable.\footnote{See also \cite{boulogeorgos2021machine} for a detailed explanation of {ML} architectures and their application in other research areas such as therapy development and nanomaterial design.}
Advancing the field, the community is beginning to embed feedforward {NN} architectures into \ac{DNA} circuits in cells \cite{nagipogu2023survey} and to investigate explainable methods for {NN} module operation~\cite {huang2021signal}.
However, only a few surveys, such as \cite{huang2021signal,rizwan2018review,gentili2024neuromorphic}, address {ML} methods suitable for developing {MC}-based schemes.
The work in \cite{huang2021signal} primarily focuses on performance comparisons between model- and data-driven detectors, the authors of \cite{rizwan2018review} primarily consider the application layer, and the work in \cite{gentili2024neuromorphic} succinctly summarizes biocompatible technologies for running {NN} modules.

\noindent
\textbf{Cross-Disciplinary Routes:} Towards the interdisciplinary understanding among researchers in communication engineering, synthetic biology, and bioengineering, a hierarchy is proposed in \cite{bi2021survey} to map communication concepts to the biological behavior of cells. 
Unlike biologists, who study natural system interactions guided by holistic views, the engineering community is developing modular designs, i.e., defining communication functionalities in layers: application, data abstraction, molecules-data interface, molecular interactions, and their propagation.
A second perspective toward interdisciplinary research lies in the recent application of information-semantic concepts to \ac{MC}-engineered systems, in which goal-oriented communication may support the design of biochemical applications~\cite{egan2023toward}.

%
\noindent
\textbf{Summary Remarks:} The variety of topics in the above-related surveys introduces a joint appeal: Scenario awareness must be part of {MC} deployments and dynamically adapt to real-world environments; see, for instance, the channel-noise dependency on data \cite[Eqs. (73)-(76)]{jamali2019channel}, adaptive coding strategies to reduce \ac{ISI} as in \cite[Sec. IV.C]{hofmann2023coding}, and the challenges to design high data-rate transmissions due to the non-stationary nature of {MC} noise, as discussed in~\cite{huang2021signal}.
A recent trend is that {NN} architectures are becoming relevant for channel estimation, synchronization, detection, and related tasks such as localization and disease detection, as illustrated in \Cref{fig_trends}.
{NNs} are primarily implemented outside the {MC} environment with digital technologies (as we discuss in \Cref{sec_communication}), although embedding {NNs} within the {MC} pipeline is a very appealing topic in the literature (as we consider later in \Cref{sec_bio_AI}).

Furthermore, three critical insights emerge from the recent literature as open aspects to research:
\begin{itemize}
    \item Hardware-Driven \ac{AI}: Unlike traditional {AI} deployments, {NN}s in the {IoBNT} must be "biocompatible" at the nanoscale.
    This requires a move beyond standard feedforward architectures toward low-energy modules that can be hosted within biological chemical networks.
    \item Explainable Artificial Intelligence architectures: There is a growing consensus that "black-box" \ac{ML} models are insufficient for in-body environments.
    Modern research is shifting toward explainable \acp{NN} that utilize domain knowledge to provide \ac{XAI}, ensuring that responses are predictable and trustworthy.
    \item Interdisciplinary Convergence: \ac{AI} has emerged as the definitive framework bridging synthetic biology and communications engineering. 
    Recent efforts to define a common language for both disciplines enable the convergence of the disciplines toward prototyping "smart" biological nodes that function as active computational entities.
\end{itemize}
In the following sections, we provide an overview of the literature groundwork on {NN} architectures, {MC} environments, and comparative performance, establishing the technical baseline for convergence between biological and digital domains in \ac{IoBNT} networks.

%

\section{Neural Networks as Enablers of IoBNT Networks}
\label{sec_communication}

This section details the reported {NN}-based solutions for the various layers in \ac{IoBNT} networks.
Different architectures have been explored, including feedforward {NN}s for channel estimation, \ac{RNN} and \ac{BiRNN}) for detection, \acp{CNN} and \acp{TCN} for sequence decoding, and more recently, transformer-based and autoencoder models for end-to-end learning.
These architectures have primarily been applied in supervised settings, e.g., for channel estimation and detection tasks, with fewer works exploring unsupervised feature learning and reinforcement learning, e.g., for synchronization and data fusion tasks.

To provide structure to the extensive body of related work, this section is organized into the following subsections: channel estimation, synchronization, detection, end-to-end learning mechanisms, and studies reported in the \ac{MAC} and upper layers.
Each subsection follows the same format: (i)~Problem description for {MC}, where the motivation for using {NN} modules is discussed, followed by (ii)~the {MC} environments where {NNs} have been deployed, (iii)~description of reported {NN} architectures and their performance, (iv)~illustrative examples of implemented code using {NN} modules, and (v)~concluding remarks.

%

\subsection{Channel Estimation} 
\label{sec_channelEst}

End-to-end {MC} channel estimators are essential to enable high-performance decoding at \ac{IoBNT} nodes.
In the literature, we find the application of {NN}-based estimators for various purposes, such as evaluating communication performance \cite{yilmaz2017machine} and designing macro-scale receivers \cite{gulec2020distance}.
We categorize these estimators into parameter estimation and channel modeling approaches.

\subsubsection{Problem Definition}
\label{sec_channel_problem}

\begin{figure}
    \centering
    \includegraphics[width=\columnwidth]{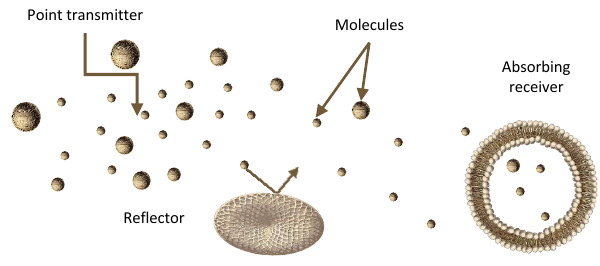}
    \caption{Conceptual illustration of a free-diffusion channel.}
    \label{fig_free_diffusion}
\end{figure}

The {CIR} in {MC} links is defined as the probability of finding a molecule at the receiver after its release from the emitter; see \cite[Def. 2]{jamali2019channel}.
The \ac{CIR} is given as a function of time, here denoted as $h(t)$ of given parameters $\{h_0,\,h_1,\dots,\, h_n\}$.
Channel estimation and modeling are the two main problems: the former concerns finding the set of parameters $\{h_i\}$, while the latter concerns finding the sequence $h(t)$.
In many situations, estimating or developing a closed-form expression for the {CIR} is not feasible due to the inherent complexity of the {MC} channel's geometry and topology.
Examples include the channel illustrated in \Cref{fig_free_diffusion},\footnote{The reflector and receiver components in the figure were generated with the assistance of ChatGPT, utilizing a Da Vinci style for the drawing.
The image for the cells in Fig. 4 was generated similarly.}
where just the inclusion of a reflector introduces a non-linearity in the channel, preventing the development of a closed-form expression for $h(t)$.

Either of these two problems unfolds into a challenge due to the unknown variables that need to be estimated, such as the communication distance, the impact of the geometry of the emitters and receivers, or the fluid velocity in vessel-like media.
Due to the difficulty of these problems, literature reports {NN}-based solutions leveraging on their capacity as universal approximators.
In the following subsections, we detail various {MC} scenarios and the reported {NN}-based estimators.

%

\subsubsection{Environments for MC Channel Estimation}

In the literature, {NN} models for channel estimation and modeling have been applied for two main {MC} environments: for simulated free-diffusion {MC} channels, as in \cite{yilmaz2017machine,lee2017machine,ozdemir2021estimating,ozdemir2024estimating,cheng2024channel,damrath2018array,cheng2025deep,ozbey2024artificial}, and for real-world testbeds for open air channels using sprayers as transmitters, as in~\cite{gulec2020distance}.
The more commonplace environment of free diffusion poses challenges in evaluating the channel model for complex topologies.
Examples include scenarios where multiple senders, reflectors, and absorbers are situated between the emitter and the receiver, which hinder the analytical development of closed-form solutions for the {CIR}; see \Cref{fig_free_diffusion}.
The {NN}-related literature considers free-diffusion environments on the ($\si{\micro\meter}$) scale, which include spherical reflectors and absorbers placed at arbitrary locations between a point emitter and a spherical absorbing receiver \cite{ozdemir2021estimating,cheng2025deep,ozdemir2024estimating,yilmaz2017machine}.

\ac{MIMO} links in {MC} using point transmitters and absorbing-receiver models are also gaining interest.
Examples are the $2\times 2$ \ac{MIMO} scheme in free-diffusion {MC} channels for \cite{lee2017machine,damrath2018array}, $4\times 4$ \ac{MIMO} in~\cite{ozbey2024artificial}, and asymmetric \ac{MIMO} channels comprising \num{2} emitters and \num{4} receivers in~\cite{cheng2024channel}.

The literature also evaluated testbed measurements that account for realistic channel effects, including the impact of droplet size, evaporation, and unsteady flows, as in~\cite{gulec2020distance}.
The testbed deploys an air-compressor sprayer to release ethyl alcohol molecules and an MQ-3 alcohol sensor to detect droplets from distances of up to $\SI{200}{\centi\meter}$.

\subsubsection{NN Models for MC Channel Estimation}
\label{sec_NN_channel}

Various models are reported for free-diffusion channels aiming to estimate {CIR} parameters, as in \cite{lee2017machine,gulec2020distance,cheng2024channel,damrath2018array,cheng2025deep}, and more broadly, to estimate the {CIR} sequence as in~\cite{ozdemir2021estimating,ozdemir2024estimating,ozbey2024artificial}.
These models implement a variety of architectures such as feed forward {NN} \cite{lee2017machine,gulec2020distance,yilmaz2017machine,cheng2024channel,damrath2018array,ozbey2024artificial}, \ac{CNN} \cite{ozdemir2021estimating}, and the popular transformers~\cite{ozdemir2024estimating,cheng2024channel,cheng2025deep}.

\noindent
\textbf{{NN}-based Solutions for {MC} Channel Estimation:}
The aforementioned {NN} architectures are used to estimate a variety of {MC} channel parameters, including distances and the orientation of emitters and receivers in the channel.
For instance, a feedforward {NN} model is developed in \cite{gulec2020distance} to estimate the distance between the emitter and the receiver.
The model employs a single hidden layer with a single node and is trained on a predefined set of features, which includes the peak of the received sequence and the rise time from low to high amplitude in the sequence of received molecules.
Results exhibit relative errors between the estimated and actual distances in the range of $\SIrange{2}{20}{\percent}$ at communication distances of $\SIrange{100}{200}{\centi\meter}$.

{NNs} are also reported in \cite{yilmaz2017machine,lee2017machine} to evaluate the parameters of the first-hitting time probability; see \cite[Eq. (2)]{lee2017machine}.
The deployment consists of single- and two- {NN} machines, trained on the recorded number of molecules at the receiver.
The single {NN} is deployed through \ac{FC} hidden layer and a total of \num{30} nodes.
The two-machine model is proposed in \cite{lee2017machine}, in which each {NN} estimates a separate set of {CIR} parameters.
Both {NNs} are implemented with a single \ac{FC} layer and \num{15} nodes each.
The results in \cite{lee2017machine} indicate that the single-machine model performs better than the two-machine ones.
These solutions achieve high accuracy with an absolute error of less than \num{e-2}.

In free-diffusion \ac{MIMO} environments, {NNs} are trained to directly estimate the parameters of a given closed-form expression for the {CIR}, as indicated in \cite{lee2017machine,damrath2018array,cheng2024channel}; see the model and the parameters in \cite[Eqs. (2) and (3)]{lee2017machine}.\footnote{Closed-form \ac{CIR} models can be derived for ideal channel geometries, e.g., comprising a point transmitter, a free-diffusion channel, and a spherical receiver; more realistic \ac{MC} channels instead require data-driven estimation.}
These solutions in \cite{lee2017machine,damrath2018array} follow a two-stage approach to train the {NN}.
In the first stage, the authors fit a {CIR} model to a $2\times 2$ \ac{MIMO} {MC} channel, using the non-linear least-squares method and samples generated via simulations.
Next, the {NN} is trained to predict the fitted \ac{CIR} using {MC} channel parameters such as distance, the molecule's diffusion coefficient, and receiver radius as inputs.


Continuing with \ac{MIMO} channels, a more typical deployment trains {NN} models using received molecule counts, as investigated in \cite{cheng2024channel,ozbey2024artificial}.
The work in \cite{cheng2024channel} compares the performance of transformer and feedforward {NN} architectures for channel estimation, with the latter achieving better performance in simulations.
The feedforward {NN} comprises one \ac{FC} layer and~\num{5} hidden layers of \numlist{10;20;40;20;10} neurons.
The encoder and decoder components of the transformer implement two \ac{FC} layers and the same number of hidden layers as the feedforward {NN}.
In particular, the work in \cite{ozbey2024artificial} estimates the distance and the rotation angle between the emitter and receiver planes of a $4\times 4$ \ac{MIMO} setup.
The solution deploys a feedforward {NN} with \num{14} connected layers, where the inputs are the numbers of molecules received and the outputs are the \ac{MC} distances and rotation angles.


\noindent
\textbf{{NN}-based Solutions for {MC} Channel Modeling:}
{NN} models are also trained to predict the {CIR} sequence in free-diffusion channels as explored in \cite{ozdemir2021estimating,ozdemir2024estimating}. 
The solution branches into two paths, one of which takes as input the red-green-blue (RGB) image of the {MC} environment.
Point transmitters in the \ac{MC} channel are depicted with dots, while absorbers, reflectors, and receivers are depicted as circles of different colors and radii. 
A \ac{CNN} is used in \cite{ozdemir2021estimating}, and a transformer (without the positional encoders) is used in \cite{ozdemir2024estimating} to extract the relevant image features.
A \ac{MLP} block is included within the architecture to normalize the sequence for the {CIR}.
The second branch of the design in \cite{ozdemir2021estimating} employs an \ac{FC} layer and an \ac{MLP} block to estimate the maximum number of received molecules.
As a result, the normalized \ac{MSE} between the ground truth and the estimated {CIR} sequence is on the order of \num{e-2}.

%

\subsubsection{Illustrative Code Example to Estimate the Distance Among Cells}
\label{sec_code_dist}

\begin{figure}
    \centering
    \includegraphics[width=\columnwidth]{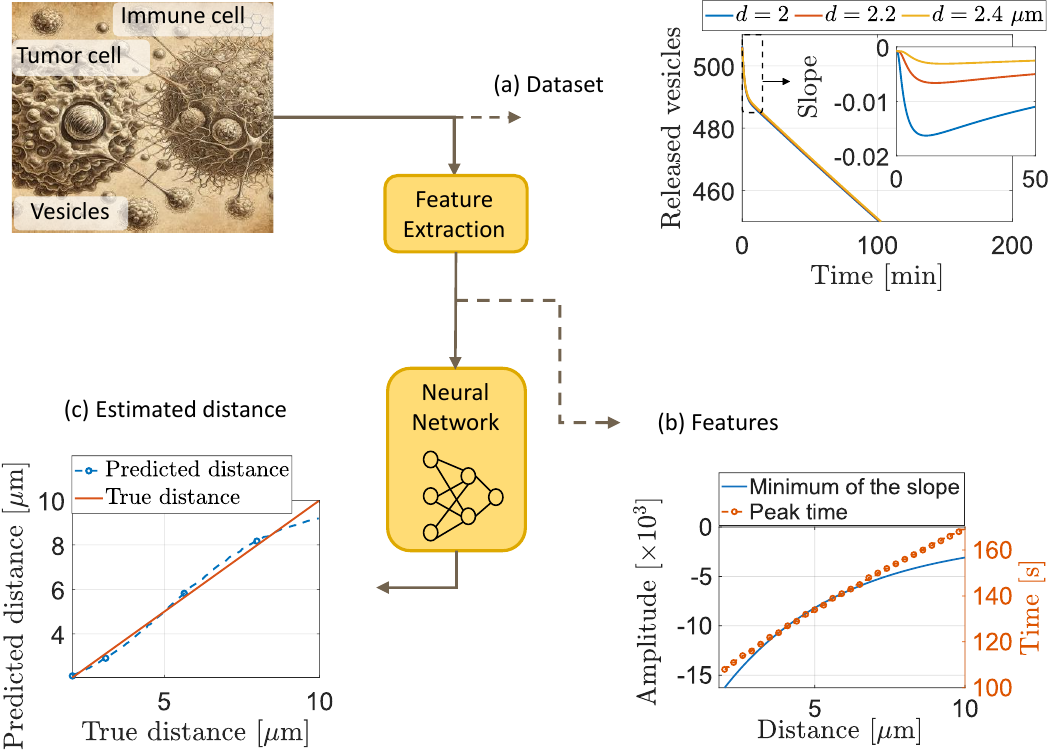}
    \caption{Schematic for estimating the distance between cells.}
    \label{fig_cell2cell}
\end{figure}

To connect the preceding discussion with implementation, we present a concise \ac{NN}-based distance-estimation example for a representative cell-to-cell \ac{MC} scenario.
This code example develops a distance estimator between immune and cancer cells based on the recorded number of vesicle molecules.
Cancer and immune cells exchange vesicles in the proximity of each other, and the concentration level of vesicles recorded at the immune cell can be readily used to estimate their distance to the cancer cell; see the sketch in \Cref{fig_cell2cell} and model details in~\cite{zoofaghari2024modeling}.
The dependence of the concentration level on the distance parameter prevents deriving a closed-form inverse relationship; therefore, a trained {NN} is employed to learn this mapping.


To model the environment, we use the code provided by the authors in \cite{zoofaghari2024modeling}, which calculates the number of vesicles released by the immune cell; see the various curves plotted over time and at different distances in \Cref{fig_cell2cell}a.
We first extract two features from the raw data to train the {NN}: (i)~The peak amplitude of the slope of the received vesicles; and (ii)~the time location of the peak (peak time), as illustrated in \Cref{fig_cell2cell}b.
We implemented a feedforward \ac{NN} in Matlab with a single layer and two nodes, reflecting the approximately linear relation between peak time and distance observed in \Cref{fig_cell2cell}b), which can be captured with two trainable coefficients.
As depicted in \Cref{fig_cell2cell}c, the {NN} accurately estimates the distance to the tumor cell in the range of $\SIrange{2}{10}{\micro\meter}$ with a relative error of $\SI{3.3}{\percent}$.

%

\subsubsection{Concluding Remarks}

The use of {NNs} for {MC} channel estimation has mostly focused on feedforward {NN} and \ac{CNN} architectures.
Feedforward {NNs} are used to estimate specific parameters in the channel, and \ac{CNN} architectures are adopted for the more ambitious goal of estimating the {CIR}.
This subject still requires further development of {NNs} to target more realistic scenarios.
In \ac{IoBNT} contexts, \ac{NN}-based channel models have not yet adequately addressed both short-range links, such as cell-to-cell communication, and long-range channels, such as those in the human circulatory system.

%
\subsection{Synchronization}

Successful communication among nodes relies on all participants being aware of the timing of the other entities involved; i.e., they must be synchronized.
Many works on {MC} assume perfect synchronization by default, thereby enabling separate analysis of receiver noise sensitivity or detection parameter optimization.
Yet in real-world systems, communicating nodes must explicitly synchronize to ensure successful communication.
The following sections will define the problem, summarize reported approaches, and describe an example implementation.

\subsubsection{Problem Definition for Synchronization}

The problem of symbol synchronization concerns the question of when a symbol is considered to start at the receiver's side.
In the simplest case, i.e., one transmitter and one receiver, the transmitter sends the symbols, and, according to the delay introduced by the {MC} channel, they arrive at the receiver.
It is the receiver's challenge to determine the start of the transmitted symbol from the received sequence.

Traditionally, symbol synchronization in {MC} is often addressed using the \ac{MLE} as an optimal method; see \cite{jamali2017symbol}.
Due to the high computational requirements for implementing \ac{MLE}, it is often not a viable option for experimental {MC} systems.
Out of this need, low-complexity versions of \ac{MLE} that rely on the use of separate signaling molecules for data transmission and synchronization were proposed in \cite{jamali2017symbol}.
Additionally, by using two different signaling molecule types, the solution in \cite{mukherjee2019synchronization} creates an artificial synchronization clock by exploiting differences in diffusion coefficients.
Due to their higher diffusion coefficient, synchronization molecules arrive at the receiver before the data-signaling molecules, thereby indicating the start of transmission.
A blind synchronization approach based on \ac{MLE} was proposed in \cite{shahmohammadian2013blind}, in which a predefined sequence of molecules is transmitted before the actual data transmission to calibrate \ac{MLE}.

In the synchronization approaches reported above, one problem is ignored: the variability of the {MC} setup.
The system is assumed to be static and well-known.
However, this assumption does not hold for real-world scenarios where transmitter-receiver distances, background flow, and the behavior of signaling molecules change over time.  
This variability triggers the need for synchronization to enable adaptation to the system's changing state. 
These problems are well-suited for {ML}-based synchronizers because they can adapt continually to fast-changing channels. 

%

\subsubsection{Relevant Environments for Synchronization}

The reported {ML}-based solutions to synchronization constitute a first step towards incorporating more complex environments into their setup and evaluating two types of dynamic transmission scenarios.
In the reported works, two extensions to the simple static {MC} channel were evaluated:

\noindent
\textbf{Mobile Setup:} 
In many expected {MC} applications, the communication participants are mobile and operate within a defined space subject to environmental drift. As a result, communication may occur from varying relative positions.
The synchronization strategy must, therefore, be able to adapt to a changing number of molecules being received over time.

\noindent
\textbf{Relay Setup}:
Depending on the environment, extending the communication range may require the use of relays.
In this case, a single receiver will likely detect molecules transmitted by the last relay node in a drift-affected channel, along with interference from previous relays.
Combined with the possible mobility of communication participants, a synchronization unit must be able to synchronize reliably in dynamically changing interference scenarios.

Still, these scenarios do not fully describe all expected application environments of {MC}.
Synchronization strategies for {MC} must additionally consider more realistic setups such as multiple transmitters and receivers or multi-path propagation.
Here, again, the synchronization method must be able to adjust to dynamic environments, and {ML} could be highly beneficial.

%

\subsubsection{NN Models for Synchronization}

As synchronization in a dynamic {MC} setup relies on continuous adaptation to an ever-changing environment, \ac{RL}, which is based on interaction with the environment, is especially fitting.
With its ability to estimate complex mathematical models \cite{luong2019applications}, \ac{RL} offers the possibility of solving the problem with little a priori knowledge.

Synchronization with the help of \ac{ML} was first employed in~\cite{debus2023reinforcement}.
An \acs{RL} agent was designed for a simulated \ac{MC} setup of an air-based \ac{MC} testbed with a mobile receiver~\cite{hofmann2022testbed}.
With rewards based on the correctness of the decoded bit, the agent learned to adjust the decoding threshold to detect a synchronization sequence.
While the agent showed potential, the necessity of knowing the correct bit value during training significantly impacted the applicability of the approach in a real-world setting.
The synchronizer has also been integrated into a relayed mobile setup consisting of a transmitter, a relay, and a receiver, as described in \cite{debus2024synchronized}.
The new \ac{RL} agent's reward was based on the difference between the counted number of molecules and the threshold set by the agent.
The synchronizer was reported to achieve a \ac{TPR} of over $\SI{80}{\percent}$ and a \ac{FPR} of below $\SI{5}{\percent}$ for all transmitted synchronization frames.
The results compared particularly well to the \ac{FBMLE} synchronizer~\cite{jamali2017symbol}, which struggled to cope with the additional interference caused by the original transmission on the link between the relay and the receiver.
In both \ac{RL}-based synchronizers, a \ac{PPO} agent was used. 
Their actor and critic networks included \ac{LSTM} layers to account for past observations and actions when evaluating the current step.

Synchronization and detection via \acp{DNN} was implemented in \cite{casaleiro2024synchronisation}.
Two setups, featuring a one-dimensional \ac{CNN} and a \ac{GRU} \ac{RNN}, were evaluated for their synchronization and decoding performance, using a padded Barker code as a synchronization frame.
The first setup used one {NN} for both synchronization and decoding, while the second setup used separate {NNs} for the two tasks.
In the evaluation, different static \ac{SNR} scenarios were considered in an unbounded free-diffusion channel.
Both setups successfully synchronized the transmissions at \acp{SNR} of \SI{45}{\decibel} or higher.

\newpage
While the reported results demonstrate the usability of {ML} to synchronize {MC} systems, they only scratch the surface of the approach's advantages.
Especially in more complex environments that might include multipath and multiple simultaneous transmissions, significant benefits are expected from employing an inherently adaptive {ML}-based synchronizer.
Similar trends have been observed in wireless communication systems, where learning-based synchronization enhances robustness to channel impairments~\cite{szott2022wifi}.
In the following section, we demonstrate how the reviewed {ML} approaches naturally extend to an {MC} setup with dynamically varying interference.
We provide an illustrative example showing how such a model can be designed, trained, and evaluated in practice.

%

\subsubsection{Illustrative Code Example to Synchronize the Receiver and Emitter Symbol Time}
\label{sec_sync_ex}

Following the example in \cite{debus2024synchronized}, we implemented an \ac{RL}-based synchronizer in Matlab and Simulink.
We utilized a particle simulation of the \ac{MM} testbed introduced in \cite{brand2022mediamodulation}, demonstrating that the reported approach can be transferred to liquid-based closed-loop {MC} scenarios.
In the testbed, the traditional transmitter-receiver structures are extended by adding an eraser positioned in the loop, after the receiver and before the transmitter.
Switchable signaling molecules \cite{brakemann2011reversibly} are used for communication in this setup.
The transmitter can turn the molecules "on"; after they pass the receiver, the eraser turns them "off" again.

\begin{figure}
    \centering
    \subfloat[\ac{RL}-based synchronizer~\cite{debus2023reinforcement}.]
    {\includegraphics[width=0.7\columnwidth]{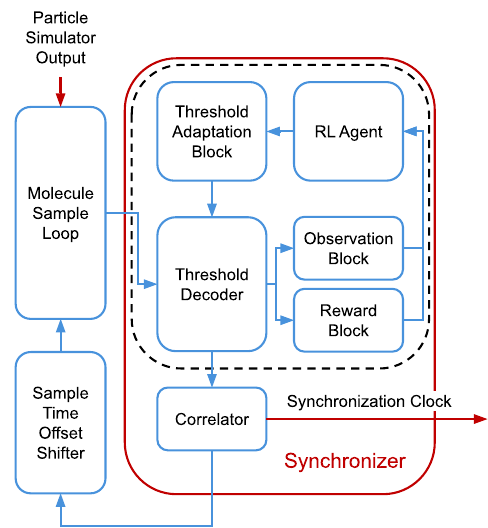}\label{fig_debus2024synchronized_environment}}\\
    \vspace{2em}
    \subfloat[\Acf{TPR}, \acf{FPR}, and \acf{STO}~\cite{debus2024synchronized}.]
    {\includegraphics[width=\columnwidth]{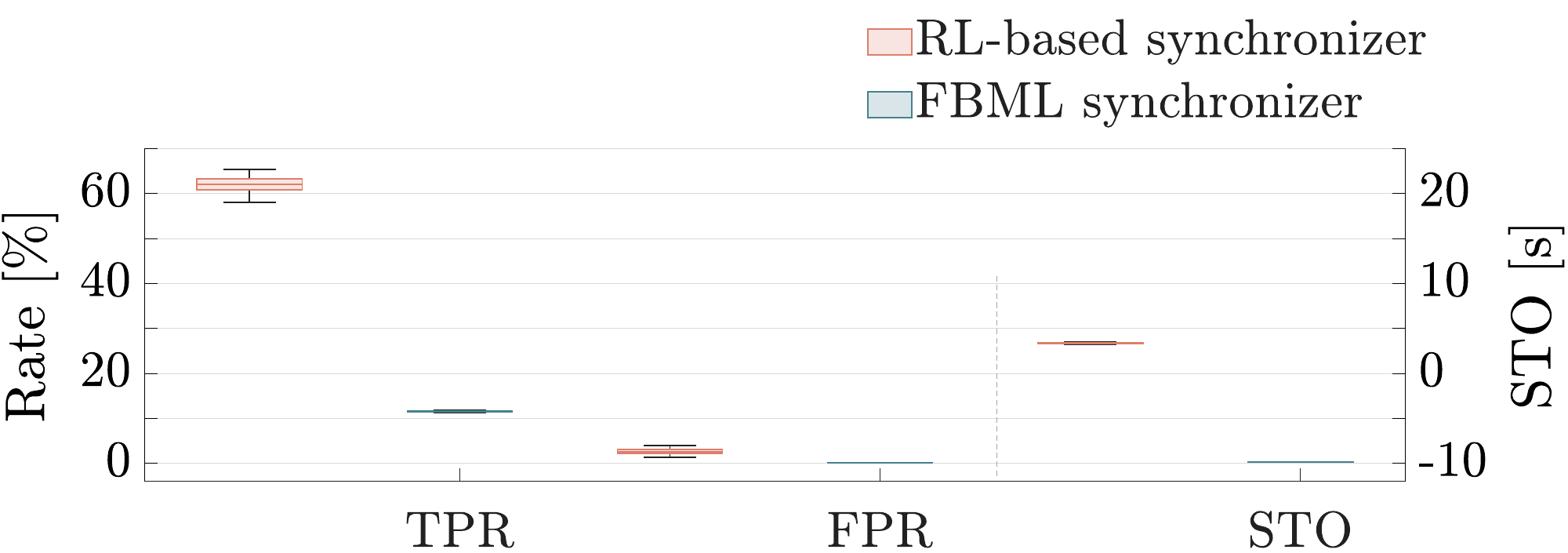}\label{fig_debus2024synchronized_results}}
    \caption{System model of the \ac{RL}-based synchronization unit and its performance.}
    \label{fig_synch_RL}
\end{figure}

Our synchronizer follows the structure displayed in \Cref{fig_debus2024synchronized_environment}.
A loop through the environment consists of: {(i)} Sampling the current number of molecules (Molecule Sample Loop), {(ii)}~decoding the current bit value (Threshold Decoder), {(iii)}~having the \ac{RL} agent adapt the threshold according to the observed state and the reward, and {(iv)} synchronizing the sampling offset if the synchronization frame was detected (Correlator and Sample Time Offset Shifter blocks).
The presented system uses a \ac{PPO} agent \cite{schulman2017proximal} with a single \ac{LSTM} layer in both the actor and critic networks.
We evaluated learning behavior across several network and layer sizes and found that the agent performed best with \num{128} cells per \ac{LSTM} layer.
Additionally, we performed hyperparameter tuning for the actor and critic learning rates, mini-batch size, experience horizon, entropy loss rate, discount factor, and reward scaling factor, as described in~\cite {andrychowicz2020what}. 
For the other parameters, we found that their influence on the agent's performance was best left at the default parameter settings.
The process of setting the parameters was based on our experience from previous work~\cite{debus2024synchronized}.

We produce data for our implementation with a particle simulator of the testbed.
Bits were transmitted using binary \ac{CSK} modulation with an instantaneous pulse releasing \num{e3} molecules to represent bit \texttt{1} and no molecules for bit \texttt{0}.
To vary the data slightly while not distorting the simulated channel behavior too much, additive white Gaussian noise was added to create an \ac{SNR} of \SI{30}{\dB}, as described in~\cite{kuran2021survey}.
The \ac{RL} agent was then trained for \num{245000} \num{5}-bit-frames in \num{35000} episodes.

We evaluate our synchronization unit by comparing it to the \ac{FBMLE} synchronization unit \cite{jamali2017symbol}.
We use its synchronization bit-sequence \texttt{[11001]} as transmitted with the \ac{CSK} modulation.
Both synchronizers ran \num{100} times \num{1000} \mbox{\num{5}-bit-frames} taken from the evaluation dataset.
To assess the accuracy of the synchronizers in detecting the transmitted synchronization frames, we evaluate the \ac{TPR}, the \ac{FPR}, and the symbol-time offset relative to the synchronization frames for both synchronizers.
The results in \Cref{fig_debus2024synchronized_results} reveal that
the \ac{RL}-based synchronizer achieves a higher \ac{TPR} than the \ac{FBMLE} synchronizer.
At the same time, it also detects more synchronization frames incorrectly, resulting in a higher \ac{FPR} than the \ac{FBMLE} approach.
Taking both detection rates together, the \ac{RL}-based synchronizer performs better.
Regarding the absolute value of the symbol time offset, both synchronizers accurately detect the symbol's start.

\subsubsection{Concluding Remarks}
The reported solutions show the potential of using {ML}-based approaches for synchronization in {MC}.
Their ability to adapt to the highly dynamic environments of {MC} channels allows smart synchronizers to outperform other synchronization approaches in real-world settings.
While existing solutions employ only \ac{RL} approaches and \ac{RNN} structures, research integrating diverse network architectures and attention mechanisms is rife with opportunities.
Besides, as we integrate {MC} into the \ac{HCS} for precision medicine applications, {ML}-based approaches offer the next step toward synchronizing workflows within the ever-changing human body environment.

%

\subsection{Detectors}

This section examines {NN} architectures for information decoding at receiver nodes, which is the most-studied topic in the literature.\footnote{The total number of references related to detection tasks is $\SI{26}{\percent}$ of the surveyed literature.}
A key benefit of {NNs} is their ability as universal approximators, enabling adaptable tuning of decoder parameters in unknown environments.
In line with Occam's razor,\footnote{Occam’s razor is the principle that, among competing element sets, the ``simplest'' one should be preferred, i.e., the set with the smallest number of elements.} {NN} models are designed to learn the minimum possible {MC} parameters of the {CIR}.
The literature reports architectures based on feedforward {NN}s, \acp{RNN}, \acp{CNN}, and transformers for decoding tasks.
In the following, we provide details on the detection problem definition, environments, deployed {NN} architectures, and an illustrative code example. 

\subsubsection{Problem Definition to Decode Incoming Symbols}
\label{sec_detect_problem}

Decoding digital transmissions refers to identifying received symbols out of an alphabet of possible ones.
Similarly to wireless communications, the decoding problem is formulated based on a probabilistic description of the {MC} channel.
{MC} channels are abstracted with the conditional probability $P_\mathrm{ch}(y_k,y_{k-1}, \dots|x_k,x_{k-1}, \dots;H)$, where $H$ refers to the set of channel parameters, while $x_k$ and $y_k$ are the transmitted and received symbols in time slot $k$, respectively.
Assuming a known probability mass function for the transmitted symbols,
denoted by $P_X(x_k, x_{k-1}, \dots)$ over a finite alphabet $\mathcal{X}$ ($x_k \in \mathcal{X}$),
the MAP estimate of the transmitted symbol $x_k$ is given by $\hat{x}_k=\underset{x_k\in \mathcal{X}}{\mathrm{argmax}}\,{P_\mathrm{ch}(y_k,y_{k-1}, \dots|x_k,x_{k-1}, \dots;H)}\times P_X(x_k,x_{k-1},\dots)$, see \cite[Sec. II]{farsad2018neural}.
%
%

Solving the above problem is typically infeasible because the conditional distribution ${P_\mathrm{ch}(y_k,y_{k-1}, \dots|x_k,x_{k-1}, \dots;H)}$ and the set of parameters $H$ are unknown.
This challenge is amplified in time-variant channels, such as those in drift fluidic environments~\cite{debus2025blood}, and in experimental testbeds where the end-to-end model depends on complex, often unmeasurable geometries~\cite{farsad2013tabletop}. 
Consequently, data-driven approaches based on \acp{NN} have been proposed to learn the channel distribution directly, avoiding explicit modeling of the end-to-end \ac{MC} channel.


Decoding may also require determining the optimal detection thresholds ($\lambda_i$) for \ac{CSK} transmissions, which is a second problem formulation stated in the literature; see \cite[Eq. (26)]{qian2019molecular}.
For optimal performance, the $\lambda_i$'s must be adjusted based on the distance between emitter and receiver, as well as the emission pattern, which is not a trivial problem, see the formulation in \cite[Eq. (13)]{qian2019molecular} for the simplest case of a zero-bit memory receiver.
Threshold-based detection requires solving the problem $\lambda_i=\underset{\lambda}{\mathrm{argmin}}{\,\mathrm{BER}(\lambda)}$, see \cite[Eq. (26)]{qian2019molecular}, where $\mathrm{BER}$ refers to the \acl{BER}.
This formulation is also solved through {NN} models due to the lack of suitable closed-form expressions.
%
%

%

\subsubsection{Environments for {NN} Detectors}

{NN} models, as detectors, are trained in {MC}- free diffusion environments, where multiple geometries are considered in the literature.
The training sequence is constructed in {MC} channels where the emitter is a point transmitter and the receiver is an absorbing sphere \cite{alshammri2018adaptive,agrawal2022neural,sharma2020deep,qian2019molecular,lu2023mcformer} or a ligand receptor \cite{kim2023machine,baydas2023estimation}.
More complex environments follow multi-hop links as in~\cite{cheng2024signal}, $5\times 5$ \ac{MIMO} channels with ligand receptors in \cite{baydas2023estimation}, and a $8\times 1$ \ac{MISO} channel as described in~\cite{kara2022molecular}.
Mobility of the transmitters and receivers is also modeled in \cite{shrivastava2021performance, shrivastava2021scaled,cheng2023signal,cheng2024informer}, which eventually distorts the received sequence.
Vessel-like channels, which have been less studied, were introduced in~\cite{kosanetzki2025demodulation}.
The communication link is established inside a pipe, where the emitter is fixed in place with a cylindrical geometry, while the receiver moves across the pipe’s cross-section and oscillates along the flow direction.

The randomness of the {MC} channels is primarily modeled using the Poisson distribution, as in \cite{qian2018receiver,qian2019molecular,farsad2018neural,alshammri2018adaptive,shrivastava2021performance,shrivastava2021scaled,agrawal2021hyperparameter}, and it is less frequently modeled with the Gaussian distribution as in \cite{chen2021selfattention,kara2022molecular,kosanetzki2025demodulation}.
More realistic {MC} models also incorporate the degradation of molecules due to chemical reactions, as evaluated in \cite{agrawal2022neural,shrivastava2021performance,shrivastava2021scaled}.

Received training sequences have been recorded from experimental testbeds in vessel-like channels.
Examples include binary emissions performed with acid and base signals in water \cite{farsad2017detection,farsad2018sliding,farsad2018neural,sun2020ctbrnn,koo2016molecular,koo2020deep,chen2021selfattention}, saline solutions and water \cite{wang2020understanding}, and using magnetic nanoparticles as in \cite{bartunik2022using,bartunik2023artificial}.
Binary transmissions are performed using \textit{E. Coli} bacteria as generators, as reported in the \textit{in vivo} testbed in \cite{vakilipoor2022hybrid,bai2023temporal}.
\textit{E. Coli} bacteria are stimulated with light to release protons and increase the medium's pH level, and a sensor measures the pH levels, serving as the receiver~\cite{vakilipoor2022hybrid}.
As another example, binary transmission is performed by releasing flagellated bacteria, which are guided by a magnetic field towards a spherical receiver, as described in \cite{bai2023temporal}.
The receiver, which was initially filled with luminescent non-motile bacteria, begins collecting the \ac{QS} molecules released by the arriving bacteria and produces light in response.
The observed intensity and the number of collected bacteria are used to subsequently decode the transmitted sequence.

To summarize, the communication methods outlined above use pulse-based modulation, i.e., \ac{OOK}.
The one bits are encoded with the number of released molecules or bacteria, and the zero bits are encoded by their absence.
Additionally, emissions occur with bit durations in the millisecond ($\si{\milli\second}$) range and the channel noise varies significantly from a considerable level ($\text{SNR}=\SI{0}{\decibel}$) to a nearly negligible one ($\text{SNR}=\SI{60}{\decibel}$).

%

\subsubsection{{NN} Architectures to Detect Incoming Symbols}
\label{sec_NN_detectors}

Reported architectures for detection in {MC} include feedforward \acp{NN}~\cite{farsad2018neural}, which exploit prior knowledge such as channel memory; \acp{CNN}~\cite{huang2021signal}, offering robustness to channel delays; and \acp{BiRNN}~\cite{farsad2018neural}, which effectively mitigate \ac{ISI} by processing both past and future samples when decoding the current symbol.
More recently, attention-based models~\cite{vaswani2017attention} have been introduced to capture long-range temporal dependencies and further improve decoding performance in channels with extended memory.


These architectures are reported to operate either on a symbol-by-symbol or on a sequence basis.
In symbol-by-symbol mode, the {NN} is trained with samples corresponding to a single symbol only, like in \cite{qian2018receiver,qian2019molecular,farsad2017detection,farsad2018detection,sharma2020deep,kim2023machine,kosanetzki2025demodulation} (simplest detection mode).
More robustly, in sequence detection, the {NN} is trained on several symbols.
In this way, the detection of the current symbol leverages observations of past and future received ones, as in \cite{qian2019molecular,farsad2018detection,sun2020ctbrnn,wang2020understanding,shrivastava2021scaled,alshammri2018adaptive}.
The architectures that follow these two approaches are described next.

\begin{figure*}
    \centering
    \subfloat[Cascade connection of feedforward {NNs}.]
    {\includegraphics[width=0.5\columnwidth]{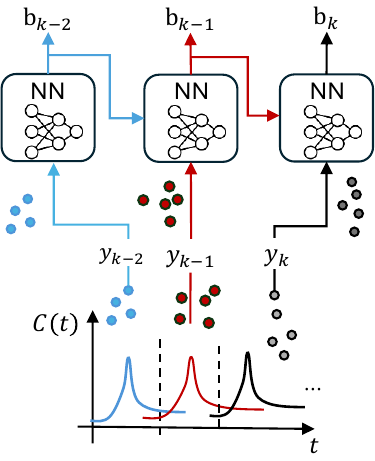}\label{fig_NN_cascade}}
    \subfloat[Cascade connection of \acp{RNN}.]
    {\includegraphics[width=0.5\columnwidth]{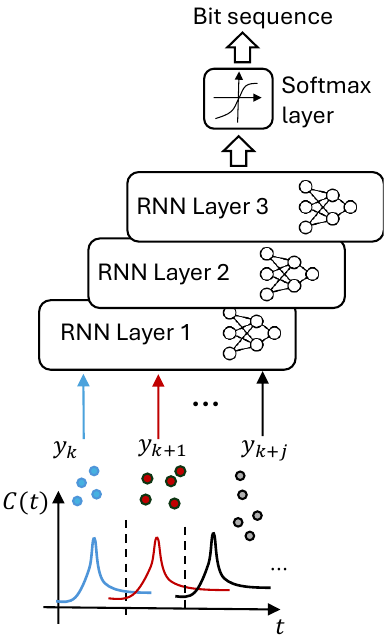}\label{fig_LSTM_detect}}
    \subfloat[Cascade connection of \acp{BiRNN}.]
    {\includegraphics[width=0.5\columnwidth]{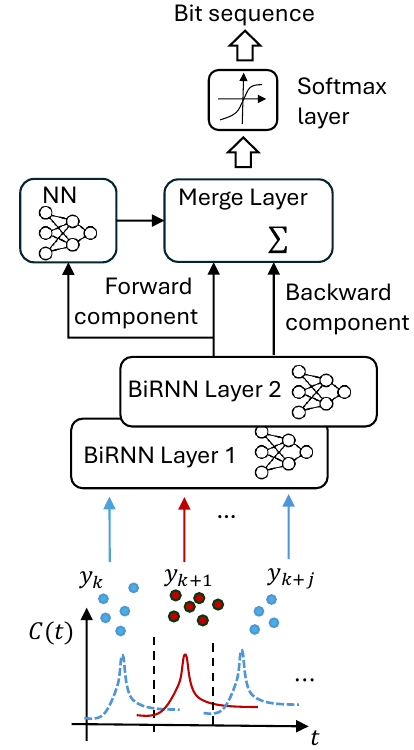}\label{fig_CoBRNN}}
    \subfloat[Sliding window detector.]
    {\includegraphics[width=0.6\columnwidth]{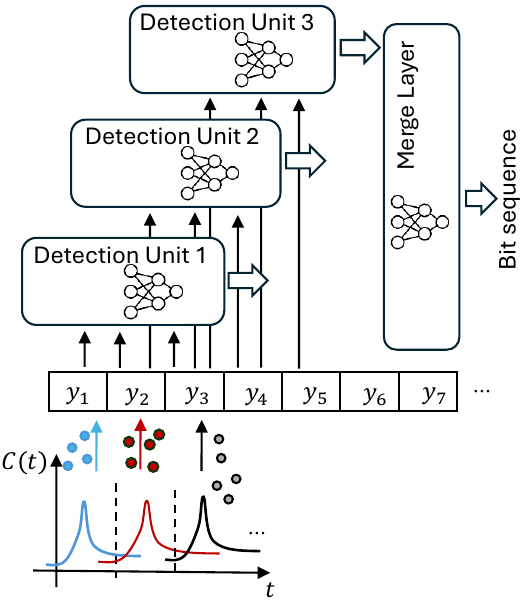}\label{fig_sliding_window}}
    \caption{Feedforward and recurrent {NN} architectures as sequence decoders.
    The \ac{RNN} modules implement \ac{LSTM} cells; see \cite{farsad2017detection,farsad2018neural}.}
    \label{fig_RNN}
\end{figure*}

\textbf{Symbol-by-Symbol Detection Mode:} In this mode, feedforward {NNs} solve the \ac{MAP} formulation stated in the previous section, as described in \cite{farsad2017detection,farsad2018detection,sharma2020deep,shrivastava2021performance,vakilipoor2022hybrid,agrawal2022neural,bai2023temporal,koo2020deep,kim2023machine}.
\acp{RNN} and \acp{CNN} architectures are also reported, as in \cite{vakilipoor2022hybrid} and \cite{kara2022molecular}, respectively.
Besides, symbol-by-symbol detectors can be trained more efficiently by using distinctive features of the input sequence.
Features are evaluated based on the concentration difference of received molecules~\cite{farsad2017detection,farsad2018neural,shrivastava2021performance,vakilipoor2022hybrid,agrawal2022neural} or by fitting a polynomial to the received concentration trace and using its coefficients as feature inputs, see \cite{vakilipoor2022hybrid}.
Other examples extract features from the received sequence using \ac{CNN} architectures comprising one \cite{farsad2017detection,vakilipoor2022hybrid,bai2023temporal} and three \cite{bartunik2022using,bartunik2023artificial,kosanetzki2025demodulation} hidden layers.

\textbf{Sequence Detection Mode:} In this mode, the reported {NN} architectures are more diverse; not only feedforward {NNs} models are used~\cite{qian2019molecular,qian2018receiver,shrivastava2021scaled,alshammri2018adaptive}, but also \acp{CNN} \cite{baydas2023estimation}, \ac{RNN}~\cite{farsad2017detection,baydas2023estimation,wang2020understanding}, and~\acp{BiRNN} \cite{farsad2018neural} encompassing \ac{LSTM} cells.
A cascade connection of the feedforward {NN} architecture is reported in \cite{qian2019molecular} operating as a threshold decoder.
This solution aims to determine the optimal threshold ($\lambda_i$) that minimizes the \ac{BER} metric, as stated in the section above.
As depicted in \Cref{fig_NN_cascade}, decoding the current bit ($b_k$) accounts for the previously decoded ones ($b_{k-1}$, $b_{k-2}$), allowing the {NN} to inherently learn the \ac{ISI} effect in the {MC} channel.

The long-term dependencies in the input sequence are more efficiently learned using \ac{LSTM} cells within \ac{RNN} architectures, as described in \cite{farsad2017detection,farsad2018neural}.
The schematic of these decoders is depicted in \Cref{fig_LSTM_detect}, where three layers are connected in cascade, each layer is of length \num{40} and trained with \num{120}-bit sequences; see \cite[Fig. 1 b)]{farsad2017detection}.
Besides, with \ac{LSTM} networks, bidirectional architectures can be more effectively deployed to exploit the correlation of past and future samples for decoding; see \cite[Fig. 1 c)]{farsad2017detection}.
Past emissions cause \ac{ISI}; therefore, knowledge of the previously transmitted bits can be used to cancel their effect on decoding the current bit.
Meanwhile, as the current emission leaks into the next, the future samples also carry information that helps decode the current one.
Exploiting input data of past and future samples, the implemented \ac{BiRNN} model in \cite{farsad2018detection,farsad2017detection} reduces the impact of \ac{ISI} and is reported as being more computationally efficient than the Viterbi decoder for long memory channels.

The above \ac{BiRNN} architecture is enhanced by adding a learning mechanism that merges the forward and backward pathways, as reported in \cite[Fig. 1b]{sun2020ctbrnn}.
As depicted in \Cref{fig_CoBRNN}, a feedforward {NN} adjusts the coefficients of the weighted sum for the merging layer.
The model is trained using the {Adam} optimization algorithm (see \cite{kingma2014adam}) and achieves a lower \ac{BER} (one order of magnitude lower) than detectors using feedforward {NNs} or \ac{CNN} architectures.

The above architectures can also be integrated into the sliding window architecture to further improve performance~\cite{farsad2018neural}; see the schematic in \Cref{fig_sliding_window}. 
Three detection units decode overlapping symbols within a sliding window, aiming to detect larger correlation lags in the input sequence.
For instance, in deciding the bit sequence within the symbol $y_3$ in \Cref{fig_sliding_window}, not only the neighbor samples $y_2$ and $y_4$ are processed (within the detection unit~\num{2}), but also $y_1$ and $y_5$ (within the detection units \num{1} and \num{3}, respectively).
In this way, the Merge Layer block is fed with a more extensive sequence when compared to the same block in \Cref{fig_CoBRNN}.
In this solution, the detection units are implemented using \acp{BiRNN} and the Merge Layer block evaluates the average of the \acp{RNN} outputs; see \cite[Eq. (10)]{farsad2018neural}.

\textbf{Attention Models in the Loop:} Attention models are increasingly referenced for improving {NN} architectures in {MC} channels; see recent examples in \cite{chen2021selfattention,cheng2023signal,cheng2024informer,cheng2024signal,lu2023mcformer}.
As their main feature, these models estimate the relevance of data within the input sequence, which improves subsequent decoding.
Although previous architectures, such as \ac{BiRNN}, inherently include this feature, they are typically limited to short-range dependencies, whereas attention models can span longer ranges within the input sequence; see~\cite{bahdanau2014neural}.
Seeking to reduce complexity, the encoder component of the transformer is implemented with a single self-attention unit, and the decoder component comprises only the {NN} unit, as described in \cite[Fig. 2]{cheng2024informer,cheng2024signal}.
Additionally, the authors in \cite{lu2023mcformer} further investigate reducing the transformer's hyperparameters (such as the encoding vector length and input size) while maintaining standard operation.
As an alternative, the encoder component of the transformer can be replaced with a three-layer \ac{CNN} module connected to the decoder component's input.
Moreover, to facilitate a more robust architecture, the sliding window scheme in \Cref{fig_sliding_window} has been extended by implementing the detection units using a transformer, as in \cite{cheng2024informer}.
Next, to illustrate the operation of {NN}s in more detail, we present a short code example that highlights how the \ac{BiRNN} architecture is designed, trained, and tested in a practical {MC} setting.

%

\subsubsection{Illustrative Code Example for Symbol Detection}
\label{sec_code_decoder}

\begin{figure*}
    \centering
    \subfloat[Experimental testbed; adapted from~\cite{debus2024synchronized}.] 
    {\includegraphics[width=0.3\linewidth]{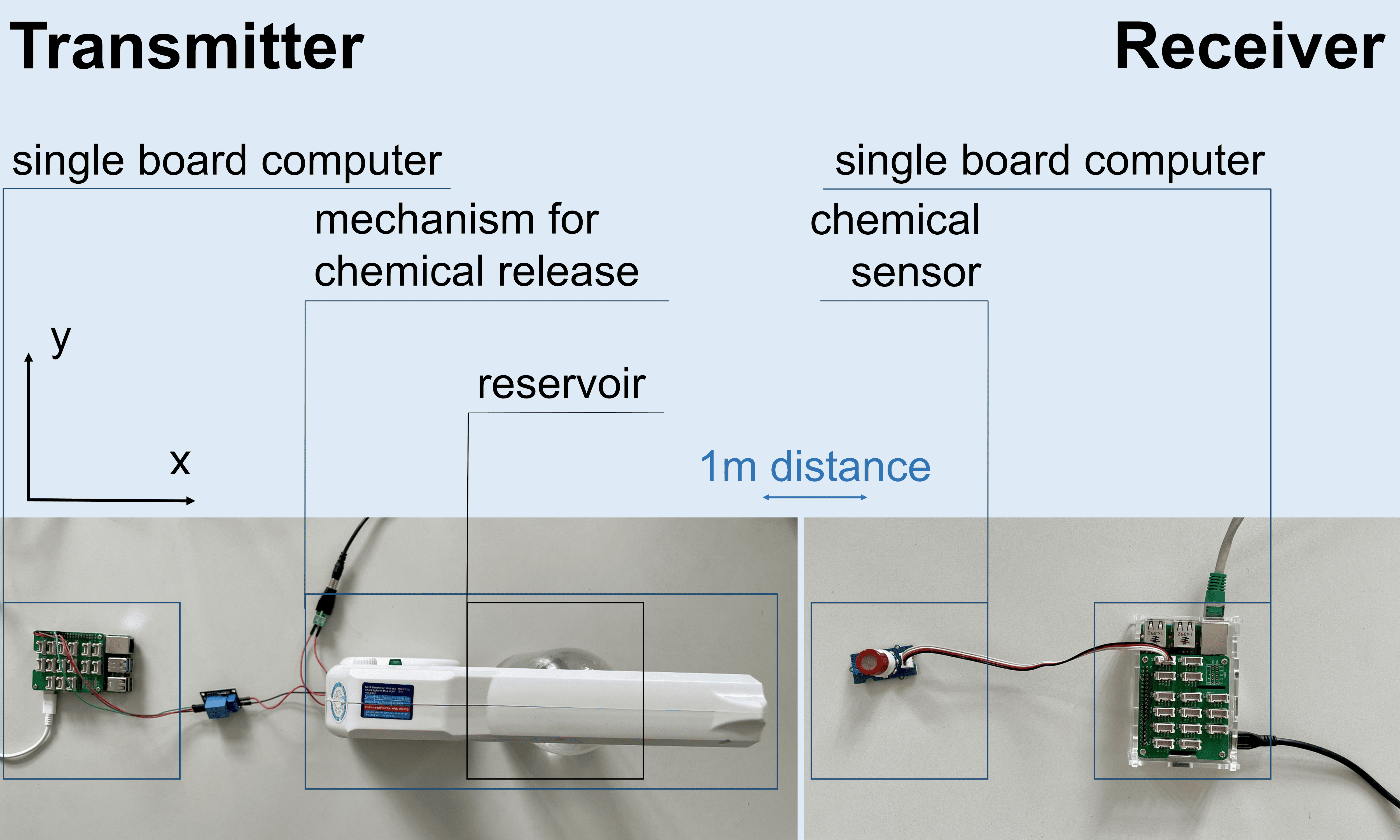}\label{fig_testbed_sprayer}}
    \subfloat[Training loss.]{\includegraphics[width=0.34\linewidth]{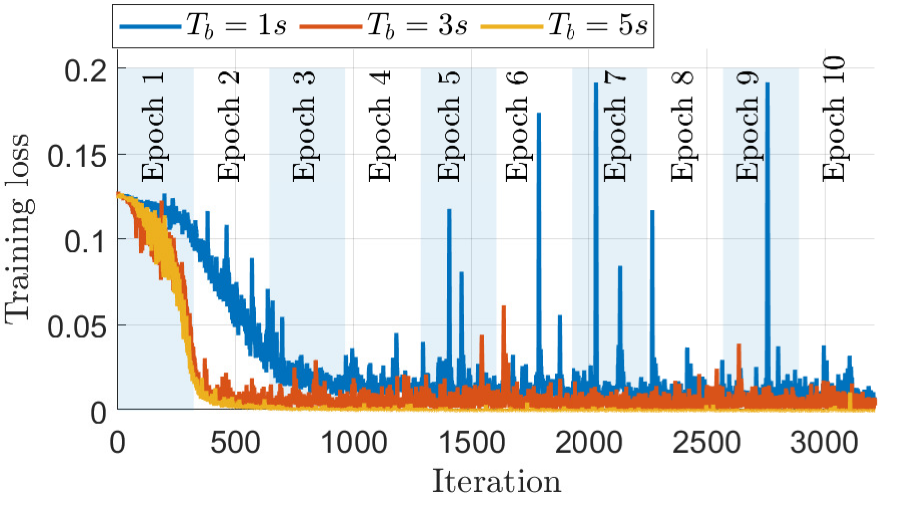}\label{fig_testbed_training}}
    \subfloat[Resulting \ac{BER}.]{\includegraphics[width=0.34\linewidth]{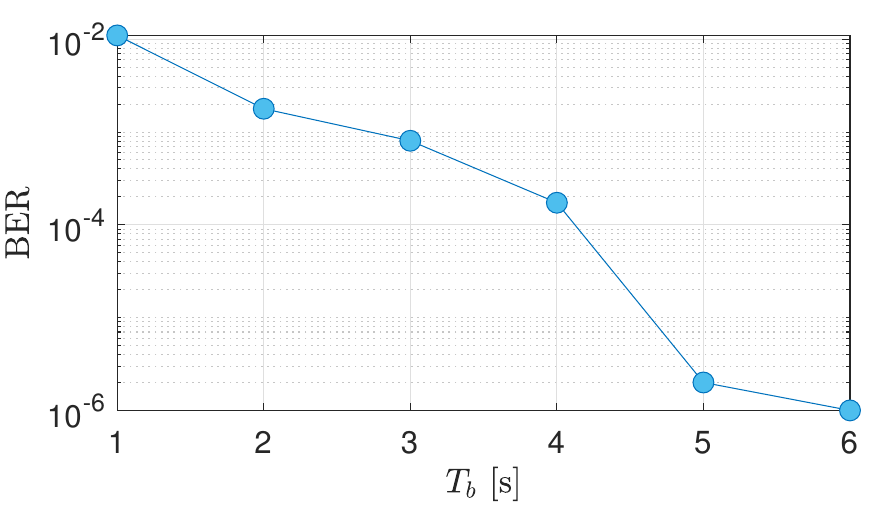}\label{fig_testbed_Rx_BER}}
    \caption{Experimental testbed used for {MC} and measurements, as developed in \cite{hofmann2022testbed}.}
    \label{fig_exp_testbed}
\end{figure*}

This section demonstrates the performance of the sliding \ac{BiRNN} architecture using samples from the experimental testbed illustrated in \Cref{fig_testbed_sprayer}.
The testbed sets a communication link over the distance of $\SI{1}{\meter}$ between the emitter sprayer and the receiver sensor.
Using ethanol molecules as carriers, \ac{OOK} emissions are performed with binary pulses of $T_b=\SI{4}{\second}$ duration each; see further details in \cite[Sec. II]{hofmann2022testbed}.
We use the recorded received pulse as the expected sequence for the emission of ones, and we model the randomness of the received sequence using the Poisson channel model; see \cite[Eq. (87)] {jamali2019channel}.
Besides, we manually (in the code) added synthetic noise molecules of concentration $\SI{10}{\milli\gram\per\liter}$, yielding an $\mathrm{SNR}$ of $\SI{27.5}{\decibel}$.
At the receiver, the observed pulse sequence is sampled with the synchronization signal, which is assumed to be known.

Due to the channel memory, consecutive emissions cause \ac{ISI}, hindering the use of low-complexity schemes, such as the threshold decoder.
For the decoder, we train the sliding \ac{BiRNN} scheme reported in~\cite{farsad2018sliding,farsad2018neural} as a sequence detector.
With the measured pulse duration from the experimental testbed ($\SI{28.6}{\second}$ in total), we evaluate the channel memory as $L=\lfloor\frac{\SI{28.6}{\second}}{T_b}\rfloor=7$, which also defines the number of hidden states of the \ac{BiRNN} model.
For training, we use the {Adam} algorithm \cite{kingma2014adam} with a learning rate of \num{e-3}, \num{10} epochs, a batch size of~\num{10} units, and an emission of \num{e5} bits.

The selected hyperparameters ensure stable learning, with the training loss decreasing over epochs and the bit duration, as illustrated in Fig. 7 b).
Specifically, we observe stability after the second epoch, and losses decrease as the bit duration increases.
The resulting \ac{BER} falls within the range \numrange{e-2}{2e-6}, where more errors occur for lower bit durations, i.e., $T_b<\SI{2}{\second}$.
Besides, as the peak of the received pulses lasts for around~$\SI{4}{\second}$, the resulting \ac{BER} dramatically decreases when transmissions are performed with the bit duration of $\SI{5}{\second}$ or more, i.e., when \ac{ISI} becomes negligible.

%

\hspace{2em}
\subsubsection{Concluding Remarks}

{NN} models have proven to successfully cope with the challenging nature of {MC} channels as detectors, achieving \acp{BER} of less than \num{e-4} in testbed environments.
The variety of environments and {NN} architectures is rich in the literature: Feedforward, recurrent neural networks, and attention models have been researched to decode sequences.
These architectures have been tested in diffusion and drift channels, using molecules and bacteria as information carriers.
However, upon reviewing the literature, little is found regarding realistic environments for precision medicine applications; the considered {MC} channels have mostly simplified geometries, such as free-diffusion point-to-point links.
In future work, diffusion models can be extended to target cell-to-cell communications, including extracellular and intracellular channel models.
Mobility models can be enhanced to replicate the complex communication environment in human vessels, presenting valuable opportunities for further {NN}-model research.
Moreover, integrating these decoders with {MC} channel estimators and noise suppressors, as in~\cite{xiang2025hybrid}, will lead to improved performance.
Future studies should aim to balance the added architectural complexity caused by this integration with its impact on performance.

%

\subsection{Autoencoders}
\label{sec_autoencoders}

End-to-end learning using \ac{AEC} is an innovative approach in {MC}, which, on the one hand, mitigates the challenges associated with {MC} channel modeling, and on the other hand, optimizes the entire {MC} system, including both the transmitter and receiver~\cite{mohamed2019modelbased,khanzadeh2023towards}.
Traditional {MC} systems typically divide the transceiver chain into distinct blocks such as modulation, channel estimation, synchronization, equalization, and demodulation.
However, greater benefits can be achieved from a design that integrates the entire communication system (end-to-end), and \ac{AEC} serves as a prime architecture to optimize the entire process \cite{aoudia2019modelfree}.

An \ac{AEC} consists of two {NN}s, known as the encoder and decoder, which are trained together in an unsupervised fashion; see a representation in \Cref{fig_AEC_architecture}.
The encoder and decoder replace the transmitter and receiver components of the communication system, and the entire communication pathway is optimized end-to-end \cite[Sec. 1.3.2 pp. 11]{eldar2022machine}.
This section summarizes the reported literature on \acp{AEC} for {MC}, defines the end-to-end learning problem in {MC}, describes the deployed {NN} architectures, and finally, provides an illustrative code example.

%

\subsubsection{Problem Definition for Autoencoders}

\begin{figure}
    \centering
    \includegraphics[width=\columnwidth]{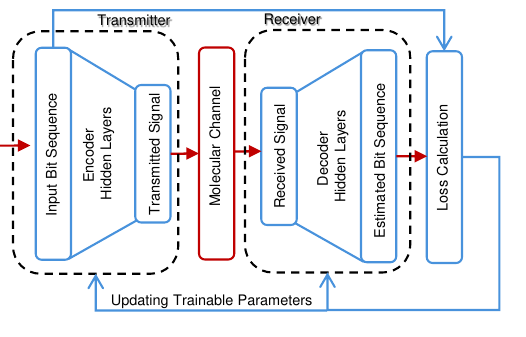}
    \caption{AEC architecture for end-to-end learning of {MC}.}
    \label{fig_AEC_architecture}
\end{figure}

The \ac{AEC} architecture comprises \acp{DNN} for the encoder and decoder, with separate trainable parameters.
The problem is to find these trainable parameters so that the information transmitted by the emitter is restored at the receiver.
Within the {MC} environment, the encoder generates an optimal number of emitted molecules (referred to as symbol generation or constellation design), ensuring that it can be accurately detected at the receiver side with minimal error, despite the stochastic nature and variations of the molecular channel.
Therefore, the ultimate goal is to train the \ac{AEC} to minimize a desired loss function; see the corresponding block (rightmost) in \Cref{fig_AEC_architecture}. 

As for the loss function, an \ac{AEC} adopts the cross-entropy as an effective metric for the end-to-end learning of systems.
This metric measures the difference between the predicted and the actual probability distributions of the transmitted information.
For symbol-based transmission, where data is encoded as individual symbols from a predefined alphabet, categorical cross-entropy is appropriate~\cite{mohamed2019modelbased}.
In contrast, binary cross-entropy is more effective for block-based transmission, which processes data in the form of larger binary sequences, as reported in ~\cite{khanzadeh2023end}.

%

\subsubsection{Environments for Autoencoders}

Advection-diffusion channels are the main type of {MC} channels considered in the literature for studying \acp{AEC}.
In this scenario, a trained point transmitter encodes information bits into molecular concentrations and controls their release accordingly.
The molecules are detected by a passive receiver, which decodes the signal and extracts the information using its trained {NN}, as reported in~\cite{khanzadeh2023towards,khanzadeh2023end,mohamed2019modelbased,angerbauer2023towards}.

\begin{figure*}
    \centering
    \subfloat[Experimental testbed.]
    {\includegraphics[width=0.24\linewidth]{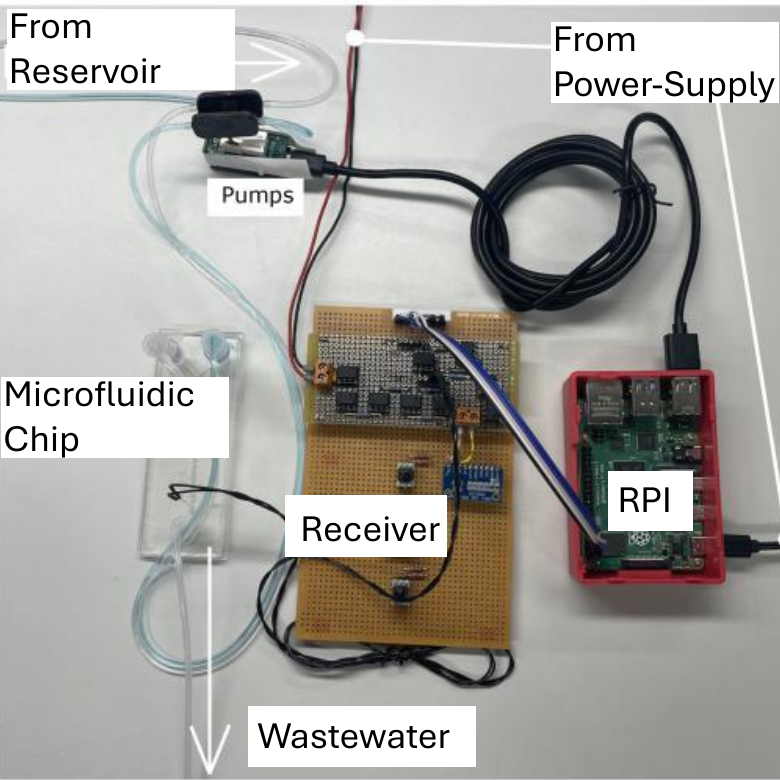}\label{fig_salinity_testbed}}
    \subfloat[Training loss for the \ac{AEC}.]
    {\includegraphics[width=0.37\linewidth]{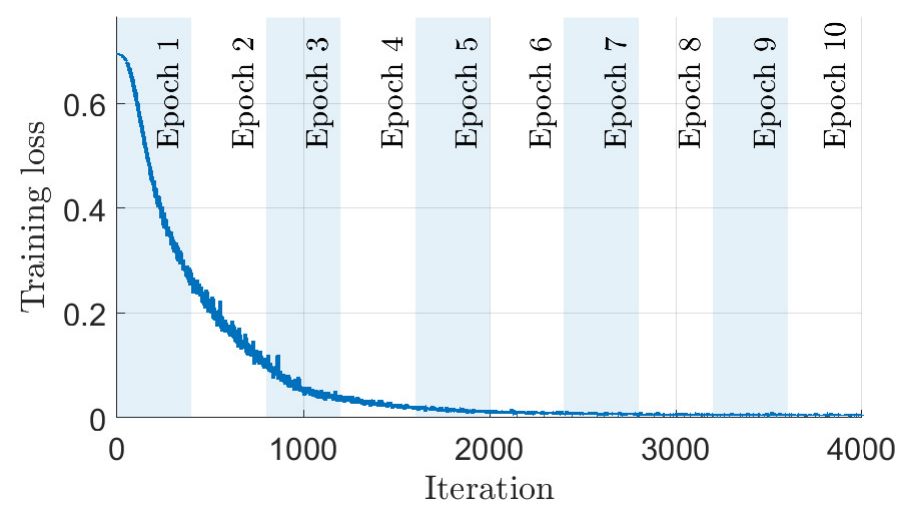}\label{fig_AEC Loss}}
    \subfloat[Resulting \ac{BER} for the \ac{AEC} when symbol duration is $\SI{300}{\milli\second}$.]
    {\includegraphics[width=0.37\linewidth]{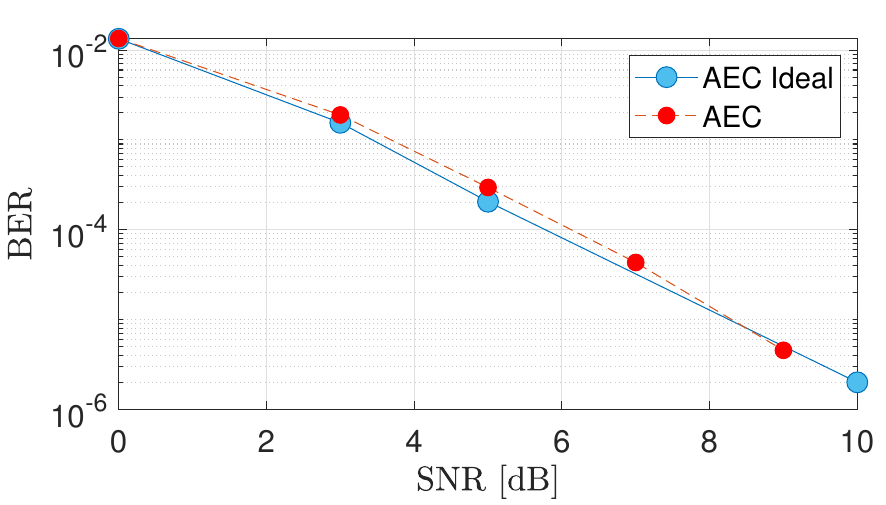}\label{fig_AEC BER}}
    \caption{Experimental testbed developed in \cite{angerbauer2023salinitybased} and the results for the trained \ac{AEC} in \cite{khanzadeh2023end}.}
    \label{fig_autoencoder_1}
\end{figure*}

%

\subsubsection{NN Models for Autoencoders}
\label{subsec_AECdeployment}

The \ac{AEC} model must first undergo training before being deployed in a practical {MC} system. The literature presents several approaches for training \acp{AEC} to facilitate end-to-end learning in MC environments, including the following.

\textbf{Model-Assumed Training:} Model-assumed training is an approach in which both the transmitter and receiver are trained using a predefined channel model.
In {MC}, this involves using the estimated {CIR} from the testbed, implemented as a fixed, non-trainable convolutional layer between the encoder and the decoder of the \ac{AEC}.
An example is given in~\cite{khanzadeh2023towards}, where the {CIR} is obtained from an experimental salinity-based testbed as reported in~\cite{angerbauer2023salinitybased}.
This approach assumes that if the model from the testbed remains stable during runtime, with minimal variations, the transmitter will identify effective strategies for encoding input during training.
Then, in the online phase, when actual transmission occurs in the experimental system, the receiver can be retrained to adapt to the slightly changed channel conditions.
At the same time, the transmitter will maintain its previously established strategies.
Thus, if the assumed model in the training phase is sufficiently accurate, the trained \ac{AEC} will also work appropriately for transmission over the experimental channel.

The solution reported in~\cite{khanzadeh2023towards} uses two \acp{CNN} as its architecture, each serving as an encoder and a decoder.
The encoder consists of three convolutional layers, with \num{16}, \num{32}, and \num{1} filters, respectively, each using a kernel size of \num{3} elements.
Each convolutional layer is followed by batch normalization and an \ac{ReLU} activation function.
A final normalization layer ensures a controlled number of transmitted molecules. 
The decoder begins with a convolutional layer featuring \num{16} filters and a kernel size proportional to the channel memory, effectively mitigating \ac{ISI}.
This is followed by an adaptive average pooling layer, a fully connected linear layer, and another convolutional layer, all of which, except the last layer, incorporate batch normalization and \ac{ReLU} activation.
The final layer employs a sigmoid activation function.
With a differentiable channel model, the \ac{AEC} is trained using backpropagation.

\textbf{Training via Data-Driven Channel Identification:} The \ac{AEC} can also be trained on a data-driven molecular channel representation, as proposed in~\cite{khanzadeh2023end}.
The training procedure involves three steps: (i) Modeling the {MC} channel using an \ac{RNN}, (ii) training the emitter and receiver components of the AEC, and (iii) fine-tuning the trained model.
A data-driven {ML} method is used in the first step to obtain a differentiable representation of molecular channels.
Specifically, an \ac{ARX} model is employed as the channel representation, implemented as an \ac{IIR} filter using a trainable \ac{RNN}.
This formulation preserves differentiability for end-to-end optimization while more effectively capturing long-term temporal dependencies in molecular propagation than feedforward models.

In the second step, both the encoder and decoder parts of the \ac{AEC} are jointly trained using backpropagation, while the \ac{RNN} representing the channel remains fixed.
Lastly, the \ac{AEC} undergoes fine-tuning to address mismatches between the approximation of the channel model and the real model.
The decoder parameters are adjusted using transmissions over the real channel, while the encoder parameters remain unchanged.

\textbf{Model-Free Training:} In this mode, the \ac{AEC} is trained without relying on any model (analytic or data-driven).
To illustrate the methodology, we consider the \ac{DRL} system with fully connected {NNs}, as proposed in \cite{mohamed2019modelbased}, where information is transmitted through a stochastic, unknown channel.
During each training iteration, the decoder is optimized while keeping the encoder parameters fixed, followed by optimizing the encoder while keeping the decoder parameters fixed.
This iterative process enhances the overall system performance.
Updating the decoder parameters is straightforward as it is a supervised task, and does not require backpropagation through the channel.
However, updating the encoder relies on backpropagation through the channel, which requires a channel model to be available.
To circumvent this challenge, the authors in~\cite{mohamed2019modelbased} employ an \ac{RL} model, treating the transmitter as an agent that receives the loss calculated at the receiver via an ideal separate feedback channel. Additionally,~\cite{khanzadeh2025end} incorporates \ac{FiLM} layers into the \ac{AEC} architecture to enable online adaptation during deployment, allowing the model to adjust to channel conditions that differ from those observed during training without retraining.

\textbf{Transmitter-Exclusive Training:} Transmitter-exclusive approaches perform the training on the transmitter side only. 
This training mode is well-suited for the {IoBNT} scenario, where the transmitter is usually easily accessible, as it is located outside the body, and the receiver is of low complexity and located inside, as specified in~\cite {angerbauer2023towards, khanzadeh2024explainable}.
The proposed \ac{AAEC} in~\cite{angerbauer2023towards, khanzadeh2024explainable} employs a \ac{CNN} encoder for binary transmission with adjustable concentration levels and a threshold-based decoder realized as a single convolutional layer.
This approach seeks to mitigate the impact of residual molecules from prior transmissions by encoding symbols to counteract lingering interference, thereby reducing \ac{ISI}.

Moreover, a \ac{DNN}-based approach for optimizing the number of molecules released by each transmitter in a mobile molecular \ac{MIMO} system is presented in~\cite{cheng2023channel}.
While \ac{AEC} is not utilized in the proposed structure, this \ac{DNN}-based method can be integrated into transmitter-exclusive training for {MC}.
The optimization is performed based on different transmitter-receiver distances to minimize the \ac{ILI}, thereby reducing the \ac{BER}.
The \ac{DNN} is trained with constraints on the molecule release count to ensure compliance with lower and upper bounds.
The results demonstrate that the \ac{DNN}-based approach outperforms the \ac{GA}-based approach, achieving a lower average BER and significantly reduced computation time.
Additionally, the \ac{DNN}-based optimization achieves a \ac{BER} comparable to that of an exhaustive search while significantly reducing computational time.

Adaptive modulation and coding are other critical tasks to dynamically adjust transmission parameters in response to changing channel conditions.
Adaptive modulation is essential for maintaining reliable and efficient communication in {MC}, where factors such as diffusion and advection strongly influence the channel.
In this context, an \ac{RL}-module has been proposed to optimize real-time transmission parameters, such as the modulation order and symbol duration, \cite{khanzadeh2024ql-based}.
A gateway device connected to the \ac{RL} model estimates channel conditions from heart rate data by leveraging a digital twin of the human circulatory system.
This solution improves the achievable raw bit rate and error performance.
Additionally, to address the limited capabilities at the nanoscale, this architecture alleviates the computational burden on resource-constrained nanodevices by offloading complexity to an external gateway.
To complement the above review, we next provide a brief illustrative example, adapted from \cite{khanzadeh2023end}, that demonstrates how this architecture is designed, trained, and evaluated in practice.

%

\subsubsection{Illustrative Example for Autoencoders}
This section complements the above architectures by illustrating the operation of the \ac{AEC} from~\cite{khanzadeh2023end} in the salinity-based testbed developed in~\cite{angerbauer2023salinitybased}.
In this testbed, information bits are mapped to salinity levels and detected based on the corresponding conductivity in water, for which a closed-form expression cannot be derived due to the lack of an accurate model of the transmitter and receiver geometries.
\Cref{fig_salinity_testbed} depicts the setup for a $\SI{3.6}{\centi\meter}$ transmitter–receiver distance.
Real microfluidic channel measurements are used to identify channels and train an \ac{RNN}.
The \ac{AEC} employs three convolutional layers with batch normalization and ReLU activation at the encoder, while the decoder uses a convolutional layer acting as a linear equalizer to mitigate \ac{ISI} and noise.
Two pooling layers (average and max) and a fully connected layer downsample features to match the transmitted bit sequence, followed by a sigmoid output for soft-input decoding.
Training is simulation-based over \num{2000} epochs with a batch size of \num{40} and \num{100} $\si{\bit}$ sequences (\Cref{fig_AEC Loss}).
Each emission comprises encoder-defined binary pulses lasting $\SI{500}{\milli\second}$.
The Adam optimizer is used with an initial learning rate of \num{0.009} and a decay of \num{0.99} every~\num{1000} iterations.
The architectural parameters, such as the layer type and number, as well as the number of neurons per layer, were empirically tuned based on observed training performance and prior experience, with domain knowledge guiding design choices.
For instance, convolutional layers with kernel sizes matching the channel memory were adopted at the receiver side.
Training hyperparameters, including learning rate, batch size, and optimizer type, were optimized through a grid search to achieve stable convergence and minimize validation loss, as described in \cite{angerbauer2023salinitybased}.
As a result, the obtained \ac{BER} ranges from $\numrange{9e-2}{9e-6}$ when the \ac{SNR} is in the interval~$\SIrange{0}{10}{\decibel}$.
These results attain the ideal case of full {MC} channel knowledge; labeled as '\ac{AEC} Ideal' in \Cref{fig_AEC BER}.

%

\subsubsection{Concluding Remarks}

\Ac{AEC} can inherently enhance the performance of {MC} by jointly optimizing the transmitter and receiver operations.
Besides, this architecture effectively addresses the complexities of dynamic and stochastic {MC} channels.
However, several practical challenges remain unresolved.
For example, real-world molecular channels are time-varying, requiring continuous fine-tuning of {NN} parameters to adapt to unseen conditions.
How to efficiently update the transmitter parameters during deployment without incurring high computational or energy costs remains an open question for future research.
Additionally, deploying resource-intensive NNs on small, biocompatible, and resource-constrained devices poses significant technical hurdles.

%

\subsection{Higher Layers}
\label{sec_resource_alloc}

The works summarized thus far mostly focus on the \ac{PHY} layer and target point-to-point communication links.
Fewer studies address higher layers, such as resource allocation (MAC layer) and localization (application layer).\footnote{Contributions related to higher layers represent $\SI{7}{\percent}$ of the surveyed literature.}
This section provides an overview of the reported research on integrating {NN} architectures into the higher layers of {IoBNT} networks.

\subsubsection{Resource Allocation}

Resource allocation problems arise when multiple users attempt to communicate using the same resource.
Such a scenario is examined in free-diffusion channels in \cite{cheng2024resource}, which investigates a transmission policy that minimizes the \ac{BER}.
The scenario considers an arbitrary number of transmitters placed at random locations in a free-diffusion channel, and restricted to a maximum number of released molecules.

The transmission policy is devised by a feedforward {NN} with three hidden layers, each containing twice as many neurons as there are transmitters, except for the first layer, which has two additional neurons, as indicated in \cite{cheng2024resource}.
The {NN} is trained using the distances from the transmitters to the receiver as inputs. The model outputs the number of molecules released by each transmitter.
As activation functions, the first layer uses the hyperbolic tangent sigmoid, the ``purelin'' (linear) in the second hidden layer, and the \ac{ReLU} for the third hidden layer.
This architecture achieves a \ac{BER} on the order of magnitude \num{e-3} with three to five transmitters positioned between $\SIrange{12.5}{14.5}{\micro\meter}$ from the receiver.
The network operates with a symbol time of $\SI{100}{\milli\second}$, a total of~\num{6e4} released molecules, and a noisy source modeled as a Gaussian distribution with variance \num{1000}.

%

\subsubsection{Localization}

Localizing diseases and self-localizing nanosensors are among the most anticipated applications of \ac{IoBNT} networks in precision medicine.
Scenarios have been developed within the dynamic environment of blood vessels as in~\cite{torres-gomez2022nanosensor2,torres-gomez2024dna-based,torres-gomez2021machine,galvan2024tailoring,jin2024transformerbased,calvobartra2024graph,hube2025set} and in the less restrictive case of free-diffusion environments as in~\cite{kose2020machine,cheng2025localizing}.

In blood vessels, prior work assumes that blood flow passively drives existing nanosensors.
In the \ac{HCS} environment, the self-localization capabilities of nanosensors are particularly challenging to develop due to the absence of a reference system.
A reference system for self-localizing nanosensors within the bloodstream can be anchored to the concentrations of neighboring nanosensors and their travel times, as proposed in~\cite{torres-gomez2022nanosensor2}.
This work assumes that nanosensors contain an internal counter and an external device that resets the counter when it travels through the heart.
Then, as the nanosensor's traveling time increases as it flows through the blood vessels, the recorded traveling time per nanosensor can be used to distinguish shorter (central body) from longer (lower body) paths, thereby self-distinguishing the body region.
Additionally, as the concentration of nanosensors depends on the specific vessel path (see the Markov model formulation in~\cite{torres-gomez2021markov,torres-gomez2021machine}), it can serve as a second metric for self-localization. 
As such, a feedforward {NN} can be trained to distinguish the traveling path of nanosensors based on these two metrics, i.e., traveling time and concentration level, as proposed in~\cite{torres-gomez2024dna-based}.
The {NN} is trained on data generated by the \ac{BVS} simulator \cite{geyer2018bloodvoyagers}, achieving approximately $\SI{85}{\percent}$ positive predictions.
Localization methods also rely on establishing a communication link among nanonodes, as described in \cite{jin2024transformerbased}.
This research assumes transmitter nodes with fixed spatial positions within the human cardiovascular system and a receiver node, which is intended to be localized.
Using the transformer architecture, the prediction accuracy for receiver coordinates ranges from $\SIrange{78}{96}{\percent}$~\cite{jin2024transformerbased}, depending on blood flow velocity and the transformer's model capacity, which is determined by its computational complexity.

A localization method using \acp{GNN} is also reported in~\cite{calvobartra2024graph}, where the \ac{HCS} is modeled as a graph using the \ac{BVS} representation with \num{94} regions.
The proposed model relies on \acp{HGT} to propagate information between region and anchor nodes, using a configuration with \num{64} hidden channels, \num{8} attention heads, \num{3} \ac{HGT} layers, and \num{4} convolutional layers. Compared with the feedforward \ac{NN} baseline in~\cite{torres-gomez2022nanosensor2,lopez2025toward}, the \ac{GNN} improves point localization accuracy by more than $\SI{30}{\percent}$ in the reported~$\SI{18}{\min}$ runtime scenario and extends localization coverage to the entire bloodstream.


Localization of nanonodes is also performed in short-range free-diffusion {MC} links, as developed in~\cite{kose2020machine} for security applications and in \cite{cheng2025localizing} for tracking a mobile emitter.
The feedforward {NN} model in~\cite{kose2020machine} is developed for a 2D free-diffusion environment and localizes an eavesdropper node with an error of less than~$\SI{4}{\micro\meter}$.
The solution in \cite{cheng2025localizing} develops a transformer architecture that tracks the position of a mobile emitter with precision exceeding \SI{80}{\percent}.

%

\subsubsection{Data Fusion}
\label{sec_fusion}

Reported research targets data fusion techniques in tasks related to detecting potential abnormalities \cite{torres-gomez2024dna-based,solak2020neural,solak2020rnn,mai2017event}. 
Applications include health-condition monitoring, environmental sensing, and the detection of toxic agents.
Commonly implemented fusion rules are the \texttt{OR}, \texttt{AND}, or \texttt{MAJORITY} rules; see for instance~\cite[Sec. IV B]{torres-gomez2024dna-based}.
However, these rules are complex to apply in practice, as they require evaluating channel-dependent parameters such as noise and interference levels, which are typically unknown.
Overcoming this impediment, {NNs} are trained and deployed to effectively fuse data from various sensor nodes, utilizing feedforward {NNs} in \cite{solak2020neural} and \ac{RNN} in~\cite{solak2020rnn}.
These two architectures are developed for free-diffusion channels, achieving a detection probability greater than \num{0.9} and a false alarm probability of \num{5e-2}.
Furthermore, the robustness of data fusion techniques is investigated in \cite{mai2017event}, where an \ac{RL} agent is trained to optimize the observation window for collecting samples at the fusion node.
The \ac{RL} agent adjusts the time slot duration to ensure a detection probability within a confidence interval and to minimize noise levels.
This work reports the fusion of binary emissions from~\num{100} nodes in free-diffusion channels, utilizing the \texttt{OR} rule at the fusion node.

%

\subsection{Concluding Remarks and Outlook}

\begin{figure}
    \centering
    \includegraphics[width=\columnwidth]{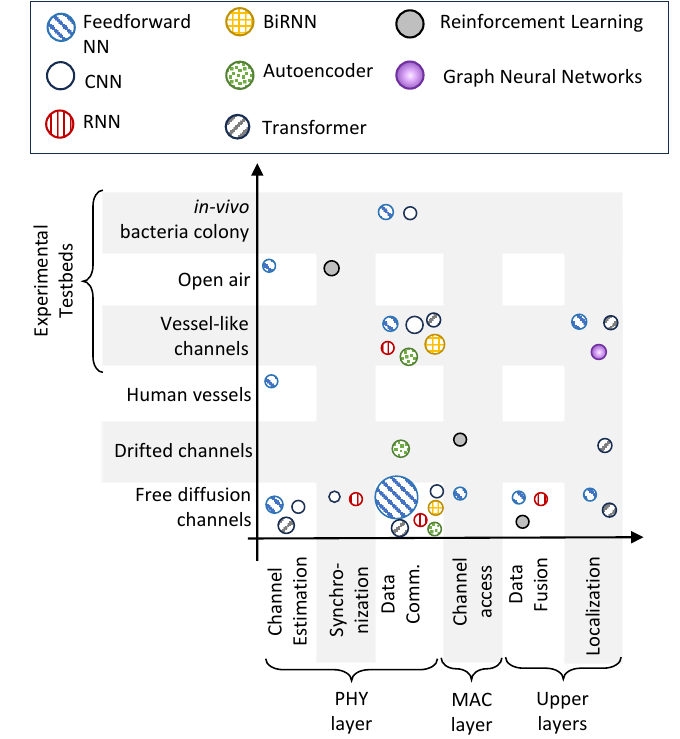}
    \caption{Bubble plot of reported {NN} architectures.}
    \label{fig_NN_architecture}
\end{figure}

Within the reported literature, the deployment of {NNs} follows two common practices: (i) The most popular models are feedforward {NNs}, and (ii) deployment focuses primarily on decoding tasks in point-to-point links.
We identify these two characteristics from the comparative summary provided in \Cref{fig_NN_architecture} and the consolidated overview in \Cref{tab_detection_parameters}.
In this figure, the size of each bubble represents the number of published works, arranged by communication layer (horizontal axis) and by {MC} scenario (vertical axis).\footnote{As a reference, the size of the bubbles in \Cref{fig_NN_architecture} reflects \num{10} publications for the largest bubble, and one publication for the smaller ones.
The generation of this plot is accessible in \url{https://github.com/tkn-tub/NN_molecular_communications/tree/main/Tables}.}
\Cref{tab_detection_parameters} reports key \ac{MC}-related parameters that describe per {MC} geometry and {NN} architecture.
This tabular overview also highlights the diversity of experimental and simulation setups used in the surveyed studies.

From \Cref{fig_NN_architecture} and \Cref{tab_detection_parameters}, we also identify several research gaps that deserve further investigation:
(i) the lack of \ac{NN}-based solutions for multipoint-to-multipoint links;
(ii) the limited attention given to upper-layer functionalities; and
(iii) the need for additional experimental validation to support the integration of \ac{NN}-based schemes.\footnote{Several physical parameters used to model {MC} channels, such as receiver radius, diffusion coefficients, and communication ranges, are often selected for analytical or simulation convenience rather than derived from specific experimental technologies. Establishing experimentally grounded parameter values remains an important direction for future developments in the field.}
Furthermore, most reported \ac{NN} models are trained offline and used solely for inference, while online and adaptive training strategies remain largely unexplored.

Another key aspect to consider is the feasibility of integrating the reported \ac{NN}-based solutions into nanodevices, which remains open for further analysis.
As nanodevices are expected to operate with limited resources, it is crucial to examine the trade-off between performance and complexity across tasks, including channel estimation, synchronization, detection, and upper-layer applications such as data fusion and localization.
Following the literature, the number of parameters reported across methods ranges from below \num{10} to several thousand, see \Cref{tab_NN_complexity}.
The least demanding architectures are feedforward and recurrent {NN}s and are used primarily for tasks related to data communication.
The most complex architectures are reported using {CNN}s and reinforcement learning methods employed for synchronization and data communication.

\newlength{\MCgeomW}
\newlength{\NNarchW}
\newlength{\AppW}
\newlength{\CommRangeW}
\newlength{\RxRadiusW}
\newlength{\DiffCoeffW}
\newlength{\ReleasedMolW}
\newlength{\SymbolDurW}
\newlength{\SnrW}
\newlength{\PerfW}
\newlength{\RefW}

\setlength{\MCgeomW}{1cm}
\setlength{\NNarchW}{1.3cm}
\setlength{\AppW}{1.6cm}
\setlength{\CommRangeW}{1.5cm}
\setlength{\RxRadiusW}{1cm}
\setlength{\DiffCoeffW}{1cm}
\setlength{\ReleasedMolW}{1cm}
\setlength{\SymbolDurW}{0.7cm}
\setlength{\SnrW}{1cm}
\setlength{\PerfW}{0.7cm}
\setlength{\RefW}{0.5cm}

\begin{table*}[!t]
\centering
\caption{Summary of {NN} architectures, applications, and parameters related to the {MC} channels and communication.}
        \label{tab_detection_parameters}
\footnotesize
\begin{tabular}{%
 p{\MCgeomW}p{\NNarchW}|p{\AppW}|p{\CommRangeW}|p{\RxRadiusW} p{\DiffCoeffW}|p{\ReleasedMolW}p{\SymbolDurW}p{\SnrW}p{\PerfW}l|p{\RefW}}

      & \multicolumn{1}{r}{} & \multicolumn{1}{l}{} &
\multicolumn{3}{l|}{End-to-end channel-related parameters} &
\multicolumn{5}{l|}{Communication-related parameters} &  \bigstrut[b]\\
\hline
\centering\mycell{\acs{MC}\\Geometry} & \multicolumn{1}{|p{\NNarchW}|}{\mycell{{NN}\\ architecture}} & \multicolumn{1}{p{\AppW}|}{\mycell{\\Application}} & \multicolumn{1}{p{\CommRangeW}}{\mycell{Communication\\ range}} & \multicolumn{1}{p{\RxRadiusW}}{\mycell{Receiver\\radius}} & \multicolumn{1}{p{\DiffCoeffW}|}{{$D$\newline{} [$\si{\nano\meter\squared\per\nano\second}$]}} & \multicolumn{1}{p{\ReleasedMolW}}{\mycell{Released\\molecules}}
& \multicolumn{1}{p{\SymbolDurW}}{\mycell{Symbol\\duration}}
& \multicolumn{1}{p{\SnrW}}{\mycell{\\\acs{SNR} [$\si{\decibel}$]}} & \multicolumn{1}{p{\PerfW}}{\mycell{\\Performance metrics}} & \multicolumn{1}{l|}{} & {\mycell{\\Ref.}} \bigstrut\\

\hline
\multirow{21}[38]{=}{\rotatebox[origin=c]{90}{\mycell{\\Point transmitter-Free diffusion-Spherical absorbing receiver}}} & \multicolumn{1}{p{\NNarchW}|}{\multirow{12}[24]{*}{\rotatebox[origin=c]{90}{\mycell{\\Feedforward \acs{NN}}}}} & \multicolumn{1}{p{\AppW}|}{\multirow{5}[10]{*}{\mycell{Channel\\ estimation}}} & \multicolumn{1}{p{\CommRangeW}}{\multirow{1.5}[4]{*}{$\SIrange[range-phrase=\text{~to~}]{2}{10}{\micro\meter}$}} & \multicolumn{1}{p{\RxRadiusW}}{\multirow{1.5}[4]{*}{$\SI{4}{\micro\meter}$}} & \multicolumn{1}{p{\DiffCoeffW}|}{\mycell{\num{33e-4},\\ \num{66e-4}}} & \multicolumn{1}{p{\ReleasedMolW}}{\multirow{1.5}[4]{*}{\num{200}}} &       &       & \multicolumn{1}{p{\PerfW}}{\multirow{1.5}[4]{*}{Error}} & \multicolumn{1}{l|}{\multirow{1.5}[4]{*}{$\SI{3.3}{\percent}$}} & \multicolumn{1}{l}{\mycell{Sec.\\ \ref{sec_code_dist}}} \bigstrut\\

\cline{4-12} &       &       & \multicolumn{1}{p{\CommRangeW}}{$\SIrange[range-phrase=\text{~to~}]{2}{11}{\micro\meter}$} & \multicolumn{1}{p{\RxRadiusW}}{ $\SIrange[range-phrase=\text{~to~}]{3}{7}{\micro\meter}$} & \multicolumn{1}{p{\DiffCoeffW}|}{\multirow{3}[6]{*}{\mycell{\num{5e-1},\\ \num{e-1}}}} & \multicolumn{1}{p{\ReleasedMolW}}{\multirow{2}[4]{*}{\num{3e3}}} &       &       & \multicolumn{1}{p{\PerfW}}{\multirow{4}[8]{*}{\acs{RMSE}}} & \multicolumn{1}{l|}{$\leq\,$\num{0.3}} & \multicolumn{1}{l}{\cite{lee2017machine} } \bigstrut\\

\cline{4-5}\cline{8-9}\cline{11-12} &       &       & \multicolumn{1}{p{\CommRangeW}}{$\SIrange[range-phrase=\text{~to~}]{2}{11}{\micro\meter}$} & \multicolumn{1}{p{\RxRadiusW}}{ $\SIrange[range-phrase=\text{~to~}]{4}{10}{\micro\meter}$} &       &       &       &       &       & \multicolumn{1}{l|}{$\leq 1$} & \multicolumn{1}{l}{\cite{yilmaz2017machine} } \bigstrut\\


\cline{4-5}\cline{7-9}\cline{11-12} &       &       & \multicolumn{1}{p{\CommRangeW}}{$\SIrange[range-phrase=\text{~to~}]{4}{12}{\micro\meter}$} & \multicolumn{1}{p{\RxRadiusW}}{ $\SIlist{3;4;5}{\micro\meter}$} &       & \multicolumn{1}{p{\ReleasedMolW}}{\num{e6}} &       &       &       & \multirow{2}[4]{*}{$\leq\,$\num{e-2}} & \multicolumn{1}{l}{\cite{cheng2024channel} } \bigstrut\\

\cline{4-9}\cline{12-12} &       &       & \multicolumn{1}{p{\CommRangeW}}{$\SIrange[range-phrase=\text{~to~}]{8}{10}{\micro\meter}$} & \multicolumn{1}{p{\RxRadiusW}}{ $\SI{4}{\micro\meter}$} & \num{7.9e-1} & \multicolumn{1}{p{\ReleasedMolW}}{\num{3e3}} &       &       &       &  & \multicolumn{1}{l}{\cite{ozbey2024artificial}} \bigstrut\\

\cline{3-12} &       & \multicolumn{1}{p{\AppW}|}{\multirow{4}[8]{*}{\mycell{Data\\communication}}} & \multicolumn{1}{p{\CommRangeW}}{\multirow{2}[4]{*}{$\SI{500}{\nano\meter}$}} & \multicolumn{1}{p{\RxRadiusW}}{\multirow{2}[4]{=}{$\SI{45}{\nano\meter}$}} & \multicolumn{1}{p{\DiffCoeffW}|}{\multirow{2}[4]{*}{\num{4.2e-1}}} &       & \multicolumn{1}{p{\SymbolDurW}}{$\SI{270}{\milli\second}$} & \multicolumn{1}{p{\SnrW}}{\numrange[range-phrase=~to~]{-5}{80}} & \multirow{4}[8]{*}{\acs{BER}} & \multicolumn{1}{l|}{$\geq$ \num{2e-2}} & \multicolumn{1}{l}{\multirow{2}[4]{*}{\cite{qian2018receiver}}} \bigstrut\\

\cline{7-9}\cline{11-11} &       &       & \multicolumn{1}{p{\CommRangeW}}{} &       &       &       & \multicolumn{1}{p{\SymbolDurW}}{$\SIlist{450}{\milli\second}$} & \multicolumn{1}{p{\SnrW}}{\numrange[range-phrase=~to~]{-5}{35}} &       & \multicolumn{1}{l|}{$\geq$ \num{2e-5}} &  \bigstrut\\

\cline{4-9}\cline{11-12} &       &       & \multicolumn{1}{p{\CommRangeW}}{\multirow{1.5}[4]{*}{$\SI{5}{\micro\meter}$}} & \multicolumn{1}{p{\RxRadiusW}}{\multirow{1.5}[4]{*}{$\SI{50}{\nano\meter}$}} & \multirow{1.5}[4]{*}{\num{1.01}} & \multicolumn{1}{p{\ReleasedMolW}}{\mycell{\num{4e4} to\\ \num{4.5e4}}} & \multicolumn{1}{p{\SymbolDurW}}{\multirow{1.5}[4]{*}{$\SI{100}{\milli\second}$}} & \multicolumn{1}{p{\SnrW}}{\multirow{1.5}[4]{*}{\numrange[range-phrase=~to~]{30}{56}}} &       & \multicolumn{1}{l|}{\multirow{1.5}[4]{*}{$\geq$ \num{e-4}}} & \multicolumn{1}{p{\RefW}}{%
\raggedright
\cite{shrivastava2021performance},~\cite{shrivastava2021scaled},\\
\cite{agrawal2022neural}%
} \bigstrut\\

\cline{4-9}\cline{11-12} &       &       & \multicolumn{1}{p{\CommRangeW}}{\multirow{1.5}[4]{*}{$\SI{14}{\micro\meter}$}} & \multicolumn{1}{p{\RxRadiusW}}{\multirow{1.5}[4]{*}{$\SI{1}{\micro\meter}$}} & \multirow{1.2}[4]{*}{\num{6e-2}} & \multicolumn{1}{p{\ReleasedMolW}}{\num{e3}, \num{2e3} } & \multicolumn{1}{p{\SymbolDurW}}{\multirow{1.2}[4]{*}{$\SI{3}{\second}$}} & \multicolumn{1}{p{\SnrW}}{\multirow{1.2}[4]{*}{\numrange[range-phrase=~to~]{2}{20}}} &       & \multicolumn{1}{l|}{\multirow{1.2}[4]{*}{$\geq$ \num{7e-6}}} & \multirow{1.2}[4]{*}{\cite{sharma2020deep}} \bigstrut\\

\cline{3-12} &       & \multicolumn{1}{p{\AppW}|}{Localization} & \multicolumn{1}{p{\CommRangeW}}{$\SI{10}{\micro\meter}$} & \multicolumn{1}{p{\RxRadiusW}}{ $\SI{4}{\micro\meter}$} & \num{7.9e-1} & \multicolumn{1}{p{\ReleasedMolW}}{\num{e4}} & \multicolumn{1}{p{\SymbolDurW}}{$\SIlist{2;5}{\second}$} &       & Error & $<\SI{50}{\percent}$ & \cite{kose2020machine} \bigstrut\\

\cline{3-12} &       & \multicolumn{1}{p{\AppW}|}{\multirow{4}[8]{*}{\mycell{Data\\fusion}}} & \multicolumn{1}{p{\CommRangeW}}{\multirow{2}[4]{*}{$\SIrange[range-phrase=\text{~to~}]{4}{9}{\micro\meter}$}} & \multicolumn{1}{p{\RxRadiusW}}{\multirow{2}[4]{=}{$\SI{4}{\micro\meter}$}} & \multirow{4}[8]{*}{\mycell{\num{5e-1},\\ \num{7.9e-1}}} &       & \multicolumn{1}{p{\SymbolDurW}}{\multirow{2}[4]{*}{$\SI{70}{\milli\second}$}} & \multirow{2}[4]{*}{\numrange[range-phrase=~to~]{2}{5}} & $P_\mathrm{d}$ & \multicolumn{1}{l|}{$>\,$\num{0.94}} & \multirow{4}[4]{*}{\cite{solak2020neural}} \bigstrut\\

\cline{10-11}  &       &       & \multicolumn{1}{p{\CommRangeW}}{} &       &       &       &       &       & $P_\mathrm{fa}$ & \multicolumn{1}{l|}{$\leq\,$\num{7.5e-2}} &  \bigstrut\\

\cline{2-2}\cline{4-5}\cline{7-11}  & \multicolumn{1}{p{\NNarchW}|}{\multirow{2}[4]{*}{RNN}} &       & \multicolumn{1}{p{\CommRangeW}}{\multirow{2}[4]{*}{$\SIrange[range-phrase=\text{~to~}]{4}{9}{\micro\meter}$}} & \multicolumn{1}{p{\RxRadiusW}}{\multirow{2}[4]{=}{$\SI{4}{\micro\meter}$}} &       & \multirow{2}[4]{*}{} & \multicolumn{1}{p{\SymbolDurW}}{\multirow{2}[4]{*}{$\SI{70}{\milli\second}$}} & \multirow{2}[4]{*}{\numrange[range-phrase=~to~]{2}{5}} & $P_\mathrm{d}$ & \multicolumn{1}{l|}{$>\,$\num{0.96}} &  \bigstrut\\

\cline{10-11} &       &       & \multicolumn{1}{p{\CommRangeW}}{} &       &       &       &       &       & $P_\mathrm{fa}$ & \multicolumn{1}{l|}{$\leq\,$\num{7.5e-2}} &  \bigstrut\\

\cline{2-12} & \multicolumn{1}{p{\NNarchW}|}{\multirow{3}[4]{*}{\mycell{BiRNN}}} & \multicolumn{1}{p{\AppW}|}{\mycell{Channel\\estimation}} & \multicolumn{1}{p{\CommRangeW}}{\multirow{1.5}[4]{*}{$\SIrange[range-phrase=\text{~to~}]{4}{12}{\micro\meter}$}} & \multicolumn{1}{p{\RxRadiusW}}{\multirow{1.5}[4]{*}{$\SIlist{3;4;5}{\micro\meter}$}} & \num{5e-1}, \num{e-1} &       &       &       & \multirow{1.5}[4]{*}{\acs{RMSE}} & \multicolumn{1}{l|}{\multirow{1.5}[4]{*}{$\leq\,$\num{e-2}}} & \multicolumn{1}{l}{\multirow{1.5}[4]{*}{\cite{cheng2024channel}}} \bigstrut\\

\cline{3-12}  &       & \multicolumn{1}{p{\AppW}|}{\multirow{3}[6]{*}{\mycell{Data\\communication}}} & \multicolumn{1}{p{\CommRangeW}}{\multirow{1.5}[4]{*}{$\SIrange[range-phrase=\text{~to~}]{1}{50}{\micro\meter}$}} &       & \multirow{1.5}[4]{*}{\num{1}} & \multicolumn{1}{p{\ReleasedMolW}}{\mycell{\num{e4} to \\ \num{5e4}} } & \multicolumn{1}{p{\SymbolDurW}}{\multirow{1.5}[4]{*}{$\SI{10}{\milli\second}$}} & \multicolumn{1}{p{\SnrW}}{\multirow{1.5}[4]{*}{\numrange[range-phrase=~to~]{0}{60}}} & \multirow{3}[6]{*}{\acs{BER}} & \multicolumn{1}{l|}{\multirow{1.5}[4]{*}{$\geq$ \num{e-5}}} & \multirow{1.5}[4]{*}{\cite{alshammri2018adaptive}} \bigstrut\\


\cline{2-2}\cline{4-9}\cline{11-12} & \multicolumn{1}{p{\NNarchW}|}{\multirow{5}[8]{*}{Transformer}} & \multicolumn{1}{p{\AppW}|}{} & \multicolumn{1}{p{\CommRangeW}}{\multirow{1.5}[4]{*}{$\SIrange[range-phrase=\text{~to~}]{7}{12}{\micro\meter}$}} & \multicolumn{1}{p{\RxRadiusW}}{\multirow{1.5}[4]{*}{$\SI{5}{\micro\meter}$}} & \multirow{1.5}[4]{*}{\num{5}} & \multicolumn{1}{p{\ReleasedMolW}}{\mycell{\num{6e3} to\\ \num{18e3}}} & \multicolumn{1}{p{\SymbolDurW}}{\mycell{\num{35},\\ \num{100} $\si{\milli\second}$}} & \multicolumn{1}{p{\SnrW}}{\numrange[range-phrase=~to~]{0}{40}} &       & \multicolumn{1}{l|}{\multirow{1.5}[4]{*}{$\geq$ \num{4e-3}}} & \mycell{\cite{cheng2024informer},\\\cite{cheng2024signal}} \bigstrut\\

\cline{3-12} &       & \multicolumn{1}{p{\AppW}|}{\mycell{Channel\\estimation}} & \multicolumn{1}{p{\CommRangeW}}{\multirow{1.5}[4]{*}{$\SIrange[range-phrase=\text{~to~}]{4}{12}{\micro\meter}$}} & \multicolumn{1}{p{\RxRadiusW}}{\multirow{1.5}[4]{*}{$\SIlist{3;4;5}{\micro\meter}$}} & \num{5e-1}, \num{e-1} & \multicolumn{1}{p{\ReleasedMolW}}{\multirow{1.5}[4]{*}{\num{3e3}}} &       &       & \multirow{1.5}[4]{*}{\acs{RMSE}} & \multicolumn{1}{l|}{\multirow{1.5}[4]{*}{$\leq\,$\num{e-2}}} & \multicolumn{1}{l}{\multirow{1.5}[4]{*}{\cite{cheng2024channel}}} \bigstrut\\

\cline{1-1}\cline{3-12} \multirow{3.5}[6]{=}{Drifted channels} &       & \multicolumn{1}{p{\AppW}|}{Localization} & \multicolumn{1}{p{\CommRangeW}}{meter--scale} &       &       &       &       &       & Accuracy & \multicolumn{1}{l|}{$\leq\,$\SI{39}{\percent}} & \cite{hube2025set} \bigstrut\\

\cline{3-12} &       & \multicolumn{1}{p{\AppW}|}{\multirow{2.5}[4]{*}{\mycell{Data\\communication}}} & \multicolumn{1}{p{\CommRangeW}}{$\SI{10}{\micro\meter}$} & \multicolumn{1}{p{\RxRadiusW}}{ $\SI{1.5}{\micro\meter}$} & \num{0.79} & \num{4e3} & \multicolumn{1}{p{\SymbolDurW}}{$\SI{200}{\milli\second}$} & \multicolumn{1}{p{\SnrW}}{\numrange[range-phrase=~to~]{10}{40}} & \multirow{2}[4]{*}{\acs{BER}} & \multicolumn{1}{l|}{$\geq$ \num{4e-3}} & \cite{lu2023mcformer} \bigstrut\\

\cline{2-2}\cline{4-9}\cline{11-12} & \multicolumn{1}{p{\NNarchW}|}{\multirow{1.5}[4]{*}{Autoencoder}} &       & \multicolumn{1}{p{\CommRangeW}}{\multirow{1.5}[4]{*}{$\SI{38}{\milli\meter}$}} &       & \multirow{1.5}[4]{*}{\num{1.24e5}} &       & \multicolumn{1}{p{\SymbolDurW}}{\mycell{\num{100} to\\ \num{600} $\si{\milli\second}$}} & \multicolumn{1}{p{\SnrW}}{\multirow{1.5}[4]{*}{\numrange[range-phrase=~to~]{0}{10}}} &       & \multicolumn{1}{l|}{\multirow{1.5}[4]{*}{$\geq$ \num{2e-6}}} & \multirow{1.5}[4]{*}{\cite{khanzadeh2023end}} \bigstrut\\
\hline
\end{tabular}%
\end{table*}
\normalsize

\begin{figure*}
    \centering
    \includegraphics[width=0.9\linewidth]{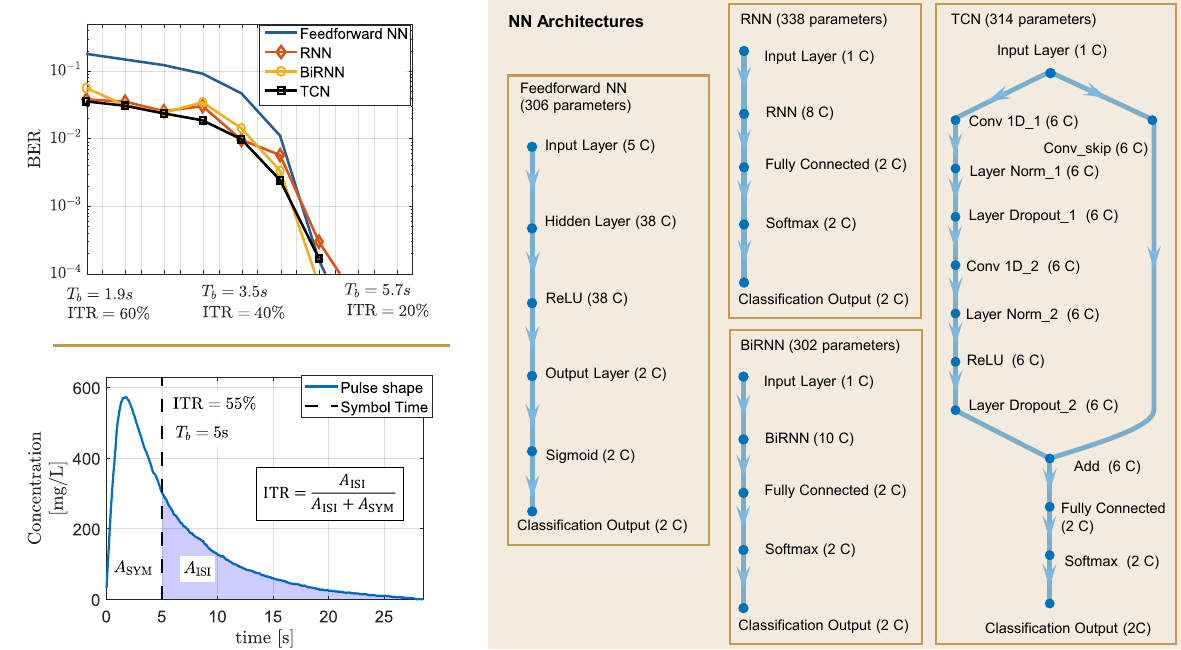}
    \caption{Graph representation and performance of various {NN} architectures.
    'C' denotes the number of input channels per layer, $A_\mathrm{SYM}$ and $A_\mathrm{ISI}$ evaluate the area below the curve.}
    \label{fig_BER_vs_ITR}
\end{figure*}

To shed more light on this topic, we examine a variety of architectures for the data communication task.
We evaluated the trade-off between detection performance and complexity across various \ac{NN} architectures, and present a comparative summary in \Cref{fig_BER_vs_ITR}.
This figure illustrates the performance of the various architectures, evaluated in terms of the \ac{BER} as a function of the transmission rate (bit time $T_b$).\footnote{The \ac{BiRNN} consists of five \ac{LSTM} cells in both the forward and backward directions, to account for channel memory.}
In this evaluation, all architectures were implemented at a comparable complexity, following the \ac{BiRNN} reference design, and Transformer models were excluded because they performed poorly in our simulations when constrained to the same number of parameters.

The comparison in \Cref{fig_BER_vs_ITR} confirms an expected result: The \ac{BiRNN} achieves similar performance with fewer learnable parameters than the other architectures, effectively capturing interrelations among samples and improving decoding.
Moreover, the figure also displays an inferred approximation of orders of magnitude for performance–complexity trade-offs: \ac{NN} architectures with hundreds of learnable parameters are generally required in \ac{MC} channels with low \ac{ISI}, i.e., channel memory of two to five samples.
A similar analysis is still pending for other tasks such as channel estimation, synchronization, or localization.
These observations motivate several open research directions for applying \acp{NN} in \ac{IoBNT} networks.

\setlength{\MCgeomW}{0.3cm}
\setlength{\NNarchW}{1.3cm}
\setlength{\AppW}{1.3cm}
\setlength{\CommRangeW}{1.5cm}
\setlength{\RxRadiusW}{0.8cm}
\setlength{\DiffCoeffW}{0.2cm}
\setlength{\ReleasedMolW}{1cm}
\setlength{\SymbolDurW}{1.2cm}
\setlength{\SnrW}{1cm}
\setlength{\PerfW}{0.7cm}
\setlength{\RefW}{0.5cm}

\begin{table*}[!t]
\centering
\caption*{Table I -- continued from previous page}
\footnotesize
\begin{tabular}{%
p{\MCgeomW}p{\MCgeomW}|p{\NNarchW}|p{\AppW}|p{\CommRangeW}|p{\RxRadiusW} p{\DiffCoeffW}|p{\ReleasedMolW}p{\SymbolDurW}p{\SnrW}p{\PerfW}l|p{\RefW}}
      & \multicolumn{1}{r}{} & \multicolumn{1}{l}{} & \multicolumn{1}{l}{} & \multicolumn{3}{l|}{End-to-end channel-related parameters} & \multicolumn{5}{c|}{Communication-related parameters} &  \bigstrut[b]\\
\hline
\multicolumn{2}{p{\MCgeomW}|}{\mycell{\acs{MC}\\ Geometry}} & \multicolumn{1}{p{\NNarchW}|}{\mycell{NN\\architecture}} & \multicolumn{1}{p{\AppW}|}{\mycell{\\Application}} & \multicolumn{1}{p{\CommRangeW}}{\mycell{Communication\\range}} & \multicolumn{1}{p{\RxRadiusW}}{\mycell{Receiver\\radius}} & \multicolumn{1}{p{\DiffCoeffW}|}{\mycell{$D$\\{}[$\si{\nano\meter\squared\per\nano\second}$]}} & \multicolumn{1}{p{\ReleasedMolW}}{\mycell{Released\\molecules}} & \multicolumn{1}{p{\SymbolDurW}}{Symbol\newline{}duration} & \multicolumn{1}{p{\SnrW}}{\mycell{\\\acs{SNR} [$\si{\decibel}$]}} & \multicolumn{2}{l|}{\mycell{Performance\\metrics}} & \multicolumn{1}{p{\RefW}}{\mycell{\\Ref.}} \bigstrut\\

\hline
\multicolumn{1}{p{\MCgeomW}}{\multirow{11}[22]{*}{\rotatebox[origin=c]{90}{\mycell{Experimental testbeds}}}} & \multicolumn{1}{p{\MCgeomW}|}{%
  \hspace*{-0.18cm}%
  \multirow{9}[14]{*}{\rotatebox[origin=c]{90}{\mycell{Vessel-like channels}}}%
} & \multicolumn{1}{p{\NNarchW}|}{\multirow{2}[4]{*}{\acs{RNN}}} & \multicolumn{1}{p{\AppW}|}{\multirow{2}[4]{*}{\mycell{Synchro-\\nization}}} & \multicolumn{1}{p{\CommRangeW}}{\multirow{2}[4]{*}{$\SI{50}{\centi\meter}$}} & \multirow{2}[4]{*}{} & \multicolumn{1}{p{\DiffCoeffW}|}{\multirow{2}[4]{*}{\num{0.1}}} & \multicolumn{1}{p{\ReleasedMolW}}{\multirow{2}[4]{*}{\num{e3}}} & \multicolumn{1}{p{\SymbolDurW}}{\multirow{2}[4]{*}{$\SI{20}{\second}$}} & \multirow{2}[4]{*}{\num{30}} & $P_d$ & \multicolumn{1}{l|}{\num{0.6}} & \multirow{2}[4]{*}{\mycell{Sec.\\ \ref{sec_sync_ex}}} \bigstrut\\

\cline{11-12}      &       &       &       & \multicolumn{1}{p{\CommRangeW}}{} &       &       &       &       &       & \multicolumn{1}{p{\PerfW}}{\acs{STO}} & \multicolumn{1}{l|}{{$\SI{4}{\second}$}} &  \bigstrut\\
\cline{3-13}      &       & \multicolumn{1}{p{\NNarchW}|}{Transformer} & \multicolumn{1}{p{\AppW}|}{\multirow{8}[10]{*}{\rotatebox[origin=c]{90}{\mycell{\\Data communication}}}} & \multicolumn{1}{p{\CommRangeW}}{\multirow{2}[4]{*}{$<\SI{1}{\meter}$}} &       &       &       & \multirow{2}[4]{*}{$\SI{1.2}{\second}$} &       & \multicolumn{1}{p{\PerfW}}{\multirow{3}[6]{*}{\acs{BER}}} & \multicolumn{1}{l|}{$\geq$ \num{9e-2}} & \cite{chen2021selfattention} \bigstrut\\

\cline{3-3}\cline{6-8}\cline{10-10}\cline{12-13}      &       & \multicolumn{1}{p{\NNarchW}|}{\multirow{2}[4]{*}{\acs{BiRNN}}} &       & \multicolumn{1}{p{\CommRangeW}}{} &       &       &       &       &       &       & \multicolumn{1}{l|}{$\geq$ \num{7.4e-2}} & \cite{sun2020ctbrnn} \bigstrut\\

\cline{5-10}\cline{12-13}      &       &       &       & \multicolumn{1}{p{\CommRangeW}}{\multirow{1.5}[6]{*}{$<\SI{1}{\meter}$}} &       &       &       & \mycell{\num{250} to\\ \num{500} $\si{\milli\second}$} &       &       & \multicolumn{1}{l|}{\multirow{1.2}[6]{*}{$>\,$\num{e-4}}} & \cite{farsad2017detection,farsad2018sliding} \bigstrut\\

\cline{3-3}\cline{5-13}      &       & \acs{CNN} &       & \multicolumn{1}{p{\CommRangeW}}{\multirow{1.5}[6]{*}{$\SI{5}{\centi\meter}$}} &       &       &       & \num{0.25}, \num{0.5}, \num{1} $\si{\second}$ &       & \multicolumn{1}{p{\PerfW}}{Accuracy} & $\geq\SI{33}{\percent}$ & \cite{bartunik2022using} \bigstrut\\

\cline{3-3}\cline{5-13}      &       & \multirow{2}[4]{*}{\mycell{Feedforward\\ NN}} &       & \multicolumn{1}{p{\CommRangeW}}{} &       &       &       & \mycell{\num{0.5}, \num{1},\\ \num{2}, \num{3} $\si{\second}$} &       & \multicolumn{1}{p{\PerfW}}{\multirow{1.2}[6]{*}{\acs{BER}}} & \multicolumn{1}{l|}{\multirow{1.2}[6]{*}{$\geq$ \num{4e-1}}} & \multirow{1.2}[6]{*}{\cite{koo2020deep}} \bigstrut\\

\cline{2-2}\cline{4-13}      & \multirow{5}[4]{*}{\rotatebox[origin=c]{90}{Open air}} &       & \multicolumn{1}{p{\AppW}|}{\mycell{Channel\\ estimation}} & \multicolumn{1}{p{\CommRangeW}}{\multirow{1.5}[6]{*}{$\SIrange[range-phrase=\text{~to~}]{1}{2}{\meter}$}} &       &       &       & \num{250}, \num{500}, \num{750} $\si{\milli\second}$ &       & \multicolumn{1}{p{\PerfW}}{\multirow{1.2}[6]{*}{\acs{RMSE}}} & \multicolumn{1}{l|}{\multirow{1.2}[6]{*}{\num{20.1}}} & \multirow{1.2}[6]{*}{\cite{gulec2020distance}} \bigstrut\\

\cline{3-13}      &       & \multicolumn{1}{p{\NNarchW}|}{\acs{BiRNN}} & \multicolumn{1}{p{\AppW}|}{\multirow{4}[6]{*}{\rotatebox[origin=c]{90}{\mycell{Data\\communication}}}} & \multicolumn{1}{p{\CommRangeW}}{\multirow{1.2}[6]{*}{$\SI{1}{\meter}$}} &       & \multicolumn{1}{p{\DiffCoeffW}|}{\multirow{1.2}[6]{*}{\num{0.84}}} & \multirow{1.2}[6]{*}{\num{4.9e23}} & \multirow{1.2}[6]{*}{$\SIrange{1}{6}{\second}$} & \multirow{1.2}[6]{*}{\num{27.5}} & \multicolumn{1}{p{\PerfW}}{\multirow{3}[6]{*}{\acs{BER}}} & \multicolumn{1}{l|}{\multirow{1.2}[6]{*}{$\geq$ \num{e-6}}} & \mycell{Sec.\\\ref{sec_code_decoder}} \bigstrut\\

\cline{2-3}\cline{5-10}\cline{12-13}\multicolumn{2}{p{\dimexpr 2\MCgeomW + 2\tabcolsep\relax}|}{\multirow{2}[4]{*}{\mycell{\textit{in-vivo}\\ bacteria\\colony}}} & \multicolumn{1}{p{\NNarchW}|}{\multirow{3}[4]{*}{\acs{CNN}}} &       & \multicolumn{1}{p{\CommRangeW}}{\multirow{1.2}[6]{*}{$\SI{100}{\micro\meter}$}} & \multicolumn{1}{p{\RxRadiusW}}{\multirow{1.2}[6]{*}{$\SI{6.7}{\micro\meter}$}} & \multicolumn{1}{p{\DiffCoeffW}|}{\multirow{1.2}[6]{*}{\num{0.75}}} & \mycell{\num{e5}\\bacteria} & \multirow{1.2}[6]{*}{$\SI{4}{\second}$} & \multirow{1.2}[6]{*}{\numrange[range-phrase=~to~]{10}{35}} &       & \multicolumn{1}{l|}{\multirow{1.2}[6]{*}{$\geq$ \num{e-4}}} & \multirow{1.2}[6]{*}{\cite{bai2023temporal}} \bigstrut\\

\cline{5-10}\cline{12-13}\multicolumn{2}{p{\dimexpr 2\MCgeomW + 2\tabcolsep\relax}|}{} &       &       & \multicolumn{1}{p{\CommRangeW}}{} &       &       &       & $\SI{1}{\minute}$ &       &       & \multicolumn{1}{l|}{$\geq$\num{e-2}} & \cite{vakilipoor2022hybrid} \bigstrut\\

\hline
\multicolumn{2}{p{\dimexpr 2\MCgeomW + 2\tabcolsep\relax}|}{\multirow{2}[4]{*}{\rotatebox[origin=c]{90}{\mycell{Human\\ vessels}}}} & \multirow{2}[4]{*}{\mycell{Feedforward\\ NN}} & \multicolumn{1}{p{\AppW}|}{\mycell{Channel\\ estimation}} & \multicolumn{1}{p{\CommRangeW}}{} &       &       &       &       &       & \multirow{1.2}[6]{*}{\acs{RMSE}} & \multicolumn{1}{l|}{\multirow{1.2}[6]{*}{$\geq,$\num{5e-2}}} & \cite{mohamed2021biocyber} \bigstrut\\

\cline{4-13}\multicolumn{2}{p{\dimexpr 2\MCgeomW + 2\tabcolsep\relax}|}{} &       & Detection & \multicolumn{1}{p{\CommRangeW}}{$\approx\SI{2.5}{\meter}$} &       &       & \num{e3} &       &       & Accuracy & $<\SI{85}{\percent}$ & \cite{ torres-gomez2024dna-based} \bigstrut\\

\hline
\end{tabular}%
\begin{tablenotes}
    \item Notes:
    \item $\bullet$ In the Ref. column of the table, the entries referring to "Sec." refer to the corresponding section in the present manuscript.
    \item $\bullet$ Missing entries in the table are not specified in the corresponding reference.
    \item $\bullet$ The column $D\,[\si{\nano\meter\squared\per\nano\second}]$ refers to the diffusion coefficient of molecules.
    \item $\bullet$ The column "Released molecules" refers to the number of molecules released at the emitter.
    \item $\bullet$ Few contributions report the channel memory.
    The work in \cite{qian2018receiver} reports a channel length of $6$ symbols.
    The work in \cite{chen2021selfattention} reports \numlist{3;5;13}.
    The work in \cite{koo2020deep} reports \num{1}.
    Our solution in Sec. III.C.4 reports a channel memory of \num{7} symbols.
    \item $\bullet$ All reported "Data communication" schemes uses the \ac{OOK} modulation except for \cite{bartunik2022using} which report multilevel modulation as well.
    \item $\bullet$ The acronyms \acs{RMSE} in the table refers to \acl{RMSE}.
    \item $\bullet$ The entry table for reference in \cite{yilmaz2017machine} also develops a \ac{MC} geometry comprised of a volumetric transmitter.
    Besides, we evaluated the \acs{RMSE} metric by inspecting the printed Fig. 3 on the same paper.
    \item $\bullet$ The flow velocity corresponding to the entry in the Table "Drifted channel" is $\SI{30}{\micro\meter\per\second}$ as follows from reference \cite{lu2023mcformer},  $\SI{5.5}{\centi\meter\per\second}$ as in \cite{khanzadeh2023end}, and $\SI{10}{\centi\meter\per\second}$ as in the entry for Sec.III.B.4.
    Additionally, the entry table "Open air" also specifies a flow velocity of $\SI{3.5}{\meter\per\second}$ for Sec. III.C.4 in the primary document.
    \item $\bullet$ The "\acs{RNN}" and "\ac{BiRNN}" entries in the table implement \ac{LSTM} cells.
    \item $\bullet$ The entry table for the references \cite{shrivastava2021performance,shrivastava2021scaled,agrawal2022neural} refer to a mobile scenario where the diffusion coefficient of the emitter is $\SI{4.74e-5}{\nano\meter\squared\per\nano\second}$ and for the receiver is $\SIlist{2.31e-3}{\nano\meter\squared\per\nano\second}$.
    Similarly, the entry table for the reference in \cite{cheng2023signal,cheng2024informer,cheng2024signal} define a diffusion coefficient of the receiver as $\SI{e-5}{\nano\meter\squared\per\nano\second}$.
    %
    %
    \item $\bullet$ The number of released molecules within the entry table of reference \cite{torres-gomez2024dna-based} refers to the number of nanosensors instead.
    \end{tablenotes}
\end{table*}
\normalsize

%

\textbf{1) Alternative NN Architectures:} Recent literature has primarily considered feedforward \acp{NN} and \acp{RNN} architectures for \ac{MC} systems; see \Cref{fig_NN_architecture}.
In contrast, other architectures that have gained significant traction in the broader \ac{NN} literature—most notably transformers and reinforcement learning—have only recently begun to attract attention in the context of \ac{MC}.
Transformer-based attention mechanisms have been highly successful in domains such as language processing, enabling joint alignment and decoding by capturing long-range dependencies in sequential data \cite{vaswani2017attention}.
However, despite initial examples discussed in \Cref{sec_NN_detectors}, current transformer implementations are not yet tailored to the specific interdependencies induced by \ac{ISI} in \ac{MC} channels.
As a result, transformer-based architectures require substantial adaptation to fully exploit \ac{ISI}-driven sequence structure in \ac{MC} decoding, as also noted in \cite{ma2024survey}.


Furthermore, {NNs} remain ill-suited to nonlinear scenarios, e.g., turbulent flows, and require vast amounts of data to converge, often with poor accuracy. 
Nevertheless, given that the underlying \emph{equations} governing the system are known, physics-informed {NN} models arise as a promising approach.
This architecture can include turbulence models by integrating Reynolds-averaged Navier-Stokes or Fokker-Planck equations into the learning algorithm, referring to examples in \cite{wu2018physicsinformed, chen2021solving}.

%

\newpage
\textbf{2) Hyperparameter Tuning:} Within the {MC} literature, only a few references \cite{lu2023mcformer,kosanetzki2025demodulation,agrawal2022neural} investigate the minimization of the \ac{BER} by optimizing the training and architecture-related hyperparameters.
The performance of neural network models in {MC} systems strongly depends on the selection of the training hyperparameters.
These include parameters such as the learning rate, batch size, optimizer type, number of training epochs, and regularization strategies, which directly influence convergence behaviour and generalization capability.
Looking after the optimization methods, the authors of \cite{lu2023mcformer} report using a grid search method to find hyperparameters related to the dimension of the \ac{NN}'s input and the total number of encoders and decoders in the transformer architecture.
A similar procedure is illustrated in \cite{agrawal2022neural} to find the number of layers and nodes per layer in a feedforward \ac{NN}.
More exhaustively, the authors in \cite{kosanetzki2025demodulation} use Bayesian optimization to find training parameters, including learning and dropout rates, as well as the number of layers in the \ac{CNN} and \ac{BiRNN} architectures.
However, the choice of these hyperparameters across the large number of reported papers is empirical and varies significantly with the {MC} scenario, dataset size, and network architecture.
To provide a consolidated overview of configurations adopted in prior studies, \Cref{tab_NN_hyperparameters_rew2} summarizes the reported hyperparameter settings used to train neural network models for {MC} applications.
The table highlights the diversity of training strategies and illustrates the lack of standardized configurations across the field.
In this regard, several promising research avenues remain open: (i) The extensive body of work on hyperparameter optimization methods has yet to be systematically applied to \ac{MC} channels (e.g., Bayesian, evolutionary, or gradient-based strategies \cite{yang2020hyperparameter});
(ii)~still existing studies mainly focus on the detection task, leaving channel estimation, synchronization, and higher-layer processes largely unexplored; and
(iii) current approaches are mostly agnostic to the underlying \ac{MC} link, optimizing for \ac{BER} without accounting for the physical characteristics and dynamics of molecule propagation.

%

\textbf{3) Channel Access in MC-based IoBNT Networks:} Channel access strategies face the problem that the number of nodes in the link is unknown, which is the case in nanonetwork formation in {MC} scenarios.
For instance, effective disease treatment relies on coordinated nanosensor actions, enabling the controlled release of drugs toward targets such as cancer cells~\cite{mosayebi2019early}.
Protocols have been designed to enable these applications, as seen in \cite{felicetti2014tcplike}; however, the dynamics of {MC} channels in fluidic environments hinder their straightforward deployment (see \cite{debus2025blood}).
Cognitive strategies developed for node clustering in computer networks, such as those in \cite{lai2011cognitive}, can be repurposed in {MC} environments for nanosensor cluster formation as well.

%

\textbf{4) Leveraging Cognitive Radio Concepts within the \ac{IoBNT} Framework:} Cognitive radio, originally developed for adaptive wireless systems \cite{haykin2005cognitive}, can be naturally extended to {MC} environments.
As discussed in \Cref{sec_communication}, early research already hints at this integration but remains incomplete.
Advancing cross-layer and cooperative strategies could substantially enhance {MC} performance.
At the system level, cognitive units may adapt sensing and transmission to support goal-oriented tasks, such as mimicking bioprocesses in cellular-scale digital twins~\cite{torres-gomez2023fine-tune}, thereby requiring unified cognitive control across all {MC} layers.

%

\section{Enabling Biocomputing in IoBNT Networks}
\label{sec_bio_AI}

\begin{figure*}[!t]
    \centering
    \includegraphics[width=\textwidth]{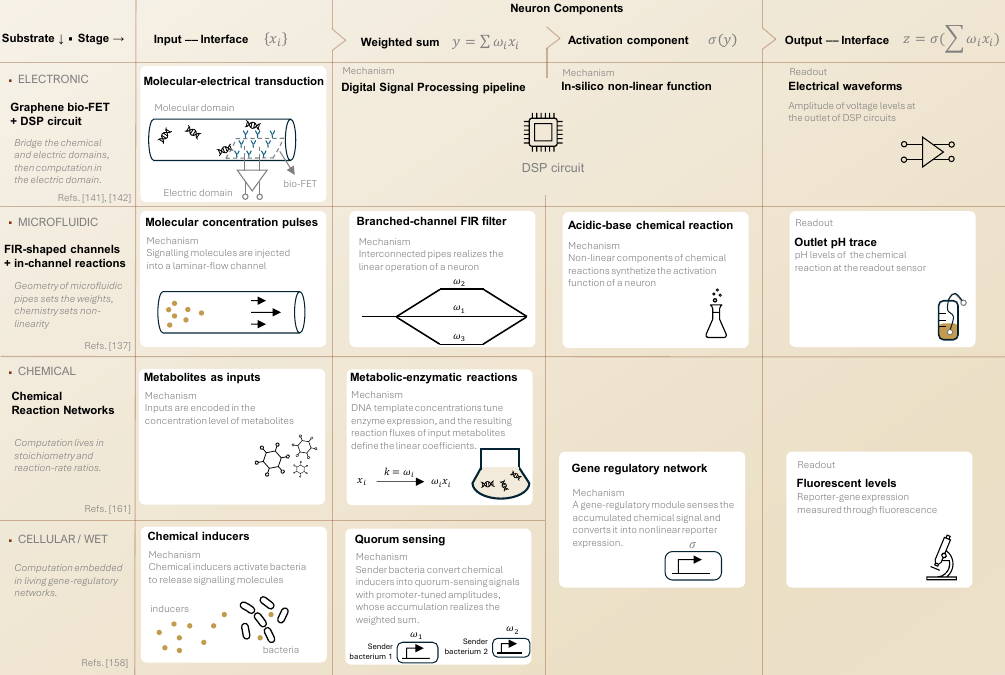}
    \caption{Summary of biocompatible technologies for {NN}s deployment.}
    \label{fig_bio}
\end{figure*}

Having reviewed the main methodological and algorithmic approaches, we now shift our focus toward their practical deployment.
In particular, this section discusses how the previously described techniques translate into experimental platforms, emerging bio-nanotechnological systems, and real-world applications.
This trend motivates the search for biocompatible computing substrates, as future \ac{IoBNT} nanonodes are expected to support smart processing tasks while remaining compatible with biological environments~\cite{marzo2019nanonetworks}.
To guide this discussion, \Cref{fig_bio} provides a conceptual overview of promising substrates for realizing \ac{NN} components in bio-nano systems, organized according to the main stages of a neuron: molecular or electrical input interfacing, weighted summation, nonlinear activation, and output readout.
Bio-inspired micro/nano computer design is a relatively young research field, and the literature reflects varying levels of readiness for individual concepts, which we summarize in this section.
Developments in this area lay the foundation to support the intelligent capabilities of the nanocomponents of \ac{IoBNT} networks.

As summarized in \Cref{fig_bio}, already investigated technologies are developed along three main directions:
\begin{itemize}
    \item[(i)] Extending the biological role of \ac{DNA}-based chemical reactions within the cell for computing; see the evolution in \cite[Fig. 1]{nagipogu2023survey} and \cite[Fig. 2a]{haluan2024synthetic},
    \item[(ii)] developing \acp{CRN} \cite{agiza2024phcontrolled}, and
    \item[(iii)] engineering microfluidic circuits for computing \cite{wang2013modular}.
\end{itemize}
Towards the development of {NN} models, a major trend in the literature is the use of engineered \acp{GRN} (see \Cref{sec_biocomp_gen}) due to their programmability for complex calculations \cite[Sec. 4.3]{nagipogu2023survey}.
Primarily developed within the communication-engineering community, a second trend involves integrating chemical reactions within microfluidic circuits. 
Following these research streams, we trace the early developments in digital logic and \ac{DSP} operations in biocompatible substrates for the development of {NN} modules.

%

\subsection{Biocomputing in the Digital Domain}


%


In the digital domain, reported research emulates neurons through chemical implementation and integration of logic gates.
Basically, a set of excitatory and inhibitory chemical reactions is connected by common substrates to realize logical operations such as \texttt{AND}, \texttt{OR}, and  \texttt{NOT}.\footnote{A \ac{CRN} is an abstraction of the dynamics of a system of chemical reactions given by a finite set of differential equations; see \cite[Sec. 3.1]{nagipogu2023survey}.}
The first description of \acp{CRN}, realizing neuron components of McCulloch–Pitts neurons,\footnote{McCulloch–Pitts neurons have only two possible states, either firing or quiescent.} dates back to the work in \cite{hjelmfelt1991chemical}, where a neuron was prototyped through an enzyme reaction system in free-diffusion environments.
Subsequent work develops neurons by connecting logic gates, in which a \num{2}-neuron feedforward {NN} is realized in \cite{agiza2024phcontrolled} to classify black-and-white images.
The linear neuron operation results from mixing encoded bit-1s and bit-0s, analogous to the mixing of acid and base compounds.
The neuron’s non-linearity is implemented via pH-based thresholding.
The classifier identifies $8\times 8 $ resolution images of handwritten digits with an accuracy larger than $\SI{95}{\percent}$.

The literature also reports models for \ac{DSP} operations in microfluidic channels, where the analytic formulation is valid under laminar flow conditions~\cite {bicen2013systemtheoretic}.
A \ac{FIR} filter is modeled in \cite[Sec. V]{bicen2013systemtheoretic} to design lowpass, stopband, and bandpass filtering of molecule concentration levels.\footnote{Calculations are reported in \cite{bicen2013systemtheoretic} to be accurate whenever the flow is laminar, i.e., at small Reynolds numbers, see \cite[Eq. (13)]{jamali2019channel}.}
The impulse response (in the time domain) and transfer function (in the frequency domain) are derived for combined microfluidic straight channels of arbitrary length, turns, bifurcations, and junctions.
Although it was not originally conceived for implementing neurons, the \ac{FIR}-based microfluidic circuit design provides a basis for realizing neuron-like operations in \ac{NN} architectures.
For instance, a 1D \ac{CNN} operation, $\sum_i \omega_i x_{(i-n)}$, maps naturally onto an \ac{FIR} structure by treating the filter coefficients $\omega_i$ as learnable parameters.
The neuron's nonlinearity, corresponding to the activation function, may then be implemented via chemical reactions embedded in the microfluidic channel, thereby completing the weighted-sum/activation pipeline illustrated in the microfluidic row of \Cref{fig_bio}; related nonlinear chemical mechanisms are reported in~\cite{bi2020chemical,walter2023real,amerizadeh2021bacterial}.

Complementing purely microfluidic realizations, the electrical domain provides a mature substrate for implementing the \ac{DSP} components of \acp{NN}, such as adders, multipliers, and nonlinear functions.
In this approach, molecular concentration patterns are first transduced into electrical waveforms, allowing subsequent weighted-sum and nonlinear operations to be exe-
%

%

\clearpage       
\onecolumn       
\newpage
\begin{landscape}
\footnotesize
        \begin{xltabular}{\linewidth}{p{1.7em}p{2em}|p{4.5em}|l|l|p{2.5em}|p{2.5em}|l|ll|lll|p{4.5em}|p{4.4em}}
        \caption{Complexity of reported \acs{NN} architectures.}
        \label{tab_NN_complexity}\\
\multicolumn{2}{c|}{\multirow{4}[6]{*}{{\acs{MC} Geometry}}} & \multicolumn{1}{c|}{\multirow{4}[6]{*}{{\mycell{{NN}\\ architecture}}}} & \multicolumn{1}{c|}{\multirow{1}[2]{*}{\multirow{4}[6]{*}{\mycell{Application}}}} & \multicolumn{1}{c|}{\multirow{2}[1]{*}{\multirow{3}[6]{*}{\mycell{Input\\ length}}}} & \multicolumn{3}{c|}{Fully connected layers} & \multicolumn{2}{c|}{\acs{RNN} layers} & \multicolumn{3}{c|}{\acs{CNN} layers} & \multicolumn{1}{c|}{\multirow{4}[6]{*}{\mycell{Number of\\parameters}}} & \multicolumn{1}{c}{\multirow{4}[6]{*}{Ref.}} \bigstrut[b]\\

\cline{6-13}\multicolumn{2}{c|}{} &       &       &       & \mycell{FC\\layers} & \mycell{Hidden\\layers} & \multicolumn{1}{p{3.5em}|}{\mycell{Nodes per\\hidden layer}} & \multicolumn{1}{p{3em}|}{\mycell{Number\\of layers}} & \multicolumn{1}{p{4.355em}|}{\mycell{\acs{LSTM} cells\\ per layers}} & \multicolumn{1}{p{3.43em}|}{\mycell{Number\\of layers}} & \multicolumn{1}{p{3.93em}|}{\mycell{Number of \\ filters}} & \multicolumn{1}{p{1.5em}|}{\mycell{Filter\\ size}} &       &  \bigstrut\\

\hline
\multicolumn{2}{c|}{\multirow{10}[20]{*}{\rotatebox[origin=c]{90}{\mycell{Point transmitter-Free diffusion-\\Spherical absorbing receiver}}}} & \multirow{9}[18]{*}{\rotatebox[origin=c]{90}{\mycell{\\Feedforward \acs{NN}}}} & \multicolumn{1}{l|}{\multirow{1.2}[4]{*}{\mycell{Channel\\estimation}}} & \num{2} & \multicolumn{1}{l|}{\num{1}} & \multicolumn{1}{l|}{\num{1}} & \num{2} &       &       &       &       &       & \multicolumn{1}{l|}{\num{6}} & \mycell{\Cref{sec_code_dist}} \bigstrut\\

\cline{5-15}\multicolumn{2}{c|}{} &       &       & \num{3} & \multicolumn{1}{l|}{\num{1}} & \multicolumn{1}{l|}{\num{5}} &  \numlist{10;20;40;20;10} &       &       &       &       &       & \multicolumn{1}{l|}{\num{2163}} & \multicolumn{1}{l}{\cite{cheng2024channel}} \bigstrut\\

\cline{4-15}\multicolumn{2}{c|}{} &       & Localization & \num{e3} & \multicolumn{1}{l|}{\num{1}} & \multicolumn{1}{l|}{\num{5}} & \num{400} &       &       &       &       &       & \num{1041603} & \cite{kose2020machine} \bigstrut\\

\cline{4-15}\multicolumn{2}{c|}{} &       & Detection & \num{16} & \multicolumn{1}{l|}{\num{1}} & \multicolumn{1}{l|}{\num{1}} & \num{16} &       &       &       &       &       & \num{272} & \cite{solak2020neural} \bigstrut\\

\cline{4-15}\multicolumn{2}{c|}{} &       & \multicolumn{1}{c|}{\multirow{5}[10]{*}{\mycell{Data\\communication}}} & \num{6} & \multicolumn{1}{l|}{\num{5}} & \multicolumn{1}{l|}{\num{10}} & \num{4} &       &       &       &       &       & \num{837} & \multicolumn{1}{p{3.355em}}{\cite{qian2019molecular}} \bigstrut\\

\cline{5-15}\multicolumn{2}{c|}{} &       &       & \num{11} & \multicolumn{1}{l|}{\num{1}} & \multicolumn{1}{l|}{\num{2}} & \num{20} &       &       &       &       &       & \num{722} & \cite{shrivastava2021performance,shrivastava2021scaled} \bigstrut\\

\cline{5-15}\multicolumn{2}{c|}{} &       &       & \num{11} & \multicolumn{1}{l|}{\num{1}} & \multicolumn{1}{l|}{\num{12}} & \num{10} &       &       &       &       &       & \num{1242} & \cite{agrawal2022neural} \bigstrut\\

\cline{5-15}\multicolumn{2}{c|}{} &       &       & \num{120} & \multicolumn{1}{l|}{\num{1}} & \multicolumn{1}{l|}{\num{2}} &  \numlist{70;10} &       &       &       &       &       & \num{9111} & \cite{sharma2020deep} \bigstrut\\

\cline{5-15}\multicolumn{2}{c|}{} &       &       & \num{12} & \multicolumn{1}{l|}{\num{8}} & \multicolumn{1}{l|}{\num{1}} &  \numlist{200;5;1} &       &       & \multicolumn{1}{p{3.43em}|}{\num{3}} & \multicolumn{1}{p{3.93em}|}{\num{3}} & \numlist{10;3;1} & \num{15688} & \cite{cheng2024informer} \bigstrut\\

\cline{3-15}\multicolumn{2}{c|}{} & \acs{RNN} & Detection & \num{32} & \multicolumn{1}{l|}{\num{1}} & \multicolumn{1}{l|}{\num{1}} & \num{32} & \multicolumn{1}{l|}{\num{1}} & \num{32} &       &       &       & \num{417} & \cite{solak2020rnn} \bigstrut\\

\hline
\multicolumn{1}{c|}{\multirow{5}[10]{*}{\rotatebox[origin=c]{90}{\mycell{Experimental\\testbeds}}}} & \mycell{Open air} & \multirow{3}[6]{*}{\acs{BiRNN}} & \multirow{3}[6]{*}{\mycell{Data\\communication}} & \num{8} & \multicolumn{1}{l}{} & \multicolumn{1}{l}{} &       & \multicolumn{1}{l|}{\num{6}} & \num{8} &       &       &       & \num{576} & \Cref{sec_code_decoder} \bigstrut\\

\cline{2-2}\cline{5-15}\multicolumn{1}{c|}{} & \multicolumn{1}{l|}{\multirow{4}[8]{*}{\rotatebox[origin=c]{90}{\mycell{Vessel-like\\channels}}}} &       &       & \num{5} & \multicolumn{1}{l|}{\num{2}} & \multicolumn{1}{l|}{\numlist{1;1}} & \num{25} & \multicolumn{1}{l|}{\num{2}} & \num{5} &       &       &       & \num{540} & \cite{sun2020ctbrnn} \bigstrut\\

\cline{5-15}\multicolumn{1}{c|}{} &       &       &       & \num{40} & \multicolumn{1}{l}{} & \multicolumn{1}{l}{} &       & \multicolumn{1}{l|}{\num{32}} & \num{40} &       &       &       & \num{2880} & \cite{farsad2018sliding} \bigstrut\\

\cline{3-15}\multicolumn{1}{c|}{} &       & \acs{RNN} & Synchronization & \num{1} & \multicolumn{1}{l|}{\num{2}} & \multicolumn{1}{l|}{\numlist{1;1}} & \numlist{256;256} & \multicolumn{1}{l|}{\num{1}} & \num{128} &       &       &       & \num{35328} & \Cref{sec_sync_ex} \bigstrut\\

\cline{3-15}\multicolumn{1}{c|}{} &       & \multirow{3}[6]{*}{\acs{CNN}} & \multirow{3}[6]{*}{\mycell{Data\\communication}} & \num{128} & \multicolumn{1}{l|}{\num{3}} & \multicolumn{1}{l|}{\num{1}} & \numlist{4096;4096;6} &       &       & \multicolumn{1}{l|}{\num{3}} & \multicolumn{1}{l|}{\numlist{64;2;2}} & \numlist{7;5;3} & \num{33587738} & \cite{bartunik2022using} \bigstrut\\

\cline{2-2}\cline{5-15}\multicolumn{2}{c|}{\multirow{2}[4]{*}{\mycell{\textit{in-vivo} bacteria colony}}} &       &       & $15\times60\times2$ & \multicolumn{1}{l|}{\num{2}} & \multicolumn{1}{l|}{\num{1}} & \numlist{32;2} &       &       & \multicolumn{1}{l|}{\num{1}} & \multicolumn{1}{l|}{\num{16}} & \multicolumn{1}{l|}{\num{15}} & \num{462144} & \cite{vakilipoor2022hybrid} \bigstrut\\

\cline{5-15}\multicolumn{2}{c|}{} &       &       & \num{5} & \multicolumn{1}{l}{} & \multicolumn{1}{l}{} & \multicolumn{1}{l}{} &       &       & \multicolumn{1}{l|}{\num{2}} & \multicolumn{1}{l|}{\num{2}} & \num{3} & \num{30} & \cite{bai2023temporal} \bigstrut\\

\hline
\multicolumn{2}{l|}{\multirow{1.2}[6]{*}{\mycell{Human vessels}}} & \mycell{Feedforward\\ NN} & \multirow{1.2}[6]{*}{Detection} & \multirow{1.2}[6]{*}{\num{2}} & \multicolumn{1}{l|}{\multirow{1.2}[6]{*}{\num{1}}} & \multicolumn{1}{l|}{\multirow{1.2}[6]{*}{\num{1}}} & \multirow{1.2}[6]{*}{\num{6}} &       &       &       &       &       & \multirow{1.2}[6]{*}{\num{18}} & \multirow{1.2}[6]{*}{\cite{torres-gomez2024dna-based}} \bigstrut\\

\hline
\multirow{1.2}[6]{*}{\mycell{Drifted channels}} &       & \multirow{1.2}[6]{*}{Autoencoder} & \mycell{Data\\communication} & \multirow{1.2}[6]{*}{\num{1}} & \mycell{Enc: \num{3}\\ Dec: \num{5}} & \mycell{Enc: \num{1}\\ Dec: \num{1}} &       &       &       &       &       &       & \multirow{1.2}[6]{*}{\num{244}} & \multirow{1.2}[6]{*}{\cite{khanzadeh2023end}} \bigstrut\\

\hline
\end{xltabular}
\begin{tablenotes}
    \item Notes:
    %
    %
    %
    %
    %
    %
    %
    %
    \item $\bullet$ The calculation of the number of elements per layer is evaluated as follows:
    \item - Feedforward \acs{NN}: We compute the total of parameters by adding the number of coefficients and bias per layer.
    Per layer, the total of coefficients is $n_\mathrm{in}\times n_\mathrm{out}$, and the total of biases is $n_\mathrm{out}$, where $n_\mathrm{in}$ is the number of inputs, and $n_\mathrm{out}$ is the number of outputs for the given layer.
    The number of nodes in the output layer is \num{1}, except for the entry in \cite{agrawal2022neural}, where the nodes in the output layer are \num{2}.
    \item - \ac{RNN} and \acs{BiRNN} architectures: We compute the number of parameters based on the gates within the \ac{LSTM} cells, as all reported methods deploy them.
    The \ac{LSTM} cell comprises four gates: input, forget, cell, and output.
    Each gate implements a separate set of weights for the input, hidden states, and biases, with eight weights and four biases, totaling \num{12} parameters per cell.
    In total, the amount of parameters for $n_\mathrm{cells}$ cells will be $n_\mathrm{cells}\times(12)$.
    In the case of the \ac{BiRNN}, calculations are twice the number as for the \ac{RNN}, as each layer implements both a backward and forward direction.
    \item - \ac{CNN} architecture: The filter size gives the number of parameters.
    A filter of size $K$ will implement $K$ weights and a single bias.
    Additionally, the total number of filters is determined by the comparative sizes of the input and output.
    For instance, if the input is a \num{1}D vector as $128\times1$ and the output is a \num{2}D matrix as $128\times64$, the number of filters is \num{64}.
    If the input and output are \num{2}D matrices, given by $64\times64$ and $64\times128$, respectively, then two filters would be needed, as the output dimension is twice the input dimension.
    \item - Transformer architecture: The complexity for this architecture is evaluated based on the code published by the authors in \cite{cheng2024informer}.
    In this entry, the number of nodes per hidden layer, \num{200}, refers to the six fully connected layers implemented in the transformer architecture, while the other values, \numlist{5;1}, refer to the remaining fully connected layers implemented in the decoder.
    This architecture also deploys three \acp{CNN} at the Transformer input to extract features.
 \end{tablenotes}
\normalsize
\newpage
\footnotesize
        \begin{xltabular}{\linewidth}{ll|l|l|l|p{4.775em}|l|l|l|l}
        \caption{Summary of hyperparameters related to the training of \acs{NN} architectures.}
        \label{tab_NN_hyperparameters_rew2}\\
\multicolumn{2}{p{9.36em}|}{\mycell{\\ \acs{MC} Geometry}} & \multicolumn{1}{p{4.09em}|}{\mycell{\acs{NN}\\architecture}} & \multicolumn{1}{p{5.955em}|}{\mycell{\\Application}} & \multicolumn{1}{p{6.41em}|}{\mycell{Optimizer\\algorithm}} & \mycell{Learning\\rate} & \multicolumn{1}{p{8em}|}{\mycell{Number\\of samples training}} & \multicolumn{1}{p{4.18em}|}{\mycell{Number\\of epochs}} & \multicolumn{1}{p{3.545em}|}{\mycell{Batch\\size}} & \multicolumn{1}{p{4.045em}}{\mycell{\\Ref.}} \bigstrut[b]\\

\hline
\multicolumn{2}{c|}{\multirow{10}[20]{*}{\rotatebox[origin=c]{90}{\mycell{Point transmitter-Free diffusion-\\Spherical absorbing receiver}}}} & \multirow{9}[18]{*}{\rotatebox[origin=c]{90}{\mycell{\\Feedforward NN}}} & \multicolumn{1}{l|}{\mycell{Channel\\estimation}} & \multicolumn{1}{l|}{\multirow{1.5}{*}{BFGS}} & \multicolumn{1}{l|}{} & \multicolumn{1}{l|}{\multirow{1.5}{*}{$\SI{481}{\samples}$}} & \multicolumn{1}{l|}{\multirow{1.5}{*}{\num{15}}} &       & \multicolumn{1}{l}{\multirow{1.5}{*}{\Cref{sec_code_dist}}} \bigstrut\\

\cline{4-10}\multicolumn{2}{c|}{} &       & Localization & \multicolumn{1}{l|}{\multirow{2}[4]{*}{Adam}} & \num{e-2} &       & \num{100} & \num{256} & \cite{kose2020machine} \bigstrut\\
\cline{4-4}\cline{6-10}\multicolumn{2}{c|}{} &       & Detection &       & \num{e-3} & \multicolumn{1}{l|}{$\SI{e5}{\samples}$} & \num{100} & \num{10} & \cite{solak2020neural} \bigstrut\\
\cline{4-10}\multicolumn{2}{c|}{} &       & \multirow{6}[12]{*}{\mycell{Data\\ communication}} & \multicolumn{1}{l|}{\multirow{3}[6]{*}{LM}} & \multirow{2}[4]{*}{\num{e-2}} & \multicolumn{1}{l|}{$\SI{e3}{\bit}$} & \multicolumn{1}{c|}{\multirow{2}[4]{*}{\num{200}}} & \num{50} & \multicolumn{1}{l}{\cite{qian2018receiver}} \bigstrut\\
\cline{7-7}\multicolumn{2}{c|}{} &       &       &       & \multicolumn{1}{l|}{} & \multicolumn{1}{l|}{$\SI{5e4}{\bit}$} &       & \num{1e3} & \multicolumn{1}{l}{\cite{qian2019molecular}} \bigstrut\\
\cline{6-10}\multicolumn{2}{c|}{} &       &       &       & \multirow{2}[4]{*}{\num{e-3}} & \multicolumn{1}{l|}{\multirow{2}[4]{*}{$\SI{5e4}{\bit}$}} & \numlist{50;55;125} &       & \multirow{2}[4]{*}{\cite{shrivastava2021performance,shrivastava2021scaled}} \bigstrut\\
\cline{5-5}\cline{8-9}\multicolumn{2}{c|}{} &       &       & \multicolumn{1}{l|}{\multirow{2}[4]{*}{BFGS}} & \multicolumn{1}{l|}{} &       & \numlist{184;188;215} &       &  \bigstrut\\
\cline{6-10}\multicolumn{2}{c|}{} &       &       &       & \multicolumn{1}{l|}{} & \multicolumn{1}{l|}{$\SI{e4}{\bit}$} &       &       & \cite{agrawal2022neural} \bigstrut\\
\cline{6-9}\multicolumn{2}{c|}{} &       &       &       & \num{e-3} &       & \num{500} &       & \cite{sharma2020deep} \bigstrut\\
\cline{3-10}\multicolumn{2}{c|}{} & \acs{RNN} & Detection & \multicolumn{1}{l|}{Gradient descent} & \num{e-3} & \multicolumn{1}{l|}{$\SI{8e3}{\samples}$} & \num{100} & \num{10} & \cite{solak2020rnn} \bigstrut\\
\hline
\multicolumn{1}{c}{\multirow{5}[10]{*}{\mycell{Experimental\\testbeds}}} & Open air & \multirow{3}[6]{*}{\acs{BiRNN}} & \multicolumn{1}{c|}{\multirow{3}[6]{*}{\mycell{Data\\ communication}}} & \multirow{3}[6]{*}{Adam} & \multicolumn{1}{l|}{\num{e-3}} & $\SI{e6}{\samples}$ & \num{10} & \num{10} & \Cref{sec_code_decoder} \bigstrut\\
\cline{2-2}\cline{6-10}      & \multicolumn{1}{l|}{\multirow{4}[8]{*}{\mycell{Vessel-like\\channels}}} &       &       &       & \multicolumn{1}{l|}{} & $\SI{8e4}{\bit}$ & \num{2e4} &       & \cite{sun2020ctbrnn} \bigstrut\\
\cline{6-10}      &       &       &       &       & \multicolumn{1}{l|}{\num{e-3}} & $\SI{120}{\bit}$ & \num{200} & \num{10} & \cite{farsad2018sliding} \bigstrut\\

\cline{3-10}      &       & \acs{RNN} & Synchronization & BFGS  & \multicolumn{1}{l|}{\num{2e-4}} & $\SI{1.2e6}{\bit}$ & \num{35e3} &       & \Cref{sec_sync_ex} \bigstrut\\

\cline{3-10}      &       & \acs{CNN} & \multirow{2}[4]{*}{\mycell{Data\\ communication}} & Adam  & \multicolumn{1}{l|}{\num{e-3}} &       & \num{20} & \num{64} & \cite{bartunik2022using} \bigstrut\\

\cline{2-3}\cline{5-10}\multicolumn{2}{l|}{\mycell{\textit{in vivo} bacteria colony}} & \multicolumn{1}{l|}{\acs{CNN}} &       & SGDM  & \multicolumn{1}{l|}{} &       &       &       & \cite{vakilipoor2022hybrid} \bigstrut\\

\hline
\multicolumn{2}{l|}{\multirow{1.5}[4]{*}{Human vessels}} & \mycell{Feedforward\\ NN} & \multirow{1.5}[4]{*}{Detection} &       & \multicolumn{1}{l|}{} &       & \multirow{1.5}[4]{*}{\num{4e4}} &       & \multirow{1.5}[4]{*}{\cite{torres-gomez2024dna-based}} \bigstrut\\

\hline
\multirow{1.5}[4]{*}{Drifted channels} &       & \multirow{1.5}[4]{*}{Autoencoder} & \mycell{Data\\ communication} &       & \multirow{1.5}[4]{*}{\num{9e-3}} &       & \multirow{1.5}[4]{*}{\num{2e3}} & \multirow{1.5}[4]{*}{\num{40}} & \multirow{1.5}[4]{*}{\cite{khanzadeh2023end}} \bigstrut\\
\bottomrule
\end{xltabular}
\begin{tablenotes}
    \item Notes:
    \item $\bullet$ In the Ref. column of the table, the entries referring to ``Sec.'' are pointing to the corresponding section in the main document.
    \item $\bullet$ The abbreviations LM, BFGS, and SGDM introduced in the column ``Optimizer algorithm'' refer to Lavenberg-Marquardt, Broyden-Fletcher-Goldfarb-Shanno (see \cite{sun2020survey}), and stochastic gradient descent with momentum.
    \item $\bullet$ Missing entries in the table are not reported in the corresponding manuscript.
\end{tablenotes}

\normalsize
\end{landscape}
\clearpage       
\twocolumn       

%

\noindent
cuted by conventional electronic circuits.
Graphene biofield-effect transistors (bio-FET) can provide this molecular--electrical interface by detecting \ac{DNA} strands flowing in microfluidic channels and converting them into electrical signals~\cite{kuscu2021fabrication,abdali2024frequencydomain}.
This pipeline is illustrated in the first row of \Cref{fig_bio}; although experimental demonstrations in \ac{IoBNT} networks remain unexplored, bio-FET sensing technologies represent one of the more mature routes for molecular--electrical interfacing.

%

\subsection{Biocomputing in the Analog Domain}

Unlike \textit{in silico} technologies, where {NNs} are deployed digitally, the integration of {NNs} with chemical-based technologies is more naturally pursued in the analog domain.
This is mainly because chemically wiring pre-existing logic gates in fluidic media to implement more complex arithmetics remains challenging.
The following paragraphs summarize several reported {NN} implementations that are based on compartment-based models, \ac{GRN}, and biological \textit{in vivo} systems.

\subsubsection{Wet Neuromorphic Computing}
\label{sec_biocomp_gen}

Computing can be realized in the analog domain using the already-in-place neuromorphic capabilities of biological systems \cite{perera2025wetneuromorphic,somathilaka2025analyzing}.\footnote{Neuromorphic computing denotes computing principles inspired by biological nervous systems, in which information processing emerges from distributed interactions among interconnected units and memory is embedded in the system dynamics, rather than being separated from computation as in conventional von Neumann architectures.}
For instance, within the cell, memory and computation are inherently integrated into the \ac{GRN} (which refers to neuromorphic computing), enabling regression and pattern recognition tasks, as illustrated in \cite{somathilaka2025internet}.
The concept also extends to offloading computation into tissues and entire organisms; see the most vivid examples in \cite{cai2023brain,kagan2022invitro}.

Within the cell, arithmetic can be realized in the analog domain by engineering \ac{DNA} circuits, as summarized in \cite{purcell2014synthetic,balasubramaniam2023realizing,nagipogu2023survey}.
Examples include adders, multipliers, subtractors, power-laws, and dividers realized by combining \ac{DNA} circuits, as illustrated in the early work in \cite{daniel2013synthetic}.
To a greater extent, the literature focuses on implementing feedforward {NN} architectures that leverage the natural interactions among genes within the cell.
For instance, {NNs} are identified with genes as the computing units, transcription factors as the stimulus, and the degree of influence of transcription factors in the genes as the weights \cite{somathilaka2023revealing,somathilaka2024wet,kim2004neural}.
Using these analogies between \ac{DNA}-chemical components and {NN} functions, the potential of operating {NN} structures already within the gene regulatory circuits is extensive --as a result of mining the connection in the \ac{GRN}, more than a hundred single-layer {NNs} have been identified, see~\cite[Fig.~3]{balasubramaniam2023realizing}.

The first work demonstrating the potential of \ac{DNA}-based engineering relied on the \ac{DNA} strand-displacement technique as developed in \cite{qian2011neural}.\footnote{\ac{DNA} strand displacement regulates the gene expression and is capable of creating universal computation \cite{soloveichik2010dna}.}
In this technique, neurons are implemented via the reaction of input strands and toeholds (\ac{DNA} segments) with three gates.
The multiplier gates output a number of \ac{DNA} strands in proportion to the weight, i.e., the operation $\omega_i \times x_i$, where $\omega_i$ is the weight and $x_i$ is the input.
The integration gate is chemically related to the output strands from the multiplier gates in a way that a reaction product reflects a strand as the sum of the inputs, i.e., $\sum \omega_i \times x_i$.
Finally, thresholding is implemented by a dedicated threshold gate that releases output strands only when the accumulated signal exceeds a prescribed level; these output strands are then detected through fluorescence from \ac{DNA} reporter gates.
Using \num{112} \ac{DNA} strands assembled into \num{72} initial \ac{DNA} species, this technology implemented a four-neuron Hopfield associative memory and demonstrated pattern recall through a ``read your mind'' game with four stored patterns.


On a larger scale, other examples include combining multiple engineered cells~\cite{sarkar2021single,li2021synthetic} and the striking performance of brain organoids cultured cells \cite{smirnova2023organoid}.
The research in \cite{sarkar2021single} conceives a single-layer {NN} via cell-to-cell communication in a diffusive medium.
The weights of the {NN} are set by the diffusive properties of the information molecules.
The required nonlinearity is achieved through gene expression in an engineered \textit{E. Coli} cell.
Integrating \numrange{2}{6} bacterial populations, each population acting as a \num{2}-input neuron, this work demonstrated the implementation of multiplexers, majority rules (data fusion), and decoding, among other operations.   
A different approach is developed in \cite{li2021synthetic}, in which multiple bacteria serve as transmitters and release molecules proportional to the magnitudes of the learnable coefficients of the {NN}.
Another bacterium serves as a receiver and implements the nonlinear function; see an illustration of the complete pipeline in the bottom row in \Cref{fig_bio}.
The concept's applicability to \ac{ML} was demonstrated in~\cite{li2021synthetic} through a pattern-recognition task, in which an engineered bacterial population, performing as a neuron of \num{9} learnable coefficients, classified chemical-inducer input patterns arranged as $3\times3$-bit images.
In both implementations \cite{sarkar2021single,li2021synthetic}, the biochemical computation is transduced into fluorescent reporter expression, which serves as the measurable readout of the \ac{NN} output.


Brain organoids, which develop in cultures of human stem cells, are known to perform complex tasks in real-time.
The work in \cite{kagan2022invitro} illustrates training living neuron cells to play the arcade game 'Pong'.
The learning capability of these networks relies on neuronal neuroplasticity.
The experimental work illustrates the response of surrounding cells to an electric stimulus and the development of new neural connections during the learning phase.
This technology offers two key advantages related to scalability: (i) reduced energy consumption, with the human brain serving as a reference for highly connected yet energy-efficient computation at approximately $\SI{20}{\watt}$, and (ii) faster training; it takes about \num{4} epochs to match the performance of \ac{LSTM} cells, which require \num{50} epochs. 

%

\subsubsection{Analog Computation with \acp{CRN}}

\ac{DNA} regulatory circuits are also in the domain of \acp{CRN}, as they provide the substrate for the reaction pathways.\footnote{Substrates refer to reactant compounds in a chemical reaction.}
However, other chemical solutions implement the computation without engineering \ac{DNA} strands.
These solutions aim to overcome limitations such as temperature sensitivity, long computation times, and the use of a large number of chemical components, as stated in \cite{agiza2023digital}.
For instance, the metabolic circuits of cells are engineered in~\cite{pandi2019metabolic} to implement \num{2} and~\mbox{\num{4}--input} {NNs} of a single layer.
This work achieved positive weights with concentration level in a cell-free framework and negative weights with attenuating reactions.
However, due to the large number of required reactions, these techniques are often combined with \ac{DNA}-computing \cite{sawlekar2021survey,cherry2018scaling,rodriguez2021loser} to facilitate implementation.
Notably, \acp{CRN} were used to implement a chemical {NN} that even allows backpropagation and, thus, online in-situ {NN} training~\cite{nagipogu2024neuralcrns}.

%

\subsubsection{Compartment-Based Models}

Nanoscale computing units can be realized through the propagation of molecules between connected compartments and the chemical reactions occurring within them~\cite{angerbauer2024novel}.
Matrix weights are mapped by adjusting compartment volumes, forming the basis of a compartment-based reaction–diffusion computer, which has been experimentally demonstrated using phase transition, precipitation, and acid–base reactions~\cite{angerbauer2024investigation}.
These experiments achieved reasonable accuracy, indicating feasibility for microscale implementations.
Further theoretical work extends this concept to molecular nano-NNs and has been validated through classification and regression tasks~\cite{angerbauer2023molecular}.

\subsubsection{Reservoir Computing in MC Channels}

Reservoir computing has recently been introduced in the \ac{MC} domain to realize recurrent architectures~\cite{uzun2025molecular}, also via brain organoids~\cite{cai2023brain}.
This architecture employs a static \ac{RNN} in a reservoir, with learning occurring only in the output layer~\cite{stepney2024physical}.
The reservoir is implemented physically as a point-transmitter free-diffusion \ac{MC} channel with ligand–receptor interactions, which represents the state update equation of the \ac{RNN}~\cite[Eq.~(1)]{uzun2025molecular}, or using brain organoids as described in \cite{cai2023brain}.
The output layer is interfaced with a silicon substrate and trained as a weighted vector of bound-molecule samples over time (see ~\cite[Eq.~(9)]{uzun2025molecular}).

\subsection{Resource Demand and Feasibility of Bio-AI Unit Implementation}

The required arithmetic operations for {NN} operation, such as adders, multipliers, and non-linear functions, are highly resource-demanding to be implemented by logic units within cells or microfluidic devices.
For instance, an eight-bit adder would require \num{16} \texttt{AND}, \num{8} \texttt{OR} and \num{8} \texttt{XOR} two-input gates; see \cite[Sec. 5.2.1]{brown2009fundamentals}.
The reported {NN}-based decoder architectures in \Cref{sec_communication} require around \num{300} learnable parameters (see \Cref{fig_BER_vs_ITR}) and nearly the same number of adders, yielding a total of $300\times(16+8+8)=9600$ logic gates to be implemented.
Enforcing a gate per cell or per microfluidic circuit imposes a physical dimension on the millimeter scale, preventing low-dimensional developments at the micrometer or nanometer scale.\footnote{We assume the dimension of a cell is in the range of $\SIrange{1}{2}{\micro\meter}$ as for \textit{E.~Coli} \cite{elhajj2015how} and for microfluidic circuits in the range of $\SIrange{1}{10}{\micro\meter}$ \cite{prakash2007microfluidic}.}
Furthermore, hundreds of learnable parameters lead to a high number of connections, imposing a greater challenge for the chemical wiring of the corresponding logic units.

By contrast, \acp{GRN} within the cell allows a higher computational density.
Following the entries in \cite[Fig. 3]{balasubramaniam2023realizing} related to feedforward {NN} implementation, we can find networks of~\num{26} learnable parameters maximum.\footnote{We perform this calculation assuming the number of hidden nodes follows the number of inputs.
}
Therefore,~\num{14} connected cells can achieve the complexity reported for decoders (approximately \num{300} in \Cref{fig_BER_vs_ITR}) and still fit within the micrometer range.
However, the limited reliability of DNA circuits \cite{nagipogu2023survey} and their long processing times (on the order of hours \cite[Fig. 4]{cherry2018scaling}) remain significant drawbacks.
Beyond purely genetic circuits, brain organoids \cite{smirnova2023organoid} and reservoir computing \cite{uzun2025molecular} are candidate technologies for developing {NN} architectures at the micrometer scale.
Reported designs are in the range of $\SIrange{10}{100}{\micro\meter}$, although not limited to further scaling down their dimensions.
These advancements demonstrate the practicality of {NN} architectures for {IoBNT} applications; yet, a design for {NN} module interfaces (both biological and electrical) remains unavailable.

A comparative summary of the technologies presented in this section is provided in \Cref{tab_bio}, which evaluates their capabilities in terms of weight density, latency, scalability, and \ac{TRL}.\footnote{For a definition of readiness levels, see \cite[pag. 14]{EuropeanCommission2025}.}
To ensure a fair comparison across diverse substrates, we evaluate computational capabilities using spatiotemporal weight density (number of weights per unit volume per unit time).
This metric is necessary because a purely spatial evaluation tends to overestimate the efficiency of silicon technologies.
That is, in contrast to chemical-based technologies, silicon enables iterative hardware reuse through code loops. 
By accounting for the reusability capacity of \textit{in~silico} technologies, this evaluation provides a realistic sense of the physical scale required to synthesize {NN} models.
The values are extracted from the experimental studies cited in the last column of the table, with specific evaluation details for weight counts and volume, as provided in the table notes.

As a result of these calculations, the highest \ac{TRL} is achieved with bioFET technology, see also a summary of similar biosensors on the market in \cite{lau2025using}.
Chemical technologies correspond to lower \ac{TRL} levels, where prototypes have been developed, and some have been tested in relevant scenarios.
We find that among the chemical-based technologies, compartment-based technologies have the highest gain, although it is six orders of magnitude lower than that of silico technologies.
We also remark that the lower densities of the chemical technologies are primarily due to the reported experimental volumes being on the order of tens of $\si{\micro\liter}$, which understate their weight-density capacities.
These contributions illustrated the affordability of developing {NNs} with bio-compounds, but less attention was paid to their efficiency in terms of size and performance.
This motivates the development of more size-efficient proofs of concept.


\subsection{Concluding Remarks and Future Outlook}

In summary, there are numerous approaches to implementing {NNs} in biological and/or chemical contexts. 
A key challenge lies in the limited scalability of these concepts, constrained by geometrical~\cite{angerbauer2024investigation}, chemical \cite{nagipogu2024neuralcrns}, or biological \cite{rizik2022synthetic} constraints.
While numerous concepts have been shown to work in practice, a unified framework for the design and standardization of biological or chemical {NNs} is currently missing. 
The development of such a unified scheme is anticipated, as current research in the field is advancing.

Besides, we identify the following open research directions related to the implementation of {NNs} in the biological domain:

\textbf{1. {MC} Networks as {ML} Platforms:} Inspired by paradigms in WiFi and {ML} joint development~\cite{szott2022wifi}, {MC} networks can also constitute a running platform for {ML} development (see \cite[Chap. 12]{eldar2022machine}).
Various examples exist in wireless networks where intelligence is distributed among nodes.
Such ideas can be extended to {MC} networks for deploying large {NN} architectures challenging to fit on a single node.

\textbf{2. Deployment at the Nanoscale Level:} The current integration of \ac{MC} and \ac{ML} primarily relies on macro-scale external computers hosting the \ac{NN} inference (\Cref{sec_communication}).
The micro-scale implementations illustrated in this section still remain limited in complexity.
A promising direction for reducing resource consumption is to develop logic based on a one-bit digital representation of molecular signals, such as delta-sigma modulation schemes~\cite[Sec. 12.3]{carlson2002communication}.
This approach could lead to digital-like biocomputers based on molecular analog-to-digital converters and low-bit-precision logic elements—such as adders and multipliers—implemented via chemical or biological reactions~\cite{angerbauer2024molecular,holt2020protease,huynh2013chemical,mokon2022computational,lau2021efficient}.

\section{Early Work on Explainable Neural Networks}
\label{sec_XAI}

Over the past few decades, {MC} has evolved from theoretical concepts to practical implementations, leveraging advances in nanotechnology and bioengineering \cite{lotter2023experimental,lotter2023experimental2}.
Concurrently, the integration of {AI} has transformed {MC} by enhancing the efficiency and robustness of signal processing, encoding, transmission, and decoding processes, as discussed in~\cite{huang2021signal}.
Despite these advancements, the complexity and opacity of {AI} models pose a significant barrier to their full realization in {MC} systems, as proof of correctness is lacking.
The directions presented in this section address these challenges by introducing methods to make the application of {NNs} transparent.
This section outlines the motivation behind this research direction, followed by a brief description of the fundamental tools, a summary of applied \ac{XAI} in the context of \ac{IoBNT} networks, and an illustrative code example.

%

\subsection{Motivation}

The application of {NNs} in {MC} has traditionally focused on optimizing various aspects of the communication process, see examples in \Cref{sec_communication}.
However, the ``black-box'' nature of these models often leads to a lack of transparency and interpretability.
Hence, the traditional application of {NN}s can be limiting in critical applications where understanding the decision-making process is crucial.
This is where \acf{XAI} comes into play, offering a suite of techniques designed to elucidate the inner workings of {AI} models, making them more transparent and comprehensible.

The motivation behind \ac{XAI} research in {MC} is twofold~\cite{huang2021signal}.
Firstly, there is a pressing need to bridge the gap between the sophisticated {NN} algorithms used in {MC} and the requirement for transparency and explainability, particularly in healthcare applications.
\ac{XAI} can help provide evidence supporting the correctness of deployed \ac{NN}-based models, thereby increasing confidence in their operation.
Moreover, \ac{XAI} can facilitate interdisciplinary collaboration by making {NN} models more accessible to researchers and practitioners across biology, chemistry, and engineering.
Interpreting the operation of {NN} modules is ultimately a translation of abstract neural weights into physically meaningful insights rooted in

%

\scriptsize
\begin{table*}[htbp]
  \centering
  \caption{Comparison of reported bio-computing technologies.}
  \label{tab_bio}%
    \begin{tabular}{p{0.8cm}p{0.4em}p{1.5cm}p{3em}p{2cm}p{1.5cm}p{1.7cm}p{1.7cm}p{1cm}p{1.2cm}}
      & \multicolumn{2}{l}{\mycell{\\Technology}} & \multicolumn{1}{c}{\mycell{\\Domain}} & {\mycell{Weight Density \\$[\si{\w\per\cubic\milli\meter\per\second}]$}} & \multicolumn{1}{c}{\mycell{\\Latency}} & \multicolumn{1}{c}{\mycell{Scalability\\ (approx.)}} & \mycell{Bio-\\compatibility} & \multicolumn{1}{l}{\mycell{\\TRL}} & \multicolumn{1}{l}{\mycell{\\Ref.}} \bigstrut[b]\\

\hline
\mycell{Digital\\ only} & \multicolumn{2}{l}{\multirow{1.2}[4]{*}{Graphene bio-FET}} & \multirow{1.2}[4]{*}{Silico} & \multirow{1.2}[4]{*}{\num{e5}} & \multirow{1.2}[4]{*}{milli seconds} & \multirow{1.2}[4]{*}{High}  & \multirow{1.2}[4]{*}{Medium} & \multirow{1.2}[4]{*}{TRL \num{6}-\num{7}} & \multirow{1.2}[4]{*}{\cite{kuscu2021fabrication},~\cite{abdali2024frequencydomain}} \bigstrut\\

\hline
\mycell{Digital\\\& Analog} & \multicolumn{2}{l}{\multirow{1.2}[4]{*}{\Acl{CRN}}} & \multirow{1.2}[4]{*}{Chemical} & \multirow{1.2}[4]{*}{\num{e-4}} & \multirow{1.2}[4]{*}{minutes} & \multirow{1.2}[4]{*}{Low-Medium} & \multirow{1.2}[4]{*}{Medium–high} & \multirow{1.2}[4]{*}{TRL \num{3}-\num{5}} & \mycell{\cite{pandi2019metabolic}~\cite{cherry2018scaling},\\ \cite{agiza2023digital}} \bigstrut\\

\hline
\multirow{5}[6]{=}{\mycell{Analog\\only}} & \multicolumn{2}{l}{Compartment-based} & Chemical & \num{e-1} & seconds & Medium-High & Medium–high & TRL \num{4}-\num{5} & \cite{angerbauer2024novel},~\cite{angerbauer2023molecular} \bigstrut\\

\cline{2-10}      & \multicolumn{1}{c}{\multirow{3.5}[4]{*}{\mycell{Wet\\ Neuromorphic\\ Systems}}} & DNA Strand Displacement & \multirow{1.2}[4]{*}{Chemical} & \multirow{1.2}[4]{*}{\num{e-4}} & \multirow{1.2}[4]{*}{hours} & \multirow{1.2}[4]{*}{Medium} & \multirow{1.2}[4]{*}{Very high} & \multirow{1.2}[4]{*}{TRL \num{3}-\num{4}} & \multirow{1.2}[4]{*}{\cite{qian2011neural}} \bigstrut\\

\cline{3-10}      &       & Bacteria Colony & \multirow{1.2}[4]{*}{Cellular} & \multirow{1.2}[4]{*}{\num{e-6}} & \multirow{1.2}[4]{*}{hours} & \multirow{1.2}[4]{*}{Medium} & \multirow{1.2}[4]{*}{Very high} & \multirow{1.2}[4]{*}{TRL \num{2}-\num{3}} & \multirow{1.2}[4]{*}{\cite{sarkar2021single},~\cite{li2021synthetic}}
\bigstrut\\
\bottomrule
    \end{tabular}%
\begin{tablenotes}
    \item Notes:
    \item $\bullet$ The "Computational Density" column denotes the areal concentration of {NN} weights that can be physically hosted per unit volume and evaluated per unit of time.
    To evaluate the computational density per entry, we followed the remarks below
    \begin{itemize}            
        \item \textbf{Graphene bio-FET:} In this case, we assume that a digital back-end circuit is attached to the graphene bio-FET, which serves as the interface between the molecular and electrical domains. 
        This digital circuit incorporates an \ac{ADC} for signal conversion and a microcontroller unit that hosts the on-device \ac{NN} model.
        Following TinyML solutions targeting healthcare applications, as reported in~\cite{diab2022embedded}, we selected the work in~\cite{faraone2020convolutionalrecurrent} as an illustrative high-complexity example.
        This solution deploys a \ac{CNN} of \num{197596} parameters within the  microcontroller nRF52832 of dimension \SI{3}{\milli\meter}\,\(\times\)\,\SI{3.2}{\milli\meter}~\,\(\times\)\,\SI{0.225}{\milli\meter}\cite{noauthor_nrf52832_nodate}.
        The reported processing time is $\SI{97.8}{\milli\second}$ \cite[Sec.IV.C]{faraone2020convolutionalrecurrent}, resulting in a parameter density of $
        \frac{197596}{\SI{3}{\milli\meter}\times \SI{3.2}{\milli\meter} \times \SI{0.225}{\milli\meter}\times \SI{97.8}{\milli\second}} \approx \SI{9.2}{\kelvin\w\per\cubic\milli\meter\per\second}$.
        \item \textbf{\Acl{CRN}:} The reference in \cite{agiza2023digital} reports three fully connected neurons to process input images of size $8 \times 8$, $12 \times 12$, $16 \times 16$, and $28 \times 28$, yielding a total of weights per image dimension as \numlist{64;144;256;784} (\cite[Tab. 2-41, Supplement]{agiza2023digital}).
        The reaction volume is reported as $2\times \SI{2.4}{\micro\liter}$ per weight, yielding a total volume per image dimension as $\SIlist{307.2;691.2;1228.8;3763,2}{\micro\liter}$.
        The reaction times per image dimension as $\SIlist{14;32;57;174}{\min}$.
        Following these values, the density per image dimension yields  $\SIlist{2.5e-4;1.1e-4;6.1e-4;2e-5}{\w\per\cubic\milli\meter\per\second}$.
        
        %
        \item \textbf{Compartment-based:} The contribution in \cite{angerbauer2024novel}, refers to a total of $6$ parameters ($3\times 2$ matrix multiplication in \cite[Eq. (71)]{angerbauer2024novel}) synthesized in a total volume of~$\SI{25}{\cubic\micro\meter}$ (see \cite[Tab. II]{angerbauer2024novel}).
        The latency to develop the matrix multiplication yields approximately $3\times \SI{0.4}{\second}=\SI{1.2}{\second}$, see the \cite[Fig. 14]{angerbauer2024novel}.
        This figures yield a density of $\frac{4}{\SI{20}{\cubic\milli\meter}\SI{1.2}{\second}}\approx \SI{0.16}{\w\per\cubic\milli\meter\per\second}$.
        Similar order of magnitude holds for the work in \cite{angerbauer2023molecular}, where the number of parameters is  $8$ \cite[Eq. (23)]{angerbauer2023molecular}, the reaction volume is $\SI{15}{\micro\liter}$ (\cite[Tab. II]{angerbauer2023molecular}) and latency is approximately $\SI{2.5}{\second}$ (\cite[Fig. 3]{angerbauer2023molecular}), yielding a density of  $\SI{0.2}{\w\per\cubic\milli\meter\per\second}$.
        \item \textbf{\ac{DNA} Strand Displacement:} We base our calculations on the experimental design of the four-neuron Hopfield associative network introduced in \cite[Fig. 3]{qian2011neural}.
        A total of \num{12} weights are synthesized with \ac{DNA} strands \cite[Supp. S4]{qian2011neural} in a reaction volume of $\SI{1}{\cubic\milli\meter}$.
        The circuit operates with a latency of $\SI{8}{\hour}$, see \cite[Fig. 3 d]{qian2011neural}.
        These figures yield a density of $\SI{4.2e-4}{\w\per\cubic\milli\meter\per\second}$.
        %
        \item \textbf{Bacteria Colony:} The contribution in \cite{sarkar2021single} develops a bacteria colony, where \num{9} weights are synthetized using \num{5} bacteria, see \cite[Fig. 5c]{sarkar2021single}.
        The processing time takes up to $\SI{16}{\hour}$, see \cite[Supp. So]{sarkar2021single}.
        Experiments are reported on \num{96}-well microplate, which we assume with the minimum work volume of$\SI{25}{\micro\liter}$, see \cite{noauthor_96well_nodate}.
        Following these figures, the calculations for the density yields $
        \frac{9}{5\times\SI{25}{\cubic\micro\meter} \times \SI{16}{\hour}} \approx \SI{1.3e-6}{\kelvin\w\per\cubic\milli\meter\per\second}$.
        A similar order of magnitude holds for the report in \cite{li2021synthetic}, where \num{8} weights are synthesized in a \num{96}-well microplate and the calculation span in the range $\SIrange{2}{5}{\hour}$.
        \item \textbf{Reservoir Computing:} These contributions develop a reservoir computing architecture.
        Following the contribution in \cite{uzun2025molecular}, the learning phase is given by the vector $\textbf{W}_\text{out}$ in \cite[Eq. 2]{uzun2025molecular}, which boils down to performing a weighted sum.
        As such, the density for this approach relies on the means to develop a vector multiplication through any of the listed entries in this table.
        Similar calculations account for the contribution in \cite{cai2023brain}.
        As such, we avoid listing Reservoir Computing as a separate row in the table.
        \item \textbf{Brain Organoids:} We avoid listing the calculations for the contribution in \cite{kagan2022invitro} as there isn't an explicit association of learnable coefficients to the connection among cells. 
        \item \textbf{GRN:} We are not listing \ac{GRN} as we couldn't find literature that follows an experimental design.   
        \end{itemize}
\end{tablenotes}
\end{table*}
\normalsize
%

\noindent
biology and chemistry, a process \textit{per se} that requires multidisciplinary expertise.
%

\subsection{Explainable and Interpretable Molecular Communication}

The interpretability of a model is limited by a balance between its complexity and transparency requirements \cite{adadi2018peeking}.
Models with fewer parameters and near-linearity become more interpretable, albeit with limited performance.
Performance and explainability are counteracting; balancing these two aspects poses significant challenges in designing high-performance methods for sensitive use cases.
Within the scope of this challenge, a variety of metrics are used to evaluate the effectiveness of explanations, as described next.

\subsubsection{Explainability Metrics}
The following metrics evaluate to what degree {NN}-driven predictions can be trusted and understood within the intricate dynamics of nanoscale interactions \cite{nauta2023from}:

\textbf{Fidelity:} Fidelity quantifies how accurately an explanation with a surrogate model reflects the behavior of the original model. 
A surrogate model serves to explain the primary model's predictions, and fidelity measures the alignment between them.
A high-fidelity explanation implies that the surrogate model accurately captures the core decision-making processes of the original one.
This is particularly crucial in {MC}, where understanding the molecular mechanisms that drive predictions can shed light on the interactions at the nanoscale level.

\textbf{Sparsity:} Sparsity reflects the simplicity of an explanation by quantifying the number of features involved.
Higher sparsity corresponds to fewer features being included, leading to more concise and interpretable explanations.
In the realm of {MC}, where interactions at the nanoscale level are inherently complex and multifaceted, sparse explanations are particularly valuable to identify the most relevant physical processes that influence predictions.

\textbf{Stability:} Stability measures the consistency of an explanation when the input is subject to small perturbations.
In {MC}, where environmental factors and molecular interactions are dynamic, stable explanations ensure that the insights provided by the {NN} model are reliable and robust across different scenarios.

\textbf{Causality:} Causality assesses whether the features identified in the explanation are genuinely responsible for the model's output.
In the context of {MC}, this involves identifying whether specific signaling molecules or interactions directly influence the predicted cellular responses.
This metric is particularly valuable for validating experimental designs and ensuring the reliability of insights in complex molecular networks.

\textbf{Comprehensibility:} Comprehensibility reflects how easily a human can interpret the explanation. 
Although difficult to quantify directly, it can be associated with the simplicity of the explanation.
For example, a shorter rule-based explanation with fewer conditions is generally easier to understand than a complex one.

%

\subsubsection{Explainability Techniques}

Benefiting from the above-mentioned metrics, there are several key techniques for providing explanations in {NN} models, as summarized below:

\textbf{Feature Attribution Methods:} These methods assign an importance score to each feature.
The most common techniques include \ac{LIME}, \ac{SHAP}, \ac{LRP}, and \ac{DeepLIFT} \cite{adadi2018peeking}.
\Ac{LIME} operates by perturbing the input data and observing the changes in prediction to create a local, interpretable model.
A weighted linear model is then trained with the perturbed data, aiming to approximate the original model; see details in \cite{ribeiro2016why}.

\Ac{SHAP} values are derived from cooperative game theory, providing a unique solution that allocates payoffs to players in the fairest manner (features).
The \ac{SHAP} value for a given feature is evaluated as in \cite[Eq. (4)]{lundberg2017unified}.
This evaluation considers all possible combinations of input features and computes the marginal contribution of each feature to the model’s prediction.
In essence, \ac{SHAP} estimates how much a particular feature, when added to different subsets of features, changes the prediction, and then averages these effects to yield a global contribution score.
This mechanism allows for a consistent and locally accurate explanation of feature influence.
We develop an illustrative example for this method in \Cref{sec_xai_code}.

\Ac{LRP} is a method designed to attribute an {NN}’s prediction back to its input features \cite{binder2016layerwise}.
This is achieved by backpropagating relevance from the output layer through the network to the input features, with relevance determined by distributing contributions in proportion to the weighted connections between neurons.
In the context of {MC}, \ac{LRP} helps to identify which nanoscale signals or features play the most critical roles in the prediction, enabling better interpretability of complex {NN} models.

\newpage
\Ac{DeepLIFT} extends \ac{LRP} by comparing the actual input to a reference input or baseline, quantifying the difference in model outputs; see further details in~\cite{shrikumar2017learning}.
By explicitly accounting for deviations from a baseline, \ac{DeepLIFT} is particularly useful in scenarios where the relative importance of nanoscale molecular changes can provide crucial insights into the system's dynamics.
For instance, in {MC} systems where slight variations in vesicle concentration, timing, or signal shape may signify different biological states or disease progressions, \ac{DeepLIFT} can highlight which input perturbations most significantly influence the model's prediction.
This enables researchers to trace model decisions back to specific molecular behaviors, such as an unexpected peak in a molecule's release rate or a delay in arrival time, thereby improving interpretability in AI-driven diagnostic or monitoring systems.
Unlike gradient-based methods that may suffer from vanishing signals, \ac{DeepLIFT} preserves contribution scores even in saturated or nonlinear regions, which are common in biochemical signaling cascades.

\textbf{Saliency Maps:} Saliency maps, commonly used in computer vision, highlight regions of an input image that are most important for a model's prediction \cite{mundhenk2019efficient}.
This can be helpful in {MC}, as it visually highlights the most influential features in the {NN} model's decision-making process.
This technique generates visual representations that show which areas of input data contribute most to the model's output, allowing researchers to quickly identify key molecular factors or patterns driving the communication process. 

\textbf{Counterfactual Explanations:} These methods provide insights by answering "what-if" questions \cite{wachter2017counterfactual}.
For a given input, a counterfactual explanation identifies the smallest input change such that the model output shifts to a target value.
Where molecular interactions dictate system responses, counterfactual explanations can help researchers understand the causal relationships between molecular signals and communication outcomes.
For instance, counterfactual explanations can reveal how these changes would affect the model’s prediction of system behavior by modifying the concentration of a particular molecule or altering a specific signaling pathway.
This technique enhances the interpretability of {AI} models by providing actionable insights into the conditions under which {MC} might behave differently, helping researchers better understand and control nanoscale interactions.

\textbf{Manual Permutation Importance:} The \ac{MPI} method is a model-agnostic technique used to quantify the contribution of individual features to a {ML} model's predictive accuracy.
It measures feature importance by randomly shuffling a specific feature's values and observing the resulting increase in model error, thereby effectively disrupting its relationship with the target variable.
This approach was first introduced in the context of Random Forests \cite{breiman2001random}, where it was used to assess variable significance by measuring the impact of permuting a variable on prediction accuracy.
For {MC}, where factors such as molecular diffusion, receptor binding, and molecular degradation introduce stochastic variability, explainability methods such as \ac{MPI} ensure that {AI}-driven models remain not only accurate but also biologically meaningful.
By identifying which features most significantly affect model performance, \ac{MPI} aids in understanding the underlying biological processes and enhances the interpretability of complex {MC} models. 
We develop an illustrative example for this method in \Cref{sec_xai_code}.

%

\subsection{Reported Research}

Traditional communication schemes in {MC} often struggle to evaluate complex end-to-end channel models and minimize the \ac{BER}.
Although reported {NN} models are a promising alternative to conventional detection methods (see those reported in \Cref{sec_communication}), a critical limitation is their inherent lack of transparency \cite{huang2021signal}.
As summarized below, research studies have begun to address these limitations, with a primary focus on feature-attribution methods to improve interpretability.
Existing approaches largely focus on PHY-layer symbol detection.
Published research work aims to provide a proof of correctness for \ac{NN}-based approaches by mapping them to conventional detectors, such as threshold, slope, or linear decoders, see e.g. \cite{li2023explainability,torres-gomez2023explainability,khanzadeh2024explainable,huang2025demystifying}.

\Ac{XAI} methods have been applied to elucidate the inner workings of {NN}-based symbol detectors for a $2\times2$ \ac{MIMO} {MC} channel in \cite{li2023explainability}.
To interpret the {NN}'s behavior, the authors employed \ac{XAI} techniques such as \ac{LIME}, \ac{PDP}, and the \ac{ICE} method.
Their application revealed that the trained {NN} effectively operates as a threshold detector and slope detector in the low and high \ac{ISI} regimes, respectively.
This association of the {NN} with a threshold or slope detector is the model's interpretation and also provides proof of the {NN}-based detector's correctness.

In a similar direction, researchers generated synthetic data from real testbed measurements to train an {NN} for binary symbol detection in \cite{torres-gomez2023explainability}.
The key objective was to demystify the ``black box'' nature of {NN}s and provide assurance of their correctness in symbol detection tasks.
To achieve this, they employed \ac{XAI} techniques, such as \ac{LIME} and \ac{ICE} plots.
The analysis revealed that the trained {NN}'s decision-making process closely mirrored that of standard peak and slope detectors in low and high \ac{ISI} regimes, respectively.
The {NN} behaved similarly to traditional methods used in {MC} for symbol detection based on signal features such as amplitude peaks and signal slopes.

Another notable development in {MC} is the integration of the {IoBNT} framework, where powerful external transmitters communicate with computationally constrained internal receivers.
To address this asymmetry, the authors of \cite{khanzadeh2024explainable} introduced an \ac{AAEC} architecture for end-to-end learning in {MC} systems and analyzed the transmitter using a surrogate model, showing that it behaves similarly to a zero-forcing precoder in low and moderate \ac{ISI} regimes.
This correspondence enables anticipation of the {NN}'s behavior using well-known linear precoding principles while improving the transparency of the learned transmitter.


By integrating \ac{SHAP} into {MC} detector design, the authors in~\cite{huang2025demystifying} identified crucial feature points in the received molecular signal.
\ac{SHAP}-driven analysis highlights the most informative segments of molecular signal waveforms, providing a systematic explanation of how different {NN} models arrive at detection decisions.
Previous work in {MC} has demonstrated the potential of combining model-based methods with {ML} to compensate for unknown channel parameters; however, the inherent opacity of deep NNs remains a persistent barrier to practical deployment.
This approach maintains the adaptability and high accuracy of data-driven detectors while enhancing explainability, ultimately facilitating safer and more reliable MC applications in areas such as targeted drug delivery and \textit{in vivo} biochemical monitoring.

%

\subsection{Illustrative Code Example and Results}
\label{sec_xai_code}

As a code example, we use the model already developed in \Cref{sec_code_dist}, which employs a \num{2}-node {NN} to estimate the distance between immune and cancer cells.
The model first extracts two features from the number of vesicles in immune cells: the amplitude and the time coordinate (peak time) of the minimum slope in this sequence; see the block diagram in \Cref{fig_cell2cell}.
The {NN} inputs are these two features, and the output is the predicted distance, as illustrated in \Cref{fig_cell2cell}c.
Following our previous work in \cite{basaran2025xai-enhanced}, we implemented the \ac{MPI} method to evaluate the significance of each feature for the {NN}'s output, and also included the \ac{SHAP} value for further illustration.

\begin{figure}
    \centering
    \subfloat[Permutation importance.]
    {\includegraphics[width=0.6\columnwidth]{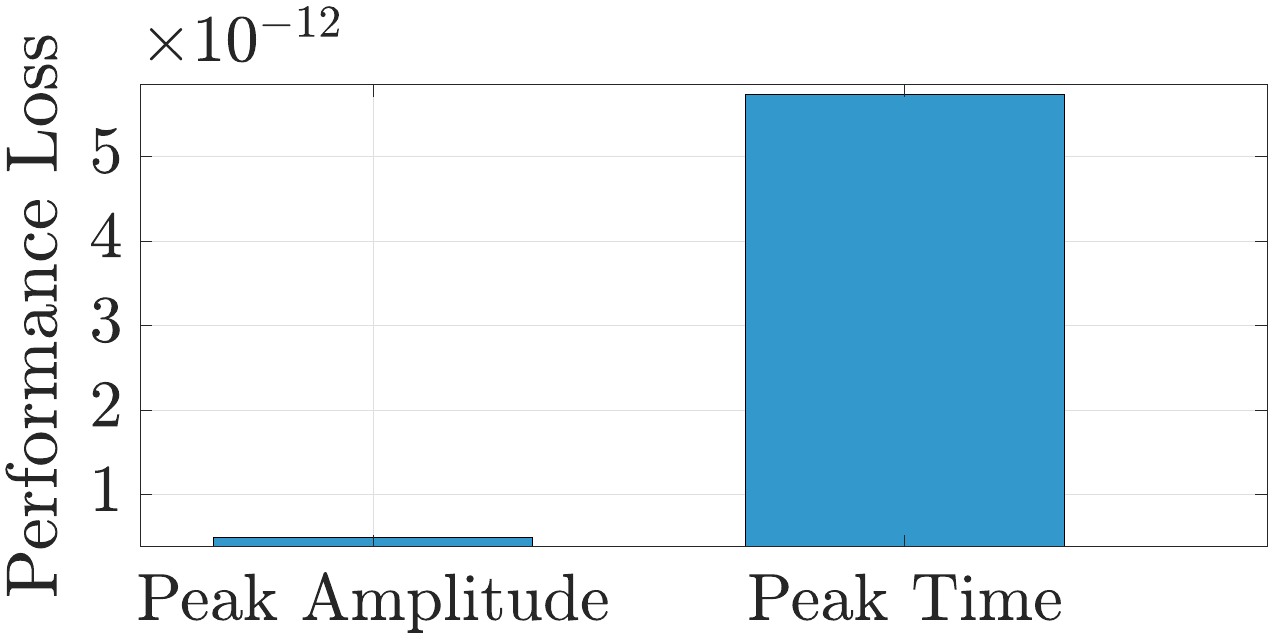}\label{fig_xaia}}
    \vspace{2em}
    \subfloat[Shapley summary.]{\includegraphics[width=0.75\columnwidth]{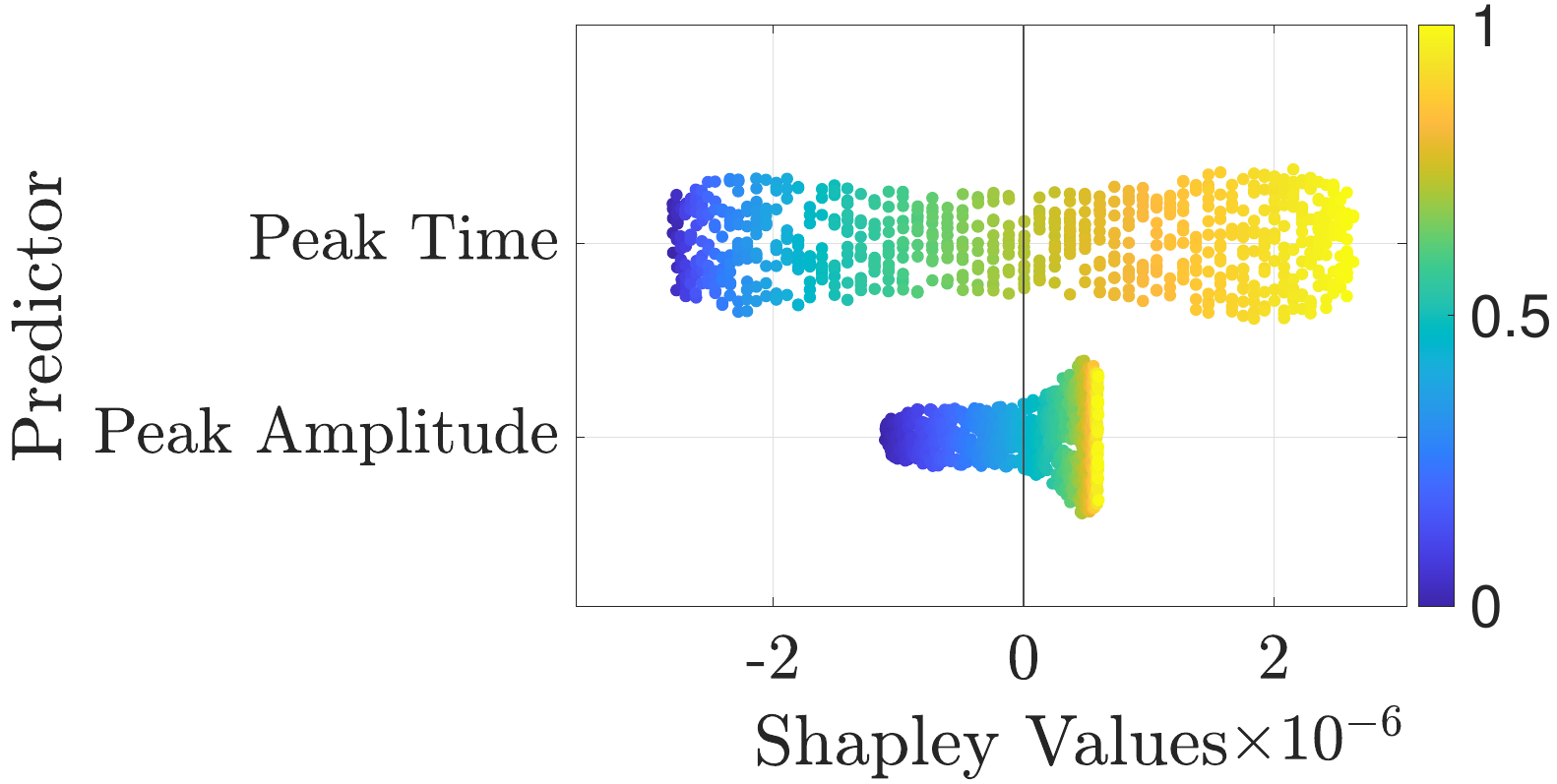}\label{fig_xaib}}
    \caption{Feature importance analysis using \acl{MIP}.}
    \label{fig_xai}    
\end{figure}

The results of applying the \ac{MPI} method are illustrated in \Cref{fig_xaia}.
The plot highlights the dominant role of the peak position, resulting in a substantially larger performance loss than the peak-amplitude feature.
This indicates a stronger correlation with distance and suggests that the peak time encodes key characteristics of vesicle-mediated communication.
Complementing this, the \ac{SHAP} analysis in \Cref{fig_xaib} shows that peak time spans a broader range of \ac{SHAP} values---reflecting both stronger and more variable influence on predictions---whereas peak amplitude remains tightly centered around zero.
This behavior aligns with the physics of free-diffusion channels, where the travel time encoded in the peak position follows the relation $t_\mathrm{peak} \propto d^2 / D$~\cite[below Eq.~(2.8)]{berg1993random}, where $d$ is the distance between cells and $D$ is the diffusion coefficient of the released vesicles.
Overall, the \ac{NN} architecture effectively focuses on peak timing to capture this fundamental relationship between peak time and distance.

\subsection{Concluding Remarks and Outlook}

The transparent operation of technologies is crucial in sensitive applications such as healthcare and medicine, where understanding the inner workings of deployed {NNs} may determine their adoption within the {IoBNT} framework.
Currently, this aspect of {NNs} in {IoBNT} is explored less than their functional applications.

As summarized in \Cref{tab_XAI}, the literature remains scarce and focuses on a limited set of {NN} architectures, mostly feedforward {NN}s.
Preliminary research interprets the operation of {NNs} using classical methods, including threshold, slope detectors, and zero-forcing precoders.
In the context of {MC}, \ac{XAI} remains nascent, and the interpretability of NN modules across other layers of {IoBNT} networks remains an open research challenge.


\footnotesize
\begin{table*}[htbp]
  \centering
  \caption{Summary of \acs{XAI} research in molecular communication.}
  \label{tab_XAI}%
    \begin{tabular}{p{6em}p{1.5cm}p{3cm}p{8em}p{6cm}p{0.7cm}}
\toprule
XAI Method & {NN}-model & {MC}-Geometry & Focus & \mycell{Description \& Key findings} & Ref. \bigstrut[b]\\

\hline
\multirow{2}[6]{=}{LIME, ICE} & \multirow{7}[8]{=}{\rotatebox[origin=c]{90}{\mycell{\\Feedforward {NN}}}} & Air-based free diffusion channel & \multirow{3}[6]{=}{Symbol detection} & \multirow{3.}[6]{=}{Interpret the {NN} operation as a receiver based on the identification of the most critical samples in the received sequence over {MC} channels. As a result, the \ac{NN} is interpreted to operate as a peak detector or as a slope detector based on the \ac{ISI} level of transmissions.} & \multirow{1.2}[4]{*}{\cite{torres-gomez2023explainability}} \bigstrut\\

\cline{3-3}
\cline{6-6}
\multicolumn{1}{c}{} & \multicolumn{1}{c}{} & $2\times 2$ \acs{MIMO} in free diffusion channel & \multicolumn{1}{l}{} & \multicolumn{1}{l}{} & \multirow{1.2}[4]{*}{\cite{li2023explainability}} \bigstrut\\

\cline{1-1}\cline{3-3}\cline{6-6} \multirow{1}[2]{=}{ \acs{SHAP}} & \multicolumn{1}{c}{} & Point transmitter - Free diffusion - Spherical receiver & \multicolumn{1}{l}{} & \multicolumn{1}{l}{} & \multirow{1.2}[4]{*}{\cite{huang2025demystifying}} \bigstrut\\

\cline{1-1} \cline{3-6}
\multirow{3}[2]{=}{ \acs{SHAP}, \acs{MPI}} &  & \multirow{2.2}[4]{*}{\mycell{Extracellular\\matrix}} & \multirow{2}[6]{=}{Distance estimation} & Quantifies the contribution of peak amplitude and timing features to the predicted cell-to-cell distance. The {NN} is interpreted as an estimator based on the peak time of received pulses. & \multirow{2}[4]{*}{\cite{basaran2025xai-enhanced}} \bigstrut\\

\hline
\mycell{Linear\\modeling} & \multirow{1.2}[4]{*}{Autoencoder} & \multirow{1.2}[4]{*}{Drifted channels} & End-to-end learning of {MC} channels & Develop a surrogate model to interpret the transmitter's operation as a zero-forcing precoder. & \multirow{1.2}[4]{*}{\cite{khanzadeh2024explainable}} \bigstrut\\

\bottomrule
    \end{tabular}%
\end{table*}%
\normalsize


Lastly, a particularly promising direction is to incorporate physical laws directly into the training process via \acp{PINN}.
Unlike purely data-driven approaches, \acp{PINN} can embed the governing \acp{PDE} of molecular transport (e.g., Fick’s diffusion law, reaction-diffusion kinetics, or fluid dynamics in vesicle-mediated signaling) as additional loss terms within the training objective.
Not only do these models exhibit better generalization to previously unseen conditions, but they also offer greater interpretability of the solution space, as latent layers must satisfy fundamental conservation laws and boundary conditions intrinsic to molecular transport.

\section{The Backbone of Neural Networks: Training Data}
\label{sec_data}

Having discussed the modeling approaches and learning methodologies for {MC} systems in the previous sections, it becomes apparent that their performance critically depends on the availability and quality of data. 
Data is the backbone of {NN} training, and this section provides an overview of existing {MC} datasets.
We include datasets that have not yet been exploited in the context of {NNs}.
Furthermore, the datasets are discussed with respect to their generation and accessibility, and some remarks are made on their documentation. 
Finally, current limitations and future perspectives on synthetic data generation are outlined.

Many datasets, particularly those derived from simulations and published in papers, i.e., data used for plotting, are not publicly available. 
This lack of accessibility undermines the reproducibility and validation of the findings, limiting the research community's ability to build on existing research. 
By focusing on publicly available and accessible datasets, such as \cite{tuccitto2023dataset,tucitto2023dataset_zenodo,hofmann2023dataset_macroscale}, we aim to provide a comprehensive summary that researchers across domains can use.

%

\subsection{General Considerations}

Data can be generated based on observations of physical or virtual processes, such as wet-lab experiments or computer simulations.
Data generation can be time-consuming and resource-intensive.
Avoiding repeating the process, datasets shared among different research groups can {(i)} facilitate reproducible research and {(ii)} provide an abstraction layer for the actual process of interest, e.g., {MC} between synthetic cells, that allows other researchers to develop models and algorithms without the need to (re-)run and observe the process.

Both these goals hinge on the {\em availability} and the {\em usability} of the datasets.
Also, a thorough {\em documentation} of the shared data is required.
These aspects, which relate primarily to data and metadata handling, will be covered later in this section.
However, in determining which data to collect in the first place, the properties of the process that generates the data and the primary purpose of data collection play a key role.

The purpose of collecting data can be to validate a specific theoretical model.
In this case, it is essential to control the experimental conditions as best as possible and obtain a representative set of data samples under these conditions.
The main focus of data collection ito confirm or reject a specific, oftenas is the case, e.g., for {NN} training quantitative, hypothesis on the process, e.g., the average received \ac{SNR}, that originates from the theoretical model.\footnote{The theoretical model can also be an {NN}, in which case the hypothesis results from the inference step.}
On the other hand, data can also be used to build a model, as in {NN} training.
In this case, data should be collected to avoid training-set imbalances and overfitting to a narrow subset of experimental conditions.

Training data can be distinguished into two types: {(i)} Synthetic and {(ii)} experimental data.
In this survey, synthetic data refers to data artificially generated through algorithms or simulations that mimic real-world scenarios.
Experimental data is collected from experimental samples, alongside the challenges of building environments in which various parameters can be precisely controlled.

%

\subsection{Synthetic Data Generation}

\subsubsection{Existing Datasets}
\label{subsubsec:data}

Existing datasets primarily address two environments: Pipe-like and free-diffusion channels in fluid and air-based media.
For pipe-like channels, the authors in~\cite{hofmann2024openfoam} introduce a cylindrical water-filled duct designed to study steady-state flow conditions across various channel lengths.
The authors benchmark their simulation results against \ac{PBS} and an analytical channel modeling approach, as accessible in~\cite{hofmann2023dataset_simulation}.

The above dataset has also been extended in~\cite{zhou2024github} to cover not only the flow-dominated regime but also the dispersion and the mixed regimes.
Furthermore, targeting healthcare applications for advanced plaque modeling~\cite{wietfeld2024advanced}, the work in~\cite{hofmann2024molecular} develops a dataset accessible in \cite{hofmann2024dataset_plaque}, using the \ac{CFD} solver OpenFOAM.
A previously published dataset in~\cite{hofmann2023dataset_simulation} provides a framework for {MC} specific post-processing, i.e., utilizing Python to analyze the OpenFOAM simulation output and evaluate the {CIR} of the \ac{MC} channel.

The dataset in~\cite{gulec2024_zenodo} includes results from \ac{CFD} simulations performed using the proprietary ANSYS software to study the transmission of airborne pathogens in turbulent airflow environments. 
The simulations examine the spread of pathogen-laden droplets and aerosols, ``mimicking scenarios such as coughing-induced flows''~\cite{gulec2024_zenodo}.
An overview and summary of all simulation results, along with a detailed explanation, can be found in~\cite{gulec2023computational}.

While the datasets discussed so far mostly contain data from fluid dynamics simulations, some other datasets reproduce time sequences as a result of processing algorithms.
This is the case in~\cite{gulec2020github}, where a macroscale \ac{MC} system—using an electric alcohol sprayer and sensor without forced airflow—is provided as Matlab files and raw measurement data.
The estimated distance is produced by {ML}-based methods, see~\cite{gulec2020distance}, and the Fluid Dynamics-based Distance Estimation (FDDE) algorithm, see~\cite{gulec2021fluid}.

The dataset in~\cite{gao2023dataset} includes, in addition to the dataset itself, the code for creating it in Matlab, which is published on a university server, making the results from \cite{gao2021molecular,gao2023typespread} accessible to other researchers.
The authors of \cite{gao2021molecular,gao2023typespread} address the development of multiple-access schemes and detection strategies for diffusion-based {MC}, focusing on mitigating \ac{ISI} and \ac{MAI}.
Furthermore, \cite{gao2021molecular} and \cite{gao2023typespread} introduce the type-spread \ac{MoSK} modulation technique, which uses additional molecule types to reduce the \ac{ISI} effect.
Type-spread \ac{MoSK} is expanded for a multiple-access system to support multiple nanomachines.
In addition, a molecular code-division multiple-access scheme is proposed in~\cite{gao2021molecular,gao2023typespread}, relying on two molecule types.

Published datasets can also describe the parameters used in a publication~\cite{das2022dataset}.
For example, in~\cite{das2022received}, an improved channel model for {MC} with a partially absorbing receiver is proposed.
The partially absorbing receiver has four parameters determined by \ac{PSO}.
The optimum of the determined parameters, as mentioned in~\cite{das2022received}, can be found in the corresponding dataset in~\cite{das2022dataset}.


%

\subsubsection{Dataset Generation Tools}

In addition to the published datasets, simulators and software tools are also available to generate synthetic datasets based on users' requirements and system model parameters. 
Generally, in some {MC} simulators, the medium flow is simulated explicitly, whereas other tools use analytical models to approximate or neglect it~\cite{hamidovic2024microfluidic}.
The latter set of tools focuses mainly on Monte Carlo simulation and is referred to as {\em flow-agnostic simulators}.
In contrast, the physical fluid flow characteristics are considered in the former one, referred to as {\em flow-aware simulators}.
Most of the tools are publicly available, except for the commercial simulators COMSOL Multiphysics, ANSYS Fluent, and Matlab, for which a license is required to utilize the full range of available functions. 
Further requirements also arise for the {MC} \ac{GPU} simulator (also known as ``parallel simulation framework for nanonetworking''~\cite{gokarslan2020github}), as it requires an Nvidia \ac{GPU} supporting the Nvidia \ac{CUDA}~\cite{gokarslan2020github}.

The flow-agnostic simulators encompass tools designed for various applications, such as bacterial {MC} (nanoNS3~\cite{jian2017nanons3}, BNSim~\cite{wei2013efficient}), reaction-transport modeling (URDME~\cite{drawert2012urdme}), and Brownian motion-based simulations (Smoldyn~\cite{andrews2010detailed,andrews2016smoldyn}, N3Sim~\cite{llatser2011exploring,llatser2014n3sim}, AcCoRD~\cite{noel2017simulating}). 
Diverse computational frameworks support these simulators, including Matlab, COMSOL, and NS3.
In contrast, fluid dynamics simulators used in the context of {MC}, such as OpenFOAM, COMSOL Multiphysics, and ANSYS Fluent, are primarily based on the concept of \ac{CFD} and numerical solutions of the Navier-Stokes equation in general.
Availability varies significantly, with several simulators offering freely accessible source codes (e.g., OpenFOAM, Munich microfluidic toolkit, and NS3). 
In contrast, others, like ANSYS Fluent and COMSOL Multiphysics, are commercially licensed. 
This highlights the trade-off between cost and accessibility across the simulators, with open-source tools typically targeting niche research applications and commercially available software catering to broader engineering applications.

In addition to the simulators, other existing analytical simulation codes, data generation codes, and models have been published on GitHub as repositories for code development.
This is the case in~\cite{qian2020github}, which models a free-diffusion {MC} channel and implements an {NN}-based detector.
Three other GitHub projects have been created as part of a lecture series in which the results in~\cite{qian2019molecular} were reproduced.\footnote{These code projects are
{(i)} Sangani~\cite{sangani2021github}, {(ii)} Patel~\cite{patel2021github} (also improved the code in terms of the {CIR} of a passive receiver or enzymatic degradation), and {(iii)} Shastri in~\cite{shastri2020github}.}
Furthermore, a trained \ac{ANN} model is published on Matlab Central~\cite{birkan2017ANN} to evaluate the number of received molecules for a spherical reflecting transmitter and a spherical absorbing receiver using an \ac{ANN} approach~\cite{yilmaz2017machine}.

Matlab code is also developed to simulate the random walk and the signal reconstruction process at the receiver in free-diffusion channels \cite{gulec2018github,atakan2019signal}.
Introducing the \ac{MoHANET} concept in which {MC} principles and analysis are applied to pathogen-laden droplets, code developments also encompass airborne pathogen transmission, integrating insights from epidemiology, biology, medicine, and fluid dynamics~\cite{gulec2022github,gulec2022mobile}.
These code files represent proof-of-concept results validated using empirical COVID-19 data from the reference~\cite{dong2020interactive}.
Finally, the published code also accounts for the modeling of a stochastic biofilm formation model based on bacterial \ac{QS}~\cite{gulec2023stochastic,gulec2023stochastic2,gulec2022github_stochastic,gulec2022github_stochastic2}.

The code in \cite{zhang2024informer_github} includes an {NN} model to realize a signal sequence detector for a mobile {MC} system based on the Informer model in~\cite{cheng2024informer}.
Considering a diffusion-based environment, the mobile {MC} system comprises a point transmitter and a spherical passive receiver~\cite{cheng2024informer}.
The signal sequence detector computes the autocorrelation coefficient of the input sequence to determine the optimal sequence length. 
Numerically obtained results in~\cite{cheng2024informer} demonstrate that the performance of the Informer-based model is better than that of the Transformer-based model in terms of detection ability.
However, the dataset is not publicly available.

In~\cite{scazzoli2024molecular}, a ``model for generating synthetic data for a biological {MC}'' for training an {NN} for the discrimination of transmitted bits is presented. 
Testbed measurement data from~\cite{grebenstein2018biological} is used as a benchmark for the synthetically generated data.
The corresponding dataset is available upon request.
Table~\ref{tab_MC_simulators} reviews the existing dataset generation tools. To assess their accessibility, we introduce three complementary metrics that characterize code availability and their accesbility; see table notes. 

%

\subsection{Experimental Data Generation}

Experimental data generation from testbeds can generally be distinguished into air-based~\cite{hofmann2023dataset_analog,hofmann2023dataset_macroscale} and liquid-based~\cite{tuccitto2023dataset,tucitto2023dataset_zenodo,bartunik2023dataset_channel,cali2022interfacial,cali2022dataset_interfacial,abbaszadeh2019dataset_signal,abbaszadeh2020dataset_signal,grebenstein2019dataset_molecular,walter2023dataset_realtime,scherer2025closedloop_github,scherer2025closedloop_zenodo,vakalipoor2025cam_data} data.
As listed in~\cite{lotter2023experimental}, there are more testbeds than those mentioned here, but, to the best of the authors' knowledge, no data were made publicly available for other testbeds. 

The datasets published on IEEE Dataport~\cite{tuccitto2023dataset} and Zenodo~\cite{tucitto2023dataset_zenodo} include fluorescence signal measurements (emission wavelength and intensities) of \ac{MoSK} transmissions in a liquid-based testbed, which was proposed in the related publication in~\cite{cali2024experimental}. 
In the testbed setup, \acp{GQD}, soluble in water and fluorescent, serves as a signaling molecule.
The transmitter infuses the \acp{GQD} using injection valves, and a fluorescence-based receiver detects and decodes the fluorescence signals from blue-\acp{GQD} and cyan-\acp{GQD}, which serve as the molecules to shift. 
The testbed's performance is assessed based on synchronization, detection thresholds, and symbol recognition through a \ac{PCA}, which requires a broad dataset. 

The dataset in~\cite{hofmann2023dataset_macroscale} provides experimental measurements from a macroscale, single-input, single-output, air-based {MC} testbed.
It also includes the Matlab processing code of the dataset-related publication in~\cite{torres-gomez2023explainability}.
The dataset has been reported in various research studies in the literature, including \ac{XAI} in {MC} channels \cite{torres-gomez2023explainability}, \ac{RL}-based synchronization mechanisms \cite{debus2023reinforcement,debus2024synchronized}, and adaptive detectors \cite{hofmann2022testbed}.
The dataset in~\cite{hofmann2023dataset_macroscale} was extended as a result of the research in~\cite{hofmann2023analog}. 
The new associated repository, accessible in~\cite{hofmann2023dataset_analog}, records the experimental measurement data for ethanol molecules in the air.
It also provides data sheets for the testbed components and Python code to control the sprayer, read the sensor output, and implement communication protocols.

The dataset in~\cite{bartunik2023dataset_channel} publishes a biocompatible {MC} testbed that utilizes magnetic nanoparticles as information carriers, so-called \acp{SPION}.
The \acp{SPION} are injected into a constant background flow using an injection pump, cf.~\cite[Fig.~1]{bartunik2023channel}.
The receiver detects the \acp{SPION} by a change in the inductance of the nearby fluid caused by the presence of the \acp{SPION}. 
This testbed is utilized to investigate channel parameters such as (among others) background flow, channel length, and channel diameter. 
It should be emphasized that~\cite{bartunik2023channel} discusses the dataset in detail and explains its structure, which is a limitation in other published datasets. 

\footnotesize
\begin{table*}[ht]
\newcommand{\green}{\textcolor{mygreen}{\ding{108}}} 
\newcommand{\yellow}{\textcolor{myyellow}{\ding{108}}} 
\newcommand{\red}{\textcolor{myred}{\ding{108}}} 
\centering
\caption{Overview of data-generating simulators, alphabetically sorted. URDME and MMFT denote the Unstructured Reaction-Diffusion Master Equation and the Munich MicroFluidic Toolkit, respectively. }
\label{tab_MC_simulators}
\begin{tabular}{p{1cm} p{4cm} p{4cm} p{2.5cm} p{1.2cm} p{1.2cm} p{1.2cm}}
\toprule
\textbf{Category} & \textbf{Simulator and dependence} & \textbf{Application area} & \textbf{Source code} & \textbf{Open-source} & \textbf{Availability} & \textbf{Accessibility}  \\ \hline

\multirow{27}{=}{\rotatebox[origin=c]{90}{Flow agnostic}} & \multirow{2}{=}{AcCoRD~\cite{noel2017simulating}}  & Microscopic and mesoscopic diffusion-based MC & \multirow{2}{=}{GitHub~\cite{noel2020arcord}} & \multirow{2}{=}{\green} & \multirow{2}{=}{\green} & \multirow{2}{=}{\green} \\
                                \cline{2-7}
                                & \multirow{2}{=}{BiNS~\cite{felicetti2012simulation} \& BiNS2~\cite{felicetti2013simulation}} & Diffusion-based MC in blood vessels &  & \multirow{2}{=}{\red} & \multirow{2}{=}{\yellow} & \multirow{2}{=}{\red} \\
                                \cline{2-7} 
                                & BioNetGen~\cite{harris2016bionetgen} & Biochemical systems & Website/manual \cite{bionetgen} & \green & \green & \green \\ \cline{2-7} 
                                & \mycell{blood-voyager-s~\cite{geyer2018bloodvoyagers},\\BVS-Vis~\cite{geyer2020bvs-vis},\\MEHLISSA~\cite{wendt2020mehlissa}} & Moving objects in human body & GitHub, including blood-voyager-s~\cite{wendt2020github_blood}, BVS-Vis~\cite{wendt2020github_vis}, MEHLISSA~\cite{wendt2024github_mehlissa}, & \multirow{5}{=}{\green} & \multirow{5}{=}{\green} & \multirow{5}{=}{\green} \\
                                & BVS-Net~\cite{ebner2024bvs-net} (ns-3 extensions) &  & BVS-Net~\cite{wendt2019github_net} & & & \\
                                \cline{2-7}
                                & BNSim~\cite{wei2013efficient} & Bacterial networks & Website \cite{BNSim} & \multirow{1}{=}{\green} & \multirow{1}{=}{\green} & \multirow{1}{=}{\red} \\
                                \cline{2-7} 
                                & MC GPU Simulator (Nvidia GPU supporting CUDA) & Diffusion-based MC & GitHub~\cite{gokarslan2020github} & \multirow{2}{=}{\green} & \multirow{2}{=}{\green} & \multirow{2}{=}{\green} \\
                                \cline{2-7}
                                & MesoRD~\cite{hattne2005stochastic} (visualising results in Matlab) & Reaction-diffusion simulations & SourceForge~\cite{hattne2005meso} & \multirow{2}{=}{\green} & \multirow{2}{=}{\green} & \multirow{2}{=}{\green} \\ \cline{2-7} 
                                & MolComSim & Simple active and passive MC & GitHub~\cite{morgan2015} & \green & \green & \green \\
                                \cline{2-7}
                                & \multirow{2}{=}{MUCIN~\cite{yilmaz2014simulation} (Matlab extension)} & Diffusion-based MC with drift & Matlab Central File Exchange~\cite{Birkan2024matlab} & \multirow{2}{=}{\yellow} & \multirow{2}{=}{\yellow} & \multirow{2}{=}{\yellow} \\ 
                                \cline{2-7} 
                                & Multicellular MC Simulator~\cite{saiki2022design,saiki2025general} & (Large scale) multicellular MC scenarios & GitHub~\cite{saiki2025general_github} & \multirow{2}{=}{\green} & \multirow{2}{=}{\green} & \multirow{2}{=}{\green} \\
                                \cline{2-7}
                                & nanoNS3~\cite{jian2017nanons3} (ns-3 extension) & Bacterial \ac{MC} networks & Download \cite{Download} & \green & \green & \green \\
                                \cline{2-7} 
                                & N3Sim~\cite{llatser2011exploring,llatser2014n3sim} & Diffusion-based MC & Website \cite{N3Sim} & \yellow & \yellow & \red \\ \cline{2-7} 
                                & N$^4$Sim~\cite{turgut2022n4sim} & Nervous systems and synaptic MC & GitHub~\cite{turgut2020n4sim} & \green & \green & \green \\ \cline{2-7}
                                & \multirow{3}{=}{Smartcell~\cite{ander2004smartcell}}  & \multirow{3}{=}{Cellular processes} & Available upon email request; initial download link not available & \multirow{3}{=}{\green} & \multirow{3}{=}{\yellow} & \multirow{3}{=}{\yellow} \\
                                \cline{2-7}
                                & \multirow{2}{=}{Smoldyn~\cite{andrews2010detailed}} & Diffusion, membrane interactions, and molecule reactions & Website \cite{smoldyn} & \multirow{2}{=}{\green} & \multirow{2}{=}{\green} & \multirow{2}{=}{\green} \\
                                \cline{2-7} 
                                & URDME~\cite{drawert2012urdme} (including Matlab and COMSOL interface) & Reaction-transport simulation and modeling  & GitHub~\cite{bauer2024github} & \multirow{2}{=}{\green} & \multirow{2}{=}{\green} & \multirow{2}{=}{\green} \\ 
                                 \hline
\multirow{8}{=}{\rotatebox[origin=c]{90}{Flow aware}} & \multirow{1}{=}{ANSYS Fluent}  & Wide range of fluidic scenarios & & \multirow{1}{=}{\red} & \multirow{1}{=}{\yellow} & \multirow{1}{=}{\yellow} \\                                         \cline{2-7} 
                                & \multirow{1}{=}{COMSOL Multiphysics} & Wide range of fluidic scenarios &  & \multirow{1}{=}{\red} & \multirow{1}{=}{\yellow} & \multirow{1}{=}{\yellow} \\
                                \cline{2-7} 
                                & Droplet-Based Microfluidic Simulator~\cite{grimmer2019advanced,fink2022mmft} (MMFT extension) & Droplet-based fluidic MC & Github~\cite{grimmer2019_github} & \multirow{2}{=}{\green} & \multirow{2}{=}{\green} & \multirow{2}{=}{\green} \\                                
                                \cline{2-7}
                                & \multirow{2}{=}{Matlab} & PDEs, CFD, and particle tracking for MC and fluidic scenarios & Commercially available & \multirow{2}{=}{\red} & \multirow{2}{=}{\yellow} & \multirow{2}{=}{\yellow} \\
                                \cline{2-7} 
                                & \multirow{1}{=}{OpenFOAM (Pogona~\cite{drees2020efficient,stratmann2021using})} & Wide range of fluidic scenarios & Pogona- GitHub~\cite{drees2020_github_pogona} & \multirow{1}{=}{\green} & \multirow{1}{=}{\green} & \multirow{1}{=}{\green} \\ 
                                \bottomrule
\end{tabular}%
\begin{tablenotes}
\item Notes:
\item $\bullet$ The first metric, open‑source status, assesses whether the underlying source code is publicly accessible and modifiable. Tools that are fully open‑source receive a green rating, partially accessible or account‑restricted tools receive a yellow rating, and closed‑source tools whose source code is not publicly released receive a red rating. 
\item $\bullet$ The second metric, availability, evaluates how easily a simulator can be obtained through formal release channels. Freely downloadable or straightforwardly purchasable tools receive a green rating, whereas tools requiring registration, institutional affiliation, or academic request are marked yellow. Tools that are no longer supported or unavailable receive a red rating. 
\item $\bullet$ Finally, the accessibility metric captures the hurdles users must overcome to download or install the software. Simulators offering direct, barrier‑free downloads receive a green rating; those requiring additional steps, such as email requests or account creation, receive a yellow rating; and those with broken links, discontinued repositories, or inaccessible downloads receive a red rating. 
\end{tablenotes}
\end{table*}
\normalsize

The received sequence of \ac{ISK} transmissions in a liquid-based testbed is accessible on IEEE DataPort~\cite{cali2022interfacial} and Zenodo~\cite{cali2022dataset_interfacial}.
In \ac{ISK}, the modulation of the signal exploits the effect of viscosity fingering, i.e., two miscible fluids form a (temporary) interface, given that they differ in either their viscosity or density~\cite{cali2023interfacial}.
The testbed for experimental evaluation consists of a transmitter containing an infusion pump, a six-way injection valve, and a \num{10}-way selection valve, which allows the injection of up to ten solutions~\cite[Fig~1]{cali2023interfacial}.
Fluorescent carbon nanoparticles are used as information carriers.
On the receiver side, a fluorescence detector measures the system's fluorescence output, demodulating and decoding the transmitted signal.

Images obtained from the \ac{PIV} and \ac{PLIF} tools, which are referred to for tracking and detecting fluorescent tracers in liquid, are accessible in~\cite{abbaszadeh2020dataset_signal,abbaszadeh2019dataset_signal}.
The dataset results from the research in~\cite{abbaszadeh2020molecular}, which develops two methodologies for particle tracking and detection in liquids. 
Both methods are based on a laser sheet illuminating a planar section of the medium~\cite[Figs.~2 and 3]{abbaszadeh2020molecular}, where fluorescent tracers, captured by a camera, serve as information carriers.
The repository also includes Matlab code for further image processing to track and detect the fluorescent tracers.

The open-access dataset in \cite{grebenstein2019dataset_molecular} contains experimental measurements from a biological {MC} testbed \cite{grebenstein2019molecular}.
The testbed utilizes \textit{E. Coli} bacteria, which ``express the light-driven proton pump \textit{gloeorhodopsin} from \textit{Gloeobacter violaceus}.''
When stimulated by external light, the bacteria act as transmitters, releasing protons into a liquid channel. 
The protons serve as signaling molecules, altering the system's pH, which is later detected.
The repository also provides a detailed description of the dataset and comprises two zip files containing the data along with Matlab code example for processing.

In~\cite{walter2023real}, a liquid-based microfluidic {MC} testbed is presented.
The source files for~\cite{walter2023real} are accessible in \cite{walter2023github,walter2023dataset_realtime}.
For transmitting the information, \ac{CSK} is used; in particular, information is encoded in the concentration of sodium hydroxide in that testbed.
In addition, the testbed models chemical reactions and microfluidic geometry, enabling the transmitter to shape the signal and the receiver to threshold, amplify, and detect it after propagation~\cite{walter2023real}.
Therefore, the chemical reactions are based on well-known pH-based reactants, such as hydrochloric acid and sodium chloride~\cite{walter2023real}. 
A phosphate-buffered saline solution is used for dilutions, and a spectrometer on the receiver side detects the transmitted information. 
The raw data (mainly as \texttt{csv} files) in~\cite{walter2023dataset_realtime} refers to the plotted results in~\cite{walter2023real}.
The GitHub repository in~\cite{walter2023github} contains the software for the complete automation of the testbed, including timed chemical injection using syringe pumps, measurement of the flow rate, and control of the flow rate using a proportional–integral–-derivative controller, and measurement of the UV-visible spectrum.

The freely accessible dataset in~\cite{scherer2025closedloop_zenodo} contains experimental measurements from long-term experiments using the closed-loop {MC} testbed described in~\cite {brand2024closed}.
The experiment was run for $\SI{125}{\hour}$, obtaining more than $\SI{250}{\kilo\bit}$ of transmitted data via {MC}.
The testbed utilizes a green fluorescent protein variant ``Dreiklang'' (GFPD), as the information carrier.
Using light of a specific wavelength, GFPD can be switched reversibly between two different states ~\cite{brand2024switchable}.
The testbed differs from other liquid-based testbeds because it is a closed-loop structure, not an end-to-end structure, e.g., by pumping a liquid from one reservoir to another.
Therefore, GFPD is only injected once.
The testbed contains an optical transmitter for writing information, an optical eraser for erasing information, and a receiver for reading the fluorescence state of the GFPD~\cite{brand2024closed}.
In~\cite{scherer2025closedloop_zenodo}, the raw measurement dataset (as a zip folder containing \texttt{csv} and \texttt{json} files and as a SQLite database) is published, corresponding to the plots in the publication~\cite {scherer2025closedloop}.
In addition, the Python code for synchronization and detection is available in a GitHub repository~\cite{scherer2025closedloop_github}, which includes instructions for reading the SQLite database. 

The Zenodo dataset in~\cite{vakalipoor2025cam_data} contains \num{69} experimental measurements from the study in~\cite{vakalipoor2025cam}, which introduces the first versatile three-dimensional \textit{in vivo} {MC} testbed.
Each entry includes egg and region identifiers, developmental stage, injection parameters, fluorescence intensity, estimated distributions, and corresponding error values.
This dataset offers a comprehensive experimental basis for modeling and analyzing \textit{in vivo} {MC} channels.


%

\subsection{Discussion}
\label{sec_datasets_comparison}

The datasets mentioned above exhibit different levels of accessibility, completeness, and documentation.
As a reader's guide, we develop a traffic-light system in \Cref{tab_dimension_dataset} in which green represents the highest level of dataset completeness and red represents the least along the following dimensions:
\begin{itemize}
    \item {Reproducibility (only considered for synthetic datasets):} Green dots indicate that the source code and its documentation are fully accessible.
    Yellow dots indicate that the dataset documentation is missing or that access to the source code is unavailable.
    The red dots indicate that all of the above are missing. 
    \item{Representativeness (only for experimental datasets):} Evaluated by comparing the number of replicates ($N$) of single experiments to the maximum number of repetitions over all experimental datasets. 
    For the considered set of experimental datasets, a maximum number of $N=69$~\cite{vakalipoor2025cam_data} replicates applies so that for the ranges between $N=1$ and $N=23$ replicates a red dot, between $N=24$ and $N=46$ replicates a yellow dot, and between $N=47$ and $N=69$ replicates a green dot was assigned. 
    The dataset in \cite{gulec2020github} is rated in yellow because it contains a total of $N = 55$ measurements; however, only $N = 5$ repetitions are conducted for each transmitter-receiver distance.
    The dataset in~\cite{scherer2025closedloop_zenodo,scherer2025closedloop_github} is an exception here, as it is the first long-term experimental {MC} system of its kind. The dataset size clearly stands out from the other datasets and is consequently rated green.
    \item{Usability:} Encompasses the cases ``the dataset is available'', ``the dataset processing code is available'', and ``the code for plotting is available''.
    If all three criteria are met, green follows; for two criteria (regardless of which), yellow follows; otherwise, red is applied.
    \item{Availability:} Here, we consider only two cases, either yellow (data record restricted availability, for example, behind an account wall) or green (data record freely available). 
    Datasets that are not available are not listed, so no red is assigned here.
    \item{Documentation:} Considers the completeness of the parameters' metadata and the dataset's documentation. 
    Green: Both aspects are fulfilled; Yellow: Only one of them is fulfilled; Red: None of them are fulfilled.
\end{itemize}

\begin{table}
    \newcommand{\green}{\textcolor{mygreen}{\ding{108}}} 
    \newcommand{\yellow}{\textcolor{myyellow}{\ding{108}}} 
    \newcommand{\red}{\textcolor{myred}{\ding{108}}} 
    \centering
    \caption{4D dataset clustering for synthetic and experimental data. Green, yellow, and red circles represent the highest, medium, and lowest levels, respectively.}
    \label{tab_dimension_dataset}
    \resizebox{\columnwidth}{!}{%
    \begin{tabular}{l c c c c c}
    \toprule
    \textbf{Reference} & \textbf{Reproducibility} & \textbf{Representativeness} & \textbf{Usability} & \textbf{Availability} & \textbf{Documentation} \\
    \midrule
    \cite{hofmann2024dataset_plaque} & \green & - & \yellow & \yellow & \green \\ 
    \cite{tuccitto2023dataset,tucitto2023dataset_zenodo} & - & \yellow & \red & \green & \green \\ 
    \cite{hofmann2023dataset_simulation} & \green & - & \yellow & \yellow & \green \\
    \cite{hofmann2023dataset_analog} & - & \red & \red & \yellow & \green \\ 
    \cite{hofmann2023dataset_macroscale} & - & \yellow & \green & \yellow & \green \\ 
    \cite{bartunik2023dataset_channel} & - & \red & \red & \yellow & \green \\ 
    \cite{cali2022interfacial,cali2022dataset_interfacial} & - & \red & \yellow & \green & \green \\ 
    \cite{abbaszadeh2019dataset_signal,abbaszadeh2020dataset_signal} & - & \red & \yellow & \green & \green \\ 
    \cite{grebenstein2019dataset_molecular} & - & \red & \yellow & \yellow & \green \\ 
    \cite{gao2023dataset} & \yellow & - & \green & \green & \yellow \\ 
    \cite{walter2023dataset_realtime} & - & \red & \red & \green & \green \\ 
    \cite{gulec2024_zenodo} & \yellow & - & \red & \green & \green \\ 
    \cite{das2022dataset} & \red & - & \red & \green & \yellow \\ 
    \cite{das2021dataset_2} & \red & - & \red & \green & \yellow \\  
    \cite{thakkar2020dataset} & \multicolumn{2}{c}{\red} & \red & \yellow & \red \\ 
    \cite{gulec2020github} & - & \yellow & \green & \green & \green \\ 
    \cite{lu2023github} & \green & - & \green & \green & \green \\ 
    \cite{scherer2025closedloop_zenodo,scherer2025closedloop_github} & - & \green & \yellow & \green & \green \\
    \cite{vakalipoor2025cam_data} & - & \green & \yellow & \green & \green \\
    \bottomrule
    \end{tabular}%
    }
\end{table}

Available datasets for {MC} research are diverse and accessible through multiple platforms; however, several limitations hinder broad use, cf.\ \Cref{tab_dimension_dataset}. 
One limitation is the lack of standardization in data formats and annotations, which complicates integration and comparative analysis across datasets. 
Additionally, some datasets lack detailed metadata or related publications, reducing the transparency and reproducibility of the research outcomes. 
The sizes of certain datasets are disproportionately small, limiting their applicability for machine learning or extensive simulation studies. 
Another issue is platform dependence.
While IEEE DataPort, GitHub, and Zenodo are widely used, access to some university-hosted datasets may be restricted or not well-documented.

Data generation tools face compatibility limitations, as many rely on specific frameworks such as Matlab or NS3, hindering reproducibility across software versions.
Moreover, simulator-specific modeling assumptions restrict dataset representativeness.
For example, reaction-diffusion tools such as Smoldyn or MesoRD struggle with environmental heterogeneity, whereas nanoNS3 or N3Sim may not generalize well to macroscopic systems.
Availability and usability further limit adoption, since some tools require licenses, lack active maintenance, or provide insufficient documentation for novice users.
Finally, dataset quality strongly impacts NN generalization, as label noise and class imbalance can cause overfitting and biased predictions, making high-quality, balanced data as critical as model architecture design~\cite{zhao2020role,johnson2019survey,gong2023survey}.


\footnotesize
\begin{table*}[!b]
    \centering
    \caption{Existing datasets on \ac{MC}, ordered by platform and year of publication.}
    \label{tab_datasets}
    \resizebox{1\linewidth}{!}{%
    \begin{tabular}{l  p{30em}  l l l l l l}
        \toprule
        & \textbf{Name of the dataset} & \textbf{Reference} & \textbf{Year} & \textbf{Size} & \textbf{Data type} & \textbf{Related work} & \textbf{Platform} \\
        \midrule
        (1) & Dataset for Advanced Plaque Modeling for Atherosclerosis Detection using Molecular Communication & \cite{hofmann2024dataset_plaque} & 2024 & $\sim$\SI{1.5}{\giga\byte} & Synthetic & \cite{wietfeld2025advanced} & IEEE DataPort \\
        (2) & Dataset of "Experimental Implementation of Molecule Shift Keying for Enhanced Molecular Communication" & \cite{tuccitto2023dataset,tucitto2023dataset_zenodo} & 2023 & $\sim$\SI{300}{\mega\byte} & Experimental& \cite{cali2024experimental} & IEEE DataPort~\cite{tuccitto2023dataset}, Zenodo~\cite{tucitto2023dataset_zenodo} \\
        (3) & Dataset for the Simulation of Microfluidic Molecular Communication using OpenFOAM & \cite{hofmann2023dataset_simulation} & 2023 & $\sim$\SI{40}{\giga\byte} & Synthetic & \cite{hofmann2024openfoam} & IEEE DataPort \\
        (4) & Dataset for Analog Network Coding in Molecular Communications: A Practical Implementation & \cite{hofmann2023dataset_analog} & 2023 & $\sim$\SI{0.8}{\mega\byte} & Experimental& \cite{hofmann2023analog} & IEEE DataPort  \\
        (5) & Dataset for Macroscale Molecular Communication Testbed & \cite{hofmann2023dataset_macroscale} & 2023 & $\sim$\SI{0.2}{\mega\byte} & Experimental& \cite{hofmann2022testbed,torres-gomez2023explainability,debus2023reinforcement,debus2024synchronized} & IEEE DataPort  \\
        (6) & Channel Parameter Studies with a Biocompatible Testbed for Molecular Communication & \cite{bartunik2023dataset_channel} & 2023 & $\sim$\SI{2}{\mega\byte} & Experimental& \cite{bartunik2023channel} & IEEE DataPort  \\
        (7) & The Data Related to Interfacial Shift Keying Allows a High Information Rate in Molecular Communication & \cite{cali2022interfacial,cali2022dataset_interfacial} & 2022 & $\sim$\SI{0.6}{\mega\byte} & Experimental& \cite{cal2022fluorescent,cali2023interfacial} & IEEE DataPort~\cite{cali2022interfacial}, Zenodo~\cite{cali2022dataset_interfacial} \\
        (8) & Molecular Signal Tracking and Detection Methods in Fluid Dynamic Channels (+ Method and Data) & \cite{abbaszadeh2019dataset_signal,abbaszadeh2020dataset_signal} & 2019, 2020 & $\sim$\SI{26.8}{\giga\byte} & Experimental& \cite{abbaszadeh2020molecular} & IEEE DataPort \\
        (9) & A Molecular Communication Testbed Based on Proton Pumping Bacteria & \cite{grebenstein2019dataset_molecular} & 2019 & $\sim$\SI{0.6}{\mega\byte} & Experimental& \cite{grebenstein2019molecular} & IEEE DataPort \\
         (10) & Dataset in Support of the Southampton Doctoral Thesis "Type-Spread and Multiple-Access Molecular Communications" & \cite{gao2023dataset} & 2023 & $\sim$\SI{25}{\mega\byte} & Synthetic & \cite{gao2023typespread} & University server \\ 
         (11) & Transient Indocyanine Green Distribution Measurement and Modelling in Chorioallantoic Membrane (CAM) Model & \cite{vakalipoor2025cam_data} & 2025 & $\sim \SI{220}{\mega\byte}$ & Experimental & \cite{vakalipoor2025cam} & Zenodo \\
         (12) & Closed-Loop Long-Term Experimental Molecular Communication System &  \cite{scherer2025closedloop_zenodo,scherer2025closedloop_github} & 2025 & $\sim$\SI{667}{\mega\byte} & Experimental & \cite{brand2022mediamodulation,brand2024switchable,brand2024closed,scherer2025closedloop} & Zenodo~\cite{scherer2025closedloop_zenodo}, GitHub~\cite{scherer2025closedloop_github} \\
         (13) & CFD Simulation Dataset for Airborne Pathogen Transmission in Turbulent Channels & \cite{gulec2024_zenodo} & 2024 & $\sim$\SI{506.3}{\mega\byte} & Synthetic & \cite{gulec2023computational,gulec2024github} & Zenodo \\
         (14) & Real-Time Signal Processing via Chemical Reactions for a Microfluidic Molecular Communication System & \cite{walter2023dataset_realtime} & 2023 & $\sim$\SI{2.4}{\giga\byte} & Experimental& \cite{walter2023real,walter2023github} & Zenodo \\ 
        (15) & Received Signal Modeling and BER Analysis for Molecular SISO Communications & \cite{das2022dataset} & 2022 & $\sim$\SI{3.1}{\kilo\byte} & Synthetic & \cite{das2022received} & Zenodo \\ 
        (16) & Channel Estimation and Performance Analysis of SISO Molecular Communications & \cite{das2021dataset,das2021dataset_2} & 2021 & $\sim$\SI{3.3}{\kilo\byte} & Synthetic & Not specified & Zenodo \\ 
        (17) & "molecular$\_$communication" & \cite{thakkar2020dataset} & 2020 & $\sim$\SI{1.5}{\kilo\byte} & Not specified & Not specified & Kaggle \\
        (18) & "MCFormer: A Transformer-Based Detector for Molecular Communication with Accelerated Particle-Based Solution" & \cite{lu2023github} & 2023 & $\sim$\SI{41.8}{\mega\byte} & Synthetic & \cite{lu2023mcformer} & GitHub \\
        (19) & "Distance-Estimation-in-Molecular-Communication" & \cite{gulec2020github} & 2020 & $\sim$\SI{620}{\kilo\byte} & Experimental & \cite{gulec2020distance,gulec2021fluid} & GitHub \\
        \bottomrule
    \end{tabular}
    }
\end{table*}
\normalsize

\subsection{Concluding Remarks and Outlook}

To summarize, in recent years, data generation, documentation, and publication have received attention, addressing the need for robust training datasets and reliable research outcomes.
Our survey of existing datasets also shows that eight datasets are derived from experimental testbeds. 
The use case for sharing experimental data is stronger than for synthetic data, primarily because wet-lab experiments are more challenging to reproduce than simulations.
Additionally, dataset publication still requires more extensive and comprehensive documentation practices.

%

In this topic, we also identify further research directions towards the connections between testbeds and the training and deployment of {NNs} as follows:

\textbf{1. Training in Body-like Testbeds:} The pharmaceutical industry is today testing experimental results developed in organs-on-chip.
As a proven tool for drug testing \cite{ewart2022performance}, this technology mimics organ function and fluid dynamics.
As these devices are expected to become a daily occurrence in laboratories, we anticipate they will serve as testbeds for generating data and integrating {NN}-models.

%

\textbf{2. Transferring from Simulators to Real-World Testbeds:} A broad range of simulation tools exists for {MC}; on the one hand, these simulation tools are often open-source and easy to install, so researchers use them to implement their {NN}/{ML} approaches and benchmark them against analytical approaches, such as in~\cite{pascual2024analytical}.
On the other hand, the components of experimental testbeds are often limited in their computational capabilities, as seen in the testbed presented in~\cite{hofmann2022testbed}, making the implementation of computationally intensive {NN} and {ML} approaches challenging. 
To address these limitations, recent research explores lightweight \ac{NN} architectures that lower resource demands without compromising accuracy.
Techniques such as model quantization~\cite{wei2024advances}, network pruning~\cite{he2024pruning}, and TinyML~\cite{abadade2023comprehensive} enable efficient deployment of models on resource-constrained testbeds.
However, as the {MC} community moves towards practical applications, {NN} and {ML} tools applied to {MC} simulation tools should be transferred to experimental testbeds.
In this way, experimental setups will enable benchmarking, at least in pre-defined real-world scenarios, owing to the targeted parameter control of the experimental testbeds.

%

\section{Conclusion}
\label{sec_conclu}

This survey examines the variety of {NN} architectures for the functional development of {IoBNT} networks and provides an overview of the most recent methods for integrating them into {MC} links.
A thorough evaluation of the reported research concludes that {NN} architectures are essential for handling the complex nature of {MC} connections, for which closed-form expressions for model-based solutions are often impractical to derive.
As such, {NNs} are the \textit{de facto} model for devising practical solutions within the various layers of {IoBNT} networks.
A closer look at the reported literature reveals that most of the reported developments primarily focus on the \ac{PHY} layer for decoding tasks, where the \ac{BiRNN} achieves the highest performance in \ac{ISI} channels and with less complexity compared to \acp{RNN}, feedforward {NNs}, and \acp{CNN}.
Conversely, fewer works target solutions in the upper layers, and only a few methods explore the much-needed localization and detection applications in {IoBNT} networks.
We expect further developments in {NN}-based methods towards application-oriented solutions for {IoBNT} networks.
The literature is also maturing in supplying the necessary datasets for {NN} training, and we anticipate further progress in the usability and reproducibility dimension of datasets, along with increased data generation for in-body environments.

Furthermore, studies at the intersection of \acp{NN} and \ac{IoBNT} underline two long-term research directions: (i) biocompatible deployment of \acp{NN} at the nanoscale is still in its early stages, and nanoscale architectures beyond feedforward \acp{NN} require further attention to realize the full potential of \ac{IoBNT}; and (ii) physics-informed architectures, grounded in domain knowledge, can support the development of self-explainable solutions.
We conclude with the broader perspective that \ac{AI} can provide a common conceptual and methodological framework for bridging biology and engineering, enabling interdisciplinary designs in which biological processes and communication systems are developed in closer alignment.

%

\printbibliography

@data{Birkan2024matlab,
	author = {Birkan, H.},
	title = {{MolecUlar CommunicatIoN (MUCIN) Simulator}},
	journal = {MATLAB Central File Exchange},
	url = {mathworks.com/matlabcentral/fileexchange/46066-molecular-communication-mucin-simulator},
	year = {2024},
}

@techreport{EuropeanCommission2025,
	author = {{European Commission}},
	title = {Horizon Europe Work Programme 2023-2025: 13. General Annexes},
	institution = {European Commission},
	month = May,
	note = {Annex G: Legal and financial set-up of the grant agreements},
	number = {C(2025) 2779},
	type = {Decision},
	url = {https://ec.europa.eu},
	year = {2025},
}

@article{abadade2023comprehensive,
    author = {Abadade, Youssef and Temouden, Anas and Bamoumen, Hatim and Benamar, Nabil and Chtouki, Yousra and Hafid, Abdelhakim Senhaji},
    doi = {10.1109/access.2023.3294111},
    title = {{A Comprehensive Survey on TinyML}},
    pages = {96892--96922},
    journal = {IEEE Access},
    issn = {2169-3536},
    publisher = {IEEE},
    volume = {11},
    year = {2023}
}

@incollection{abadal2019graphenebased,
	author = {Abadal, Sergi and Hosseininejad, Seyed Ehsan and Lemme, Max and Bolivar, Peter Haring and Sol{\'{e}}-Pareta, Josep and Alarc{\'{o}}n, Eduard and Cabellos-Aparicio, Albert},
	title = {{Graphene-Based Antenna Design for Communications in the Terahertz Band}},
	editor = {Vacca, John R.},
	booktitle = {Nanoscale Networking and Communications Handbook},
	edition = {1},
	isbn = {978-0-429-16304-3},
	location = {Boca Raton, FL},
	pages = {25--45},
	publisher = {CRC Press},
	year = {2019},
}

@article{abadal2024electromagnetic,
	author = {Abadal, Sergi and Han, Chong and Petrov, Vitaly and Galluccio, Laura and Akyildiz, Ian F. and Jornet, Josep Miquel},
	title = {{Electromagnetic Nanonetworks Beyond 6G: From Wearable and Implantable Networks to On-Chip and Quantum Communication}},
	doi = {10.1109/jsac.2024.3399253},
	issn = {0733-8716},
	journal = {IEEE Journal on Selected Areas in Communications},
	month = Aug,
	number = {8},
	pages = {2122--2142},
	publisher = {IEEE},
	volume = {42},
	year = {2024},
}

@data{abbaszadeh2019dataset_signal,
	author = {Abbaszadeh, Mahmoud and Atthanayake, Iresha and J. Thomas, Peter and Guo, Weisi},
	title = {Molecular Signal Tracking and Detection Methods in Fluid Dynamic Channels},
	note = {{IEEE Dataport}},
	publisher = {IEEE Dataport},
	url = {doi.org/10.21227/ynet-ss80},
	year = {2019},
}

@data{abbaszadeh2020dataset_signal,
	author = {Abbaszadeh, Mahmoud and Atthanayake, Iresha and J. Thomas, Peter and Guo, Weisi},
	title = {Molecular Signal Tracking and Detection Methods in Fluid Dynamic Channels (Method and Data)},
	month = Jan,
	note = {{IEEE Dataport}},
	publisher = {IEEE Dataport},
	url = {doi.org/10.21227/eae4-kg81},
	year = {2020},
}

@article{abbaszadeh2020molecular,
	author = {Abbaszadeh, Mahmoud and Atthanayake, Iresha and Thomas, Peter J. and Guo, Weisi},
	title = {{Molecular Signal Tracking and Detection Methods in Fluid Dynamic Channels}},
	doi = {10.1109/tmbmc.2020.3009899},
	issn = {2332-7804},
	journal = {IEEE Transactions on Molecular, Biological and Multi-Scale Communications},
	month = Nov,
	number = {2},
	pages = {151--159},
	publisher = {IEEE},
	volume = {6},
	year = {2020},
}

@article{abdali2024frequencydomain,
	author = {Abdali, Ali and Kuscu, Murat},
	title = {{Frequency-Domain Model of Microfluidic Molecular Communication Channels With Graphene BioFET-Based Receivers}},
	doi = {10.1109/tcomm.2024.3376593},
	issn = {1558-0857},
	journal = {IEEE Transactions on Communications},
	month = Aug,
	number = {8},
	pages = {4564--4576},
	publisher = {IEEE},
	volume = {72},
	year = {2024},
}

@article{adadi2018peeking,
	author = {Adadi, Amina and Berrada, Mohammed},
	title = {{Peeking Inside the Black-Box: A Survey on Explainable Artificial Intelligence (XAI)}},
	doi = {10.1109/access.2018.2870052},
	issn = {2169-3536},
	journal = {IEEE Access},
	pages = {52138--52160},
	publisher = {IEEE},
	volume = {6},
	year = {2018},
}

@article{agiza2023digital,
	author = {Agiza, Ahmed and Oakley, Kady and Rosenstein, Jacob K. and Rubenstein, Brenda M. and Kim, Eunsuk and Riedel, Marc and Reda, Sherief},
	title = {{Digital circuits and neural networks based on acid-base chemistry implemented by robotic fluid handling}},
	doi = {10.1038/s41467-023-36206-8},
	issn = {2041-1723},
	journal = {Nature Communications},
	month = Jan,
	number = {1},
	pages = {1--9},
	publisher = {Springer Nature},
	volume = {14},
	year = {2023},
}

@article{agiza2024phcontrolled,
	author = {Agiza, Ahmed and Marriott, Stephen and Rosenstein, Jacob K. and Kim, Eunsuk and Reda, Sherief},
	title = {{pH-Controlled enzymatic computing for digital circuits and neural networks}},
	doi = {10.1039/d4cp02039a},
	issn = {1463-9084},
	journal = {Physical Chemistry Chemical Physics},
	number = {31},
	pages = {20898--20907},
	publisher = {RSC},
	volume = {26},
	year = {2024},
}

@book{agrawal2021hyperparameter,
	author = {Agrawal, Tanay},
	title = {{Hyperparameter Optimization in Machine Learning: Make Your Machine Learning and Deep Learning Models More Efficient}},
	doi = {10.1007/978-1-4842-6579-6},
	isbn = {978-1-4842-6579-6},
	pages = {177},
	publisher = {Apress},
	year = {2021},
}

@inproceedings{agrawal2022neural,
	author = {Agrawal, Upendra K. and Shrivastava, Amit K. and Das, Debanjan and Mahapatra, Rajarshi},
	title = {{Neural Network Detector in Mobile Molecular Communication for Fast Varying Channels}},
	booktitle = {International Conference on Connected Systems and Intelligence (CSI 2022)},
	address = {Trivandrum, India},
	doi = {10.1109/csi54720.2022.9924143},
	isbn = {978-1-66545-815-3},
	month = Aug,
	pages = {1 -- 5},
	publisher = {IEEE},
	year = {2022},
}

@article{aktas2024odorbased,
	author = {Aktas, Dilara and Ortlek, Beyza E. and Civas, Meltem and Baradari, Elham and Kilic, Ahmet B. and Bilgen, Fatih E. and Okcu, Ayse S. and Whitfield, Melanie and Cetinkaya, Oktay and Akan, Ozgur B.},
	title = {{Odor-Based Molecular Communications: State-of-the-Art, Vision, Challenges, and Frontier Directions}},
	doi = {10.1109/comst.2024.3487472},
	issn = {1553-877X},
	journal = {IEEE Communications Surveys \& Tutorials},
	pages = {1--34},
	publisher = {IEEE},
	year = {2024},
}

@article{akyildiz2019moving,
	author = {Akyildiz, Ian F. and Pierobon, Massimiliano and Balasubramaniam, Sasitharan},
	title = {{Moving forward with molecular communication: from theory to human health applications [point of view]}},
	doi = {10.1109/jproc.2019.2913890},
	issn = {0018-9219},
	journal = {Proceedings of the IEEE},
	month = May,
	number = {5},
	pages = {858--865},
	publisher = {IEEE},
	volume = {107},
	year = {2019},
}

@article{akyildiz2020panacea,
	author = {Akyildiz, Ian F. and Ghovanloo, Maysam and Guler, Ulkuhan and Ozkaya-Ahmadov, Tevhide and Sarioglu, A. Fatih and Unluturk, Bige Deniz},
	title = {{PANACEA: An Internet of Bio-NanoThings Application for Early Detection and Mitigation of Infectious Diseases}},
	doi = {10.1109/access.2020.3012139},
	issn = {2169-3536},
	journal = {IEEE Access},
	month = Jan,
	pages = {140512--140523},
	publisher = {IEEE},
	volume = {8},
	year = {2020},
}

@inproceedings{alshammri2018adaptive,
	author = {Alshammri, Ghalib H. and Ahmed, Walid K. M. and Lawrence, Victor B.},
	title = {{Adaptive Batch Training Rule-Based Detection Scheme for On-OFF-Keying Diffusion-Based Molecular Communications}},
	booktitle = {13th IEEE Nanotechnology Materials and Devices Conference (NMDC 2018)},
	address = {Portland, OR},
	doi = {10.1109/nmdc.2018.8605873},
	issn = {2378-377X},
	month = Oct,
	pages = {1 -- 4},
	publisher = {IEEE},
	year = {2018},
}

@article{amerizadeh2021bacterial,
	author = {Amerizadeh, Atefeh and Mashhadian, Ali and Farahnak-Ghazani, Maryam and Arjmandi, Hamidreza and Rad, Maryam Alsadat and Shamloo, Amir and Vosoughi, Manouchehr and Nasiri-Kenari, Masoumeh},
	title = {{Bacterial Receiver Prototype for Molecular Communication Using Rhamnose Operon in a Microfluidic Environment}},
	doi = {10.1109/tnb.2021.3090761},
	issn = {1558-2639},
	journal = {IEEE Transactions on NanoBioscience},
	month = Oct,
	number = {4},
	pages = {426--435},
	publisher = {IEEE},
	volume = {20},
	year = {2021},
}

@article{ander2004smartcell,
	author = {Ander, M. and Tom\'as-Oliveira, I. and Ferkinghoff-Borg, J. and Beltrao, P. and Foglierini, M. and Di Ventura, B. and Serrano, L. and Lemerle, C.},
	title = {{SmartCell, a framework to simulate cellular processes that combines stochastic approximation with diffusion and localisation: analysis of simple networks}},
	doi = {10.1049/sb:20045017},
	issn = {1741-248X},
	journal = {IEEE Proceedings -- Systems Biology},
	month = Jun,
	number = {1},
	pages = {129--138},
	publisher = {IET},
	volume = {1},
	year = {2004},
}

@article{andrews2010detailed,
	author = {Andrews, Steven S. and Addy, Nathan J. and Brent, Roger and Arkin, Adam P.},
	title = {{Detailed Simulations of Cell Biology with Smoldyn 2.1}},
	doi = {10.1371/journal.pcbi.1000705},
	issn = {1553-7358},
	journal = {PLoS Computational Biology},
	month = Mar,
	number = {3},
	pages = {1--10},
	publisher = {Public Library of Science (PLoS)},
	volume = {6},
	year = {2010},
}

@article{andrews2016smoldyn,
	author = {Andrews, Steven S},
	title = {{Smoldyn: particle-based simulation with rule-based modeling, improved molecular interaction and a library interface}},
	address = {Oxford, United Kingdom},
	doi = {10.1093/bioinformatics/btw700},
	issn = {1367-4811},
	journal = {Bioinformatics},
	month = Dec,
	number = {5},
	pages = {710--717},
	publisher = {OUP},
	volume = {33},
	year = {2016},
}

@techreport{andrychowicz2020what,
	author = {Andrychowicz, Marcin and Raichuk, Anton and Sta{\'{n}}czyk, Piotr and Orsini, Manu and Girgin, Sertan and Marinier, Raphael and Hussenot, Leonard and Geist, Matthieu and Pietquin, Olivier and Michalski, Marcin and Gelly, Sylvain and Bachem, Olivier},
	title = {{What Matters In On-Policy Reinforcement Learning? A Large-Scale Empirical Study}},
	doi = {10.48550/ARXIV.2006.05990},
	institution = {arXiv},
	month = Jun,
	number = {2006.05990},
	type = {cs.LG},
	year = {2020},
}

@techreport{angerbauer2023molecular,
	author = {Angerbauer, Stefan and Haselmayr, Werner and Enzenhofer, Franz and Pankratz, Tobias and Khanzadeh, Roya and Springer, Andreas},
	title = {{Molecular Nano Neural Networks (M3N):In-Body Intelligence for the IoBNT}},
	doi = {10.36227/techrxiv.24427435.v1},
	institution = {TechRxiv},
	month = Oct,
	pages = {1--7},
	type = {preprint},
	year = {2023},
}

@article{angerbauer2023salinitybased,
	author = {Angerbauer, Stefan and Hamidovic, Medina and Enzenhofer, Franz and Bartunik, Max and Kirchner, Jens and Springer, Andreas and Haselmayr, Werner},
	title = {{Salinity-Based Molecular Communication in Microfluidic Channels}},
	doi = {10.1109/tmbmc.2023.3277391},
	issn = {2332-7804},
	journal = {IEEE Transactions on Molecular, Biological and Multi-Scale Communications},
	month = Jun,
	number = {2},
	pages = {191--206},
	publisher = {IEEE},
	volume = {9},
	year = {2023},
}

@inproceedings{angerbauer2023towards,
	author = {Angerbauer, Stefan and Khanzadeh, Roya and Enzenhofer, Franz and Springer, Andreas and Haselmayr, Werner},
	title = {{Towards Asymmetric Auto-Encoders for the IoBNT}},
	booktitle = {10th ACM International Conference on Nanoscale Computing and Communication (NANOCOM 2023)},
	address = {Coventry, United Kingdom},
	doi = {10.1145/3576781.3608733},
	month = Sep,
	pages = {166--167},
	publisher = {ACM},
	year = {2023},
}

@article{angerbauer2024investigation,
	author = {Angerbauer, Stefan and Tuccitto, Nunzio and Sfrazzetto, Giuseppe Trusso and Santonocito, Rossella and Haselmayr, Werner},
	title = {{Investigation of Different Chemical Realizations for Molecular Matrix Multiplications}},
	doi = {10.1109/tmbmc.2024.3436905},
	issn = {2332-7804},
	journal = {IEEE Transactions on Molecular, Biological and Multi-Scale Communications},
	month = Sep,
	number = {3},
	pages = {464--469},
	publisher = {IEEE},
	volume = {10},
	year = {2024},
}

@techreport{angerbauer2024molecular,
	author = {Angerbauer, Stefan and Enzenhofer, Franz and Gattringer, Hubert and Springer, Andreas and Haselmayr, Werner},
	title = {{A Molecular Analog-to-Digital Converter}},
	doi = {10.36227/techrxiv.171777814.47534874/v1},
	institution = {TechRxiv},
	month = Jun,
	pages = {1--7},
	type = {cs},
	year = {2024},
}

@article{angerbauer2024novel,
	author = {Angerbauer, Stefan and Enzenhofer, Franz and Pankratz, Tobias and Hamidovic, Medina and Springer, Andreas and Haselmayr, Werner},
	title = {{Novel Nano-Scale Computing Unit for the IoBNT: Concept and Practical Considerations}},
	doi = {10.1109/tmbmc.2024.3397050},
	issn = {2332-7804},
	journal = {IEEE Transactions on Molecular, Biological and Multi-Scale Communications},
	pages = {549--565},
	publisher = {IEEE},
	year = {2024},
}

@article{aoudia2019modelfree,
	author = {Aoudia, Fay{\c{c}}al Ait and Hoydis, Jakob},
	title = {{Model-Free Training of End-to-End Communication Systems}},
	doi = {10.1109/jsac.2019.2933891},
	issn = {0733-8716},
	journal = {IEEE Journal on Selected Areas in Communications},
	month = Nov,
	number = {11},
	pages = {2503--2516},
	publisher = {IEEE},
	volume = {37},
	year = {2019},
}

@article{atakan2019signal,
	author = {Atakan, Baris and G{\"{u}}le{\c{c}}, Fatih},
	title = {{Signal reconstruction in diffusion-based molecular communication}},
	doi = {10.1002/ett.3699},
	issn = {2161-3915},
	journal = {Transactions on Emerging Telecommunications Technologies},
	month = Dec,
	number = {12},
	pages = {1--14},
	publisher = {Wiley},
	volume = {30},
	year = {2019},
}

@techreport{bahdanau2014neural,
	author = {Bahdanau, Dzmitry and Cho, Kyunghyun and Bengio, Yoshua},
	title = {{Neural Machine Translation by Jointly Learning to Align and Translate}},
	doi = {10.48550/ARXIV.1409.0473},
	institution = {arXiv},
	month = May,
	number = {1409.0473},
	type = {cs.CL},
	year = {2014},
}

@article{bai2023temporal,
	author = {Bai, Chenyao and Zhu, Aoji and Lu, Xiwen and Zhu, Yunlong and Wang, Kezhi},
	title = {{Temporal Convolutional Network-Based Signal Detection for Magnetotactic Bacteria Communication System}},
	doi = {10.1109/tnb.2023.3262555},
	issn = {1558-2639},
	journal = {IEEE Transactions on NanoBioscience},
	month = Oct,
	number = {4},
	pages = {943--955},
	publisher = {IEEE},
	volume = {22},
	year = {2023},
}

@article{balasubramaniam2023realizing,
	author = {Balasubramaniam, Sasitharan and Somathilaka, Samitha and Sun, Sehee and Ratwatte, Adrian and Pierobon, Massimiliano},
	title = {{Realizing Molecular Machine Learning Through Communications for Biological AI}},
	doi = {10.1109/mnano.2023.3262099},
	issn = {1932-4510},
	journal = {IEEE Nanotechnology Magazine},
	month = Jun,
	number = {3},
	pages = {10--20},
	publisher = {IEEE},
	volume = {17},
	year = {2023},
}

@inproceedings{bartunik2022using,
	author = {Bartunik, Max and Keszocze, Oliver and Schiller, Benjamin and Kirchner, Jens},
	title = {{Using Deep Learning to Demodulate Transmissions in Molecular Communication}},
	booktitle = {16th IEEE International Symposium on Medical Information and Communication Technology (ISMICT 2022)},
	address = {Lincoln, NE},
	doi = {10.1109/ismict56646.2022.9828263},
	issn = {2326-8301},
	month = May,
	pages = {1 -- 6},
	publisher = {IEEE},
	year = {2022},
}

@article{bartunik2023artificial,
	author = {Bartunik, Max and Kirchner, Jens and Keszocze, Oliver},
	title = {{Artificial intelligence for molecular communication}},
	doi = {10.1515/itit-2023-0029},
	issn = {2196-7032},
	journal = {it - Information Technology (itIT)},
	month = Aug,
	number = {4-5},
	pages = {155--163},
	publisher = {De Gruyter},
	volume = {65},
	year = {2023},
}

@article{bartunik2023channel,
	author = {Bartunik, Max and Teller, Janina and Fischer, Georg and Kirchner, Jens},
	title = {{Channel Parameter Studies of a Molecular Communication Testbed With Biocompatible Information Carriers: Methods and Data}},
	doi = {10.1109/tmbmc.2023.3325405},
	issn = {2332-7804},
	journal = {IEEE Transactions on Molecular, Biological and Multi-Scale Communications},
	month = Dec,
	number = {4},
	pages = {489--498},
	publisher = {IEEE},
	volume = {9},
	year = {2023},
}

@data{bartunik2023dataset_channel,
	author = {Bartunik, Max},
	title = {Channel Parameter Studies with a Biocompatible Testbed for Molecular Communication},
	note = {{IEEE Dataport}},
	publisher = {IEEE Dataport},
	url = {doi.org/10.21227/g15d-kz12},
	year = {2023},
}

@inproceedings{basaran2025xai-enhanced,
	author = {Basaran, Osman Tugay and Torres G{\'{o}}mez, Jorge and Dressler, Falko},
	title = {{XAI-Enhanced Bilateral Molecular Communication: Revealing Cancer Microenvironment Dynamics via Extracellular Tumor Vesicles}},
	booktitle = {IEEE International Conference on Machine Learning for Communication and Networking (ICMLCN 2025)},
	address = {Barcelona, Spain},
	doi = {10.1109/ICMLCN64995.2025.11139906},
	isbn = {979-8-3315-2042-7},
	month = May,
	pages = {1--6},
	publisher = {IEEE},
	year = {2025},
}

@data{bauer2024github,
	author = {Bauer, Pavol and Engblom, Stefan and Senek, Aleksandar and Wilson, Daniel},
	title = {{URDME (Version 1.4)}},
	journal = {GitHub repository},
	publisher = {GitHub},
	url = {github.com/URDME/urdme},
	year = {2024},
}

@article{baydas2023estimation,
	author = {Baydas, O. Tansel and Cetinkaya, Oktay and Akan, Ozgur B.},
	title = {{Estimation and Detection for Molecular MIMO Communications in the Internet of Bio-Nano Things}},
	doi = {10.1109/tmbmc.2023.3252943},
	issn = {2332-7804},
	journal = {IEEE Transactions on Molecular, Biological and Multi-Scale Communications},
	month = Mar,
	number = {1},
	pages = {106--110},
	publisher = {IEEE},
	volume = {9},
	year = {2023},
}

@book{berg1993random,
	author = {Berg, Howard C.},
	title = {{Random Walks in Biology}},
	isbn = {978-0-691-00064-0},
	pages = {152},
	publisher = {Princeton University Press},
	year = {1993},
}

@article{bi2020chemical,
	author = {Bi, Dadi and Deng, Yansha and Pierobon, Massimiliano and Nallanathan, Arumugam},
	title = {{Chemical Reactions-Based Microfluidic Transmitter and Receiver Design for Molecular Communication}},
	doi = {10.1109/tcomm.2020.2993633},
	issn = {1558-0857},
	journal = {IEEE Transactions on Communications},
	month = Sep,
	number = {9},
	pages = {5590--5605},
	publisher = {IEEE},
	volume = {68},
	year = {2020},
}

@article{bi2021survey,
	author = {Bi, Dadi and Almpanis, Apostolos and Noel, Adam and Deng, Yansha and Schober, Robert},
	title = {{A Survey of Molecular Communication in Cell Biology: Establishing a New Hierarchy for Interdisciplinary Applications}},
	doi = {10.1109/comst.2021.3066117},
	issn = {1553-877X},
	journal = {IEEE Communications Surveys \& Tutorials},
	number = {3},
	pages = {1494--1545},
	publisher = {IEEE},
	volume = {23},
	year = {2021},
}

@article{bicen2013systemtheoretic,
	author = {Bicen, A. Ozan and Akyildiz, Ian F.},
	title = {{System-Theoretic Analysis and Least-Squares Design of Microfluidic Channels for Flow-Induced Molecular Communication}},
	doi = {10.1109/tsp.2013.2274959},
	issn = {1053-587X},
	journal = {IEEE Transactions on Signal Processing},
	month = Oct,
	number = {20},
	pages = {5000--5013},
	publisher = {IEEE},
	volume = {61},
	year = {2013},
}

@incollection{binder2016layerwise,
	author = {Binder, Alexander and Montavon, Gr{\'{e}}goire and Lapuschkin, Sebastian and M{\"{u}}ller, Klaus-Robert and Samek, Wojciech},
	title = {{Layer-Wise Relevance Propagation for Neural Networks with Local Renormalization Layers}},
	editor = {E.P. Villa, Alessandro and Masulli, Paolo and Pons Rivero, Antonio Javier},
	booktitle = {Artificial Neural Networks and Machine Learning},
	doi = {10.1007/978-3-319-44781-0_8},
	isbn = {978-3-319-44781-0},
	location = {Barcelona, Spain},
	pages = {63--71},
	publisher = {Springer International Publishing},
	year = {2016},
}

@data{birkan2017ANN,
	author = {{H. Birkan}},
	title = {ANN for Diffusion Channel with Reflecting Spherical to Absorbing Spherical},
	note = {{MATLAB Central File Exchange}},
	url = {mathworks.com/matlabcentral/fileexchange/61382-ann-for-diffusion-channel-with-reflecting-spherical-to-absorbing-spherical},
	year = {2017},
}

@article{boulogeorgos2021machine,
	author = {Boulogeorgos, Alexandros-Apostolos A. and Trevlakis, Stylianos E. and Tegos, Sotiris A. and Papanikolaou, Vasilis K. and Karagiannidis, George K.},
	title = {{Machine Learning in Nano-Scale Biomedical Engineering}},
	doi = {10.1109/tmbmc.2020.3035383},
	issn = {2332-7804},
	journal = {IEEE Transactions on Molecular, Biological and Multi-Scale Communications},
	month = Mar,
	number = {1},
	pages = {10--39},
	publisher = {IEEE},
	volume = {7},
	year = {2021},
}

@article{brakemann2011reversibly,
	author = {Brakemann, Tanja and Stiel, Andre C. and Weber, Gert and Andresen, Martin and Testa, Ilaria and Grotjohann, Tim and Leutenegger, Marcel and Plessmann, Uwe and Urlaub, Henning and Eggeling, Christian and Wahl, Markus C. and Hell, Stefan W. and Jakobs, Stefan},
	title = {{A Reversibly Photoswitchable GFP-like Protein with Fluorescence Excitation Decoupled from Switching}},
	doi = {10.1038/nbt.1952},
	journal = {Nature Biotechnology},
	month = Oct,
	number = {10},
	pages = {942--947},
	publisher = {Springer Nature},
	volume = {29},
	year = {2011},
}

@article{brand2022mediamodulation,
	author = {Brand, Lukas and Garkisch, Moritz and Lotter, Sebastian and Sch{\"{a}}fer, Maximilian and Burkovski, Andreas and Sticht, Heinrich and Castiglione, Kathrin and Schober, Robert},
	title = {{Media Modulation Based Molecular Communication}},
	doi = {10.1109/TCOMM.2022.3205949},
	issn = {1558-0857},
	journal = {IEEE Transactions on Communications},
	month = Nov,
	number = {11},
	pages = {7207--7223},
	publisher = {IEEE},
	volume = {70},
	year = {2022},
}

@inproceedings{brand2024closed,
	author = {Brand, Lukas and Scherer, Maike and Dieck, Teena tom and Lotter, Sebastian and Sch{\"{a}}fer, Maximilian and Burkovski, Andreas and Sticht, Heinrich and Castiglione, Kathrin and Schober, Robert},
	title = {{Closed Loop Molecular Communication Testbed: Setup, Interference Analysis, and Experimental Results}},
	booktitle = {IEEE International Conference on Communications (ICC 2024)},
	address = {Denver, CO},
	doi = {10.1109/icc51166.2024.10622231},
	issn = {1938-1883},
	month = Jun,
	pages = {4805--4811},
	publisher = {IEEE},
	year = {2024},
}

@article{brand2024switchable,
	author = {Brand, Lukas and Scherer, Maike and Lotter, Sebastian and Dieck, Teena tom and Sch{\"{a}}fer, Maximilian and Burkovski, Andreas and Sticht, Heinrich and Castiglione, Kathrin and Schober, Robert},
	title = {{Switchable Signaling Molecules for Media Modulation: Fundamentals, Applications, and Research Directions}},
	doi = {10.1109/mcom.021.2300096},
	issn = {0163-6804},
	journal = {IEEE Communications Magazine},
	month = May,
	number = {5},
	pages = {112--118},
	publisher = {IEEE},
	volume = {62},
	year = {2024},
}

@article{breiman2001random,
	author = {Breiman, Leo},
	title = {{Random Forests}},
	doi = {10.1023/A:1010933404324},
	issn = {1573-0565},
	journal = {Machine Learning},
	month = Oct,
	number = {1},
	pages = {5--32},
	publisher = {Springer},
	volume = {45},
	year = {2001},
}

@book{brown2009fundamentals,
	author = {Brown, Stephen and Vranesic, Zvonko},
	title = {{Fundamentals of Digital Logic with VHDL Design}},
	edition = {3},
	isbn = {978-0-07-352953-0},
	location = {Boston, MA},
	pages = {934},
	publisher = {McGraw-Hill},
	year = {2009},
}

@article{cai2023brain,
	author = {Cai, Hongwei and Ao, Zheng and Tian, Chunhui and Wu, Zhuhao and Liu, Hongcheng and Tchieu, Jason and Gu, Mingxia and Mackie, Ken and Guo, Feng},
	title = {{Brain organoid reservoir computing for artificial intelligence}},
	doi = {10.1038/s41928-023-01069-w},
	issn = {2520-1131},
	journal = {Nature Electronics},
	month = Dec,
	number = {12},
	pages = {1032--1039},
	publisher = {Springer},
	volume = {6},
	year = {2023},
}

@article{cal2022fluorescent,
	author = {Cal{\`{i}}, Federico and Fichera, Luca and Sfrazzetto, Giuseppe Trusso and Nicotra, Giuseppe and Sfuncia, Gianfranco and Bruno, Elena and Lanzan{\`{o}}, Luca and Barbagallo, Ignazio and Li-Destri, Giovanni and Tuccitto, Nunzio},
	title = {{Fluorescent nanoparticles for reliable communication among implantable medical devices}},
	doi = {10.1016/j.carbon.2022.01.016},
	issn = {0008-6223},
	journal = {Carbon},
	month = Apr,
	pages = {262--275},
	publisher = {Elsevier},
	volume = {190},
	year = {2022},
}

@data{cali2022dataset_interfacial,
	author = {Calì, Federico and Li-Destri, Giovanni and Tuccitto, Nunzio},
	title = {{The Data Related to Interfacial Shift Keying Allows a High Information Rate in Molecular Communication}},
	note = {{Zenodo}},
	publisher = {Zenodo},
	url = {doi.org/10.5281/zenodo.7244181},
	version = {1.0.0},
	year = {2022},
}

@data{cali2022interfacial,
	author = {Calì, Federico and Li-Destri, Giovanni and Tuccitto, Nunzio},
	title = {The Data Related to Interfacial Shift Keying Allows a High Information Rate in Molecular Communication},
	month = Oct,
	note = {{IEEE Dataport}},
	publisher = {IEEE Dataport},
	url = {doi.org/10.21227/mj9p-pt58},
	year = {2022},
}

@article{cali2023interfacial,
	author = {Cal{\`{i}}, Federico and Li-Destri, Giovanni and Tuccitto, Nunzio},
	title = {{Interfacial Shift Keying Allows a High Information Rate in Molecular Communication: Methods and Data}},
	doi = {10.1109/tmbmc.2023.3290076},
	issn = {2332-7804},
	journal = {IEEE Transactions on Molecular, Biological and Multi-Scale Communications},
	month = Sep,
	number = {3},
	pages = {300--307},
	publisher = {IEEE},
	volume = {9},
	year = {2023},
}

@article{cali2024experimental,
	author = {Cal{\`{i}}, Federico and Barreca, Salvatore and Li-Destri, Giovanni and Torrisi, Alberto and Licciardello, Antonino and Tuccitto, Nunzio},
	title = {{Experimental Implementation of Molecule Shift Keying for Enhanced Molecular Communication}},
	doi = {10.1109/tmbmc.2024.3368759},
	issn = {2332-7804},
	journal = {IEEE Transactions on Molecular, Biological and Multi-Scale Communications},
	month = Mar,
	number = {1},
	pages = {175--184},
	publisher = {IEEE},
	volume = {10},
	year = {2024},
}

@article{calvobartra2024graph,
	author = {Bartra, Gerard Calvo and Lemic, Filip and Pascual, Guillem and P{\'{e}}rez Rodas, Aina and Struye, Jakob and Delgado, Carmen and Costa-P{\'{e}}rez, Xavier},
	title = {{Graph Neural Networks as an Enabler of Terahertz-Based Flow-Guided Nanoscale Localization Over Highly Erroneous Raw Data}},
	doi = {10.1109/jsac.2024.3399257},
	issn = {0733-8716},
	journal = {IEEE Journal on Selected Areas in Communications},
	month = Aug,
	number = {8},
	pages = {1992--2008},
	publisher = {IEEE},
	volume = {42},
	year = {2024},
}

@article{canovascarrasco2020understanding,
	author = {Canovas-Carrasco, Sebastian and Asorey-Cacheda, Rafael and Garcia-Sanchez, Antonio-Javier and Garcia-Haro, Joan and Wojcik, Krzysztof and Kulakowski, Pawel},
	title = {{Understanding the Applicability of Terahertz Flow-Guided Nano-Networks for Medical Applications}},
	doi = {10.1109/access.2020.3041187},
	issn = {2169-3536},
	journal = {IEEE Access},
	pages = {214224--214239},
	publisher = {IEEE},
	volume = {8},
	year = {2020},
}

@book{carlson2002communication,
	author = {Carlson, A. Bruce and Crilly, Paul B. and Rutledge, Janet C.},
	title = {{Communication Systems: An Introduction to Signals and Noise in Electrical Communication}},
	edition = {4},
	isbn = {978-0-07-011127-1},
	location = {New York City, NY},
	pages = {850},
	publisher = {McGraw-Hill},
	year = {2002},
}

@article{casaleiro2024synchronisation,
	author = {Casaleiro, Duarte and Souto, Nuno and Silva, Jo{\~{a}}o C.},
	title = {{Synchronisation and Detection in Molecular Communication using a Deep-Learning-based Approach}},
	doi = {10.1109/access.2024.3519310},
	issn = {2169-3536},
	journal = {IEEE Access},
	pages = {192539 -- 192553},
	publisher = {IEEE},
	year = {2024},
}

@data{cell2cell2024data,
	author = {Torres G\'omez, Jorge and Khanzadeh, Roya and Angerbauer, Stefan and Debus, Lisa Y. and Hofmann, Pit and Başaran, Osman Tugay and Lotter, Sebastian and Unluturk, Bige Deniz and Abadal, Sergi and Fitzek, Frank H.P. and Haselmayr, Werner and Schober, Robert and Dressler, Falko},
	title = {{Dataset for Cell-to-Cell Communications}},
	note = {{IEEE Dataport}},
	publisher = {IEEE Dataport},
	url = {doi.org/10.21227/vv2r-qn28},
	year = {2024},
}

@article{chen2021selfattention,
	author = {Chen, Lei and Sun, Li},
	title = {{Self-Attention-Based Real-Time Signal Detector for Communication Systems With Unknown Channel Models}},
	doi = {10.1109/lcomm.2021.3082708},
	issn = {1558-2558},
	journal = {IEEE Communications Letters},
	month = Aug,
	number = {8},
	pages = {2639--2643},
	publisher = {IEEE},
	volume = {25},
	year = {2021},
}

@article{chen2021solving,
	author = {Chen, Xiaoli and Yang, Liu and Duan, Jinqiao and Karniadakis, George Em},
	title = {{Solving Inverse Stochastic Problems from Discrete Particle Observations Using the Fokker--Planck Equation and Physics-Informed Neural Networks}},
	doi = {10.1137/20m1360153},
	issn = {1095-7197},
	journal = {SIAM Journal on Scientific Computing},
	month = Jan,
	number = {3},
	pages = {B811--B830},
	publisher = {SIAM},
	volume = {43},
	year = {2021},
}

@article{cheng2023channel,
	author = {Cheng, Zhen and Sun, Jie and Zhang, Zhichao and Hu, Ping and Chi, Kaikai},
	title = {{Channel Modeling and Optimal Released Molecules for Mobile Molecular MIMO Communications Among Bionanosensors}},
	doi = {10.1109/jsen.2023.3304971},
	issn = {1530-437X},
	journal = {IEEE Sensors Journal},
	month = Oct,
	number = {19},
	pages = {22139--22152},
	publisher = {IEEE},
	volume = {23},
	year = {2023},
}

@inproceedings{cheng2023signal,
	author = {Cheng, Zhen and Zhang, Zhichao and Jiang, Ji and Sun, Jie},
	title = {{Signal Detection of Mobile Multi-user Molecular Communication System Using Transformer-Based Model}},
	booktitle = {8th International Conference on Computer and Communication Systems (ICCCS 2023)},
	address = {Guangzhou, China},
	doi = {10.1109/icccs57501.2023.10151419},
	isbn = {978-1-66545-613-5},
	month = Apr,
	pages = {85--90},
	publisher = {IEEE},
	year = {2023},
}

@article{cheng2024channel,
	author = {Cheng, Zhen and Chen, Miaodi and Liu, Heng and Xia, Ming and Gong, Weihua},
	title = {{Channel modeling for diffusion-based molecular MIMO communications using deep learning}},
	doi = {10.1016/j.nancom.2024.100543},
	issn = {1878-7789},
	journal = {Elsevier Nano Communication Networks},
	month = Dec,
	pages = {1 -- 9},
	publisher = {Elsevier},
	volume = {42},
	year = {2024},
}

@article{cheng2024informer,
	author = {Cheng, Zhen and Zhang, Zhichao and Jin, Xuancheng and Gong, Weihua and Chi, Kaikai},
	title = {{An Informer-Based Signal Sequence Detector for Mobile Molecular Communication}},
	doi = {10.1109/lcomm.2024.3381600},
	issn = {1558-2558},
	journal = {IEEE Communications Letters},
	month = Jun,
	number = {6},
	pages = {1397--1401},
	publisher = {IEEE},
	volume = {28},
	year = {2024},
}

@article{cheng2024resource,
	author = {Cheng, Zhen and Yan, Jun and Sun, Jie and Zhang, Shubin and Chi, Kaikai},
	title = {{Resource Allocation Optimization in Mobile Multiuser Molecular Communication by Deep Neural Network}},
	doi = {10.1109/tmbmc.2024.3412669},
	issn = {2332-7804},
	journal = {IEEE Transactions on Molecular, Biological and Multi-Scale Communications},
	month = Sep,
	number = {3},
	pages = {409--421},
	publisher = {IEEE},
	volume = {10},
	year = {2024},
}

@article{cheng2024signal,
	author = {Cheng, Zhen and Zhang, Zhichao and Sun, Jie},
	title = {{Signal Detection of Cooperative Multi-Hop Mobile Molecular Communication via Diffusion}},
	doi = {10.1109/tmbmc.2024.3360341},
	issn = {2332-7804},
	journal = {IEEE Transactions on Molecular, Biological and Multi-Scale Communications},
	month = Mar,
	number = {1},
	pages = {101--111},
	publisher = {IEEE},
	volume = {10},
	year = {2024},
}

@article{cheng2025deep,
	author = {Cheng, Zhen and Liu, Heng and Xu, Ziyan and Li, Jiaxin and Chi, Kaikai},
	title = {{Deep Learning-Based Estimation of Emission Time and Arrival Time in Diffusive Multi-Receiver Molecular Communication}},
	doi = {10.1109/tmbmc.2025.3546503},
	issn = {2332-7804},
	journal = {IEEE Transactions on Molecular, Biological and Multi-Scale Communications},
	month = Jun,
	number = {2},
	pages = {257--268},
	publisher = {IEEE},
	volume = {11},
	year = {2025},
}

@article{cheng2025localizing,
	author = {Cheng, Zhen and Liu, Heng and Zheng, Jianlong and Gong, Weihua and Chi, Kaikai},
	title = {{Localizing and Tracking the Transmitter Bionanosensor in Mobile Molecular Communication by Deep Learning}},
	doi = {10.1109/jsen.2025.3543552},
	issn = {1530-437X},
	journal = {IEEE Sensors Journal},
	month = Apr,
	number = {7},
	pages = {10583--10593},
	publisher = {IEEE},
	volume = {25},
	year = {2025},
}

@article{cherry2018scaling,
	author = {Cherry, Kevin M. and Qian, Lulu},
	title = {{Scaling up molecular pattern recognition with DNA-based winner-take-all neural networks}},
	doi = {10.1038/s41586-018-0289-6},
	issn = {0028-0836},
	journal = {Nature},
	month = Jul,
	number = {7714},
	pages = {370--376},
	publisher = {Springer Nature},
	volume = {559},
	year = {2018},
}

@article{damrath2018array,
	author = {Damrath, Martin and Yilmaz, Birkan and Chae, Chan-Byoung and Hoeher, Peter Adam},
	title = {{Array Gain Analysis in Molecular MIMO Communications}},
	doi = {10.1109/access.2018.2875925},
	issn = {2169-3536},
	journal = {IEEE Access},
	pages = {61091--61102},
	publisher = {IEEE},
	volume = {6},
	year = {2018},
}

@article{daniel2013synthetic,
	author = {Daniel, Ramiz and Rubens, Jacob R. and Sarpeshkar, Rahul and Lu, Timothy K.},
	title = {{Synthetic analog computation in living cells}},
	doi = {10.1038/nature12148},
	issn = {0028-0836},
	journal = {Nature},
	month = May,
	number = {7451},
	pages = {619--623},
	publisher = {Springer Nature},
	volume = {497},
	year = {2013},
}

@data{das2021dataset,
	author = {Das, Arunava and Runwal, Bharat and Cetinkaya, Oktay and Akan, Ozgur B.},
	title = {{Channel Estimation and Performance Analysis of SISO Molecular Communications - Version v1}},
	note = {{Zenodo}},
	publisher = {Zenodo},
	url = {doi.org/10.5281/zenodo.5701433},
	year = {2021},
}

@data{das2021dataset_2,
	author = {Das, Arunava and Runwal, Bharat and Cetinkaya, Oktay and Akan, Ozgur B.},
	title = {{Channel Estimation and Performance Analysis of SISO Molecular Communications - Version v2}},
	note = {{Zenodo}},
	publisher = {Zenodo},
	url = {doi.org/10.5281/zenodo.5701559},
	year = {2021},
}

@data{das2022dataset,
	author = {Das, Arunava and Runwal, Bharat and Baydas, O. Tansel and Cetinkaya, Oktay and Akan, Ozgur B.},
	title = {{Received Signal Modeling and BER Analysis for Molecular SISO Communications}},
	note = {{Zenodo}},
	publisher = {Zenodo},
	url = {doi.org/10.5281/zenodo.7036057},
	year = {2022},
}

@inproceedings{das2022received,
	author = {Das, Arunava and Runwal, Bharat and Baydas, O. Tansel and Cetinkaya, Oktay and Akan, Ozgur B.},
	title = {{Received signal modeling and BER analysis for molecular SISO communications}},
	booktitle = {9th ACM International Conference on Nanoscale Computing and Communication (NANOCOM 2022)},
	address = {Barcelona, Spain},
	doi = {10.1145/3558583.3558854},
	month = Oct,
	pages = {1--6},
	year = {2022},
}

@inproceedings{debus2023reinforcement,
	author = {Debus, Lisa Y. and Hofmann, Pit and Torres G{\'{o}}mez, Jorge and Fitzek, Frank H. P. and Dressler, Falko},
	title = {{Reinforcement Learning-based Receiver for Molecular Communication with Mobility}},
	booktitle = {IEEE Global Communications Conference (GLOBECOM 2023)},
	address = {Kuala Lumpur, Malaysia},
	doi = {10.1109/GLOBECOM54140.2023.10436754},
	month = Dec,
	pages = {558--564},
	publisher = {IEEE},
	year = {2023},
}

@article{debus2024synchronized,
	author = {Debus, Lisa Y. and Hofmann, Pit and Torres G{\'{o}}mez, Jorge and Fitzek, Frank H. P. and Dressler, Falko},
	title = {{Synchronized Relaying in Molecular Communication: An AI-based Approach using a Mobile Testbed Setup}},
	doi = {10.1109/TMBMC.2024.3420792},
	issn = {2332-7804},
	journal = {IEEE Transactions on Molecular, Biological and Multi-Scale Communications},
	month = Sep,
	number = {3},
	pages = {470--475},
	publisher = {IEEE},
	volume = {10},
	year = {2024},
}

@article{debus2025blood,
	author = {Debus, Lisa Y. and Wilhelm, Mario J. and Wolff, Henri and Wille, Luiz C. P. and Rese, Tim and Lommel, Michael and Kirchner, Jens and Dressler, Falko},
	title = {{Blood Makes a Difference: Experimental Evaluation of Molecular Communication in Different Fluids}},
	doi = {10.1109/TMBMC.2025.3602650},
	issn = {2332-7804},
	journal = {IEEE Transactions on Molecular, Biological and Multi-Scale Communications},
	month = Dec,
	number = {4},
	pages = {493--499},
	publisher = {IEEE},
	volume = {11},
	year = {2025},
}

@article{diab2022embedded,
	author = {Diab, Maha S. and Rodriguez-Villegas, Esther},
	title = {{Embedded Machine Learning Using Microcontrollers in Wearable and Ambulatory Systems for Health and Care Applications: A Review}},
	doi = {10.1109/access.2022.3206782},
	issn = {2169-3536},
	journal = {IEEE Access},
	pages = {98450--98474},
	publisher = {IEEE},
	volume = {10},
	year = {2022},
}

@article{dong2020interactive,
	author = {Dong, Ensheng and Du, Hongru and Gardner, Lauren},
	title = {{An interactive web-based dashboard to track COVID-19 in real time}},
	doi = {10.1016/s1473-3099(20)30120-1},
	issn = {1473-3099},
	journal = {The Lancet Infectious Diseases},
	month = May,
	number = {5},
	pages = {533--534},
	publisher = {Elsevier},
	volume = {20},
	year = {2020},
}

@article{dong2020recent,
	author = {Dong, Bowei and Ma, Yiming and Ren, Zhihao and Lee, Chengkuo},
	title = {{Recent progress in nanoplasmonics-based integrated optical micro/nano-systems}},
	doi = {10.1088/1361-6463/ab77db},
	issn = {1361-6463},
	journal = {Journal of Physics D: Applied Physics},
	month = Mar,
	number = {21},
	publisher = {IOP Publishing},
	volume = {53},
	year = {2020},
}

@article{drawert2012urdme,
	author = {Drawert, Brian and Engblom, Stefan and Hellander, Andreas},
	title = {{URDME: a modular framework for stochastic simulation of reaction-transport processes in complex geometries}},
	doi = {10.1186/1752-0509-6-76},
	issn = {1752-0509},
	journal = {BMC Systems Biology},
	month = Jun,
	number = {1},
	pages = {1--17},
	publisher = {Springer},
	volume = {6},
	year = {2012},
}

@data{drees2020_github_pogona,
	author = {Drees, J. P. and Stratmann, L. and Bronner, F. and Bartunik, M. and Kirchner, J. and Unterweger, H. and Dressler, F.},
	title = {{Pogona}},
	journal = {GitHub repository},
	publisher = {GitHub},
	url = {github.com/tkn-tub/pogona},
	year = {2020},
}

@inproceedings{drees2020efficient,
	author = {Drees, Jan Peter and Stratmann, Lukas and Bronner, Fabian and Bartunik, Max and Kirchner, Jens and Unterweger, Harald and Dressler, Falko},
	title = {{Efficient Simulation of Macroscopic Molecular Communication: The Pogona Simulator}},
	booktitle = {7th ACM International Conference on Nanoscale Computing and Communication (NANOCOM 2020)},
	address = {Virtual Conference},
	doi = {10.1145/3411295.3411297},
	isbn = {978-1-4503-8083-6},
	month = Sep,
	publisher = {ACM},
	year = {2020},
}

@inproceedings{ebner2024bvs-net,
	author = {Ebner, Laurenz and Torres G{\'{o}}mez, Jorge and Pal, Saswati and Ergen{\c{c}}, Do{\u{g}}analp and Wendt, Regine and Fischer, Stefan and Dressler, Falko},
	title = {{BVS-Net: A Networking Tool for Studying THz-based Intra-body Communication Links}},
	booktitle = {11th ACM International Conference on Nanoscale Computing and Communication (NANOCOM 2024), Poster Session},
	address = {Milan, Italy},
	doi = {10.1145/3686015.3691639},
	month = Oct,
	pages = {132--133},
	publisher = {ACM},
	year = {2024},
}

@article{egan2023toward,
	author = {Egan, Malcolm and Kuscu, Murat and Barros, Michael Taynnan and Booth, Michael and Llopis-Lorente, Antoni and Magarini, Maurizio and Martins, Daniel P. and Sch{\"{a}}fer, Maximilian and Stano, Pasquale},
	title = {{Toward Interdisciplinary Synergies in Molecular Communications: Perspectives from Synthetic Biology, Nanotechnology, Communications Engineering and Philosophy of Science}},
	doi = {10.3390/life13010208},
	issn = {2075-1729},
	journal = {Life},
	month = Jan,
	number = {1},
	pages = {1--14},
	publisher = {MDPI},
	volume = {13},
	year = {2023},
}

@book{eldar2022machine,
	author = {Eldar, Yonina C. and Goldsmith, Andrea and G{\"{u}}nd{\"{u}}z, Deniz and Poor, H. Vincent},
	title = {{Machine Learning and Wireless Communications}},
	doi = {10.1017/9781108966559},
	isbn = {978-1-108-83298-4},
	publisher = {Cambridge University Press},
	year = {2022},
}

@article{elhajj2015how,
	author = {El-Hajj, Ziad W. and Newman, Elaine B.},
	title = {{How much territory can a single E. coli cell control?}},
	doi = {10.3389/fmicb.2015.00309},
	issn = {1664-302X},
	journal = {Frontiers in Microbiology},
	month = Apr,
	pages = {1--12},
	publisher = {Frontiers Media SA},
	volume = {6},
	year = {2015},
}

@article{etemadi2023abnormality,
	author = {Etemadi, N. and Farahnak-Ghazani, Maryam and Arjmandi, Hamidreza and Mirmohseni, Mahtab and Nasiri-Kenari, Masoumeh},
	title = {{Abnormality Detection and Localization Schemes Using Molecular Communication Systems: A Survey}},
	doi = {10.1109/access.2022.3228618},
	issn = {2169-3536},
	journal = {IEEE Access},
	pages = {1761--1792},
	publisher = {IEEE},
	volume = {11},
	year = {2023},
}

@article{ewart2022performance,
	author = {Ewart, Lorna and Apostolou, Athanasia and Briggs, Skyler A. and Carman, Christopher V. and Chaff, Jake T. and Heng, Anthony R. and Jadalannagari, Sushma and Janardhanan, Jeshina and Jang, Kyung-Jin and Joshipura, Sannidhi R. and Kadam, Mahika M. and Kanellias, Marianne and Kujala, Ville J. and Kulkarni, Gauri and Le, Christopher Y. and Lucchesi, Carolina and Manatakis, Dimitris V. and Maniar, Kairav K. and Quinn, Meaghan E. and Ravan, Joseph S. and Rizos, Ann Catherine and Sauld, John F. K. and Sliz, Josiah D. and Tien-Street, William and Trinidad, Dennis Ramos and Velez, James and Wendell, Max and Irrechukwu, Onyi and Mahalingaiah, Prathap Kumar and Ingber, Donald E. and Scannell, Jack W. and Levner, Daniel},
	title = {{Performance assessment and economic analysis of a human Liver-Chip for predictive toxicology}},
	doi = {10.1038/s43856-022-00209-1},
	issn = {2730-664X},
	journal = {Nature Communications Medicine},
	month = Dec,
	number = {1},
	pages = {1--16},
	publisher = {Springer Nature},
	volume = {2},
	year = {2022},
}

@inproceedings{faraone2020convolutionalrecurrent,
	author = {Faraone, Antonino and Delgado-Gonzalo, Ricard},
	title = {{Convolutional-Recurrent Neural Networks on Low-Power Wearable Platforms for Cardiac Arrhythmia Detection}},
	booktitle = {2nd IEEE International Conference on Artificial Intelligence Circuits and Systems (AICAS 2020)},
	address = {Genova, Italy},
	doi = {10.1109/aicas48895.2020.9073950},
	isbn = {978-1-7281-4922-6},
	month = Aug,
	publisher = {IEEE},
	year = {2020},
}

@article{farsad2013tabletop,
	author = {Farsad, Nariman and Guo, Weisi and Eckford, Andrew W.},
	title = {{Tabletop Molecular Communication: Text Messages through Chemical Signals}},
	doi = {10.1371/journal.pone.0082935},
	issn = {1932-6203},
	journal = {PLoS ONE},
	month = Dec,
	number = {12},
	pages = {1--13},
	publisher = {Public Library of Science},
	volume = {8},
	year = {2013},
}

@techreport{farsad2017detection,
	author = {Farsad, Nariman and Goldsmith, Andrea},
	title = {{Detection Algorithms for Communication Systems Using Deep Learning}},
	doi = {10.48550/arXiv.1705.08044},
	institution = {arXiv},
	month = Jul,
	pages = {1 -- 10},
	type = {cs.LG},
	year = {2017},
}

@inproceedings{farsad2018detection,
	author = {Farsad, Nariman and Goldsmith, Andrea},
	title = {{Detection Over Rapidly Changing Communication Channels Using Deep Learning}},
	booktitle = {52nd Asilomar Conference on Signals, Systems, and Computers},
	address = {Pacific Grove, CA},
	doi = {10.1109/acssc.2018.8645187},
	isbn = {978-1-5386-9218-9},
	month = Oct,
	pages = {1 -- 5},
	publisher = {IEEE},
	year = {2018},
}

@article{farsad2018neural,
	author = {Farsad, Nariman and Goldsmith, Andrea},
	title = {{Neural Network Detection of Data Sequences in Communication Systems}},
	doi = {10.1109/tsp.2018.2868322},
	issn = {1053-587X},
	journal = {IEEE Transactions on Signal Processing},
	month = Nov,
	number = {21},
	pages = {5663--5678},
	publisher = {IEEE},
	volume = {66},
	year = {2018},
}

@inproceedings{farsad2018sliding,
	author = {Farsad, Nariman and Goldsmith, Andrea},
	title = {{Sliding Bidirectional Recurrent Neural Networks for Sequence Detection in Communication Systems}},
	booktitle = {IEEE International Conference on Acoustics, Speech and Signal Processing (ICASSP 2018)},
	address = {Calgary, Canada},
	doi = {10.1109/icassp.2018.8462140},
	issn = {2379-190X},
	month = Apr,
	pages = {2331 -- 2335},
	publisher = {IEEE},
	year = {2018},
}

@article{felicetti2012simulation,
	author = {Felicetti, Luca and Femminella, Mauro and Reali, Gianluca},
	title = {{A simulation tool for nanoscale biological networks}},
	doi = {10.1016/j.nancom.2011.09.002},
	issn = {1878-7789},
	journal = {Elsevier Nano Communication Networks},
	month = Mar,
	number = {1},
	pages = {2--18},
	publisher = {Elsevier},
	volume = {3},
	year = {2012},
}

@article{felicetti2013simulation,
	author = {Felicetti, Luca and Femminella, Mauro and Reali, Gianluca},
	title = {{Simulation of molecular signaling in blood vessels: Software design and application to atherogenesis}},
	doi = {10.1016/j.nancom.2013.06.002},
	issn = {1878-7789},
	journal = {Elsevier Nano Communication Networks},
	month = Sep,
	number = {3},
	pages = {98--119},
	publisher = {Elsevier},
	volume = {4},
	year = {2013},
}

@article{felicetti2014tcplike,
	author = {Felicetti, Luca and Femminella, Mauro and Reali, Gianluca and Nakano, Tadashi and Vasilakos, Athanasios V},
	title = {{TCP-Like Molecular Communications}},
	doi = {10.1109/jsac.2014.2367653},
	issn = {0733-8716},
	journal = {IEEE Journal on Selected Areas in Communications},
	month = Dec,
	number = {12},
	pages = {2354--2367},
	publisher = {IEEE},
	volume = {32},
	year = {2014},
}

@article{fink2022mmft,
	author = {Fink, Gerold and Costamoling, Florina and Wille, Robert},
	title = {{MMFT Droplet Simulator: Efficient Simulation of Droplet-based Microfluidic Devices}},
	doi = {10.1016/j.simpa.2022.100440},
	issn = {2665-9638},
	journal = {Software Impacts},
	month = Dec,
	pages = {100440},
	publisher = {Elsevier},
	volume = {14},
	year = {2022},
}

@inproceedings{galvan2024tailoring,
	author = {Galv{\'{a}}n, Pablo and Lemic, Filip and Calvo, Gerard and Abadal, Sergi and Costa-P{\'{e}}rez, Xavier},
	title = {{Tailoring Graph Neural Network-based Flow-guided Localization to Individual Bloodstreams and Activities}},
	booktitle = {11th ACM International Conference on Nanoscale Computing and Communication (NANOCOM 2024)},
	address = {Milan, Italy},
	doi = {10.1145/3686015.3689356},
	month = Oct,
	pages = {109--115},
	publisher = {ACM},
	year = {2024},
}

@article{gao2021molecular,
	author = {Gao, Weidong and Mak, Terrence and Yang, Lie-Liang},
	title = {{Molecular Type Spread Molecular Shift Keying for Multiple-Access Diffusive Molecular Communications}},
	doi = {10.1109/tmbmc.2020.3041182},
	issn = {2332-7804},
	journal = {IEEE Transactions on Molecular, Biological and Multi-Scale Communications},
	month = Mar,
	number = {1},
	pages = {51--63},
	publisher = {IEEE},
	volume = {7},
	year = {2021},
}

@data{gao2023dataset,
	author = {Gao, Weidong},
	title = {{Dataset in Support of the Southampton Doctoral Thesis "Type-Spread and Multiple-Access Molecular Communications"}},
	publisher = {University of Southampton},
	url = {doi.org/10.5258/SOTON/D2545},
	year = {2023},
}

@phdthesis{gao2023typespread,
	author = {Gao, Weidong},
	title = {{Type-Spread and Multiple-Access Molecular Communications}},
	advisor = {Yang, Lie-Liang and Mak, Terrence},
	institution = {School of Electronics and Computer Science},
	location = {Southampton, United Kingdom},
	month = Feb,
	school = {University of Southampton},
	type = {PhD Thesis},
	year = {2023},
}

@article{gentili2024neuromorphic,
	author = {Gentili, Pier Luigi and Zurlo, Maria Pia and Stano, Pasquale},
	title = {{Neuromorphic engineering in wetware: the state of the art and its perspectives}},
	doi = {10.3389/fnins.2024.1443121},
	journal = {Frontiers in Neuroscience},
	month = Sep,
	pages = {1--6},
	publisher = {Frontiers Media SA},
	volume = {18},
	year = {2024},
}

@inproceedings{geyer2018bloodvoyagers,
	author = {Geyer, Regine and Stelzner, Marc and B{\"{u}}ther, Florian and Ebers, Sebastian},
	title = {{BloodVoyagerS: Simulation of the Work Environment of Medical Nanobots}},
	booktitle = {5th ACM International Conference on Nanoscale Computing and Communication (NANOCOM 2018)},
	address = {Reykjav{\'{i}}k, Iceland},
	doi = {10.1145/3233188.3233196},
	isbn = {978-1-4503-5711-1},
	month = Sep,
	pages = {5:1--5:6},
	publisher = {ACM},
	year = {2018},
}

@inproceedings{geyer2020bvs-vis,
	author = {Geyer, Regine and Deter, Chris and Fischer, Stefan},
	title = {{BVS-Vis: a web-based visualizer for BloodVoyagerS}},
	booktitle = {7th ACM International Conference on Nanoscale Computing and Communication (NANOCOM 2020)},
	address = {Virtual Conference},
	doi = {10.1145/3411295.3411300},
	isbn = {978-1-4503-8083-6},
	month = Sep,
	publisher = {ACM},
	year = {2020},
}

@data{gokarslan2020github,
	author = {Gökarslan, Kerim and \c{C}a\u{g}{\i}r{\i}c{\i}, Erhan},
	title = {{MoleCom-Gpu}},
	journal = {GitHub repository},
	publisher = {GitHub},
	url = {github.com/MoleCom-Gpu/MoleCom-Gpu},
	year = {2017},
}

@article{gong2023survey,
    author = {Gong, Youdi and Liu, Guangzhen and Xue, Yunzhi and Li, Rui and Meng, Lingzhong},
    doi = {10.1016/j.infsof.2023.107268},
    title = {{A survey on dataset quality in machine learning}},
    pages = {107268},
    journal = {Information and Software Technology},
    issn = {0950-5849},
    publisher = {Elsevier},
    month = {10},
    volume = {162},
    year = {2023}
}

@inproceedings{grebenstein2018biological,
	author = {Grebenstein, Laura and Kirchner, Jens and Peixoto, Renata Stavracakis and Zimmermann, Wiebke and Wicke, Wayan and Ahmadzadeh, Arman and Jamali, Vahid and Fischer, Georg and Weigel, Robert and Burkovski, Andreas and Schober, Robert},
	title = {{Biological optical-to-chemical signal conversion interface: a small-scale modulator for molecular communications}},
	booktitle = {5th ACM International Conference on Nanoscale Computing and Communication (NANOCOM 2018)},
	address = {Reykjav{\'{i}}k, Iceland},
	doi = {10.1145/3233188.3233203},
	isbn = {978-1-4503-5711-1},
	month = Sep,
	pages = {1--6},
	publisher = {ACM},
	year = {2018},
}

@data{grebenstein2019dataset_molecular,
	author = {Grebenstein, Laura and Kirchner, Jens and Wicke, Wayan and Ahmadzadeh, Arman and Jamali, Vahid and Fischer, Georg and Weigel, Robert and Burkovski, Andreas and Schober, Robert},
	title = {A Molecular Communication Testbed Based on Proton Pumping Bacteria},
	note = {{IEEE Dataport}},
	publisher = {IEEE Dataport},
	url = {doi.org/10.21227/3zj6-pm05},
	year = {2019},
}

@article{grebenstein2019molecular,
	author = {Grebenstein, Laura and Kirchner, Jens and Wicke, Wayan and Ahmadzadeh, Arman and Jamali, Vahid and Fischer, Georg and Weigel, Robert and Burkovski, Andreas and Schober, Robert},
	title = {{A Molecular Communication Testbed Based on Proton Pumping Bacteria: Methods and Data}},
	doi = {10.1109/tmbmc.2019.2957783},
	issn = {2332-7804},
	journal = {IEEE Transactions on Molecular, Biological and Multi-Scale Communications},
	month = Oct,
	number = {1},
	pages = {56--62},
	publisher = {IEEE},
	volume = {5},
	year = {2019},
}

@data{grimmer2019_github,
	author = {Grimmer, Andreas and Hamidovi{\'{c}}, Medina and Haselmayr, Werner and Wille, Robert},
	title = {{SIMPAC-2022-234}},
	journal = {GitHub repository},
	publisher = {GitHub},
	url = {github.com/SoftwareImpacts/SIMPAC-2022-234},
	year = {2022},
}

@article{grimmer2019advanced,
	author = {Grimmer, Andreas and Hamidovi{\'{c}}, Medina and Haselmayr, Werner and Wille, Robert},
	title = {{Advanced Simulation of Droplet Microfluidics}},
	doi = {10.1145/3313867},
	issn = {1550-4840},
	journal = {ACM Journal on Emerging Technologies in Computing Systems},
	month = Apr,
	number = {3},
	pages = {1--16},
	publisher = {ACM},
	volume = {15},
	year = {2019},
}

@data{gulec2018github,
	author = {Gulec, Fatih},
	title = {{Signal-Reconstruction-in-Diffusion-based-Molecular-Communication}},
	journal = {GitHub repository},
	publisher = {GitHub},
	url = {github.com/fatihguelec/Signal-Reconstruction-in-Diffusion-based-Molecular-Communication},
	year = {2018},
}

@article{gulec2020distance,
	author = {G{\"{u}}le{\c{c}}, Fatih and Atakan, Baris},
	title = {{Distance estimation methods for a practical macroscale molecular communication system}},
	doi = {10.1016/j.nancom.2020.100300},
	issn = {1878-7789},
	journal = {Elsevier Nano Communication Networks},
	month = May,
	pages = {1 -- 15},
	publisher = {Elsevier},
	volume = {24},
	year = {2020},
}

@data{gulec2020github,
	author = {Gulec, Fatih},
	title = {{Distance-Estimation-in-Molecular-Communication}},
	journal = {GitHub repository},
	publisher = {GitHub},
	url = {github.com/fatihguelec/Distance-Estimation-in-Molecular-Communication},
	year = {2020},
}

@article{gulec2021fluid,
	author = {G{\"{u}}le{\c{c}}, Fatih and Atakan, Baris},
	title = {{Fluid dynamics-based distance estimation algorithm for macroscale molecular communication}},
	doi = {10.1016/j.nancom.2021.100351},
	issn = {1878-7789},
	journal = {Elsevier Nano Communication Networks},
	month = Jun,
	pages = {1--9},
	publisher = {Elsevier},
	volume = {28},
	year = {2021},
}

@data{gulec2022github,
	author = {Gulec, Fatih},
	title = {{Mobile-Human-Ad-Hoc-Networks}},
	journal = {GitHub repository},
	publisher = {GitHub},
	url = {github.com/fatihguelec/Mobile-Human-Ad-Hoc-Networks},
	year = {2022},
}

@data{gulec2022github_stochastic,
	author = {Gulec, Fatih},
	title = {{Stochastic-Modeling-of-Biofilm-Formation-with-Bacterial-Quorum-Sensing}},
	journal = {GitHub repository},
	publisher = {GitHub},
	url = {github.com/fatihguelec/Stochastic-Modeling-of-Biofilm-Formation-with-Bacterial-Quorum-Sensing},
	year = {2022},
}

@data{gulec2022github_stochastic2,
	author = {Gulec, Fatih},
	title = {{Stochastic-Biofilm-Disruption-Model-Based-on-Quorum-Sensing-Mimickers}},
	journal = {GitHub repository},
	publisher = {GitHub},
	url = {github.com/fatihguelec/Stochastic-Biofilm-Disruption-Model-Based-on-Quorum-Sensing-Mimickers},
	year = {2022},
}

@article{gulec2022mobile,
	author = {G{\"{u}}le{\c{c}}, Fatih and Atakan, Baris and Dressler, Falko},
	title = {{Mobile Human Ad Hoc Networks: A Communication Engineering Viewpoint on Interhuman Airborne Pathogen Transmission}},
	doi = {10.1016/j.nancom.2022.100410},
	issn = {1878-7789},
	journal = {Elsevier Nano Communication Networks},
	month = Jun,
	pages = {1--11},
	publisher = {Elsevier},
	volume = {32-33},
	year = {2022},
}

@article{gulec2023computational,
	author = {G{\"{u}}le{\c{c}}, Fatih and Dressler, Falko and Eckford, Andrew W.},
	title = {{A Computational Approach for the Characterization of Airborne Pathogen Transmission in Turbulent Molecular Communication Channels}},
	doi = {10.1109/TMBMC.2023.3273193},
	issn = {2332-7804},
	journal = {IEEE Transactions on Molecular, Biological and Multi-Scale Communications},
	month = Jun,
	number = {2},
	pages = {124--134},
	publisher = {IEEE},
	volume = {9},
	year = {2023},
}

@inproceedings{gulec2023stochastic,
	author = {Gulec, Fatih and Eckford, Andrew},
	title = {{Stochastic Modeling of Biofilm Formation with Bacterial Quorum Sensing}},
	booktitle = {IEEE International Conference on Communications (ICC 2023)},
	address = {Rome, Italy},
	doi = {10.1109/icc45041.2023.10278566},
	isbn = {978-1-5386-7463-5},
	issn = {1938-1883},
	month = May,
	pages = {4470--4475},
	publisher = {IEEE},
	year = {2023},
}

@article{gulec2023stochastic2,
	author = {Gulec, Fatih and Eckford, Andrew},
	title = {{A Stochastic Biofilm Disruption Model Based on Quorum Sensing Mimickers}},
	doi = {10.1109/tmbmc.2023.3292321},
	issn = {2332-7804},
	journal = {IEEE Transactions on Molecular, Biological and Multi-Scale Communications},
	month = Sep,
	number = {3},
	pages = {346--350},
	publisher = {IEEE},
	volume = {9},
	year = {2023},
}

@data{gulec2024_zenodo,
	author = {Gulec, Fatih},
	title = {{CFD Simulation Dataset for Airborne Pathogen Transmission in Turbulent Channels}},
	note = {{Zenodo}},
	publisher = {Zenodo},
	url = {doi.org/10.5281/zenodo.13793238},
	year = {2024},
}

@data{gulec2024github,
	author = {Gulec, Fatih},
	title = {{CFD-Approach-for-the-Characterization-of-Airborne-Transmission-in-Turbulent-MC-Channels}},
	journal = {GitHub repository},
	publisher = {GitHub},
	url = {github.com/fatihguelec/CFD-Approach-for-the-Characterization-of-Airborne-Transmission-in-Turbulent-MC-Channels},
	year = {2024},
}

@article{haluan2024synthetic,
	author = {Halužan Vasle, Ana and Moškon, Miha},
	title = {{Synthetic biological neural networks: From current implementations to future perspectives}},
	doi = {10.1016/j.biosystems.2024.105164},
	issn = {0303-2647},
	journal = {Biosystems},
	month = Mar,
	pages = {1--11},
	publisher = {Elsevier},
	volume = {237},
	year = {2024},
}

@article{hamidovic2024microfluidic,
	author = {Hamidovi{\'{c}}, Medina and Angerbauer, Stefan and Bi, Dadi and Deng, Yansha and Tugcu, Tuna and Haselmayr, Werner},
	title = {{Microfluidic Systems for Molecular Communications: A Review From Theory to Practice}},
	doi = {10.1109/tmbmc.2024.3368768},
	issn = {2332-7804},
	journal = {IEEE Transactions on Molecular, Biological and Multi-Scale Communications},
	month = Mar,
	number = {1},
	pages = {147--163},
	publisher = {IEEE},
	volume = {10},
	year = {2024},
}

@article{harris2016bionetgen,
	author = {Harris, Leonard A. and Hogg, Justin S. and Tapia, Jos\'e-Juan and Sekar, John A. P. and Gupta, Sanjana and Korsunsky, Ilya and Arora, Arshi and Barua, Dipak and Sheehan, Robert P. and Faeder, James R.},
	title = {{BioNetGen 2.2: advances in rule-based modeling}},
	address = {Oxford, United Kingdom},
	doi = {10.1093/bioinformatics/btw469},
	issn = {1367-4811},
	journal = {Bioinformatics},
	month = Jul,
	number = {21},
	pages = {3366--3368},
	publisher = {OUP},
	volume = {32},
	year = {2016},
}

@data{hattne2005meso,
	author = {Hattne, Johan and Fange, David and Elf, Johan},
	title = {{MesoRD}},
	journal = {SourceForge repository},
	publisher = {SourceForge},
	url = {sourceforge.net/projects/mesord/},
	year = {2005},
}

@article{hattne2005stochastic,
	author = {Hattne, J. and Fange, D. and Elf, J.},
	title = {{Stochastic reaction-diffusion simulation with MesoRD}},
	address = {Oxford, United Kingdom},
	doi = {10.1093/bioinformatics/bti431},
	issn = {1367-4811},
	journal = {Bioinformatics},
	month = Apr,
	number = {12},
	pages = {2923--2924},
	publisher = {OUP},
	volume = {21},
	year = {2005},
}

@article{haykin2005cognitive,
	author = {Haykin, Simon},
	title = {{Cognitive radio: brain-empowered wireless communications}},
	doi = {10.1109/JSAC.2004.839380},
	issn = {0733-8716},
	journal = {IEEE Journal on Selected Areas in Communications},
	month = Feb,
	number = {2},
	pages = {201--220},
	publisher = {IEEE},
	volume = {23},
	year = {2005},
}

@article{he2024pruning,
	author = {He, Yang and Xiao, Lingao},
	title = {{Structured Pruning for Deep Convolutional Neural Networks: A Survey}},
	doi = {10.1109/TPAMI.2023.3334614},
	journal = {IEEE Transactions on Pattern Analysis and Machine Intelligence},
	month = Nov,
	number = {5},
	pages = {2900--2919},
	volume = {46},
	year = {2024},
}

@article{hjelmfelt1991chemical,
	author = {Hjelmfelt, A. and Weinberger, E. D. and Ross, J.},
	title = {{Chemical implementation of neural networks and Turing machines}},
	address = {Washington, D.C.},
	doi = {10.1073/pnas.88.24.10983},
	issn = {1091-6490},
	journal = {Proceedings of the National Academy of Sciences (PNAS)},
	month = Dec,
	number = {24},
	pages = {10983--10987},
	publisher = {National Academy of Sciences},
	volume = {88},
	year = {1991},
}

@inproceedings{hofmann2022testbed,
	author = {Hofmann, Pit and Torres G{\'{o}}mez, Jorge and Dressler, Falko and Fitzek, Frank H. P.},
	title = {{Testbed-based Receiver Optimization for SISO Molecular Communication Channels}},
	booktitle = {5th IEEE International Balkan Conference Communications and Networking (BalkanCom 2022)},
	address = {Sarajevo, Bosnia and Herzegovina},
	doi = {10.1109/BalkanCom55633.2022.9900720},
	isbn = {978-1-66548-764-1},
	month = Aug,
	pages = {120--125},
	publisher = {IEEE},
	year = {2022},
}

@inproceedings{hofmann2023analog,
	author = {Hofmann, Pit and Cabrera, Juan A. and Bassoli, Riccardo and Fitzek, Frank H. P.},
	title = {{Analog Network Coding in Molecular Communications: A Practical Implementation}},
	booktitle = {IEEE Global Communications Conference (GLOBECOM 2023)},
	address = {Kuala Lumpur, Malaysia},
	doi = {10.1109/globecom54140.2023.10437513},
	month = Dec,
	pages = {571--576},
	publisher = {IEEE},
	year = {2023},
}

@article{hofmann2023coding,
	author = {Hofmann, Pit and Cabrera, Juan A. and Bassoli, Riccardo and Reisslein, Martin and Fitzek, Frank H. P.},
	title = {{Coding in Diffusion-Based Molecular Nanonetworks: A Comprehensive Survey}},
	doi = {10.1109/access.2023.3243797},
	issn = {2169-3536},
	journal = {IEEE Access},
	pages = {16411--16465},
	publisher = {IEEE},
	volume = {11},
	year = {2023},
}

@data{hofmann2023dataset_analog,
	author = {Hofmann, Pit and Cabrera, Juan A. and Bassoli, Riccardo and Fitzek, Frank H.P.},
	title = {Dataset for Analog Network Coding in Molecular Communications: A Practical Implementation},
	note = {{IEEE Dataport}},
	publisher = {IEEE Dataport},
	url = {doi.org/10.21227/57z0-9q64},
	year = {2023},
}

@data{hofmann2023dataset_macroscale,
	author = {Hofmann, Pit and Torres Gómez, Jorge and Frank H.P., Frank H.P. and Dressler, Falko},
	title = {Dataset for Macroscale Molecular Communication Testbed},
	note = {{IEEE Dataport}},
	publisher = {IEEE Dataport},
	url = {doi.org/10.21227/ytkm-xp81},
	year = {2023},
}

@data{hofmann2023dataset_simulation,
	author = {Hofmann, Pit and Zhou, Pengjie and Lee, Changmin and Reisslein, Martin and Fitzek, Frank H.P. and Chae, Chan-Byoung},
	title = {Dataset for the Simulation of Microfluidic Molecular Communication using OpenFOAM},
	note = {{IEEE Dataport}},
	publisher = {IEEE Dataport},
	url = {doi.org/10.21227/b71c-4286},
	year = {2023},
}

@data{hofmann2024dataset_plaque,
	author = {Hofmann, Pit and Wietfeld, Alexander and Fuchtmann, Jonas and Zhou, Pengjie and Zheng, Ruifeng and Cabrera, Juan and Fitzek, Frank H.P. and Kellerer, Wolfgang},
	title = {Dataset for Advanced Plaque Modeling for Atherosclerosis Detection using Molecular Communication},
	note = {{IEEE Dataport}},
	publisher = {IEEE Dataport},
	url = {doi.org/10.21227/mn8d-0h55},
	year = {2024},
}

@article{hofmann2024molecular,
	author = {Hofmann, Pit and Schmidt, Sebastian and Wietfeld, Alexander and Zhou, Pengjie and Fuchtmann, Jonas and Fitzek, Frank H. P. and Kellerer, Wolfgang},
	title = {{A Molecular Communication Perspective on Detecting Arterial Plaque Formation}},
	doi = {10.1109/tmbmc.2024.3423005},
	issn = {2332-7804},
	journal = {IEEE Transactions on Molecular, Biological and Multi-Scale Communications},
	month = Sep,
	number = {3},
	pages = {458--463},
	publisher = {IEEE},
	volume = {10},
	year = {2024},
}

@article{hofmann2024openfoam,
	author = {Hofmann, Pit and Zhou, Pengjie and Lee, Changmin and Reisslein, Martin and Fitzek, Frank H. P. and Chae, Chan-Byoung},
	title = {{OpenFOAM Simulation of Microfluidic Molecular Communications: Method and Experimental Validation}},
	doi = {10.1109/access.2024.3438243},
	issn = {2169-3536},
	journal = {IEEE Access},
	pages = {109494--109512},
	publisher = {IEEE},
	volume = {12},
	year = {2024},
}

@article{holt2020protease,
	author = {Holt, Brandon Alexander and Kwong, Gabriel A.},
	title = {{Protease circuits for processing biological information}},
	doi = {10.1038/s41467-020-18840-8},
	issn = {2041-1723},
	journal = {Nature Communications},
	month = Oct,
	number = {1},
	pages = {1--12},
	publisher = {Springer Nature},
	volume = {11},
	year = {2020},
}

@article{huang2021signal,
	author = {Huang, Yu and Ji, Fei and Wei, Zhuangkun and Wen, Miaowen and Guo, Weisi},
	title = {{Signal Detection for Molecular Communication: Model-Based vs. Data-Driven Methods}},
	doi = {10.1109/mcom.001.2000957},
	issn = {0163-6804},
	journal = {IEEE Communications Magazine},
	month = May,
	number = {5},
	pages = {47--53},
	publisher = {IEEE},
	volume = {59},
	year = {2021},
}

@article{huang2022survey,
	author = {Huang, Xinyu and Fang, Yuting and Yang, Nan},
	title = {{A survey on estimation schemes in molecular communications}},
	doi = {10.1016/j.dsp.2021.103163},
	issn = {1051-2004},
	journal = {Elsevier Digital Signal Processing},
	month = May,
	pages = {1--13},
	publisher = {Elsevier},
	volume = {124},
	year = {2022},
}

@article{huang2025demystifying,
	author = {Huang, Yu and Luo, Min and Huang, Xinyu and Wen, Miaowen and Chae, Chan-Byoung},
	title = {{Demystifying Molecular Data-driven Detection with Explainable Artificial Intelligence}},
	doi = {10.1109/lwc.2025.3554889},
	issn = {2162-2337},
	journal = {IEEE Wireless Communications Letters},
	publisher = {IEEE},
	year = {2025},
}

@techreport{hube2025set,
	author = {Hube, Mika Leo and Lemic, Filip and Shitiri, Ethungshan and Bartra, Gerard Calvo and Abadal, Sergi and Costa-P{\'{e}}rez, Xavier},
	title = {{Set Transformer Architectures and Synthetic Data Generation for Flow-Guided Nanoscale Localization}},
	doi = {10.48550/ARXIV.2508.16200},
	institution = {arXiv},
	month = Aug,
	number = {2508.16200},
	type = {cs.ET},
	year = {2025},
}

@article{huynh2013chemical,
	author = {Huynh, Toan and Sun, Bing and Li, Liang and Nichols, Kevin P. and Koyner, Jay L. and Ismagilov, Rustem F.},
	title = {{Chemical Analog-to-Digital Signal Conversion Based on Robust Threshold Chemistry and Its Evaluation in the Context of Microfluidics-Based Quantitative Assays}},
	doi = {10.1021/ja4062882},
	issn = {1520-5126},
	journal = {Journal of the American Chemical Society},
	month = Sep,
	number = {39},
	pages = {14775--14783},
	publisher = {ACS},
	volume = {135},
	year = {2013},
}

@article{jamali2017symbol,
	author = {Jamali, Vahid and Ahmadzadeh, Arman and Schober, Robert},
	title = {{Symbol Synchronization for Diffusion-Based Molecular Communications}},
	doi = {10.1109/tnb.2017.2782761},
	issn = {1558-2639},
	journal = {IEEE Transactions on NanoBioscience},
	month = Dec,
	number = {8},
	pages = {873--887},
	publisher = {IEEE},
	volume = {16},
	year = {2017},
}

@article{jamali2019channel,
	author = {Jamali, Vahid and Ahmadzadeh, Arman and Wicke, Wayan and Noel, Adam and Schober, Robert},
	title = {{Channel Modeling for Diffusive Molecular Communication - A Tutorial Review}},
	doi = {10.1109/jproc.2019.2919455},
	issn = {0018-9219},
	journal = {Proceedings of the IEEE},
	month = Jul,
	number = {7},
	pages = {1256--1301},
	publisher = {IEEE},
	volume = {107},
	year = {2019},
}

@article{jian2017nanons3,
	author = {Jian, Yubing and Krishnaswamy, Bhuvana and Austin, Caitlin M. and Bicen, A. Ozan and Einolghozati, Arash and Perdomo, Jorge E. and Patel, Sagar C. and Fekri, Faramarz and Akyildiz, Ian F. and Forest, Craig R. and Sivakumar, Raghupathy},
	title = {{nanoNS3: A network simulator for bacterial nanonetworks based on molecular communication}},
	doi = {10.1016/j.nancom.2017.01.004},
	issn = {1878-7789},
	journal = {Elsevier Nano Communication Networks},
	month = Jun,
	pages = {1--11},
	publisher = {Elsevier},
	volume = {12},
	year = {2017},
}

@article{jin2024transformerbased,
	author = {Jin, Xuancheng and Cheng, Zhen and Chen, Miaodi and Liu, Heng and Gong, Weihua and Chi, Kaikai},
	title = {{Transformer-Based Receiver Localization in Vessel-Like and Flow-Induced Molecular Communication via Diffusion}},
	doi = {10.1109/lcomm.2024.3432146},
	issn = {1558-2558},
	journal = {IEEE Communications Letters},
	month = Oct,
	number = {10},
	pages = {2283--2287},
	publisher = {IEEE},
	volume = {28},
	year = {2024},
}

@article{johnson2019survey,
	author = {Johnson, Justin M. and Khoshgoftaar, Taghi M.},
	title = {{Survey on Deep Learning with Class Imbalance}},
	DOI = {10.1186/s40537-019-0192-5},
	journal = {Journal of Big Data},
	month = Mar,
	number = {1},
	publisher = {Springer},
	volume = {6},
	year = {2019},
}

@article{jornet2023nanonetworking,
	author = {Jornet, Josep Miquel and Sangwan, Amit},
	title = {{Nanonetworking in the Terahertz Band and Beyond}},
	doi = {10.1109/mnano.2023.3262105},
	issn = {1932-4510},
	journal = {IEEE Nanotechnology Magazine},
	month = Jun,
	number = {3},
	pages = {21--31},
	publisher = {IEEE},
	volume = {17},
	year = {2023},
}

@article{kagan2022invitro,
	author = {Kagan, Brett J. and Kitchen, Andy C. and Tran, Nhi T. and Habibollahi, Forough and Khajehnejad, Moein and Parker, Bradyn J. and Bhat, Anjali and Rollo, Ben and Razi, Adeel and Friston, Karl J.},
	title = {{\textit{In vitro} neurons learn and exhibit sentience when embodied in a simulated game-world}},
	doi = {10.1016/j.neuron.2022.09.001},
	issn = {0896-6273},
	journal = {Neuron},
	month = Dec,
	number = {23},
	pages = {3952--3969},
	publisher = {Elsevier},
	volume = {110},
	year = {2022},
}

@article{kara2022molecular,
	author = {Kara, Ozgur and Yaylali, Gokberk and Pusane, Ali E. and Tugcu, Tuna},
	title = {{Molecular index modulation using convolutional neural networks}},
	doi = {10.1016/j.nancom.2022.100420},
	issn = {1878-7789},
	journal = {Elsevier Nano Communication Networks},
	month = Dec,
	pages = {1 -- 8},
	publisher = {Elsevier},
	volume = {34},
	year = {2022},
}

@inproceedings{khanzadeh2023end,
	author = {Khanzadeh, Roya and Angerbauer, Stefan and Springer, Andreas and Haselmayr, Werner},
	title = {{End-to-End Learning of Communication Systems with Novel Data-Efficient IIR Channel Identification}},
	booktitle = {57th Asilomar Conference on Signals, Systems, and Computers},
	address = {Pacific Grove, CA},
	doi = {10.1109/ieeeconf59524.2023.10476924},
	issn = {2576-2303},
	month = Oct,
	pages = {40--46},
	publisher = {IEEE},
	year = {2023},
}

@inproceedings{khanzadeh2023towards,
	author = {Khanzadeh, Roya and Angerbauer, Stefan and Enzenhofer, Franz and Springer, Andreas and Haselmayr, Werner},
	title = {{Towards End-to-End Learning for Salinity-based Molecular Communication}},
	booktitle = {7th Workshop on Molecular Communications (WMC 2023)},
	address = {Erlangen, Germany},
	month = Apr,
	pages = {1--2},
	year = {2023},
}

@inproceedings{khanzadeh2024explainable,
	author = {Khanzadeh, Roya and Angerbauer, Stefan and Torres G{\'{o}}mez, Jorge and Hofmann, Pit and Dressler, Falko and Fitzek, Frank H. P. and Springer, Andreas and Haselmayr, Werner},
	title = {{Explainable Asymmetric Auto-Encoder for End-to-End Learning of IoBNT Communications}},
	booktitle = {IEEE International Conference on Machine Learning for Communication and Networking (ICMLCN 2024)},
	address = {Stockholm, Sweden},
	doi = {10.1109/ICMLCN59089.2024.10624774},
	month = May,
	pages = {412--418},
	publisher = {IEEE},
	year = {2024},
}

@article{khanzadeh2024ql-based,
	author = {Khanzadeh, Roya and Angerbauer, Stefan and Torres G{\'{o}}mez, Jorge and Springer, Andreas and Dressler, Falko and Haselmayr, Werner},
	title = {{QL-based Adaptive Transceivers for the IoBNT Communications}},
	doi = {10.1109/TMBMC.2024.3420749},
	issn = {2332-7804},
	journal = {IEEE Transactions on Molecular, Biological and Multi-Scale Communications},
	month = Sep,
	number = {3},
	pages = {476--480},
	publisher = {IEEE},
	volume = {10},
	year = {2024},
}

@inproceedings{khanzadeh2025end,
	author = {Khanzadeh, Roya and Angerbauer, Stefan and Springer, Andreas and Haselmayr, Werner},
	title = {End-to-End Learning for Time-Varying Non-Binary Molecular Communications},
	booktitle = {2025 IEEE International Conference on Machine Learning for Communication and Networking (ICMLCN)},
	organization = {IEEE},
	pages = {1--6},
	year = {2025},
}

@inproceedings{kim2004neural,
	author = {Kim, Jongmin and Hopfield, John J. and Winfree, Erik},
	title = {{Neural network computation by in vitro transcriptional circuits}},
	booktitle = {18th International Conference on Neural Information Processing Systems (NIPS 2004)},
	address = {Vancouver, Canada},
	month = Dec,
	pages = {681--688},
	publisher = {MIT Press},
	year = {2004},
}

@article{kim2023machine,
	author = {Kim, Su-Jin and Singh, Pankaj and Jung, Sung-Yoon},
	title = {{A machine learning-based concentration-encoded molecular communication system}},
	doi = {10.1016/j.nancom.2022.100433},
	issn = {1878-7789},
	journal = {Elsevier Nano Communication Networks},
	month = Mar,
	pages = {1 -- 12},
	publisher = {Elsevier},
	volume = {35},
	year = {2023},
}

@techreport{kingma2014adam,
	author = {Kingma, Diederik P. and Ba, Jimmy},
	title = {{Adam: A Method for Stochastic Optimization}},
	doi = {10.48550/arXiv.1412.6980},
	institution = {arXiv},
	month = Dec,
	number = {1412.6980},
	pages = {1--15},
	type = {cs.LG},
	year = {2014},
}

@article{koo2016molecular,
	author = {Koo, B. H. and Lee, C. and Yilmaz, H. Birkan and Farsad, N. and Eckford, Andrew W. and Chae, C. B.},
	title = {{Molecular MIMO: From Theory to Prototype}},
	doi = {10.1109/JSAC.2016.2525538},
	issn = {0733-8716},
	journal = {IEEE Journal on Selected Areas in Communications},
	month = Mar,
	number = {3},
	pages = {600--614},
	publisher = {IEEE},
	volume = {34},
	year = {2016},
}

@inproceedings{koo2020deep,
	author = {Koo, Bon-Hong and Kim, Ho Joong and Kwon, Jang-Yeon and Chae, Chan-Byoung},
	title = {{Deep Learning-based Human Implantable Nano Molecular Communications}},
	booktitle = {IEEE International Conference on Communications (ICC 2020)},
	address = {Virtual Conference},
	doi = {10.1109/icc40277.2020.9148818},
	isbn = {978-1-7281-7441-9},
	issn = {2474-9133},
	month = Jun,
	pages = {1 -- 7},
	publisher = {IEEE},
	year = {2020},
}

@article{koo2021mimo,
	author = {Koo, Bon-Hong and Lee, Changmin and Pusane, Ali E. and Tugcu, Tuna and Chae, Chan-Byoung},
	title = {{MIMO Operations in Molecular Communications: Theory, Prototypes, and Open Challenges}},
	doi = {10.1109/mcom.110.2000984},
	issn = {0163-6804},
	journal = {IEEE Communications Magazine},
	month = Sep,
	number = {9},
	pages = {98--104},
	publisher = {IEEE},
	volume = {59},
	year = {2021},
}

@inproceedings{kosanetzki2025demodulation,
	author = {Kosanetzki, Dorian and Kesz{\"{o}}cze, Oliver and Kroll, Lukas and Thiem, J{\"{o}}rg and Kirchner, Jens},
	title = {{Demodulation with Deep Neural Networks for Time-Dependent Flow-Based Molecular Communication Channels}},
	booktitle = {IEEE International Conference on Machine Learning for Communication and Networking (ICMLCN 2025)},
	address = {Barcelona, Spain},
	doi = {10.1109/icmlcn64995.2025.11140309},
	isbn = {979-8-3315-2042-7},
	month = May,
	pages = {1--6},
	publisher = {IEEE},
	year = {2025},
}

@article{kose2020machine,
	author = {Kose, Oyku Deniz and Gursoy, Mustafa Can and Saraclar, Murat and Pusane, Ali E. and Tugcu, Tuna},
	title = {{Machine Learning-Based Silent Entity Localization Using Molecular Diffusion}},
	doi = {10.1109/lcomm.2020.2968319},
	issn = {1558-2558},
	journal = {IEEE Communications Letters},
	month = Apr,
	number = {4},
	pages = {807--810},
	publisher = {IEEE},
	volume = {24},
	year = {2020},
}

@article{kuran2021survey,
	author = {Kuran, Mehmet Sukru and Yilmaz, H. Birkan and Demirkol, Ilker and Farsad, Nariman and Goldsmith, Andrea},
	title = {{A Survey on Modulation Techniques in Molecular Communication via Diffusion}},
	doi = {10.1109/comst.2020.3048099},
	issn = {1553-877X},
	journal = {IEEE Communications Surveys \& Tutorials},
	month = Jan,
	number = {1},
	pages = {7--28},
	publisher = {IEEE},
	volume = {23},
	year = {2021},
}

@article{kuscu2021fabrication,
	author = {Kuscu, Murat and Ramezani, Hamideh and Dinc, Ergin and Akhavan, Shahab and Akan, Ozgur B.},
	title = {{Fabrication and microfluidic analysis of graphene-based molecular communication receiver for Internet of Nano Things (IoNT)}},
	doi = {10.1038/s41598-021-98609-1},
	issn = {2045-2322},
	journal = {Scientific Reports},
	month = Oct,
	number = {1},
	pages = {1--20},
	publisher = {Springer Nature},
	volume = {11},
	year = {2021},
}

@article{kuscu2021internet,
	author = {Kuscu, Murat and Unluturk, Bige Deniz},
	title = {{Internet of Bio-Nano Things: A review of applications, enabling technologies and key challenges}},
	doi = {10.52953/chbb9821},
	journal = {ITU Journal on Future and Evolving Technologies},
	month = Dec,
	number = {3},
	pages = {1--24},
	publisher = {International Telecommunication Union},
	volume = {2},
	year = {2021},
}

@article{lagasse2023future,
	author = {Lagasse, Eric and Levin, Michael},
	title = {{Future medicine: from molecular pathways to the collective intelligence of the body}},
	doi = {10.1016/j.molmed.2023.06.007},
	issn = {1471-4914},
	journal = {Trends in Molecular Medicine},
	month = Sep,
	number = {9},
	pages = {687--710},
	publisher = {Elsevier},
	volume = {29},
	year = {2023},
}

@article{lai2011cognitive,
	author = {Lai, Lifeng and El Gamal, Hesham and Jiang, Haiyang and Poor, H. Vincent},
	title = {{Cognitive Medium Access: Exploration, Exploitation, and Competition}},
	doi = {10.1109/tmc.2010.65},
	issn = {1536-1233},
	journal = {IEEE Transactions on Mobile Computing},
	month = Feb,
	number = {2},
	pages = {239--253},
	publisher = {IEEE},
	volume = {10},
	year = {2011},
}

@article{lau2021efficient,
	author = {Lau, Florian-Lennert Adrian and Wendt, Regine and Fischer, Stefan},
	title = {{Efficient in-message computation of prevalent mathematical operations in DNA-based nanonetworks}},
	doi = {10.1016/j.nancom.2021.100348},
	issn = {1878-7789},
	journal = {Elsevier Nano Communication Networks},
	month = Jun,
	pages = {1--17},
	publisher = {Elsevier},
	volume = {28},
	year = {2021},
}

@article{lau2025using,
	author = {Lau, Florian-Lennert Adrian and Prange, Lara Josephine and Wendt, Regine and Scheer, Sarah and Hyttrek, Christian and Pal, Saswati and Torres G{\'{o}}mez, Jorge and Dressler, Falko and Fischer, Stefan},
	title = {{Using Off-the-Shelf Biosensors to Implement Gateways for Alarm-System Nanonetworks}},
	doi = {10.1016/j.nancom.2025.100584},
	issn = {1878-7789},
	journal = {Elsevier Nano Communication Networks},
	month = Sep,
	pages = {100584},
	publisher = {Elsevier},
	volume = {45},
	year = {2025},
}

@inproceedings{lee2017machine,
	author = {Lee, Changmin and Yilmaz, H. Birkan and Chae, Chan-Byoung and Farsad, Nariman and Goldsmith, Andrea},
	title = {{Machine learning based channel modeling for molecular MIMO communications}},
	booktitle = {18th IEEE International Workshop on Signal Processing Advances in Wireless Communications (SPAWC 2017)},
	address = {Sapporo, Japan},
	doi = {10.1109/spawc.2017.8227765},
	issn = {1948-3252},
	month = Jul,
	pages = {1 -- 5},
	publisher = {IEEE},
	year = {2017},
}

@article{lemic2021survey,
	author = {Lemic, Filip and Abadal, Sergi and Tavernier, Wouter and Stroobant, Pieter and Colle, Didier and Alarcon, Eduard and Marquez-Barja, Johann M. and Famaey, Jeroen},
	title = {{Survey on Terahertz Nanocommunication and Networking: A Top-Down Perspective}},
	doi = {10.1109/jsac.2021.3071837},
	issn = {0733-8716},
	journal = {IEEE Journal on Selected Areas in Communications},
	month = Jun,
	number = {6},
	pages = {1506--1543},
	publisher = {IEEE},
	volume = {39},
	year = {2021},
}

@article{li2021synthetic,
	author = {Li, Ximing and Rizik, Luna and Kravchik, Valeriia and Khoury, Maria and Korin, Netanel and Daniel, Ramez},
	title = {{Synthetic neural-like computing in microbial consortia for pattern recognition}},
	doi = {10.1038/s41467-021-23336-0},
	issn = {2041-1723},
	journal = {Nature Communications},
	month = May,
	number = {1},
	pages = {1--12},
	publisher = {Springer Nature},
	volume = {12},
	year = {2021},
}

@phdthesis{li2023explainability,
	author = {Li, Xin},
	title = {{Explainability of NN-based Detectors in MIMO Molecular Channels}},
	advisor = {Torres G{\'{o}}mez, Jorge},
	institution = {School of Electrical Engineering and Computer Science},
	location = {Berlin, Germany},
	month = Jul,
	referee = {Dressler, Falko and Sikora, Thomas},
	school = {TU Berlin},
	type = {Master's Thesis},
	year = {2023},
}

@inproceedings{llatser2011exploring,
	author = {Llatser, Ignacio and Pascual, Inaki and Garralda, Nora and Cabellos-Aparicio, Albert and Pierobon, Massimiliano and Alarcon, Eduard and Sol{\'{e}}-Pareta, Josep},
	title = {{Exploring the Physical Channel of Diffusion-Based Molecular Communication by Simulation}},
	booktitle = {IEEE Global Communications Conference (GLOBECOM 2011)},
	address = {Houston, TX},
	doi = {10.1109/glocom.2011.6134028},
	issn = {1930-529X},
	month = Dec,
	pages = {1--5},
	publisher = {IEEE},
	year = {2011},
}

@article{llatser2014n3sim,
	author = {Llatser, Ignacio and Demiray, Deniz and Cabellos-Aparicio, Albert and Altilar, D. Turgay and Alarc{\'{o}}n, Eduard},
	title = {{N3Sim: Simulation framework for diffusion-based molecular communication nanonetworks}},
	doi = {10.1016/j.simpat.2013.11.004},
	issn = {1569-190X},
	journal = {Elsevier Simulation Modelling Practice and Theory},
	month = Mar,
	pages = {210--222},
	publisher = {Elsevier},
	volume = {42},
	year = {2014},
}

@article{lopez2025toward,
	author = {L{\'{o}}pez, Arnau Brosa and Lemic, Filip and Calvo, Gerard and P{\'{e}}rez, Aina and Struye, Jakob and Torres G{\'{o}}mez, Jorge and Municio, Esteban and Delgado, Carmen and Dressler, Falko and Alarc{\'{o}}n, Eduard and Famaey, Jeroen and Abadal, Sergi and Costa-P{\'{e}}rez, Xavier},
	title = {{Toward Standardized Performance Evaluation of Flow-guided Nanoscale Localization}},
	doi = {10.1109/TMBMC.2024.3523428},
	issn = {2332-7804},
	journal = {IEEE Transactions on Molecular, Biological and Multi-Scale Communications},
	month = Mar,
	number = {1},
	pages = {116--127},
	publisher = {IEEE},
	volume = {11},
	year = {2025},
}

@article{lotter2023experimental,
	author = {Lotter, Sebastian and Brand, Lukas and Jamali, Vahid and Sch{\"{a}}fer, Maximilian and Loos, Helene M. and Unterweger, Harald and Greiner, Sandra and Kirchner, Jens and Alexiou, Christoph and Drummer, Dietmar and Fischer, Georg and Buettner, Andrea and Schober, Robert},
	title = {{Experimental Research in Synthetic Molecular Communications -- Part I}},
	doi = {10.1109/mnano.2023.3262100},
	issn = {1932-4510},
	journal = {IEEE Nanotechnology Magazine},
	month = Jun,
	number = {3},
	pages = {42--53},
	publisher = {IEEE},
	volume = {17},
	year = {2023},
}

@article{lotter2023experimental2,
	author = {Lotter, Sebastian and Brand, Lukas and Jamali, Vahid and Sch{\"{a}}fer, Maximilian and Loos, Helene M. and Unterweger, Harald and Greiner, Sandra and Kirchner, Jens and Alexiou, Christoph and Drummer, Dietmar and Fischer, Georg and Buettner, Andrea and Schober, Robert},
	title = {{Experimental Research in Synthetic Molecular Communications -- Part II}},
	doi = {10.1109/mnano.2023.3262377},
	issn = {1932-4510},
	journal = {IEEE Nanotechnology Magazine},
	month = Jun,
	number = {3},
	pages = {54--65},
	publisher = {IEEE},
	volume = {17},
	year = {2023},
}

@article{lu2020wireless,
	author = {Lu, Yi and Ni, Rui and Zhu, Qian},
	title = {{Wireless Communication in Nanonetworks: Current Status, Prospect and Challenges}},
	doi = {10.1109/tmbmc.2020.3004304},
	issn = {2332-7804},
	journal = {IEEE Transactions on Molecular, Biological and Multi-Scale Communications},
	month = Nov,
	number = {2},
	pages = {71--80},
	publisher = {IEEE},
	volume = {6},
	year = {2020},
}

@data{lu2023github,
	author = {Lu, Xiwen},
	title = {{MCFormer}},
	journal = {GitHub repository},
	publisher = {GitHub},
	url = {github.com/Xiwen-Lu/MCFormer},
	year = {2023},
}

@article{lu2023mcformer,
	author = {Lu, Xiwen and Bai, Chenyao and Zhu, Aoji and Zhu, Yunlong and Wang, Kezhi},
	title = {{MCFormer: A Transformer-Based Detector for Molecular Communication With Accelerated Particle-Based Solution}},
	doi = {10.1109/lcomm.2023.3303091},
	issn = {1558-2558},
	journal = {IEEE Communications Letters},
	month = Oct,
	number = {10},
	pages = {2837--2841},
	publisher = {IEEE},
	volume = {27},
	year = {2023},
}

@inproceedings{lundberg2017unified,
	author = {Lundberg, Scott M. and Lee, Su-In},
	title = {{A Unified Approach to Interpreting Model Predictions}},
	booktitle = {31st International Conference on Neural Information Processing Systems (NIPS 2017)},
	address = {Long Beach, CA},
	isbn = {978-1-5108-6096-4},
	month = Dec,
	pages = {4768--4777},
	publisher = {Curran Associates Inc.},
	year = {2017},
}

@article{luong2019applications,
	author = {Luong, Nguyen Cong and Hoang, Dinh Thai and Gong, Shimin and Niyato, Dusit and Wang, Ping and Liang, Ying-Chang and Kim, Dong In},
	title = {{Applications of Deep Reinforcement Learning in Communications and Networking: A Survey}},
	doi = {10.1109/comst.2019.2916583},
	issn = {1553-877X},
	journal = {IEEE Communications Surveys \& Tutorials},
	number = {4},
	pages = {3133--3174},
	publisher = {IEEE},
	volume = {21},
	year = {2019},
}

@article{ma2024survey,
	author = {Ma, Qianli and Liu, Zhen and Zheng, Zhenjing and Huang, Ziyang and Zhu, Siying and Yu, Zhongzhong and Kwok, James T.},
	title = {{A Survey on Time-Series Pre-Trained Models}},
	doi = {10.1109/tkde.2024.3475809},
	issn = {1041-4347},
	journal = {IEEE Transactions on Knowledge and Data Engineering},
	pages = {1--20},
	publisher = {IEEE},
	year = {2024},
}

@article{mai2017event,
	author = {Mai, Trang C. and Egan, Malcolm and Duong, Trung Q. and Di Renzo, Marco},
	title = {{Event Detection in Molecular Communication Networks With Anomalous Diffusion}},
	doi = {10.1109/lcomm.2017.2669315},
	issn = {1558-2558},
	journal = {IEEE Communications Letters},
	month = Jun,
	number = {6},
	pages = {1249--1252},
	publisher = {IEEE},
	volume = {21},
	year = {2017},
}

@article{marzo2019nanonetworks,
	author = {Marzo, Jose Luis and Jornet, Josep Miquel and Pierobon, Massimiliano},
	title = {{Nanonetworks in Biomedical Applications}},
	doi = {10.2174/1389450120666190115152613},
	issn = {1389-4501},
	journal = {Current Drug Targets},
	month = May,
	number = {8},
	pages = {800--807},
	publisher = {Bentham Science Publishers Ltd.},
	volume = {20},
	year = {2019},
}

@article{mohamed2019modelbased,
	author = {Mohamed, Soha and Jiang, Dong and Junejo, A. R. and Zuo, De Cheng},
	title = {{Model-Based: End-to-End Molecular Communication System Through Deep Reinforcement Learning Auto Encoder}},
	doi = {10.1109/access.2019.2916701},
	issn = {2169-3536},
	journal = {IEEE Access},
	pages = {70279--70286},
	publisher = {IEEE},
	volume = {7},
	year = {2019},
}

@article{mohamed2021biocyber,
	author = {Mohamed, Soha and Dong, Jian and El-Atty, Saied M. Abd and Eissa, Mahmoud A.},
	title = {{Bio-Cyber Interface Parameter Estimation with Neural Network for the Internet of Bio-Nano Things}},
	doi = {10.1007/s11277-021-09177-6},
	journal = {Wireless Personal Communications: An International Journal},
	month = Sep,
	number = {2},
	pages = {1245--1263},
	publisher = {Springer},
	volume = {123},
	year = {2021},
}

@article{moioli2021neurosciences,
	author = {Moioli, Renan Cipriano and Nardelli, Pedro H. J. and Barros, Michael Taynnan and Saad, Walid and Hekmatmanesh, Amin and Silva, Pedro E. Goria and de Sena, Arthur Sousa and Dzaferagic, Merim and Siljak, Harun and Van Leekwijck, Werner and Melgarejo, Dick Carrillo and Latre, Steven},
	title = {{Neurosciences and Wireless Networks: The Potential of Brain-Type Communications and Their Applications}},
	doi = {10.1109/comst.2021.3090778},
	issn = {1553-877X},
	journal = {IEEE Communications Surveys \& Tutorials},
	number = {3},
	pages = {1599--1621},
	publisher = {IEEE},
	volume = {23},
	year = {2021},
}

@article{mokon2022computational,
	author = {Moškon, Miha and Pušnik, Žiga and Stanovnik, Lidija and Zimic, Nikolaj and Mraz, Miha},
	title = {{A computational design of a programmable biological processor}},
	doi = {10.1016/j.biosystems.2022.104778},
	issn = {0303-2647},
	journal = {Biosystems},
	month = Nov,
	pages = {1--12},
	publisher = {Elsevier},
	volume = {221},
	year = {2022},
}

@data{morgan2015,
	author = {Morgan, Bria},
	title = {{MolComSim}},
	journal = {GitHub repository},
	publisher = {GitHub},
	url = {github.com/calypsomatic/MolComSim},
	year = {2015},
}

@article{mosayebi2019early,
	author = {Mosayebi, Reza and Ahmadzadeh, Arman and Wicke, Wayan and Jamali, Vahid and Schober, Robert and Nasiri-Kenari, Masoumeh},
	title = {{Early Cancer Detection in Blood Vessels Using Mobile Nanosensors}},
	doi = {10.1109/tnb.2018.2885463},
	issn = {1558-2639},
	journal = {IEEE Transactions on NanoBioscience},
	month = Oct,
	number = {4},
	pages = {103--116},
	publisher = {IEEE},
	volume = {18},
	year = {2019},
}

@inproceedings{mukherjee2019synchronization,
	author = {Mukherjee, Mithun and Yilmaz, H. Birkan and Bhowmik, Bishanka Brata and Lloret, Jaime and Lv, Yunrong},
	title = {{Synchronization for Diffusion-Based Molecular Communication Systems via Faster Molecules}},
	booktitle = {IEEE International Conference on Communications (ICC 2019)},
	address = {Shanghai, China},
	doi = {10.1109/ICC.2019.8761827},
	month = May,
	pages = {1--5},
	publisher = {IEEE},
	year = {2019},
}

@techreport{mundhenk2019efficient,
	author = {Mundhenk, T. Nathan and Chen, Barry Y. and Friedland, Gerald},
	title = {{Efficient Saliency Maps for Explainable AI}},
	doi = {10.48550/ARXIV.1911.11293},
	institution = {arXiv},
	month = Mar,
	number = {1911.11293},
	type = {cs.CV},
	year = {2019},
}

@article{nagipogu2023survey,
	author = {Nagipogu, Rajiv Teja and Fu, Daniel and Reif, John H.},
	title = {{A survey on molecular-scale learning systems with relevance to DNA computing}},
	doi = {10.1039/d2nr06202j},
	journal = {Nanoscale},
	number = {17},
	pages = {7676--7694},
	publisher = {RSC},
	volume = {15},
	year = {2023},
}

@techreport{nagipogu2024neuralcrns,
	author = {Nagipogu, Rajiv Teja and Reif, John H.},
	title = {{NeuralCRNs: A Natural Implementation of Learning in Chemical Reaction Networks}},
	doi = {10.48550/ARXIV.2409.00034},
	institution = {arXiv},
	month = Aug,
	pages = {1--42},
	type = {cs.ET},
	year = {2024},
}

@article{nakano2019methods,
	author = {Nakano, Tadashi and Okaie, Yutaka and Kobayashi, Shouhei and Hara, Takahiro and Hiraoka, Yasushi and Haraguchi, Tokuko},
	title = {{Methods and Applications of Mobile Molecular Communication}},
	doi = {10.1109/jproc.2019.2917625},
	issn = {0018-9219},
	journal = {Proceedings of the IEEE},
	month = Jul,
	number = {7},
	pages = {1442--1456},
	publisher = {IEEE},
	volume = {107},
	year = {2019},
}

@article{nauta2023from,
	author = {Nauta, Meike and Trienes, Jan and Pathak, Shreyasi and Nguyen, Elisa and Peters, Michelle and Schmitt, Yasmin and Schl{\"{o}}tterer, J{\"{o}}rg and van Keulen, Maurice and Seifert, Christin},
	title = {{From Anecdotal Evidence to Quantitative Evaluation Methods: A Systematic Review on Evaluating Explainable AI}},
	doi = {10.1145/3583558},
	issn = {0360-0300},
	journal = {ACM Computing Surveys},
	month = Jul,
	number = {13s},
	pages = {1--42},
	publisher = {ACM},
	volume = {55},
	year = {2023},
}

@data{noauthor_96well_nodate,
	title = {96 Well Cell Culture Microplate CELLCOAT},
	language = {en},
	url = {shop.gbo.com/en/germany/files/27485048/655936.pdf},
}

@data{noauthor_nrf52832_nodate,
	title = {{nRF52832} {Product} {Specification}},
	language = {en},
	url = {docs.arduino.cc/resources/datasheets/nRF52832_PS_v1.1.pdf},
}

@article{noel2017simulating,
	author = {Noel, Adam and Cheung, Karen C. and Schober, Robert and Makrakis, Dimitrios and Hafid, Abdelhakim},
	title = {{Simulating with AcCoRD: Actor-based Communication via Reaction--Diffusion}},
	doi = {10.1016/j.nancom.2017.02.002},
	issn = {1878-7789},
	journal = {Elsevier Nano Communication Networks},
	month = Mar,
	pages = {44--75},
	publisher = {Elsevier},
	volume = {11},
	year = {2017},
}

@data{noel2020arcord,
	author = {Noel, Adam},
	title = {{AcCoRD}},
	journal = {GitHub repository},
	publisher = {GitHub},
	url = {github.com/adamjgnoel/AcCoRD},
	year = {2020},
}

@data{bionetgen,
	title = {{BioNetGen website}},
	publisher = {Website},
	url = {https://bionetgen.org/}
}

@data{BNSim,
	title = {{BNSim website}},
	publisher = {Website},
	url = {https://radum.ece.utexas.edu/bnsim-bacteria-network},
}

@data{Download,
	title = {{Download link}},
	publisher = {Download},
	url = {http://gnan.ece.gatech.edu/ns-allinone-3.24.zip},
}

@data{N3Sim,
	title = {N3Sim},
	publisher = {Website},
	url = {https://n3cat.upc.edu/n3sim-simulation-framework-for-diffusion-based-molecular-communication-nanonetworks/},
}

@data{smoldyn,
	title = {Smoldyn},
	publisher = {Website},
	url = {https://www.smoldyn.org/},
}

@article{ozbey2024artificial,
	author = {Ozbey, Erencem and Cicekdag, Yusuf Kagan and Yilmaz, H. Birkan},
	title = {{Artificial neural network based misorientation correction in molecular 4x4 MIMO systems}},
	doi = {10.1016/j.nancom.2024.100544},
	issn = {1878-7789},
	journal = {Elsevier Nano Communication Networks},
	month = Dec,
	pages = {1 -- 9},
	publisher = {Elsevier},
	volume = {42},
	year = {2024},
}

@inproceedings{ozdemir2021estimating,
	author = {Ozdemir, Halil Umut and Orhan, Halil Ibrahim and Turan, Meric and Buyuktas, Baris and Yilmaz, H. Birkan},
	title = {{Estimating Capture Probabilities for Complex Topologies in 2D Molecular Communication via Diffusion Channel using Artificial Neural Networks}},
	booktitle = {9th IEEE International Black Sea Conference on Communications and Networking (BlackSeaCom 2021)},
	address = {Bucharest, Romania},
	doi = {10.1109/blackseacom52164.2021.9527790},
	isbn = {978-1-66540-308-5},
	month = May,
	pages = {1 -- 6},
	publisher = {IEEE},
	year = {2021},
}

@article{ozdemir2024estimating,
	author = {Ozdemir, Halil Umut and Orhan, Halil Ibrahim and Turan, Meri{\c{c}} and B{\"{u}}y{\"{u}}ktaş, Bariş and Yilmaz, H. Birkan},
	title = {{Estimating channel coefficients for complex topologies in 3D diffusion channel using artificial neural networks}},
	doi = {10.1016/j.nancom.2024.100549},
	issn = {1878-7789},
	journal = {Elsevier Nano Communication Networks},
	month = Dec,
	pages = {1--12},
	publisher = {Elsevier},
	volume = {42},
	year = {2024},
}

@article{pandi2019metabolic,
	author = {Pandi, Amir and Koch, Mathilde and Voyvodic, Peter L. and Soudier, Paul and Bonnet, Jerome and Kushwaha, Manish and Faulon, Jean-Loup},
	title = {{Metabolic perceptrons for neural computing in biological systems}},
	doi = {10.1038/s41467-019-11889-0},
	issn = {2041-1723},
	journal = {Nature Communications},
	month = Aug,
	number = {1},
	pages = {1--13},
	publisher = {Springer Nature},
	volume = {10},
	year = {2019},
}

@inproceedings{pascual2024analytical,
	author = {Pascual, Guillem and Lemic, Filip and Delgado, Carmen and Costa-P{\'{e}}rez, Xavier},
	title = {{Analytical Modelling of Raw Data for Flow-Guided In-body Nanoscale Localization}},
	booktitle = {IEEE International Conference on Machine Learning for Communication and Networking (ICMLCN 2024)},
	address = {Stockholm, Sweden},
	doi = {10.1109/icmlcn59089.2024.10625169},
	month = May,
	pages = {428--433},
	publisher = {IEEE},
	year = {2024},
}

@data{patel2021github,
	author = {Patel, Devshree},
	title = {{Molecular-Communication}},
	journal = {GitHub repository},
	publisher = {GitHub},
	url = {github.com/devshree07/Molecular-Communications},
	year = {2020},
}

@article{perera2025wetneuromorphic,
	author = {Perera, Jeewaka and Balasubramaniam, Sasitharan and Somathilaka, Samitha and Wen, Qu and Li, Xu and Kasthurirathna, Dharshana and Roohi, Arman and Nelson, Tyler},
	title = {{Wet-Neuromorphic Computing: A New Paradigm for Biological Artificial Intelligence}},
	doi = {10.1109/mis.2025.3555551},
	issn = {1941-1294},
	journal = {IEEE Intelligent Systems},
	pages = {1--7},
	publisher = {IEEE},
	year = {2025},
}

@article{prakash2007microfluidic,
	author = {Prakash, Manu and Gershenfeld, Neil},
	title = {{Microfluidic Bubble Logic}},
	doi = {10.1126/science.1136907},
	issn = {1095-9203},
	journal = {Science},
	month = Feb,
	number = {5813},
	pages = {832--835},
	publisher = {American Association for the Advancement of Science (AAAS)},
	volume = {315},
	year = {2007},
}

@article{purcell2014synthetic,
	author = {Purcell, Oliver and Lu, Timothy K},
	title = {{Synthetic analog and digital circuits for cellular computation and memory}},
	doi = {10.1016/j.copbio.2014.04.009},
	journal = {Current Opinion in Biotechnology},
	month = Oct,
	pages = {146--155},
	publisher = {Elsevier},
	volume = {29},
	year = {2014},
}

@article{qian2011neural,
	author = {Qian, Lulu and Winfree, Erik and Bruck, Jehoshua},
	title = {{Neural network computation with DNA strand displacement cascades}},
	doi = {10.1038/nature10262},
	issn = {0028-0836},
	journal = {Nature},
	month = Jul,
	number = {7356},
	pages = {368--372},
	publisher = {Springer Nature},
	volume = {475},
	year = {2011},
}

@inproceedings{qian2018receiver,
	author = {Qian, Xuewen and Di Renzo, Marco},
	title = {{Receiver Design in Molecular Communications: An Approach Based on Artificial Neural Networks}},
	booktitle = {15th IEEE International Symposium on Wireless Communication Systems (ISWCS 2018)},
	address = {Lisbon, Portugal},
	doi = {10.1109/iswcs.2018.8491088},
	month = Aug,
	pages = {1 -- 5},
	publisher = {IEEE},
	year = {2018},
}

@article{qian2019molecular,
	author = {Qian, Xuewen and Renzo, Marco Di and Eckford, Andrew W.},
	title = {{Molecular Communications: Model-Based and Data-Driven Receiver Design and Optimization}},
	doi = {10.1109/access.2019.2912600},
	issn = {2169-3536},
	journal = {IEEE Access},
	month = Jan,
	pages = {53555--53565},
	publisher = {IEEE},
	volume = {7},
	year = {2019},
}

@data{qian2020github,
	author = {Qian, Xuewen and Di Renzo, Marco and Eckford, Andrew},
	title = {{Molecular-Communication}},
	journal = {GitHub repository},
	publisher = {GitHub},
	url = {github.com/MuskanM1/Molecular-Communication},
	year = {2020},
}

@article{qiu2024review,
	author = {Qiu, Song and Wei, Zhuangkun and Huang, Yu and Abbaszadeh, Mahmoud and Charmet, Jerome and Li, Bin and Guo, Weisi},
	title = {{Review of Physical Layer Security in Molecular Internet of Nano-Things}},
	doi = {10.1109/tnb.2023.3285973},
	issn = {1558-2639},
	journal = {IEEE Transactions on NanoBioscience},
	month = Jan,
	number = {1},
	pages = {91--100},
	publisher = {IEEE},
	volume = {23},
	year = {2024},
}

@inproceedings{ribeiro2016why,
	author = {Ribeiro, Marco Tulio and Singh, Sameer and Guestrin, Carlos},
	title = {{{``}Why Should I Trust You?{''}: Explaining the Predictions of Any Classifier}},
	booktitle = {22nd ACM SIGKDD International Conference on Knowledge Discovery and Data Mining},
	address = {San Francisco, CA},
	doi = {10.1145/2939672.2939778},
	isbn = {978-1-4503-4232-2},
	month = Aug,
	pages = {1135 -- 1144},
	publisher = {ACM},
	year = {2016},
}

@article{rizik2022synthetic,
	author = {Rizik, Luna and Danial, Loai and Habib, Mouna and Weiss, Ron and Daniel, Ramez},
	title = {{Synthetic neuromorphic computing in living cells}},
	doi = {10.1038/s41467-022-33288-8},
	issn = {2041-1723},
	journal = {Nature Communications},
	month = Sep,
	number = {1},
	pages = {1--17},
	publisher = {Springer Nature},
	volume = {13},
	year = {2022},
}

@article{rizwan2018review,
	author = {Rizwan, A. and Zoha, A. and Zhang, R. and Ahmad, W. and Arshad, K. and Ali, N. Abu and Alomainy, A. and Imran, M. A. and Abbasi, Q. H.},
	title = {{A Review on the Role of Nano-Communication in Future Healthcare Systems: A Big Data Analytics Perspective}},
	doi = {10.1109/ACCESS.2018.2859340},
	issn = {2169-3536},
	journal = {IEEE Access},
	month = May,
	pages = {41903--41920},
	publisher = {IEEE},
	volume = {6},
	year = {2018},
}

@article{rodriguez2021loser,
	author = {Rodriguez, Kellen R. and Sarraf, Namita and Qian, Lulu},
	title = {{A Loser-Take-All DNA Circuit}},
	doi = {10.1021/acssynbio.1c00318},
	issn = {2161-5063},
	journal = {ACS Synthetic Biology},
	month = Oct,
	number = {11},
	pages = {2878--2885},
	publisher = {ACS},
	volume = {10},
	year = {2021},
}

@article{saeed2021body-centric,
	author = {Saeed, Nasir and Loukil, Mohamed Habib and Sarieddeen, Hadi and Al-Naffouri, Tareq Y. and Alouini, Mohamed Slim},
	title = {{Body-Centric Terahertz Networks: Prospects and Challenges}},
	doi = {10.1109/tmbmc.2021.3135198},
	issn = {2332-7804},
	journal = {IEEE Transactions on Molecular, Biological and Multi-Scale Communications},
	month = Sep,
	number = {3},
	pages = {138--157},
	publisher = {IEEE},
	volume = {8},
	year = {2022},
}

@inproceedings{saiki2022design,
	author = {Saiki, Takanori and Nakano, Tadashi},
	title = {{Design and Implementation of a Multicellular Molecular Communication Simulator}},
	booktitle = {Joint 12th International Conference on Soft Computing and Intelligent Systems and 23rd International Symposium on Advanced Intelligent Systems (SCIS and ISIS 2022)},
	address = {Ise, Japan},
	doi = {10.1109/scisisis55246.2022.10002068},
	month = Nov,
	pages = {1--5},
	publisher = {IEEE},
	year = {2022},
}

@article{saiki2025general,
	author = {Saiki, Takanori and Imanaka, Shohei and Kobayashi, Shouhei and Nakano, Tadashi},
	title = {{A General-Purpose Simulation Platform for Multicellular Molecular Communication Systems}},
	doi = {10.1109/tmbmc.2025.3544141},
	issn = {2332-7804},
	journal = {IEEE Transactions on Molecular, Biological and Multi-Scale Communications},
	month = Jun,
	number = {2},
	pages = {152--165},
	publisher = {IEEE},
	volume = {11},
	year = {2025},
}

@data{saiki2025general_github,
	author = {Saiki, Takanori and Imanaka, Shohei},
	title = {{Multicellular Molecular Communication System Simulator}},
	journal = {GitHub repository},
	publisher = {GitHub},
	url = {github.com/ImanakaShohei/MulticellularMolecularCommunicationSystemSimulator},
	year = {2025},
}

@data{sangani2021github,
	author = {Sangani, Priyank},
	title = {{Research-in-Molecular-Communication}},
	journal = {GitHub repository},
	publisher = {GitHub},
	url = {github.com/Priyank31/Research-in-Molecular-Communication},
	year = {2021},
}

@article{sarkar2021single,
	author = {Sarkar, Kathakali and Bonnerjee, Deepro and Srivastava, Rajkamal and Bagh, Sangram},
	title = {{A single layer artificial neural network type architecture with molecular engineered bacteria for reversible and irreversible computing}},
	doi = {10.1039/d1sc01505b},
	issn = {2041-6539},
	journal = {Chemical Science},
	number = {48},
	pages = {15821--15832},
	publisher = {RSC},
	volume = {12},
	year = {2021},
}

@inproceedings{sawlekar2021survey,
	author = {Sawlekar, Rucha and Nikolakopoulos, George},
	title = {{A Survey of DNA-based Computing Devices and their Applications}},
	booktitle = {European Control Conference (ECC 2021)},
	address = {Delft, Netherlands},
	doi = {10.23919/ecc54610.2021.9654895},
	month = Jun,
	pages = {769--774},
	publisher = {IEEE},
	year = {2021},
}

@article{scazzoli2024molecular,
	author = {Scazzoli, Davide and Vakilipoor, Fardad and Magarini, Maurizio},
	title = {{Molecular communication data augmentation and deep learning based detection}},
	doi = {10.1016/j.nancom.2024.100510},
	issn = {1878-7789},
	journal = {Elsevier Nano Communication Networks},
	month = Jul,
	pages = {1--12},
	publisher = {Elsevier},
	volume = {40},
	year = {2024},
}

@techreport{scherer2025closedloop,
	author = {Scherer, Maike and Brand, Lukas and Wolf, Louis and Dieck, Teena tom and Sch{\"{a}}fer, Maximilian and Lotter, Sebastian and Burkovski, Andreas and Sticht, Heinrich and Schober, Robert and Castiglione, Kathrin},
	title = {{Closed-Loop Long-Term Experimental Molecular Communication System}},
	doi = {10.48550/ARXIV.2502.00831},
	institution = {arXiv},
	month = Feb,
	pages = {1--30},
	type = {cs.ET},
	year = {2025},
}

@data{scherer2025closedloop_github,
	author = {Scherer, Maike and Brand, Lukas and Wolf, Louis and tom Dieck, Teena and Schäfer, Maximilian and Lotter, Sebastian and Burkovski, Andreas and Sticht, Heinrich and Schober, Robert and Castiglione, Kathrin},
	title = {{Media Modulation Testbed: Code Package}},
	journal = {GitHub repository},
	publisher = {GitHub},
	url = {github.com/SyMoCADS/Media_Modulation_Testbed/},
	year = {2025},
}

@data{scherer2025closedloop_zenodo,
	author = {Scherer, Maike and Brand, Lukas and Wolf, Louis and tom Dieck, Teena and Schäfer, Maximilian and Lotter, Sebastian and Burkovski, Andreas and Sticht, Heinrich and Schober, Robert and Castiglione, Kathrin},
	title = {{Closed-Loop Long-Term Experimental Molecular Communication System}},
	month = Feb,
	note = {{Zenodo}},
	publisher = {Zenodo},
	url = {doi.org/10.5281/zenodo.13898880},
	year = {2025},
}

@book{schroedinger1944what,
	author = {Schr{\"{o}}dinger, Erwin},
	title = {{What is Life? The Physical Aspect of the Living Cell}},
	isbn = {978-1-107-68365-5},
	pages = {196},
	publisher = {Cambridge University Press},
	year = {1944},
}

@techreport{schulman2017proximal,
	author = {Schulman, John and Wolski, Rich and Dhariwal, Prafulla and Radford, Alec and Klimov, Oleg},
	title = {{Proximal Policy Optimization Algorithms}},
	doi = {10.48550/ARXIV.1707.06347},
	institution = {arXiv},
	month = Jul,
	pages = {1--12},
	type = {cs.LG},
	year = {2017},
}

@article{shahmohammadian2013blind,
	author = {ShahMohammadian, Hoda and Messier, Geoffrey G. and Magierowski, Sebastian},
	title = {{Blind Synchronization in Diffusion-Based Molecular Communication Channels}},
	doi = {10.1109/lcomm.2013.100713.131727},
	issn = {1558-2558},
	journal = {IEEE Communications Letters},
	month = Nov,
	number = {11},
	pages = {2156--2159},
	publisher = {IEEE},
	volume = {17},
	year = {2013},
}

@inproceedings{sharma2020deep,
	author = {Sharma, Sanjeev and Dixit, Dharmendra and Deka, Kuntal},
	title = {{Deep Learning based Symbol Detection for Molecular Communications}},
	booktitle = {IEEE International Conference on Advanced Networks and Telecommunications Systems (ANTS 2020)},
	address = {New Delhi, India},
	doi = {10.1109/ants50601.2020.9342782},
	isbn = {978-1-7281-9290-1},
	issn = {2153-1684},
	month = Dec,
	pages = {1 -- 6},
	publisher = {IEEE},
	year = {2020},
}

@data{shastri2020github,
	author = {Shastri, Yesha},
	title = {{Molecular-Communication---Model-based-and-Data-Driven-Receiver-Design}},
	journal = {GitHub repository},
	publisher = {GitHub},
	url = {github.com/Yesha19/Molecular-Communication---Model-based-and-Data-Driven-Receiver-Design},
	year = {2020},
}

@inproceedings{shrikumar2017learning,
	author = {Shrikumar, Avanti and Greenside, Peyton and Kundaje, Anshul},
	title = {{Learning important features through propagating activation differences}},
	booktitle = {34th International Conference on Machine Learning (ICML 2017)},
	address = {Sydney, Australia},
	month = Aug,
	pages = {3145--3153},
	publisher = {JMLR.org},
	year = {2017},
}

@article{shrivastava2021performance,
	author = {Shrivastava, Amit K. and Das, Debanjan and Mahapatra, Rajarshi},
	title = {{Performance Evaluation of Mobile Molecular Communication System Using Neural Network Detector}},
	doi = {10.1109/lwc.2021.3079522},
	issn = {2162-2337},
	journal = {IEEE Wireless Communications Letters},
	month = Aug,
	number = {8},
	pages = {1776--1779},
	publisher = {IEEE},
	volume = {10},
	year = {2021},
}

@inproceedings{shrivastava2021scaled,
	author = {Shrivastava, Amit K. and Das, Debanjan and Mahapatra, Rajarshi and Varshney, Neeraj},
	title = {{Scaled Conjugate Gradient Algorithm for Neural Network Detector in Mobile Molecular Communication}},
	booktitle = {IEEE Global Communications Conference (GLOBECOM 2021)},
	address = {Madrid, Spain},
	doi = {10.1109/globecom46510.2021.9685034},
	month = Dec,
	pages = {1 -- 6},
	publisher = {IEEE},
	year = {2021},
}

@article{smirnova2023organoid,
	author = {Smirnova, Lena and Caffo, Brian S. and Gracias, David H. and Huang, Qi and Morales Pantoja, Itzy E. and Tang, Bohao and Zack, Donald J. and Berlinicke, Cynthia A. and Boyd, J. Lomax and Harris, Timothy D. and Johnson, Erik C. and Kagan, Brett J. and Kahn, Jeffrey and Muotri, Alysson R. and Paulhamus, Barton L. and Schwamborn, Jens C. and Plotkin, Jesse and Szalay, Alexander S. and Vogelstein, Joshua T. and Worley, Paul F. and Hartung, Thomas},
	title = {{Organoid intelligence (OI): the new frontier in biocomputing and intelligence-in-a-dish}},
	doi = {10.3389/fsci.2023.1017235},
	journal = {Frontiers in Neuroscience},
	month = Feb,
	pages = {1--23},
	publisher = {Frontiers Media SA},
	volume = {1},
	year = {2023},
}

@inproceedings{solak2020neural,
	author = {Solak, Sinem Nimet and Oner, Menguc},
	title = {{Neural Network Based Decision Fusion for Abnormality Detection via Molecular Communications}},
	booktitle = {IEEE Workshop on Signal Processing Systems (SiPS)},
	address = {Coimbra, Portugal},
	doi = {10.1109/sips50750.2020.9195212},
	isbn = {978-1-7281-8099-1},
	issn = {2374-7390},
	month = Oct,
	pages = {1--5},
	publisher = {IEEE},
	year = {2020},
}

@inproceedings{solak2020rnn,
	author = {Solak, Sinem Nimet and Oner, Menguc},
	title = {{RNN based abnormality detection with nanoscale sensor networks using molecular communications}},
	booktitle = {7th ACM International Conference on Nanoscale Computing and Communication (NANOCOM 2020)},
	address = {Virtual Conference},
	doi = {10.1145/3411295.3411313},
	isbn = {978-1-4503-8083-6},
	month = Sep,
	pages = {1--6},
	publisher = {ACM},
	year = {2020},
}

@article{soloveichik2010dna,
	author = {Soloveichik, David and Seelig, Georg and Winfree, Erik},
	title = {{DNA as a universal substrate for chemical kinetics}},
	address = {Washington, D.C.},
	doi = {10.1073/pnas.0909380107},
	issn = {1091-6490},
	journal = {Proceedings of the National Academy of Sciences (PNAS)},
	month = Mar,
	number = {12},
	pages = {5393--5398},
	publisher = {National Academy of Sciences},
	volume = {107},
	year = {2010},
}

@article{somathilaka2023revealing,
	author = {Somathilaka, Samitha S. and Balasubramaniam, Sasitharan and Martins, Daniel P. and Li, Xu},
	title = {{Revealing gene regulation-based neural network computing in bacteria}},
	doi = {10.1016/j.bpr.2023.100118},
	issn = {2667-0747},
	journal = {Biophysical Reports},
	month = Sep,
	number = {3},
	pages = {1--21},
	publisher = {Elsevier},
	volume = {3},
	year = {2023},
}

@techreport{somathilaka2024wet,
	author = {Somathilaka, Samitha and Ratwatte, Adrian and Balasubramaniam, Sasitharan and Vuran, Mehmet Can and Srisa-an, Witawas and Li{\`{o}}, Pietro},
	title = {{Wet TinyML: Chemical Neural Network Using Gene Regulation and Cell Plasticity}},
	doi = {10.48550/ARXIV.2403.08549},
	institution = {arXiv},
	month = Mar,
	pages = {1--7},
	type = {cs.NE},
	year = {2024},
}

@article{somathilaka2025analyzing,
	author = {Somathilaka, Samitha and Balasubramaniam, Sasitharan and Martins, Daniel P.},
	title = {{Analyzing Wet-Neuromorphic Computing Using Bacterial Gene Regulatory Neural Networks}},
	doi = {10.1109/tetc.2025.3546119},
	issn = {2168-6750},
	journal = {IEEE Transactions on Emerging Topics in Computing},
	month = Jul,
	number = {3},
	pages = {902--918},
	publisher = {IEEE},
	volume = {13},
	year = {2025},
}

@article{somathilaka2025internet,
	author = {Somathilaka, Samitha and Elayan, Hadeel and Atkinson, Joshua T. and Jornet, Josep Miquel and Balasubramaniam, Sasitharan},
	title = {{The Internet of Biofilm Living AI Devices}},
	doi = {10.1109/mcom.001.2400651},
	issn = {0163-6804},
	journal = {IEEE Communications Magazine},
	pages = {1--7},
	publisher = {IEEE},
	year = {2025},
}

@article{stepney2024physical,
	author = {Stepney, Susan},
	title = {{Physical reservoir computing: a tutorial}},
	doi = {10.1007/s11047-024-09997-y},
	issn = {1572-9796},
	journal = {Natural Computing},
	month = Nov,
	number = {4},
	pages = {665--685},
	publisher = {Springer},
	volume = {23},
	year = {2024},
}

@article{stratmann2021using,
	author = {Stratmann, Lukas and Drees, Jan Peter and Bronner, Fabian and Dressler, Falko},
	title = {{Using Vector Fields for Efficient Simulation of Macroscopic Molecular Communication}},
	doi = {10.1109/TMBMC.2021.3054930},
	issn = {2332-7804},
	journal = {IEEE Transactions on Molecular, Biological and Multi-Scale Communications, Special Section - Advances in Molecular Communication},
	month = Jun,
	number = {2},
	pages = {73--77},
	publisher = {IEEE},
	volume = {7},
	year = {2021},
}

@article{sun2020ctbrnn,
	author = {Sun, Li and Wang, Yuwei},
	title = {{CTBRNN: A Novel Deep-Learning Based Signal Sequence Detector for Communications Systems}},
	doi = {10.1109/lsp.2019.2953673},
	journal = {IEEE Signal Processing Letters},
	pages = {21 -- 25},
	publisher = {IEEE},
	volume = {27},
	year = {2020},
}

@article{sun2020survey,
	author = {Sun, Shiliang and Cao, Zehui and Zhu, Han and Zhao, Jing},
	title = {{A Survey of Optimization Methods From a Machine Learning Perspective}},
	doi = {10.1109/tcyb.2019.2950779},
	issn = {2168-2275},
	journal = {IEEE Transactions on Cybernetics},
	month = Aug,
	number = {8},
	pages = {3668--3681},
	publisher = {IEEE},
	volume = {50},
	year = {2020},
}

@article{szott2022wifi,
	author = {Szott, Szymon and Kosek-Szott, Katarzyna and Gawłowicz, Piotr and Torres G{\'{o}}mez, Jorge and Bellalta, Boris and Zubow, Anatolij and Dressler, Falko},
	title = {{Wi-Fi Meets ML: A Survey on Improving IEEE 802.11 Performance with Machine Learning}},
	doi = {10.1109/COMST.2022.3179242},
	issn = {1553-877X},
	journal = {IEEE Communications Surveys \& Tutorials},
	month = Jul,
	number = {3},
	pages = {1843--1893},
	publisher = {IEEE},
	volume = {24},
	year = {2022},
}

@data{thakkar2020dataset,
	author = {Thakkar, Mithilesh},
	title = {{molecular\_communication}},
	doi = {},
	note = {{Kaggle}},
	publisher = {Kaggle},
	url = {kaggle.com/datasets/mithilesh16/molecular-communication},
	year = {2020},
}

@inproceedings{torres-gomez2021machine,
	author = {Torres G{\'{o}}mez, Jorge and Kuestner, Anke and Pitke, Ketki and Simonjan, Jennifer and Unluturk, Bige Deniz and Dressler, Falko},
	title = {{A Machine Learning Approach for Abnormality Detection in Blood Vessels via Mobile Nanosensors}},
	booktitle = {19th ACM Conference on Embedded Networked Sensor Systems (SenSys 2021), 2nd ACM International Workshop on Nanoscale Computing, Communication, and Applications (NanoCoCoA 2021)},
	address = {Coimbra, Portugal},
	doi = {10.1145/3485730.3494037},
	month = Nov,
	pages = {596--602},
	publisher = {ACM},
	year = {2021},
}

@inproceedings{torres-gomez2021markov,
	author = {Torres G{\'{o}}mez, Jorge and Wendt, Regine and Kuestner, Anke and Pitke, Ketki and Stratmann, Lukas and Dressler, Falko},
	title = {{Markov Model for the Flow of Nanobots in the Human Circulatory System}},
	booktitle = {8th ACM International Conference on Nanoscale Computing and Communication (NANOCOM 2021)},
	address = {Virtual Conference},
	doi = {10.1145/3477206.3477477},
	isbn = {978-1-4503-8710-1},
	month = Sep,
	pages = {1--7},
	publisher = {ACM},
	year = {2021},
}

@article{torres-gomez2022nanosensor2,
	author = {Torres G{\'{o}}mez, Jorge and Kuestner, Anke and Simonjan, Jennifer and Unluturk, Bige Deniz and Dressler, Falko},
	title = {{Nanosensor Location Estimation in the Human Circulatory System using Machine Learning}},
	doi = {10.1109/TNANO.2022.3217653},
	issn = {1941-0085},
	journal = {IEEE Transactions on Nanotechnology},
	month = Oct,
	pages = {663--673},
	publisher = {IEEE},
	volume = {21},
	year = {2022},
}

@article{torres-gomez2023explainability,
	author = {Torres G{\'{o}}mez, Jorge and Hofmann, Pit and Fitzek, Frank H. P. and Dressler, Falko},
	title = {{Explainability of Neural Networks for Symbol Detection in Molecular Communication Channels}},
	doi = {10.1109/TMBMC.2023.3297135},
	issn = {2332-7804},
	journal = {IEEE Transactions on Molecular, Biological and Multi-Scale Communications},
	month = Sep,
	number = {3},
	pages = {323--328},
	publisher = {IEEE},
	volume = {9},
	year = {2023},
}

@inproceedings{torres-gomez2023fine-tune,
	author = {Torres G{\'{o}}mez, Jorge and Spicher, Nicolai and Gonz{\'{a}}lez Rios, Jorge Luis and Dressler, Falko},
	title = {{Fine-tune Circuit Representation of Human Vessels through Reinforcement Learning: A Novel Digital Twin Approach for Hemodynamics}},
	booktitle = {10th ACM International Conference on Nanoscale Computing and Communication (NANOCOM 2023)},
	address = {Coventry, United Kingdom},
	doi = {10.1145/3576781.3608717},
	month = Sep,
	pages = {46--52},
	publisher = {ACM},
	year = {2023},
}

@article{torres-gomez2024dna-based,
	author = {Torres G{\'{o}}mez, Jorge and Unluturk, Bige Deniz and Lau, Florian-Lennert Adrian and Simonjan, Jennifer and Wendt, Regine and Fischer, Stefan and Dressler, Falko},
	title = {{DNA-Based Nanonetwork for Abnormality Detection and Localization in the Human Body}},
	doi = {10.1109/TNANO.2024.3495541},
	issn = {1941-0085},
	journal = {IEEE Transactions on Nanotechnology},
	month = Nov,
	pages = {794--808},
	publisher = {IEEE},
	volume = {23},
	year = {2024},
}

@data{tuccitto2023dataset,
	author = {Tuccitto, Nunzio},
	title = {Dataset of "Experimental Implementation of Molecule Shift Keying for Enhanced Molecular Communication"},
	note = {{IEEE Dataport}},
	publisher = {IEEE Dataport},
	url = {doi.org/10.21227/krbw-ge80},
	year = {2023},
}

@data{tucitto2023dataset_zenodo,
	author = {Nunzio, Tuccitto},
	title = {{Dataset of "Experimental Implementation of Molecule Shift Keying for Enhanced Molecular Communication"}},
	month = Dec,
	note = {{Zenodo}},
	publisher = {Zenodo},
	url = {doi.org/10.5281/zenodo.10390855},
	year = {2023},
}

@data{turgut2020n4sim,
	author = {Turgut, Nafi A.},
	title = {{N4Sim}},
	journal = {GitHub repository},
	publisher = {GitHub},
	url = {github.com/nafiturgut/N4Sim},
	year = {2020},
}

@article{turgut2022n4sim,
	author = {Turgut, Nafi A. and Bilgin, Bilgesu A. and Akan, Ozgur B.},
	title = {{N$^4$ Sim: The First Nervous NaNoNetwork Simulator With Synaptic Molecular Communications}},
	doi = {10.1109/tnb.2021.3118851},
	issn = {1558-2639},
	journal = {IEEE Transactions on NanoBioscience},
	month = Oct,
	number = {4},
	pages = {468--481},
	publisher = {IEEE},
	volume = {21},
	year = {2022},
}

@techreport{uzun2025molecular,
	author = {Uzun, Mustafa and Ikiz, Kaan Burak and Kuscu, Murat},
	title = {{Molecular Communication Channel as a Physical Reservoir Computer}},
	doi = {10.48550/ARXIV.2504.17022},
	institution = {arXiv},
	month = Apr,
	pages = {1--8},
	type = {cs.ET},
	year = {2025},
}

@article{vakalipoor2025cam,
	author = {Vakilipoor, Fardad and Ettner-Sitter, Andreas and Brand, Lukas and Lotter, Sebastian and Aung, Thiha and Harteis, Silke and Schober, Robert and Schäfer, Maximilian},
	title = {{The CAM Model: An in Vivo Testbed for Molecular Communication Systems}},
	doi = {10.1109/TMBMC.2025.3601432},
	journal = {IEEE Transactions on Molecular, Biological, and Multi-Scale Communications},
	month = Aug,
	note = {early access},
	number = {},
	volume = {},
	year = {2025},
}

@data{vakalipoor2025cam_data,
	author = {Vakilipoor, Fardad and Ettner-Sitter, Andreas and Schober, Robert and Haerteis, Silke and Schäfer, Maximilian},
	title = {{Transient Indocyanine Green Distribution Measurement and Modelling in Chorioallantoic Membrane (CAM) Model}},
	month = Mar,
	note = {{Zenodo}},
	publisher = {Zenodo},
	url = {doi.org/10.5281/zenodo.14626107},
	year = {2025},
}

@inproceedings{vakilipoor2022hybrid,
	author = {Vakilipoor, Fardad and Scazzoli, Davide and Ratti, Francesca and Scalia, Gabriele and Magarini, Maurizio},
	title = {{Hybrid deep learning-based feature-augmented detection for molecular communication systems}},
	booktitle = {9th ACM International Conference on Nanoscale Computing and Communication (NANOCOM 2022)},
	address = {Barcelona, Spain},
	doi = {10.1145/3558583.3558859},
	isbn = {978-1-4503-9867-1},
	month = Oct,
	pages = {1 -- 6},
	publisher = {ACM},
	year = {2022},
}

@inproceedings{vaswani2017attention,
	author = {Vaswani, Ashish and Shazeer, Noam and Parmar, Niki and Uszkoreit, Jakob and Jones, Llion and Gomez, Aidan N. and Kaiser, Łukasz and Polosukhin, Illia},
	title = {{Attention is all you need}},
	booktitle = {31st International Conference on Neural Information Processing Systems (NIPS 2017)},
	address = {Long Beach, CA},
	isbn = {978-1-5108-6096-4},
	month = Dec,
	pages = {6000--6010},
	publisher = {Curran Associates Inc.},
	year = {2017},
}

@techreport{wachter2017counterfactual,
	author = {Wachter-Zeh, Antonia and Mittelstadt, Brent and Russell, Chris},
	title = {{Counterfactual Explanations without Opening the Black Box: Automated Decisions and the GDPR}},
	doi = {10.48550/ARXIV.1711.00399},
	institution = {arXiv},
	month = Mar,
	pages = {1--52},
	type = {cs.AI},
	year = {2017},
}

@data{walter2023dataset_realtime,
	author = {Walter, Vivien and Bi, Dadi and Salehi-Reyhani, Ali and Deng, Yansha},
	title = {{Real-Time Signal Processing via Chemical Reactions for a Microfluidic Molecular Communication System}},
	note = {{Zenodo}},
	publisher = {Zenodo},
	url = {doi.org/10.5281/zenodo.8422465},
	year = {2023},
}

@data{walter2023github,
	author = {Walter, Vivien and Bi, Dadi and Deng, Yansha},
	title = {{MolCommUI}},
	journal = {GitHub repository},
	publisher = {GitHub},
	url = {github.com/kcl-yansha/MolCommUI},
	year = {2023},
}

@article{walter2023real,
	author = {Walter, Vivien and Bi, Dadi and Salehi-Reyhani, Ali and Deng, Yansha},
	title = {{Real-time signal processing via chemical reactions for a microfluidic molecular communication system}},
	doi = {10.1038/s41467-023-42885-0},
	issn = {2041-1723},
	journal = {Nature Communications},
	number = {1},
	pages = {1--14},
	publisher = {Springer Nature},
	volume = {14},
	year = {2023},
}

@article{wang2013modular,
	author = {Wang, Baojun and Barahona, Mauricio and Buck, Martin},
	title = {{A modular cell-based biosensor using engineered genetic logic circuits to detect and integrate multiple environmental signals}},
	doi = {10.1016/j.bios.2012.08.011},
	issn = {0956-5663},
	journal = {Biosensors and Bioelectronics},
	month = Feb,
	number = {1},
	pages = {368--376},
	publisher = {Elsevier},
	volume = {40},
	year = {2013},
}

@inproceedings{wang2020understanding,
	author = {Wang, Jiaming and Hu, Dongyin and Hassanieh, Haitham},
	title = {{Understanding and Embracing the Complexities of the Molecular Communication Channel in Liquids}},
	booktitle = {26th ACM International Conference on Mobile Computing and Networking (MobiCom 2020)},
	address = {Virtual Conference},
	doi = {10.1145/3372224.3419191},
	month = Sep,
	year = {2020},
}

@article{wei2013efficient,
	author = {{G. Wei} and {P. Bogdan} and {R. Marculescu}},
	title = {{Efficient Modeling and Simulation of Bacteria-Based Nanonetworks with BNSim}},
	doi = {10.1109/JSAC.2013.SUP2.12130019},
	issn = {0733-8716},
	journal = {IEEE Journal on Selected Areas in Communications},
	month = Dec,
	number = {12},
	pages = {868--878},
	publisher = {IEEE},
	volume = {31},
	year = {2013},
}

@article{wei2024advances,
	author = {Wei, Lu and Ma, Zhong and Yang, Chaojie and Yao, Qin},
	title = {{Advances in the Neural Network Quantization: A Comprehensive Review}},
	DOI = {10.3390/app14177445},
	journal = {Applied Sciences},
	month = Aug,
	note = {{Art. no. 7445}},
	number = {17},
	publisher = {MDPI AG},
	volume = {14},
	year = {2024},
}

@data{wendt2019github_net,
	author = {Wendt, Regine and Ebner, Laurenz and Gomez, Jorge Torres},
	title = {{BVS\_Net}},
	journal = {GitHub repository},
	publisher = {GitHub},
	url = {github.com/tkn-tub/BVS_Net},
	year = {2019},
}

@data{wendt2020github_blood,
	author = {Wendt, Regine},
	title = {{blood-voyager-s}},
	journal = {GitHub repository},
	publisher = {GitHub},
	url = {github.com/RegineWendt/blood-voyager-s},
	year = {2020},
}

@data{wendt2020github_vis,
	author = {Wendt, Regine},
	title = {{BVS-Vis}},
	journal = {GitHub repository},
	publisher = {GitHub},
	url = {github.com/RegineWendt/BVS-Vis},
	year = {2020},
}

\end{document}